\documentclass[12pt]{article}

\usepackage{appendix}
\usepackage{natbib}
\usepackage{graphicx}
\graphicspath{ {../../figures/} }

\usepackage{amsmath,amsthm,amssymb}
\usepackage{enumerate}
\usepackage{url} 
\usepackage{bm}
\usepackage{bbm} 
\usepackage{ulem}

\usepackage{multirow,booktabs}
\usepackage{caption}
\usepackage{xr}
\usepackage{chklref}

\newtheorem{theorem}{Theorem}[section]
\newtheorem{lemma}{Lemma}[section]
\newtheorem{proposition}{Proposition}[section]
\newtheorem{corollary}{Corollary}[section]
\newtheorem{example}{Example}[section]
\newtheorem{remark}{Remark}[section]
\newtheorem{definition}{Definition}[section]
\newtheorem{assumption}{Assumption}[section]

\def\Prob{\mathbb{P}}
\def\E{\mathbb{E}}

\def\Cov{\mathrm{cov}}

\def\t{\boldsymbol{\tau}}
\def\th{\boldsymbol{\hat{\tau}}}
\def\tt{\boldsymbol{\hat{\theta}}}

\def\T{\boldsymbol{T}}
\def\Th{\boldsymbol{\hat{T}}}

\def\S{\boldsymbol{\Sigma}}
\def\Sh{\boldsymbol{\hat{\Sigma}}}

\def\Sb{\boldsymbol{\bar\Sigma}}

\def\G{\boldsymbol{\Gamma}}

\def\I{\boldsymbol{I}}
\def\B{\boldsymbol{B}}
\def\U{\boldsymbol{U}}
\def\X{\boldsymbol{X}}
\def\Z{\boldsymbol{Z}}
\def\R{{\boldsymbol{R}}}

\def\SS{\bs{A}} 
\def\SA{\bs{S}} 
\def\Sset{\mathcal{A}_p}

\def\cth{\vartheta}
\def\bcth{{\boldsymbol{\vartheta}}}
\def\h{\boldsymbol{h}}

\providecommand{\bs}[1]{\boldsymbol{#1}}
\DeclareMathOperator*{\argmin}{\arg\,\min}

\usepackage{xcolor}

\newcommand{\blind}{1}

\begin{document}

\def\spacingset#1{\renewcommand{\baselinestretch}%
{#1}\small\normalsize} \spacingset{1}


\if1\blind
{
  \title{\bf Hypothesis tests for structured rank correlation matrices}
  \author{Samuel Perreault$^{1,3}$\thanks{
    The authors gratefully acknowledge the Natural Sciences and Engineering Research Council of Canada (RGPIN-2015-06801 for JGN and RGPIN-2016-05883 for TD); the Canadian Institute of Statistical Sciences; the Fonds de recherche du Qu\'ebec -- Nature et technologies; and the Institut de valorisation des données (PRF-2019-3055954398).}		\quad
	Johanna G. Ne\v{s}lehov\'{a}$^2$
	\quad
    Thierry Duchesne$^{3}$\\
    \normalsize$^1$Department of Statistical Sciences, University of Toronto\\
	\normalsize$^2$Department of Mathematics and Statistics, McGill University\\
	\normalsize$^3$D{\'e}partement de math{\'e}matiques et de statistique, Universit{\'e} Laval}
  \maketitle
} \fi

\if0\blind
{
  \bigskip
  \bigskip
  \bigskip
  \begin{center}
    {\LARGE\bf Title}
\end{center}
  \medskip
} \fi

\smallskip 
\begin{abstract}
Joint modeling of a large number of variables often requires dimension reduction strategies that lead to structural assumptions of the underlying correlation matrix, such as equal pair-wise correlations within subsets of variables. The underlying correlation matrix is thus of interest for both model specification and model validation.
In this paper, we develop tests of the hypothesis that the entries of the Kendall rank correlation matrix are linear combinations of a smaller number of parameters.
The asymptotic behavior of the proposed test statistics is investigated both when the dimension is fixed and when it grows with the sample size.
We pay special attention to the restricted hypothesis of partial exchangeability, which contains full exchangeability as a special case. We show that under partial exchangeability, the test statistics and their large-sample distributions simplify, which leads to computational advantages and better performance of the tests. We propose various scalable numerical strategies for implementation of the proposed procedures, investigate their 
behavior through simulations and power calculations under local alternatives, and demonstrate their use on a real dataset of mean sea levels at various geographical locations.
\end{abstract}

\noindent%
{\it Keywords:}  block structure, Kendall's tau, exchangeability, structure learning.
\vfill

\newpage
\spacingset{1.25} 
\section{Introduction}

Modeling the dependence between the components of a random vector $\bs{X}=(X_1,\ldots, X_d)$ is of central interest in multivariate statistics and in many applications. This task is particularly challenging when the dimension $d$ is large; the model needs to be flexible, yet parsimonious and feasible to fit. This requires dimension reduction strategies that typically lead to structural assumptions or sparsity of the underlying correlation matrix.
The form of the latter matrix is thus important, particularly for model specification and validation.

In this paper, we develop tests of the hypothesis that the entries of a correlation matrix are linear combinations of a small number of parameters, say $L$. To make the methodology broadly applicable, even in situations when the dependence is not Normal or when the marginal distributions are heavy-tailed, we focus on the matrix $\T$ of pair-wise Kendall's $\tau$. The null hypothesis considered here can then be formulated as
\begin{align} \label{eq:H0}
	H_0:\ \t_p = \B \bs{\beta} \quad \text{for some} \quad \bs{\beta} \in [-1,1]^L,
\end{align}
where  $\t_p$ is a $p= d(d-1)/2$ dimensional vectorization of the above-diagonal entries of $\T$ and $\B$ is a known $p \times L$ matrix of rank $L$, where $L < p$.
Suitable choices of $\B$ correspond to a wide variety of patterns. When $\B$ is the vector of ones, $\T$ is an equicorrelation matrix, which is core to, e.g, shrinkage techniques in genetics \citep{Schafer/Strimmer:2005}. When $B_{r \ell} = \mathbbm{1}(\tau_{p,r} = \beta_\ell)$ for all $r \in \{1,\ldots, p\}$ and $\ell \in \{1,\ldots, L\}$, $\t_p$ possesses only $L$ distinct entries and $\T$ has a block structure, which arises, e.g., in the block DECO model of \cite{Engle/Kelly:2012} or in partially exchangeable copula models  considered by \cite{Perreault/Duchesne/Neslehova:2019}.
Other choices of $\B$ can make $\T$  a Toeplitz matrix as in, e.g., AR models or in the shrinkage technique of \cite{Zhang/Zhou/Li:2019}, or a banded matrix ensuing, e.g., in ante-dependence models for longitudinal data, viz.\ \cite{Zimmermanetal:2010}.

The fact that $H_0$ in \eqref{eq:H0} is more general than most existing proposals is an advantage in applications as complex correlation patterns are often observed. For example, tests of hypotheses that are more general than equicorrelation or partial exchangeability are required to find a suitable dependence model for sea level measurements considered herein.

The main contribution of this paper is the development of distribution-free tests
of the general hypothesis in \eqref{eq:H0} for continuous random vectors.
Such comprehensive methodology had so far been lacking, although special hypotheses concerning the entries of the Pearson or Spearman correlation matrices have
received some attention in the literature. Under the assumption that $\bs{X}$ is multivariate Normal, the hypothesis that the Pearson correlation matrix is an equicorrelation matrix was considered by several authors, see, e.g., \cite{Aitkin/Nelson/Reinfurt:1968} and references therein; while \cite{Wang/Daniels:2014} test for a banded matrix. 
\cite{Gaisser:2010} developed tests of the hypothesis that the Spearman rank correlation matrix is an equicorrelation matrix in a fixed dimension $d$.  A special case of $H_0$ in \eqref{eq:H0} arises when $\T$ is the identity matrix. This pattern is implied by pair-wise independence of the components of $\X$, but not vice versa. Still, off-diagonal entries of $\T$ have been used as a basis of tests of independence, most recently in high-dimensional settings \citep{Han/Chen/Liu:2017,Mao:2018,Leung/Drton:2018}.   

In this paper, we propose two classes of statistics to test $H_0$ in \eqref{eq:H0} for any fixed matrix $\B$ and examine their asymptotic properties under the null and under local alternatives. Gaussian approximations under $H_0$ in the high-dimensional setting rely on the results for $U$-statistics of \cite{Chen:2018} which utilize the bounds in \cite{Chernozhukov/Chetverikov/Kato:2017}. Our results differ from those of \cite{Han/Chen/Liu:2017}, \cite{Leung/Drton:2018}, and \cite{Drton/Han/Shi:2020} in that no distributional requirements on $\X$ are made other than continuity of its margins; in particular, we do not assume Normality of $\X$ or independence of its margins.

We further focus on the special case of \eqref{eq:H0} when $\B$ induces full or partial
exchangeability coined by \cite{Perreault/Duchesne/Neslehova:2019}. The latter paper develops a learning algorithm to find the blocks of $\T$, but does not consider testing $H_0$ in \eqref{eq:H0}. Here, we find a suite of original results, such as analytic expressions for the eigenvalues of the variance
matrix of the empirical estimator of $\t_p$, that simplify the test statistics and their null distributions under partial exchangeability, and lead to scalable numerical strategies for real data implementation.

The methodology developed here is particularly relevant for model specification, validation and structure learning. Indeed,  $\T$ has a particular structure
in many copula models, which are widely used to capture non-Normal dependence. Examples are certain vines \citep{Czado:2019}, factors \citep{Krupskii/Joe:2015,Oh/Patton:2017} or nested and hierarchical models \citep{Mai/Scherer:2012,Brechmann:2014, Joe:2015, Hofert/Huser/Prasad:2018}. The literature on structure learning for these models is emerging. For example, algorithms are being developed for hierarchical Archimedean copulas, viz. \citep{Okhrin/Okhrin/Schmid:2013,Segers/Uyttendaele:2014,Gorecki/Hofert/Holena:2016,Gorecki/Hofert/Holena:2017,Cossette/Gadouri/Marceau/Robert:2019} and references therein. The tests proposed here can serve as a basis for learning algorithms for more complex structures; first steps in this direction are made in \cite{Perreault:2020}.



The paper is organized as follows. Section \ref{sec:preliminary-notions} contains results about
the empirical estimator of $\t_p$
in standard and high-dimensional asymptotic regimes. Section \ref{sec:tests} introduces statistics to test $H_0$ in \eqref{eq:H0} while Section \ref{sec:exchangeability}
treats the special case of partial exchangeability. Asymptotic properties of the tests under the null are derived in Section \ref{sec:null}. 
The performance of the tests is assessed via an extensive simulation study summarized in Section~\ref{sec:sim-study}; behavior under local alternatives is considered in Section~\ref{sec:local-alternatives}. A data illustration is provided in Section \ref{sec:application} and Section \ref{sec:conclusion} concludes.
The online Supplementary Material contains proofs, auxiliary results, and details of all simulations and numerical implementations.

\section{Preliminary considerations} \label{sec:preliminary-notions}

\subsection{Notation}

Vectors in $\mathbb{R}^p$ are denoted by bold symbols such as $\boldsymbol{x}$ or $\boldsymbol{y}$, and operations between vectors, such as $\boldsymbol{x} + \boldsymbol{y}$, are understood as component-wise operations. Matrices are denoted by bold capital letters.
Furthermore, $\bs{0}_p$ and $\bs{1}_p$ are zero and one vectors in $\mathbb{R}^p$, respectively; $\I_p$ denotes the  $ p \times p$ identity matrix and $\bs{J}_p$ stands for  the $p \times p$ matrix of ones.  For $\bs{v} \in \mathbb{R}^p$ and $c \in \mathbb{R}$, $c + \bs{v}$ means $c\bs{1}_p + \bs{v}$; similarly, for a $p\times p$ matrix $\mathbf{M}$ and $c \in \mathbb{R}$,  $c + \mathbf{M}$ means $c\bs{J}_p + \mathbf{M}$.
For $\boldsymbol{x} \in \mathbb{R}^p$, the Euclidean and maximum norms are denoted by  $\| \boldsymbol{x} \|_2 = \sqrt{\bs{x}^\top \bs{x}}$ and $\|\boldsymbol{x} \|_{\infty} = \max(|x_1|,\ldots, |x_p|)$, respectively. Whenever possible, we use the same index notation for sets that appear often: $i,j\in\{1,\dots,d\}$; $r,s\in\{1,\dots,p\}$; $\nu,\eta \in \{1,\ldots, n\}$.

Let $\bs{X} = (X_1,\ldots,X_d)^{\top}$ be a $d$-dimensional random vector with distribution function $F$ and continuous univariate margins  $F_i$, $i\in\{1,\dots,d\}$.  From the work of \cite{Sklar:1959} it is known that there exists a unique copula $C$, i.e., a distribution function with standard uniform marginals, such that for all $x_1,\ldots, x_d \in \mathbb{R}$, $F(x_1,\dots,x_d) = C\{F_1(x_1),\dots,F_d(x_d)\}$.
In fact, $C$ is the distribution function of $\bs{U} = (U_1,\ldots, U_d)$ with components $U_i= F_i(X_i)$, $i \in \{1,\ldots, d\}$, see, e.g., \cite{Nelsen:2006}. For any  $i \neq j \in \{1,\ldots, d\}$, $C_{ij}$ denotes the distribution function of $(U_i, U_j)$, i.e., the copula of the distribution function $F_{ij}$ of $(X_i,X_j)$.

For any $i \neq j \in \{1,\ldots, d\}$, Kendall's tau of the  random pair $(X_i,X_j)$ is defined as
\begin{align*} 
	\tau(X_i,X_j) 
	= \Prob\{(X_i - X_i^*)(X_j - X_j^*) > 0\} - \Prob\{(X_i - X_i^*)(X_j - X_j^*) < 0\},
\end{align*}
{where $\bs{X}^*$ is an independent copy of $\bs{X}$. In other words,} $\tau(X_i,X_j) $ is the difference between the probabilities of concordance and discordance of $(X_i,X_j)$ and $(X_i^*, X_j^*)$. Kendall's tau, going back to \cite{Kendall:1938} and \cite{Hoeffding:1947}, is more robust and can better capture non-linear dependencies than Pearson correlation \citep{McNeil/Frey/Embrechts:2015}. This is because it depends only on the copula, viz. $\tau(X_i,X_j) = -1 + 4\ \E \left\{ C_{ij}(U_i,U_j) \right\}$ \citep{Nelsen:2006}.

Let $\T$ be the $d \times d$ {matrix of Kendall's taus} with $T_{ij}=\tau(X_i,X_j)$. Let also $\t_p$ be the vector of length
$p = d(d-1)/2$ obtained by stacking the entries above the main diagonal of $\T$ column-wise; for example when $d=4$,
$\t_6 = (T_{12},T_{13},T_{23},T_{14},T_{24},T_{34})$.
Implicit in the definition of $\t_p$ is a bijection $\iota$ from the indices of $\t_p$, $\{1,\ldots,p\}$, to those of $\T$, $\{(i,j): 1 \leqslant i < j \leqslant {d}\}$, defined so that for each $r \in \{1,\ldots, p\}$, $r \mapsto \iota(r) = (i_r, j_r)$, where
\begin{equation}\label{eq:bijection}
\quad \binom{j_r -1}{2} < r \leqslant \binom{j_r}{2} , \quad i_r = r - \binom{j_r-1}{2}
\end{equation}
with the convention that $\binom{1}{2}  =0 $. Note that the inverse of $\iota$ satisfies $\iota^{-1}(i_r,j_r)  = i_r + \binom{j_r-1}{2}$ for all $r \in \{1,\ldots, p\}$.  For this form of stacking, $\iota(r)$ does not depend on $d$, whenever $r \leqslant  p$.



\subsection{Estimation of Kendall's tau} \label{subsec:2.2}

Let $\X_\nu = (X_{\nu 1},\ldots, X_{\nu d})$, $\nu \in\{ 1,\ldots, n\}$, be a random sample from $\X$, based on which we {wish to make inference about the Kendall's tau} matrix $\T$ of $\bs{X}$. As is well-known, the entries of the empirical estimator $\Th_n$ of $\T$ are, for all $i\neq j \in \{1,\ldots, d\}$,
$\hat{T}_{ij} = 2\{n(n-1)\}^{-1} \sum_{1 \leqslant \nu < \eta \leqslant n	} h_{ij}(\bs{X}_\nu,\bs{X}_\eta)$,
where {$h_{ij} : \mathbb{R}^d \times \mathbb{R}^d \to \mathbb{R}$  is given, for all $\bs{x}, \bs{y} \in \mathbb{R}^d$, by $h_{ij}(\bs{x},\bs{y})= 2 \times \mathbbm{1}\{(x_{i} - y_{i})(x_{j} - y_{j}) > 0\} - 1$.
{In analogy to $\t_p$, let $\th_{np}$ be the vectorized version of the entries above the main diagonal of $\Th_n$. For $h: \mathbb{R}^d \times \mathbb{R}^d \to \mathbb{R}^p$} such
that $h=(h_1,\ldots,h_p)^\top$, where for $r\in\{1,\ldots, p\}$, $h_r = h_{i_rj_r}$ with $\iota(r) = (i_r, j_r)$, we have that
\begin{align} \label{eq:tau-hat}
\th_{np}=  \frac{2}{n(n-1)} \sum\limits_{1 \leqslant \nu < \eta \leqslant n} h(\bs{X}_\nu,\bs{X}_\eta).
\end{align}
From \eqref{eq:tau-hat}, $\th_{np}$ is a vector-valued $U$-statistic of order 2 and hence, for a fixed dimension $d$, it is an unbiased and asymptotically normal estimator of $\t_p$ \citep{Hoeffding:1948,Cliff/Charlin:1991,ElMaache/Lepage:2003,Genest/Neslehova/BenGhorbal:2011}. Specifically, as $n \to \infty$,
\begin{align} \label{eq:asymptotic-tau-n}
	\sqrt{n}(\th_{np}-\t_p) \rightsquigarrow {\boldsymbol{Z}_p},
\end{align}
where $\rightsquigarrow$ denotes convergence in distribution and $\boldsymbol{Z}_p \sim  {\mathcal{N}_p(\bs{0}_p,\S_{p})}$. The proof of \eqref{eq:asymptotic-tau-n} uses the  H\'ajek projection $H_n$ of $\th_{np}-\t_p$, i.e., the leading term of the Hoeffding decomposition,
\begin{align} \label{eq:Hajek}
H_n = \sum_{\nu=1}^n \E(\th_{np}-\t_p | \X_\nu) = \frac{2}{n} \sum_{\nu=1}^n g(\X_\nu),
\end{align}
where $g(\bs{x}) = \E\{h(\bs{x}, \bs{X})\} - \t_p$. Hence, for each $r \in \{1,\ldots, p\}$, $g_r(\bs{x}) = 2\{F_{i_rj_r}(x_{i_r},x_{j_r}) + \bar F_{i_rj_r}(x_{i_r},x_{j_r})\}-1-\tau_r$, where $(i_r, j_r) = \iota(r)$ and $\bar F_{i_rj_r}$ denotes the survival function of $F_{i_rj_r}$. As is well known, $\sqrt{n} (\th_{np}-\t_p-H_n)\to 0$ in probability \citep{vanderVaart:1998}, so that \eqref{eq:asymptotic-tau-n} follows from the Central Limit Theorem and $\S_p=4\Cov\{g(\bs{X})\}$. 
Expressions for the finite-sample covariance matrix {$\S_{np}$} of $\th_{np}$ are given in \cite{Genest/Neslehova/BenGhorbal:2011}. 

When both $n\to \infty$ and $d\to \infty$, the asymptotic behavior of $\th_{np}$ can be derived from the following theorem, which is a slight extension of Theorem~2.1 of \cite{Chen:2018} and needed here later on. Its proof may be found in Appendix~\ref{app:proofs-prelim} in the Supplementary Material.
\begin{theorem}\label{thm:Chen}
Let $\bs{P}_p$ be a $p \times p$ matrix whose entries may depend on $p$ but not on $n$. Assume that there exists a constant $\underline{b} \in (0,\infty)$ and a sequence $(B_n)$ of real numbers possibly tending to infinity with $B_n \geqslant 1$ for all $n$, such that the following inequalities
\begin{align*}
	{\rm(M.1)}\ \mathrm{diag}(\bs{P}_p \S_{p} \bs{P}_p^\top) \geqslant \underline{b} \bs{1}_p, \qquad {\rm (M.2)}\ \max_{1 \leqslant r \leqslant p} \sum_{s=1}^p |\bs{P}_{rs}| \leqslant B_n
\end{align*}
hold componentwise. Suppose also that there exists a constant $\bar b > 0$ independent of $n$ and $p$ with $\log p < \bar b n$.
Then there exists a constant $\kappa(\underline{b},\bar{b})$ independent of $n$ and $p$ such that
\begin{equation}\label{eq:ChenThm}
\sup_{E \in \mathcal{E}_{p}}|\Prob\{\sqrt{n}\bs{P}_{p}( \th_{np}-\t_p) \in E\} - \Prob(\bs{Z}_{p} \in E)| \leqslant  \kappa(\underline{b},\bar{b}) \left\{\frac{B_n^{4} \log^7(np)}{n}\right\}^{1/6},
\end{equation}
where $\bs{Z}_{p} \sim \mathcal{N}_p(\boldsymbol{0}_p,\bs{P}_{p} \S_{p} \bs{P}_{p}^\top)$ and $ \mathcal{E}_p$ is the set of all hyper-rectangles $\{ [\bs{a}, \bs{b}] : \bs{a}, \bs{b} \in \overline{\mathbb{R}}^p  \}$.
\end{theorem}
By Corollary~2.2 in \cite{Chen:2018} and the fact that the kernel $h$ of $\th_{np}$ is bounded, it follows from the conditions of Theorem \ref{thm:Chen} that if there exists a constant $\underline{b}>0$ as in Theorem~\ref{thm:Chen} and a constant $\bar b >0$ so that $B_n^4\log^7(np) \leqslant \bar b n^{1-\lambda}$ for some $\lambda \in (0,1)$, then there exists a constant $\kappa(\underline{b},\bar b) > 0$ so that the rate in \eqref{eq:ChenThm} is bounded above by $\kappa(\underline{b},\bar b) \times n^{-\lambda/6}$. 
Although it is not known whether the rate in Eq.~\eqref{eq:ChenThm} is optimal, it is argued in Remark~1 in \cite{Chen:2018} that it seems unimprovable in $n$. Note also that when the components of $\bs{X}$ are independent, \cite{Leung/Drton:2018} show that suitably scaled sums of $U$-statistics, including of $\th_{np}$, converge to a standard Normal as $n,d\to\infty$.

\section{Test statistics} \label{sec:tests}

Let $\B$ be some fixed, known  $p \times L$ matrix of rank $L$, where $L < p$ and consider testing the hypothesis $H_0$ in \eqref{eq:H0}. With the Moore--Penrose pseudoinverse $\B^{+}$  of $\B$ and
$$
	\mathcal{T}_p = \{\bs{\theta} \in \mathbb{R}^p : \bs{\theta} = \B \bs{\beta}, \bs{\beta} \in [-1,1]^L \}= \{\bs{\theta} \in \mathbb{R}^p : \bs{\theta}=\B\B^+ \bs{\theta}\},
$$
the hypothesis to be tested can alternatively be formulated as $H_0 : \bs{\tau}_p \in \mathcal{T}_p$.  To test $H_0$, it is thus natural to focus on the distance between the empirical estimator $\th_{np}$ and $\mathcal{T}_p$. To this end, we endow $\mathbb{R}^p$ with the Mahalanobis norm given, for any $\bm{x}\in \mathbb{R}^p$, by $\sqrt{\bm{x}^\top \SA^{-1} \bm{x}}$ for some positive definite $p \times p$ matrix $\SA$; this norm is induced by the scalar product $\bm{x}^\top \SA^{-1} \bm{y}$. Note that $\bm{x}^\top \SA^{-1} \bm{x} = \| \SA^{-1/2} \bm{x}\|_2^2$ where $\| \cdot \|_2$ denotes the Euclidean norm and $\SA^{-1/2}$ is the inverse of the principal square root of $\SA$. The latter is given by $\SA^{1/2} = \bs{V} \bs{\Delta}^{1/2} \bs{V}^{\top}$, where the columns of $\bs{V}$ are $p$ orthonormal eigenvectors of $\SA$ and $\bs{\Delta} = \mathrm{diag}(\bs{\lambda})$ is the diagonal matrix of eigenvalues of $\SA$ so that for all $r \in \{1,\ldots, p\}$, the $r$-th column of $\bs{V}$ is an eigenvector associated with $\lambda_r$. Also, $\SA^{-1/2} = \bs{V} \bs{\Delta}^{-1/2} \bs{V}^{\top}$ is the unique symmetric and positive definite matrix such that $\SA^{-1/2} \SA^{-1/2} = \SA^{-1}$ \citep[Thm.~7.2.6]{Horn/Johnson:2012}.

Because $\mathcal{T}_p$ is a complete subspace of $\mathbb{R}^p$, the distance between $\th_{np}$ and $\mathcal{T}_p$ equals the distance between 
 $\th_{np}$ and its unique orthogonal projection on $\mathcal{T}_p$ given by 
$\bs{\hat \theta}_{np} = \argmin_{\bs{\theta} \in \mathcal{T}_p } (\th_{np} - \bs{\theta})^\top \SA^{-1} (\th_{np}-\bs{\theta}).$
It is well known that 
 $\bs{\hat \theta}_{np} = \Gamma(\SA) \th_{np}$, where $ \Gamma(\SA)$ is the projection matrix given by
\begin{equation}\label{eq:Gamma}
\Gamma(\SA) =\B (\B^\top \SA^{-1} \B)^{-1} \B^\top \SA^{-1}.
\end{equation}
The projection $\bs{\hat \theta}_{np}$ can be viewed as a constrained estimator of $\t_p$ under the null hypothesis. Because  $\th_{np}$ and $\bs{\hat \theta}_{np}$ are both consistent estimators of $\t_p$ under $H_0$, it makes sense to choose
\begin{equation}\label{eq:euclidean}
		E_{np}= \|\SA^{-1/2}(\th_{np}-\bs{\hat \theta}_{np}) \|_2^2 = (\th_{np}-\bs{\hat \theta}_{np})^\top \SA^{-1}(\th_{np}-\bs{\hat \theta}_{np}).
\end{equation}	
to test $H_0$. An alternative statistic, this time without a geometric interpretation, is
\begin{equation}\label{eq:supremum}
	M_{np}  = \|\SA^{-1/2}(\th_{np}-\bs{\hat \theta}_{np}) \|_\infty =\max_{1 \leqslant r \leqslant p} \big| \{\SA^{-1/2}(\th_{np}-\bs{\hat \theta}_{np})\}_r \big|.
\end{equation}
Suitable choices of $\SA$ are the scaled identity matrix $(1/n)\I_p$ or, in view of the asymptotic normality of $\th_{np}$, an estimator of $\S_{p}$. The latter could be either the jackknife estimator  
\begin{equation} \label{eq:sigma-jackknife}
	\Sh_{np}^{\rm J} = \frac{4}{\{n(n-1)\}^2} \sum_{\nu=1}^n \sum_{\mu \neq \nu} \sum_{\xi \neq \nu} \{ h(\bs{X}_\nu,\bs{X}_\mu) - {\th}_{np} \}\{ h(\bs{X}_\nu,\bs{X}_\xi) - {\th}_{np}\}^\top,
\end{equation}
of \cite{Chen:2018} or the plug-in estimator $\Sh_{np}^{\rm P}$ of  \cite{Genest/Neslehova/BenGhorbal:2011}. For convenience, details on these estimators are provided in
Appendix~\ref{app:Sigma} of the Supplementary Material.
\begin{remark}
The statistics \eqref{eq:euclidean} and \eqref{eq:supremum} can be compared to the procedures $\mathcal{T}_{n,1}$, $\mathcal{T}_{n,2}$, $\mathcal{T}_{n,3}$, and $\mathcal{T}_{n,4}$ of \cite{Gaisser:2010}. Although these authors focus exclusively on Spearman's rho, their hypothesis that all pair-wise Spearman's rank correlations are equal is akin to $H_0$ with $\B = \bs{1}_p$. Their statistics $\mathcal{T}_{n,1}$ and  $\mathcal{T}_{n,3}$ are not convenient here because they depend on the way the rank correlation matrix is vectorized. However, $\mathcal{T}_{n,2}$ is an analogue of $p E_{np}$ when $\SA= (1/n)\I_p$,
while
$\mathcal{T}_{n,4}$
leads
to $M_{np}' = \sqrt{n}\sup_{1 \leqslant r,s \leqslant p} |\hat\tau_{np,r} - \hat\tau_{np,s}|$, which, in contrast to $M_{np}$ with $\SA = (1/n)\I_p$, does not involve $\tt_{np}$. 
Now consider a variant of $M_{np}$ in which $\bs{\hat \theta}_{np}$ is replaced by $\argmin_{\bs{\theta} \in \mathcal{T}_p } \|\SA^{-1/2}(\th_{np}-\bs{\theta}) \|_\infty$. When $\B = \bs{1}_p$ and $\SA = (1/n)\I_p$, this statistic becomes $M_{np}'/2$. However,  the norm given for all $\bm{x} \in \mathbb{R}^p$ by $\| \SA^{-1/2} \bm{x}\|_\infty$ is not induced by a scalar product. This means that the projection of $\th_{np}$ on $\mathcal{T}_p$ may be neither unique nor explicit unless in special cases.  We thus refrain from pursuing this idea here.
\end{remark}

\section{Special case of partial exchangeability} \label{sec:exchangeability}


\subsection{Hypotheses of full and partial exchangeability}\label{sec:2.3}

For structural learning in various copula models, $\T$ with a block structure is of particular interest because it can have substantially fewer distinct entries. Such a $\T$ arises, e.g., when the copula $C$ of $\X$ is partially exchangeable. This notion goes back to \cite{Perreault/Duchesne/Neslehova:2019} and is defined below for convenience.
\begin{definition}
A copula $C$ is said to be partially exchangeable with respect to a partition $\mathcal{G} = \{ \mathcal{G}_1,\dots,\mathcal{G}_K\}$ of $\{1,\dots,d\}$ with $K < d$, if for any $u_1,\dots, u_d \in [0,1]$,
$C(u_1, \dots, u_d) = C(u_{\pi(1)}, \dots, u_{\pi(d)})$
for any permutation $\pi$ of $1,\dots, d$ such that for all $i \in\{1,\dots, d\}$ and all $k \in \{1,\dots, K\}$, $i \in \mathcal{G}_k$ if and only if $\pi(i) \in \mathcal{G}_k$.
The set of all copulas that are partially exchangeable with respect to a given partition $\mathcal{G}$ is denoted by $\mathcal{C}_{\mathcal{G}}$.
\end{definition}

The reason why partial exchageability merits special attention in this paper is that when $C$ is partially exchageable, $\S_{np}$ and $\S_p$ have a specific block structure; Proposition~A.1 of \cite{Perreault/Duchesne/Neslehova:2019} shows that $\S_{np}$ and $\S_p$ belong to the set of matrices $\mathcal{S}_{\mathcal{G}}$ defined on p. 413 in Appendix~A.2 of the latter paper. As we shall see shortly, this is important for the construction of hypothesis tests. Alongside $H_0$, we thus consider the hypothesis of partial exchangeability with respect to a given partition $\mathcal{G}$, viz.
\begin{equation} \label{eq:H0-star}
H_0^* : C \in \mathcal{C}_{\mathcal{G}}.
\end{equation}
When $H_0^*$ holds, $\T$ contains only $L = K(K+1)/2 - \sum_{k=1}^K \mathbbm{1}(|\mathcal{G}_k| = 1)$ distinct off-diagonal entries and $H_0$ holds with the so-called block membership matrix $\B$. The latter is defined on page~403 of \cite{Perreault/Duchesne/Neslehova:2019}; as explained in Section~2 therein, the entries of $\B$ are either $0$ or $1$ and such that each row has exactly one non-zero entry.


The rest of this section considers the special case when $C$ is fully exchangeable, as happens, e.g., for Archimedean and elliptical copulas with an equicorrelation matrix.  Full exchangeability means that $H_0^*$ holds with $\mathcal{G} = \{\{1,\ldots, d\}\}$. In this case,  $K=L=1$, $\B = \bs{1}_p$;  furthermore, by Lemma~\ref{lem:JGN} in the Supplementary Material, $\mathcal{S}_{\mathcal{G}}$ reduces to the set of $p\times p$ matrices $\Sset=\{\SS_p(s_0,s_1, s_2) : s_0, s_1, s_2 \in \mathbb{R}\}$, where for arbitrary $s_0, s_1, s_2 \in \mathbb{R}$,  $\SS_p(s_0,s_1, s_2)$ is defined as follows. For any $r, s \in \{1,\ldots, p\}$, let $(i_r,j_r) = \iota(r)$ and $(i_s, j_s) = \iota(s)$, with $\iota$ as in \eqref{eq:bijection}. The $(r,s)$-th entry of $\SS_p(s_0,s_1, s_2)$ is then given by
\begin{multline}\label{eq:S-cal}
s_0 \times \mathbbm{1}(|\mathcal{I}_{rs}| =0) + s_1 \times \mathbbm{1}(|\mathcal{I}_{rs}| =1) + s_2 \times \mathbbm{1}(|\mathcal{I}_{rs}| =2), \quad \mathcal{I}_{rs}=\{ i_r, j_r\} \cap \{i_s, j_s\}.
\end{multline}

Because the matrices $\S_{np}$ and $\S_p$ are elements of $\Sset$ under full exchangeability, we will need several results about the matrices in $\Sset$ to develop tests of $H_0^*$.
The first finding concerns
their eigenvalues. Its proof may be found in Appendix~\ref{app:exchangeability} in the Supplementary Material, along with Remark~\ref{rem:eigen-Sigma-2} which discusses the cases $d = 2$ and $3$.
\begin{proposition} \label{prop:eigen-Sigma}
Let $d\geqslant 4$ and $\SS_p=\SS_p(s_0,s_1,s_2)$ for some $s_2,s_1,s_0\in\mathbb{R}$, as defined in Eq.~\eqref{eq:S-cal}.
Then $\SS_p$ has three real eigenvalues given by
$\delta_{1,d}(s_0,s_1,s_2) = s_2 +2(d-2)s_1 + (p-2d+3)s_0$, $\delta_{2,d}(s_0,s_1,s_2) = s_2 + (d-4)s_1 - (d-3)s_0$ and $\delta_{3}(s_0,s_1,s_2) = s_2 -2s_1 + s_0$,
with respective geometric multiplicities $1$, $d-1$ and $p-d$.
\end{proposition}

\begin{remark} \label{rem:eigen-Sigma}
Note that depending on the values of $s_0, s_1, s_2$, some (or all) of the eigenvalues $\delta_{1,d}$, $\delta_{2,d}$ and $\delta_{3}$ may coincide. When this happens, the geometric multiplicities add up, because the eigenspaces are orthogonal; this follows from the proof of Proposition~\ref{prop:eigen-Sigma}. Also apparent from the proof is that the eigenvectors do not depend on $s_0,s_1, s_2$.
\end{remark}

The next result concerns the inverses of the matrices in $\Sset$.
Its proof, given in Appendix~\ref{app:exchangeability} of the Supplementary Material, relies on the findings of \cite{Perreault/Duchesne/Neslehova:2019}.

\begin{proposition}\label{prop:inverse}
Suppose that $\SS_p = \SS_p(s_0,s_1,s_2) \in \Sset$ is invertible. Then $\SS_p^{-1} \in \Sset$, that is, there exist $t_0,t_1,t_2 \in \mathbb{R}$ such that $\SS_p^{-1} = \SS_p(t_0, t_1,t_2)$. The eigenvalues of $\SS_p^{-1}$ are given by $\delta_{\ell,d}(t_0,t_1,t_2) = 1/\delta_{\ell,d}(s_0,s_1,s_2)$ for  $\ell\in\{1,2\}$ and $\delta_{3}(t_0,t_1,t_2) = 1/\delta_{3}(s_0,s_1,s_2)$.
\end{proposition} 

It follows directly from Proposition~\ref{prop:inverse} that when $C$ is exchangeable and $\S_p$ and $\S_{np}$ are invertible, the inverses of these matrices have the same block structure. Their entries generally depend on $d$ and can be calculated from \eqref{eq:S-cal} and Propositions~\ref{prop:eigen-Sigma} and \ref{prop:inverse}.

Finally, in order to treat the situation when $d\to \infty$,
we now consider, for any fixed $s_0,s_1, s_2 \in \mathbb{R}$, the sequence of matrices $\SS_p(s_0,s_1,s_2)$ with $p = d(d-1)/2$, for $d=2,3, \ldots$.
The next proposition gives the conditions under which each member of such a sequence is positive definite. The proof is in Appendix~\ref{app:exchangeability} in the Supplementary Material.
\begin{proposition}\label{prop:pd}
Let $s_0,s_1, s_2 \in \mathbb{R}$, $d \geqslant 4$, $p=d(d-1)/2$ and $\SS_p=\SS_p(s_0,s_1,s_2)$. Then $\SS_p$ is positive definite for all $d \geqslant 4$ if and only if $s_1 \geqslant s_0 \geqslant 0$ and $s_2 -s_1 > s_1 -s_0$.
\end{proposition}
From Proposition~\ref{prop:pd}, a necessary (but not sufficient) condition for $\SS_p(s_0,s_1,s_2)$ to be positive definite for all $d \geqslant 4$ is that $s_2 > s_1 \geqslant s_0 \geqslant 0$. Moreover, the following holds.
\begin{corollary} 
Let $d \geqslant 4$ and $s_0,s_1, s_2 \in \mathbb{R}$ be such that $s_1 \geqslant s_0 \geqslant 0$ and $s_2 -s_1 > s_1 -s_0$. Then the eigenvalues $\delta_{1,d},\delta_{2,d},\delta_{3}$ 
in Proposition~\ref{prop:eigen-Sigma}
satisfy
$\delta_{1,d} \geqslant \delta_{2,d} \geqslant \delta_{3} \geqslant 0$.
\end{corollary}

\subsection{Test statistics for testing $H_0^*$} \label{subsec:estimation-exchangeability}

The statistics in Eqs.~\eqref{eq:euclidean} and \eqref{eq:supremum}  are suitable for testing $H_0^*$ as well, given that under $H_0^*$, $H_0$ holds for the block membership matrix $\B$. However, for the purpose of testing $H_0^*$, it makes sense to pick $\SA \in \mathcal{S}_{\mathcal{G}}$. This is always the case when $\SA=(1/n)\I_p$. As explained in Appendix~\ref{app:Sigma} of the Supplementary Material, the plug-in or the jackknife estimators of $\S_{np}$ can be modified to lie in $\mathcal{S}_{\mathcal{G}}$ by suitable averaging of their entries; this leads to the so-called structured estimators $  \Sb_{np}^{\rm J} $ and  $ \Sb_{np}^{\rm P} $. When $\mathcal{G}$ is coarse, this can significantly improve estimation of $\S_{np}$; in the case of full exchangeability, we show in Proposition~\ref{prop:sigma-exch} that $\Sb_{np}^{\rm J}  = \SS_p(\hat\sigma_{n0}^{\rm J},\hat\sigma_{n1}^{\rm J},\hat\sigma_{n2}^{\rm J})$, where $\hat\sigma_{n0}^{\rm J}$, $\hat\sigma_{n1}^{\rm J}$, $\hat\sigma_{n2}^{\rm J}$ are explicit and easy to calculate. 

Choosing $\SA \in \mathcal{S}_{\mathcal{G}}$ also leads to simplifications of the test statistics. First, the projection matrix
in Eq.~\eqref{eq:Gamma} becomes $\Gamma(\SA)=\B\B^+$ and thus does not depend on $\SA$ \citep[Theorem~1]{Perreault/Duchesne/Neslehova:2019}. Further simplifications are possible in the case of full exchangeability. To see this, note first that $\SA \in \Sset$ implies $\Gamma(\SA) = \G =  (1/p)\bs{J}_p$
so that
$\tt_{np} = \G\th_{np} = \bar \tau_{np} \bs{1}_p$, where $\bar\tau_{np} = (1/p)(\hat \tau_{np,1} + \ldots + \hat \tau_{np,p})$ is the average of the entries of $\th_{np}$. Now introduce 
\begin{align} \label{eq:Gamma-star}
\G^* = \SS_p\left(-\frac{2}{(d-1)(d-2)}, \frac{d-3}{(d-1)(d-2)}, \frac{2}{d-1} \right)
\end{align}
and set $\tt^*_{np} = \G^* \th_{np}$. For each  $r \in \{1,\ldots, p\}$, the $r$-th entry of $\tt^*_{np}$ is related to the column means of $\hat \T_n$. Indeed, set $(i_r,j_r) = \iota(r)$, where $\iota$ is as in Eq. \eqref{eq:bijection} and let, for $i \in \{1,\ldots, d\}$,
\begin{align} \label{eq:tau-bar}
\bar T_{ni} = \frac{1}{d-1}\sum_{r \in \mathcal{R}_i} \hat\tau_{np,r}, \quad \mathcal{R}_i = \{\iota^{-1}(i,j) : 1 \leqslant i < j \} \cup \{\iota^{-1}(j,i) : j <  i \leqslant d \},
\end{align}
be the average of the off-diagonal entries of the $i$-th column of $\hat \T_n$. Then
\begin{align}\label{eq:tau-star}
(d-2) \hat\theta^*_{np,r} =  (d-1)(\bar T_{ni_r} + \bar T_{nj_r} )-d\bar\tau_{np},
\end{align}
so that $\hat\theta^*_{np,r} \to \tau$ in probability under $H_0^*$ as $n\to \infty$.  
%
More importantly, we can show that $(\th_{np} - \tt_{np}^*)^\top(\tt_{np}^* - \tt_{np}) = 0$, which leads to the following result, proved in Appendix~\ref{app:exchangeability}. This finding makes the tests of $H_0^*$ substantially easier computationally.

\begin{proposition}\label{prop:decomposition} 
Suppose that $d \geqslant 4$ and that $\SS_p \in \Sset$ is positive definite with eigenvalues $\delta_{1,d}$, $\delta_{2,d}$, and $\delta_{3}$, as given in 
Proposition~\ref{prop:eigen-Sigma}.
Then the following hold.
\begin{enumerate}
\item[(a)] $\SS_p^{-1/2} (\tt_{np}^* - \tt_{np}) = \delta_{2,d}^{-1/2} (\tt_{np}^* - \tt_{np})$  and $\SS_p^{-1/2} (\th_{np} - \tt_{np}^*) = \delta_{3}^{-1/2} (\th_{np} - \tt_{np}^*)$.
\item[(b)] $(\th_{np} - \tt_{np})^\top \SS_p^{-1} (\th_{np} - \tt_{np}) = \delta_{3}^{-1} (\th_{np} - \tt_{np}^*)^\top(\th_{np} - \tt_{np}^*) + \delta_{2,d}^{-1} (\tt_{np}^* - \tt_{np})^\top(\tt_{np}^* - \tt_{np})$.
\end{enumerate}
\end{proposition}

\section{Asymptotic null distributions of the test statistics} \label{sec:null}

\subsection{Fixed $d$, large $n$ asymptotics} \label{subsec:n-asymptotics}

In this section, we derive the asymptotic null distributions of the test statistics in Eqs~\eqref{eq:euclidean} and \eqref{eq:supremum}, and specify how they simplify under the restricted hypothesis $H_0^*$.
We first consider the case when the sample size $n$ grows while the dimension $d$ of $\X$ is fixed.
All asymptotic results for this case are consequences of the asymptotic normality of $\th_{np}$ stated in Eq.~\eqref{eq:asymptotic-tau-n}. The first is the following theorem, proved in Appendix~\ref{app:proofs-null} in the Supplementary Material.
\begin{theorem} \label{thm:asymptotic-n}
Suppose that $H_0$ in \eqref{eq:H0} holds and that $n\SA_{np}$ and $\bs{P}_{np}$ converge in probability, as $n\to\infty$ to some $p \times p$ matrices $\SA_{p}$ and $\bs{P}_p$, respectively.
Assume that $\SA_{p}$ is positive definite,  and that for all $n$, $\SA_{np}$ is positive definite and $\bs{P}_{np}\t_p = \bs{0}_p$.
Then, as $n\to \infty$,
	$\SA^{-1/2}_{np} \bs{P}_{np}\th_{np} \rightsquigarrow \Z_p$,
where $\Z_p \sim {N}_{p}(\bs{0}_p,\SA_p^\dag)$ with $\SA_p^\dag = \SA_{p}^{-1/2} \bs{P}_p \S_{p} \bs{P}_p^\top \SA_{p}^{-1/2}$.
Furthermore,
\begin{align} \label{eq:asymptotic-n}
	\|\SA^{-1/2}_{np} \bs{P}_{np}\th_{np}\|_{2}^2 \rightsquigarrow \sum_{k=1}^{m} \lambda_{k} \chi_{\upsilon_{k}}^2 \qquad \text{and} \qquad  \|\SA^{-1/2}_{np} \bs{P}_{np}\th_{np}\|_{\infty} \rightsquigarrow \|\Z_{p}\|_{\infty}
\end{align}
as $n \to \infty$, where $\lambda_{k}$ is the $k$-th of the $m$ distinct non-zero eigenvalues of $\SA_p^\dag$ and $\upsilon_{k}$ is the geometric multiplicity of $\lambda_k$.
\end{theorem}

Let us now focus on the Euclidean statistic $E_{np}$ of Eq.~\eqref{eq:euclidean}. If $\SA = \SA_{np}$, the latter statistic can be expressed as $E_{np}  =\|\SA^{-1/2}_{np} \bs{P}_{np}\th_{np}\|_{2}^2$ for  $\bs{P}_{np} = \I_p-\Gamma(\SA_{np})$.
When $n\SA_{np}$ is a consistent estimator of $\S_{p}$, $\SA_p^\dag$ in Theorem \ref{thm:asymptotic-n} is idempotent. The mixture of chi-square distributions in Eq.~\eqref{eq:asymptotic-n} thus reduces to a single chi-square distribution. 
If $\SA = (1/n)\I_p$, then neither $\tt_{np}$ nor $E_{np}$ require the estimation of $\S_{np}$.
The price to pay is that the asymptotic distribution of $E_{np}$ under $H_0$ does not simplify to a single chi-square distribution,
but rather to a mixture of chi-squares whose weights do depend on $\S_p$. This is  formally recorded below and proved in Appendix~\ref{app:proofs-null} in the Supplementary Material.
\begin{proposition} \label{prop:asymptotic-euclidean}
Assume that $H_0$ in \eqref{eq:H0} holds and consider $E_{np}$ in Eq.~\eqref{eq:euclidean}.
Then $E_{np} \rightsquigarrow \| \Z_p \|_2^2$ as $n \to \infty$, where $\bs{Z}_p \sim \mathcal{N}_p(\bs{0}_p,\SA_p^\dag)$, whenever one of the following assumptions hold.
\begin{itemize}
\item[(a)] $\SA_{np} = \Sh_{np}$, where $n\Sh_{np}$ is a consistent estimator of $\S_{p}$. In this case, $\SA_p^\dag = \I_p - \S_p^{-1/2} \Gamma(\S_p) \S_p^{1/2}$, where $\Gamma(\S_p)$ is given by Eq. \eqref{eq:Gamma}, and $\| \Z_p \|_2^2 \sim \chi_{p-L}^2$.
\item[(b)] $\SA_{np} = (1/n)\I_{p}$. In this case, $\SA_p^\dag = (\I_p - \B\B^+) \S_p (\I_p - \B\B^+)$  and $\| \Z_p \|_2^2 \sim \sum_{k=1}^{m} \lambda_{k} \chi_{\upsilon_{k}}^2$,
where $\lambda_{k}$ is the $k$-th of the $m$ distinct non-zero eigenvalues of $\S_{p} (\I_p - \B \B^+)$ and $\upsilon_{k}$ is the geometric multiplicity of $\lambda_k$.
\end{itemize}
\end{proposition}


In the special case of full exchangeability, the coefficients $\lambda_k$ and the degrees of freedom $\upsilon_k$ can be computed explicitly. This leads to the following corollary to Proposition~\ref{prop:asymptotic-euclidean} (b), proved in  Appendix~\ref{app:proofs-null} in the Supplementary Material.
\begin{corollary} \label{cor:asymptotic-euc-exchangeable}
When $H_0^*$ in \eqref{eq:H0-star} holds with $\mathcal{G}=\{\{1,\ldots,d\}\}$ and $\SA = (1/n)\I_p$ in Eq. \eqref{eq:euclidean}, then $E_{np} \rightsquigarrow \delta_{3} \chi_{p-d}^2 + \delta_{2,d} \chi_{d-1}^2$, where $\delta_3$ and $\delta_{2,d}$ are the eigenvalues of $\S_p$ 
given in Proposition \ref{prop:eigen-Sigma}.
\end{corollary}

We next derive the asymptotic behavior of the supremum norm statistic $M_{np}$. Similarly to $E_{np}$, if $\SA = \SA_{np}$, $M_{np}  =\|\SA^{-1/2}_{np} \bs{P}_{np}\th_{np}\|_{\infty}$ with  $\bs{P}_{np} = \I_p-\Gamma(\SA_{np})$. The two choices of $\SA$ lead to the following result, proved in Appendix~\ref{app:proofs-null} in the Supplementary Material.
\begin{proposition} \label{prop:asymptotic-supremum}
Assume that $H_0$ in \eqref{eq:H0} holds and consider $M_{np}$ in Eq.~\eqref{eq:supremum}. Then $M_{np} \rightsquigarrow \|\Z_p \|_{\infty}$ as $n\to \infty$, where $\bs{Z}_p \sim \mathcal{N}_p(\bs{0}_p,\SA_p^\dag)$, whenever one of the following assumptions hold.
\begin{itemize}
\item[(a)] $\SA_{np} = \Sh_{np}$, where $n\Sh_{np}$ is a consistent estimator of $\S_{p}$. In this case, $\SA_p^\dag = \I_p - \S_p^{-1/2} \Gamma(\S_p) \S_p^{1/2}$, where $\Gamma(\S_p)$ is given by Eq. \eqref{eq:Gamma}.
\item[(b)] $\SA_{np} = (1/n)\I_{p}$. In this case, $\SA_p^\dag = (\I_p - \B\B^+) \S_p (\I_p - \B\B^+)$.
\end{itemize}
\end{proposition}

Partial and full exchangeability lead to simplifications which are recorded below and which follow directly from Lemma~B.6 of \cite{Perreault/Duchesne/Neslehova:2019}.

\begin{corollary} 
Under the assumptions of Proposition \ref{prop:asymptotic-supremum} (a) and when $H_0^*$ in \eqref{eq:H0-star} holds, one has that $\SA_p^\dag =  \I_p - \B\B^+$. When additionally $\mathcal{G}=\{\{1,\ldots,d\}\}$,   $\SA_p^\dag = \I_p - (1/p) \bs{J}_p$.
\end{corollary}


\subsection{High-dimensional asymptotics} 

As we saw in Section~\ref{subsec:n-asymptotics}, the test statistic $M_{np}$ in Eq.~\eqref{eq:supremum} with $\SA = (1/n)\I_p$ can be expressed as $\|\sqrt{n} \bs{P}_p(\th_{np} -\t_p)\|_\infty$ where $\bs{P}_p=   \I_p - \B\B^+$. This form is convenient because the asymptotic behavior under $H_0$ when both $n$ and $d$ tend to infinity follows from Theorem~\ref{thm:Chen}, as we now show. The corollary below follows directly from the latter theorem and parallels Corollary~2.2 of \cite{Chen:2018}. 
\begin{corollary}\label{cor:Chen2}
Assume that $H_0$ in \eqref{eq:H0} holds and consider $M_{np}$ in Eq.~\eqref{eq:supremum} with $\SA = (1/n)\I_p$.
Further let $\bs{P}_p = \I_p - \B\B^+$ and suppose that there exist constants $\underline{b},\bar{b} > 0$, $B_n = c \geqslant 1$ and $\lambda \in (0,1)$ such that (M.1) and (M.2) in Theorem~\ref{thm:Chen}  hold and that $\log^7(np) \leqslant \bar{b}n^{1-\lambda}$.
Then
$\sup_{E \in \mathcal{E}_{1}}| \mathbb{P}(M_{np} \in E) - \mathbb{P}(\|\bs{Z}_{p}\|_{\infty} \in E)| \leqslant  \kappa(\underline{b},\bar{b}) n^{-\lambda/6}$,
where $\bs{Z}_{p} \sim \mathcal{N}_p(\boldsymbol{0}_p,\bs{P}_p \S_{p}\bs{P}_p)$ and $ \mathcal{E}_1$ is the set of all intervals $ [a, b]$, $a,b \in \bar{\mathbb{R}}$.
\end{corollary}


Specific choices of $\B$ lead to situations in which the conditions (M.1) or (M.2) or both always hold. This is detailed in the following two propositions, which are proved in Appendix~\ref{app:proofs-null} in the Supplementary Material. We shall begin with (M.2), which is easier because it only pertains to the entries of $\bs{P}_p$.

\begin{proposition} \label{prop:P-sum-less-1}
Let $\bs{P}_p = \I_p - \B\B^+$. Assume that there exist $0 < a < c$ such that for all $r \in \{1,\dots,p\}$ and $\ell \in \{1,\dots,L\}$,  $|B_{r\ell}| \in \{0\} \cup [a,c]$. Further suppose that $\B$ has exactly one non-zero entry per row. Then for all $r \in \{1,\dots,p\}$, $1 \leqslant \sum_{s=1}^p |P_{rs}| < 1+(c/a)^2$. In particular, (M.2) holds with $B_n = 1 + (c/a)^2$.
\end{proposition}
Proposition~\ref{prop:P-sum-less-1} implies that (M.2) holds with $B_n=2$ when testing $H_0^*$ because each of the rows of the block membership matrix $\B$ has exactly one non-zero entry equal to $1$.

The validity of (M.1) is more difficult to verify because it depends not only on the entries of $\bs{P}_p$, but also on the covariance matrix $\S_p$. 
Interestingly, assuming that $\S_p$ is positive definite does not guarantee (M.1). For example,
take $d=3$ and consider $\B$ to be the $3\times 2$ matrix with columns $(1,0,0)^\top$ and $(0,1,1)^\top$. The
orthogonal projection $\bs{P}_3=\I_3 - \B\B^+$ has zeros in its first row, so that 
$(\bs{P}_3 \S_3 \bs{P}_3^\top)_{11} = \bs{0}_3^\top \S_3 \bs{0}_3 = 0$ irrespectively of $\S_3$ and (M.1) is never fulfilled. However, the validity of (M.1) can be checked more easily when testing $H_0^*$ with $\mathcal{G} = \{ \{ 1 ,\dots,d\}\}$ because then $\S_p \in \Sset$. This leads to the following result.

\begin{proposition} \label{prop:M1-exchangeable}
Suppose that  $H_0^*$ in \eqref{eq:H0-star} holds with $\mathcal{G} = \{ \{ 1 ,\dots,d\}\}$. Let $\bs{\sigma}=(\sigma_0,\sigma_1, \sigma_2)$ be such that  $\S_p = \SS_p(\bs{\sigma})$ and let $\delta_{1,d}(\bs{\sigma})$ be an eigenvalue of $\S_p$ as in 
Proposition~\ref{prop:eigen-Sigma}.
Set $\bs{P}_p = \I_p - \B\B^+=  \I_p-(1/p)\bs{J}_p$.
Then
$
	\mathrm{diag}(\bs{P}_p \S_p \bs{P}_p) = \{\sigma_2 - (1/p)\delta_{1,d}(\bs{\sigma})\} \bs{1}_p \geqslant (2/3)(\sigma_2 - \sigma_1)\bs{1}_p.
$
If $\S_p$ is positive definite for all $d \geqslant 4$, (M.1) holds with $\underline{b} = (2/3)(\sigma_2 - \sigma_1) > 0$.\end{proposition}

\section{Simulation study} \label{sec:sim-study}

\subsection{Study design} 

We now study the power and size of the tests presented in this paper in finite samples. To carry out the procedures, three user-specific choices first need to be made: the type of the statistic (either $E_{np}$ or $M_{np}$); the matrix $\SA$  (either $(1/n)\I_p$ or $\Sh_{np}$); and the estimator $\Sh_{np}$ of $\S_{np}$. For the latter, we consider $\Sh_{np}^{\rm P}$ or $\Sh_{np}^{\rm J}$ when testing $H_0$ and $\Sb_{np}^{\rm P}$ or $\Sb_{np}^{\rm J}$ when testing $H_0^*$; see Sections~\ref{sec:tests} and \ref{subsec:estimation-exchangeability} and Appendix~\ref{app:Sigma} in the Supplementary Material for details. In dimensions $d\geqslant 50$, only the jackknife estimators $\Sh_{np}^{\rm J}$ and $\Sb_{np}^{\rm J}$ are used.  \textcolor{blue}{We approximate the $p$-values using the asymptotic results in Section~\ref{sec:null}, relying either on Monte Carlo draws from the limiting distribution when needed or on the bootstrap method of \cite{Chen:2018}. Full details are given in Appendix~\ref{app:p-values} in the Supplementary Material; the number of replicates is set to $N=5000$.}
All tests are at the $5$\% level and executed on random samples of size $n \in \{50,100,150,250\}$ and varied dimensions.  All simulation results are reported in Appendix~\ref{app:sim} in the Supplementary Material; the number of simulation runs is either $1000$ or $2500$ as indicated therein.  Here we provide a summary of the key findings.

\subsection{Size investigations} \label{subsec:sim-study-level}

To investigate the size of the tests, we pick $\tau \in \{0,.3,.6\}$ and set
\begin{equation}\label{eq:T-equi-null}
\T = (1-\tau) \I_d + \tau \bs{J}_d,
\end{equation}
so that $H_0$ and $H_0^*$ hold with $\B = \bs{1}_p$ and $\mathcal{G}=\{\{1,\ldots,d\}\}$, respectively; this allows us to compare the performance of all the tests developed herein. The copula under the null is taken to be Normal, Student $t_4$, Gumbel, or Clayton with Kendall's tau matrix $\T$; in each case, there is a one-to-one correspondence between the copula parameter and $\tau$ .

Tables~\ref{tab:sim-level-normal}--\ref{tab:sim-level-clayton} display rejection rates of the tests of $H_0$ based on $E_{np}$ or $M_{np}$ for various choices of $\SA$ and $\Sh_{np}$.  The results are similar for all dependence structures considered.  First, we can conclude that when $\SA = \Sh_{np}$, the tests are excessively liberal with increasing $d$ and decreasing $n$ whatever the test statistics, the estimator of $\S_{np}$, and the value of $\tau$. The observed levels can reach up to $100$\% already for $d=15$; they only get close to $5$\% when $d=5$ and $n=250$.  Due to this poor behavior, we discard these tests henceforth. When $\SA = (1/n)\I_{p}$, the tests become increasingly conservative with increasing $d$; for any fixed $d$, the levels worsen with increasing $\tau$ and improve with increasing $n$.  The choice of estimator has only a minor effect, although the jackknife estimator leads to slightly more conservative tests. Interestingly, for any given sample size, the observed size drops to $0$ much more quickly with increasing $d$ when the statistic $E_{np}$ is used. 

When testing $H_0^*$, using structured estimators of $\S_{np}$ improves the results considerably; viz.\ Tables~\ref{tab:sim-level-star-normal}--\ref{tab:sim-level-star-clayton}. All tests now hold the level reasonably well when $M_{np}$ is used, and this across all values of $d$ and $\tau$. Increasing $d$ and $\tau$ still leads to conservative tests when $E_{np}$ is used, and slightly more so when $\SA=\Sh_{np}$, but far less so compared to Tables~\ref{tab:sim-level-normal}--\ref{tab:sim-level-clayton}.  The most striking difference is that observed sizes are now acceptable when $\SA=\Sh_{np}$. 

Finally, we investigate the size of  the tests 
when $\T$ has the block structure given by
\begin{equation}\label{eq:T-block}
	\T = (\bs{C}_{k\ell})_{1 \leqslant k,\ell \leqslant 3},
\end{equation}
where, for any $k \neq \ell \in \{1,2,3\}$, $\bs{C}_{k\ell} = \bs{C}_{\ell k}^\top$ are $d_k \times d_\ell$ matrices with all entries equal to $c_{k\ell}$, while the diagonal blocks $\bs{C}_{kk}$ are $d_k \times d_k$ matrices of the form described in Eq.~\eqref{eq:T-equi-null} with $\tau = c_{kk}$, $k \in \{1,2,3\}$. We set $c_{k\ell} = .4 - (.15)|k-\ell|$ for all $k,\ell \in \{1,2,3\}$. Such a matrix $\T$ satisfies $H_0^*$ with
$\mathcal{G} = \{\{1,\ldots, d_1\}, \{d_1+1,\ldots, d_1+d_2\}, \{d_1+d_2+1,\ldots, d\}\}$
and hence also $H_0$ with $\B$ being the block membership matrix described in Section~\ref{sec:2.3}. We consider balanced blocks with  $d_1=d_2=d_3=d/3$ as well as unbalanced blocks with $d_1=d_2/2=d_3/3=d/6$. 
The results, reported in Tables~\ref{tab:sim-level-blocks-normal}--\ref{tab:sim-level-blocks-t4} and \ref{tab:sim-level-star-blocks-normal}--\ref{tab:sim-level-star-blocks-t4}, lead to broadly similar conclusions as when $\T$ is as in Eq.~\eqref{eq:T-equi-null}.

\subsection{Power study} \label{subsec:sim-study-power}

To assess the power of the tests, we consider departures $\bs{T}_\Delta$ from $\bs{T}$ in Eq.~\eqref{eq:T-equi-null} with entries
\begin{align} \label{eq:departure}
	(a) \quad (\T_{\Delta})_{ij}= \T_{ij} + \Delta\times  \mathbbm{1}\{(i,j) = (1,2)\} \qquad 
	(b) \quad (\T_{\Delta})_{ij} = \T_{ij} + \Delta\times \mathbbm{1}(1 \not\in \{i,j\})
\end{align}
which are termed single (a) and column (b) departures. The distribution of $\bs{X}$ under the alternative is Normal or Student $t_4$ with a correlation matrix specified by $\T_{\Delta}$ \citep[Proposition 7.43]{McNeil/Frey/Embrechts:2015}, or a nested Archimedean copula with two nesting levels and two child copulas which are either Clayton or Gumbel  \citep{Hofert/Pham:2013}.


The first study examines the power of the tests of $H_0$ with $\B = \bs{1}_p$ based on $E_{np}$ or $M_{np}$ with $\SA = (1/n)\I_{p}$ and various estimators of $\S_{np}$;   $\SA=\Sh_{np}$ is excluded since the tests do not hold the level, viz.\ Section~\ref{subsec:sim-study-level}. The results for $\Delta\in \{0.1,0.2\}$ and single and column departures are reported in Tables~\ref{tab:sim-power-1-normal}--\ref{tab:sim-power-2-clayton}. Regardless of the dependence structure, the power of each test increases with $n$, $\tau$, and $\Delta$, and decreases with $d$. The drop in power with increasing $d$ is greater when $E_{np}$ is used; for example, the test based on $E_{np}$ and the jackknife estimator has no power at all when $d=100$ for all sample sizes. The lack of power of the tests based on $E_{np}$ for large $d$ resonates with their observed size of nearly $0$; see Tables~\ref{tab:sim-level-normal}--\ref{tab:sim-level-clayton}.  The tests based on $M_{np}$  suffer much less from the curse of dimensionality; the estimated power when the jackknife estimator is used is close to $100$\% when $\tau =0.6$, $\Delta \geqslant0.1$ and $n \geqslant150$.  The effect of the estimator of $\S_{np}$ seems to be minor; the jackknife estimator leads to slightly smaller power but is more feasible computationally. Furthermore, column departures are easier to detect than single departures for the same value of $\Delta$.  Finally, the tests based on $M_{np}$ have higher power than the tests based on $E_{np}$ to detect single departures, but column departures are often better detected by a test based on $E_{np}$ in small dimensions. These observations are illustrated in the top row of Figure~\ref{fig:power}, which compares the power of the tests based on $E_{np}$ and $M_{np}$ as a function of $d$ for single (left panel) and column (right panel) departures when $n=100$, $\tau=0.3$, $\Delta=0.1$, and $\Sh_{np}=\Sh_{np}^{\rm J}$. It transpires from the same figure that the power generally depends on the underlying copula, with power being the largest when the copula is Normal and smallest when it is $t_4$ or Clayton; this is also true for other values of $n$, $\tau$, and $\Delta$.

\begin{figure}[t!]
	\centering
	\includegraphics[width=.81\textwidth]{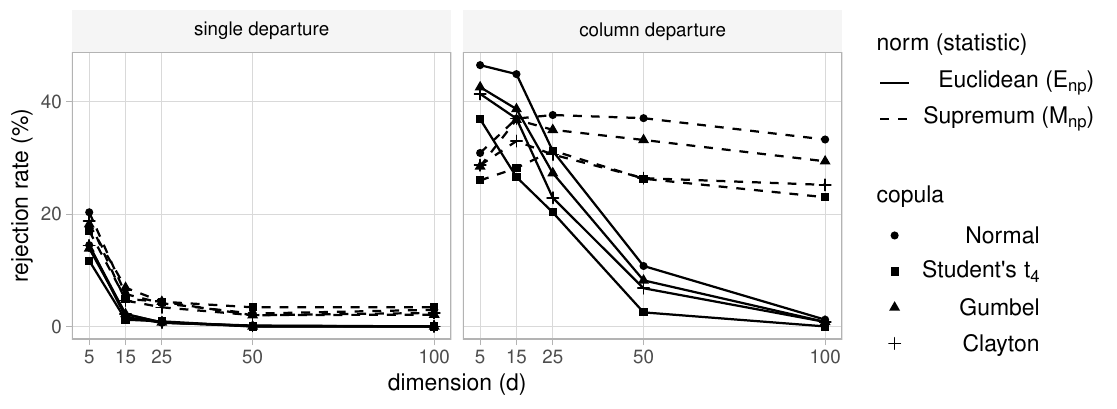}
	
	\smallskip
	\includegraphics[width=.81\textwidth]{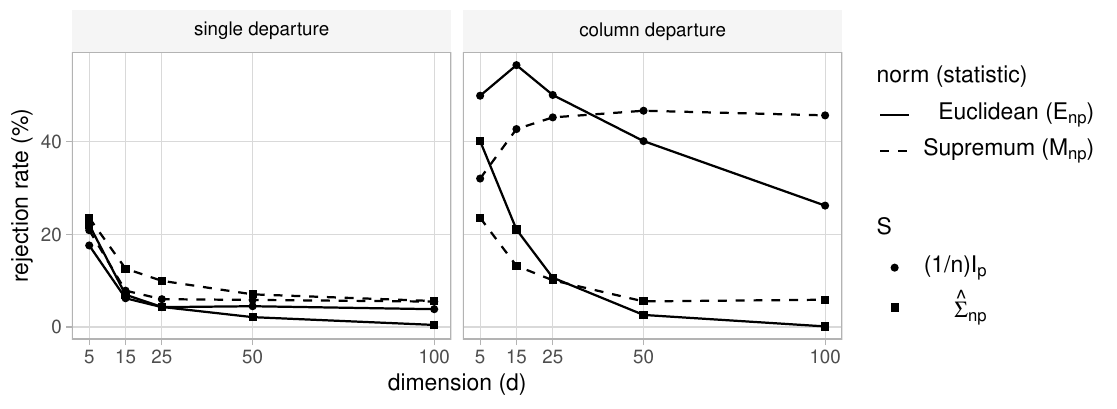}
	\vspace{-5pt}
	\caption{Observed rejection rates for single and column departures of the tests based on $E_{np}$ and $M_{np}$ as a function of $d$ when $n=100$, $\tau=0.3$, $\Delta=0.1$, and the jackknife estimator of $\S_{np}$ is used. The top row pertains to tests of $H_0$ with $\B = \bs{1}_p$ for various copula models; the bottom row corresponds to tests of
 $H_0^*$ with $\mathcal{G}=\{\{1,\ldots,d\}\}$ for the Normal distribution and various choices of $\SA$.}
\label{fig:power}
\end{figure}

The second study investigates tests of $H_0^*$ with $\mathcal{G}=\{\{1,\ldots,d\}\}$. The results, reported in Tables~\ref{tab:sim-power-star-single-1-normal}--\ref{tab:sim-power-star-column-2-clayton}, now include the tests with $\SA=\Sh_{np}$ since they hold their level well. Irrespective of the underlying copula, we can again conclude that the power of all tests increases with $n$, $\tau$, and $\Delta$, decreases with $d$, and that column departures are easier to detect than single departures. The drop in power when $d$ increases is again much greater for $E_{np}$ compared to $M_{np}$.  The preferred choice of $\SA$ depends on the alternative: $\SA=(1/n)\I_p$ leads to higher power to detect column departures, particularly for larger values of $d$. In contrast,  $\SA=\Sh_{np}$ is slightly preferable for single departures, especially when $d \leqslant 25$ and $\tau \geqslant 0.3$. The choice of the estimator $\Sh_{np}$ has a minor effect overall. Compared to $E_{np}$, $M_{np}$ has slightly better power to detect single departures in small dimensions, and substantially higher power to identify column departures in high dimensions. Power also depends on the distribution of $\X$ in a similar way as when testing $H_0$, albeit less so. Some of these observations are highlighted in the bottom row of Figure~\ref{fig:power}, which shows observed rejection rates as a function of $d$ for various choices of test statistics and $\SA$ when $\X$ is Normal.


The final study considers single departure alternatives with $\Delta=0.1$ for the case when $\T$ has the more elaborate block structure~\eqref{eq:T-block} with parameters as in Section~\ref{subsec:sim-study-level}. The results for the tests of $H_0$ and $H_0^*$ are summarized in Tables~\ref{tab:sim-power-blocks-normal}--\ref{tab:sim-power-blocks-t4} and \ref{tab:sim-power-star-blocks-normal}--\ref{tab:sim-power-star-blocks-t4}, respectively. Again, power increases with $n$ and decreases with $d$, and is higher when $\X$ is Normal than when it is $t_4$. Specific to this study is the observation that single departures are easier to detect in the unbalanced than in the balanced block design. Interestingly, when testing $H_0^*$, $\SA=\Sh_{np}$ leads to more powerful tests than $\SA=(1/n)\I_p$ this time.


\section{Power under local alternatives}\label{sec:local-alternatives}

Since the margins of $\X$ are continuous, there are no ties in the sample almost surely. Hence, 
\begin{equation}\label{eq:tau-u}
\sqrt{n}(\th_{np}-\t_p) = \frac{2}{n} \sum_{\nu=1}^n g^*(\U_\nu) + o_P(1)
\end{equation}
where $\U_\nu =  (F_1(X_{\nu 1}),\ldots, F_d(X_{\nu d}))$ is distributed as the copula $C$ of $\X$ and $g^* = (g^*_1,\ldots, g_p^*)$ with 
$
g_r^*(\bs{u}) = 2\{C_{i_rj_r}(u_{i_r},u_{j_r}) + \bar C_{i_rj_r}(u_{i_r},u_{j_r})\}-1-\tau_r,
$
where $\bar C_{i_rj_r}$ denotes the survival function of $C_{i_rj_r}$. This means that we only need to consider local copula alternatives.

\begin{assumption}\label{ass:qm}
Suppose that $\mathcal{C}=\{C_\bcth, \bcth \in \Theta \subseteq \mathbb{R}^k\}$ is a family of $d$-dimensional copulas with Lebesgue densities $c_\bcth$, $\bcth \in \Theta$. Suppose that  $\Theta$ is an open subset of $\mathbb{R}^k$.  Assume that  $\mathcal{C}$ is differentiable at $\bcth\in\Theta$ in quadratic mean, that is, there exists a measurable function $\dot \ell_\bcth=(\dot \ell_{\bcth 1},\ldots,\dot \ell_{\bcth k})$ so that as $\boldsymbol{h} \in \mathbb{R}^k$ tends to $\boldsymbol{0}$,
\begin{equation*} 
\int_0^1 \dotsi \int_0^1 \Bigl\{ \sqrt{c_{\bcth + \h}(\bs{u}) } - \sqrt{c_\bcth(\bs{u})}  - \frac{1}{2} \h^\top \dot \ell_\bcth (\bs{u})  \sqrt{c_\bcth(\bs{u}) } \Bigr\}^2 d u_1 \dotsm d u_d = o(\| \h \|^2).
\end{equation*}
\end{assumption}

Under Assumption~\ref{ass:qm}, an analogue of Theorem~\ref{thm:asymptotic-n}  holds under local alternatives.

\begin{theorem}
\label{thm:asymptotic-n-loc}
Suppose that $\mathcal{C}$ is a family of copulas that satisfies Assumption \ref{ass:qm}. Consider the null  $H_0: \bcth = \bcth_0$ and the sequence of local alternatives $H_{1n}:  \bcth = \bcth_0 + \h_n/\sqrt{n}$ where $\h_n \to \h$ as $n\to \infty$.  Suppose that under $H_0$, $n\SA_{np}$ and $\bs{P}_{np}$ converge in probability, as $n\to\infty$ to some $p \times p$ matrices $\SA_{p}$ and $\bs{P}_p$, respectively.
Assume that $\SA_{p}$ is positive definite,  and that for all $n$, $\SA_{np}$ is positive definite and $\bs{P}_{np}\t_p = \bs{0}_p$.
Then under $H_{1n}$, 
$\SA^{-1/2}_{np} \bs{P}_{np}\th_{np} \rightsquigarrow \bs{Z}_p$
as $n\to \infty$, where $\bs{Z}_p \sim \mathcal{N}_p(\bs{\zeta}_{\h}, \SA_p^\dag)$ with $\SA_p^\dag = \SA_{p}^{-1/2} \bs{P}_p \S_{p} \bs{P}_p^\top \SA_{p}^{-1/2}$ and $\bs{\zeta}_{\h} = \SA^{-1/2}_{p} \bs{P}_{p}\bs{a}$, where for each $r \in \{1,\ldots, p\}$, $a_r = E\{2g_r^*(\U) \h^\top \dot \ell_{\bcth_0} (\U)\}$ with $\U \sim C_{\bcth_0}$.
\end{theorem}

The proof of Theorem~\ref{thm:asymptotic-n-loc} relies on Le Cam's Third Lemma and may be found in Appendix~\ref{app:local-alternatives} in the Supplementary Material. As in Section~\ref{subsec:n-asymptotics}, Theorem~\ref{thm:asymptotic-n-loc} implies that under $H_{1n}$ and Assumptions (a) or (b) in Propositions~\ref{prop:asymptotic-euclidean} and \ref{prop:asymptotic-supremum}, $E_{np} \rightsquigarrow \| \Z_p \|_2^2$ and $M_{np} \rightsquigarrow \|\Z_p \|_{\infty}$ as $n\to \infty$, respectively, where $\bs{Z}_p \sim \mathcal{N}_p(\bs{a},\SA_p^\dag)$. However, the distribution of $ \| \Z_p \|_2^2$ is no longer chi-square  or a mixture thereof but has a rather complicated representation in terms of independent Normal variables \citep[Eq.~(4.1.2)]{Mathai/Provost:1992}.

The distributions of the test statistics under local alternatives allow us to calculate asymptotic power curves. To illustrate, we consider a special case of the full exchangeability hypothesis and assume that all entries of $\t$ are equal to a given $\tau_0$. We further focus on the Normal and Student $t_4$  families whose parameters are smooth functions of $\t$. This allows us to consider local alternatives of the form $\t + \h/\sqrt{n}$ with $\h = \Delta \mathbf{e}$, where $\Delta > 0$ and for $ r \in \{1,\ldots, p\}$, $\mathbf{e}_r$ equals $\mathbbm{1}\{(i_r,j_r) = (1,2)\}$ for single and $\mathbbm{1}(1 \not\in \{i_r,j_r\})$ for column departures in Eq. \eqref{eq:departure}, respectively. Detailed calculations of the drift $\boldsymbol{a}$ for these copula families may be found in Appendix~\ref{app:local-alternatives-examples} in the Supplementary Material. The resulting asymptotic power curves when $\tau_0=0.3$ are shown in Figure~\ref{fig:power-local}. For dimensions $d\in\{5,25\}$, the plots lead to similar conclusions as the simulation study: The power of the tests generally worsens with $d$; $M_{np}$ is preferable for single departures while $E_{np}$ performs better for column departures; $\SA=(1/n)\I_p$ and $\SA=\Sh_{np}$ leads to higher power for column and single departures, respectively, whatever the test statistic. It can also be seen that power is generally larger when the copula family is Normal compared to Student $t_4$. 

\begin{figure}[t]
	\centering
	\includegraphics[width=.95\textwidth]{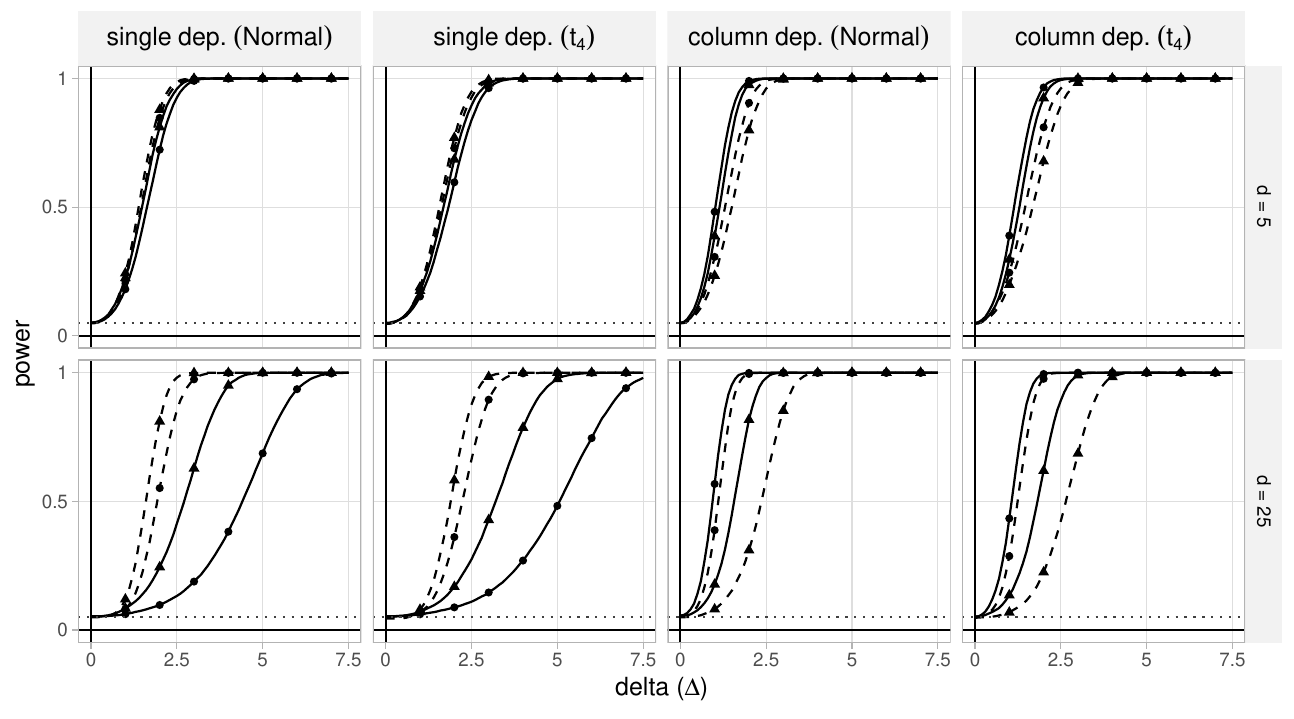}
	\vspace{-5pt}
	\caption{Asymptotic power curves of the tests based on $E_{np}$ (full) and $M_{np}$ (dashed) for the Normal and Student $t_4$ copula alternatives with single and column departures when $\tau_0=0.3$. The choice of $\SA$ is $(1/n)\I_p$ (circles) or $\Sh_{np}$ (triangles) and $d$ is set to either $5$ (top row) or $25$ (bottom row). } 
\label{fig:power-local}
\end{figure}

\section{Application} \label{sec:application}

\subsection{Data description and preprocessing} 
\defcitealias{PSMSL:2020}{PSMSL, 2020}
\defcitealias{IBC:2015}{IBC, 2015}

We used the methodology developed in this paper to analyze the dependence between average sea levels measured in the month of February at $d = 18$ different coastal stations from $1954$ to $2018$, inclusively ($n = 65$).
All stations are located in the continental United States, except for one in Hawaii, two in Alaska, two in Canada and one in Panama.
The data were retrieved on March 2, 2020 from the Permanent Service for Mean Sea Level (\citealp{Holgate/al:2013}; \citetalias{PSMSL:2020}).
The station names, as given in the original dataset, are listed in Appendix~\ref{app:application} in the Supplementary Material.

Figure~\ref{fig:stations} displays the $18$ stations. By looking at the map, it seems reasonable to believe that sea levels at certain stations are related and that geographical proximity plays a role.
Accounting for this spatial dependence is of interest in the context of flood insurance, for example.
In order to evaluate the financial risk associated with floods, an insurer needs to model not only the probability of such events occurring at specific locations, but also how likely it is that many of these locations will be flooded simultaneously.
In countries like Canada, where the private sector has been offering homeowners flood insurance products for less than a decade (\citetalias{IBC:2015}), flood data are rare and insurers usually rely on modeling water flows or water levels as a first step in estimating the financial risk underlying these products.
A common approach consists of using such models to generate synthetic water levels from which the financial losses are then estimated.
As the number of locations considered can be large, it often appears convenient to use, e.g., factor models, or similar constructions from which it is relatively simple to generate synthetic observations. Such models however require the factor structure as input. The tests developed here are useful for identifying and validating such a structure, as we now illustrate.

\begin{figure}[t]
	\centering
	\includegraphics[width=.78\textwidth]{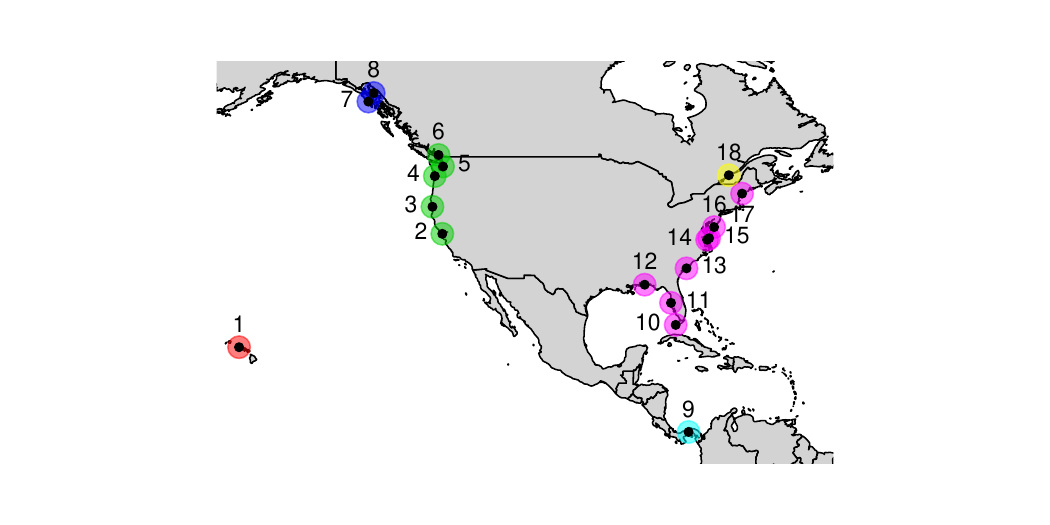}
	\vspace{-20pt}
	\caption{Location and indexation of the $d=18$ coastal stations. The colours indicate hypothesized clusters.} \label{fig:stations}
\end{figure}

Figure~\ref{fig:stations} shows six clusters of stations grouped by the geographical region in which they are located. All stations in the same cluster are highlighted with the same colour: $\mathcal{G}_1 = \{1\}$, $\mathcal{G}_2 = \{ 2,\dots,6\}$, $\mathcal{G}_3 = \{ 7,8\}$, $\mathcal{G}_4 = \{9\}$, $\mathcal{G}_5 = \{ 10,\dots,17\}$ and $\mathcal{G}_6 = \{ 18\}$. Station \#18 forms a single cluster because it is not located on the Atlantic coast but in the delta of the Saint Lawrence river.
The left panel of Figure~\ref{fig:Taus} displays the matrix of empirical Kendall's taus; the diagonal entries are highlighted with the same colour as the stations in Figure~\ref{fig:stations}. The above clusters seem to induce a block structure, which may be exploited when building a joint model in order to reduce its dimensionality, such as a factor model, for example.

Because the measurements at many stations show a monotone trend in time, a phenomenon widely attributed to global warming \citep{Oppenheimer/al:2019}, the data were preprocessed as explained in Appendix~\ref{app:application} in the Supplementary Material.
The methodology developed in this paper was then applied to regression residuals; although these are not i.i.d, the work of \cite{Cote/Genest/Omelka:2019} shows that the asymptotic results derived here in the fixed $d$ setting still apply to residuals from regression models with Normal errors.

\begin{figure}[t]
	\centering
	\begin{minipage}{.0747\textwidth}
		\centering
		\includegraphics[width=1\textwidth]{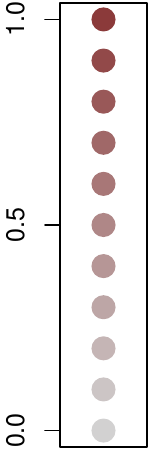}
		{\ }
	\end{minipage}
	\qquad
	\begin{minipage}{.225\textwidth}
		\centering
		\includegraphics[width=1\textwidth]{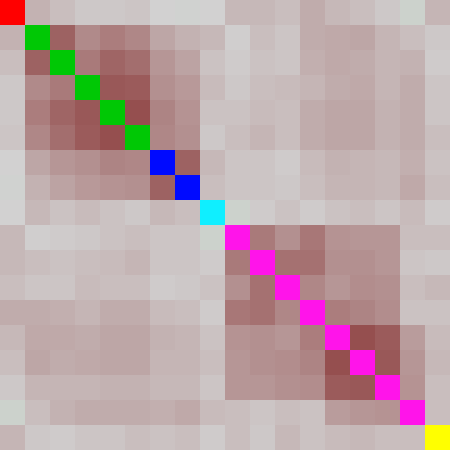}
		(a)
	\end{minipage}
	\qquad
	\begin{minipage}{.225\textwidth}
		\centering
		\includegraphics[width=1\textwidth]{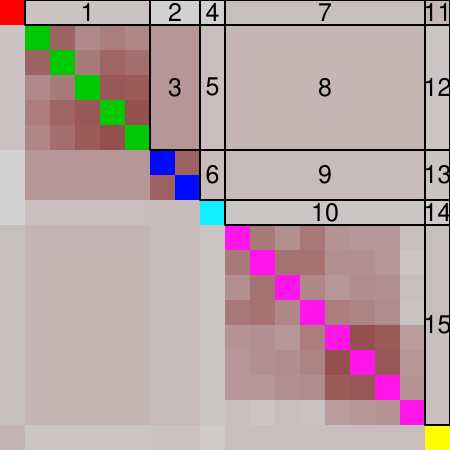}
		(b)
	\end{minipage}
	\vspace{-5pt}
	\caption{Matrix of empirical Kendall's tau (a) and its block structured equivalent (b). The diagonal entries of the matrices are such that they match the colour of the station they refer to in Figure~\ref{fig:stations}. The indexation indicates the column of $\B$ that encodes the constraints associated to each block.} \label{fig:Taus}
\end{figure}

\subsection{Application of the proposed methodology} \label{subsec:application-results}

To show how the proposed tests can help point towards a suitable block structure of the population matrix $\bs{T}$ of Kendall's taus, we tested three hypotheses. The first is $H_0^*$ with $\mathcal{G}=\{\mathcal{G}_1,\ldots, \mathcal{G}_6\}$. This hypothesis of partial exchangeability of the underlying copula induces the following block structure in $\bs{T}$: there are six diagonal blocks, three of which are of size $1$, and $15$ off-diagonal blocks, shown in the right panel of Figure~\ref{fig:Taus}. This means that under $H_0^*$, $\T$ contains only $18$ distinct entries off the main diagonal, which is considerably less than the total number $p = d(d-1)/2 = 153$ of pair-wise Kendall's taus. This reduction in the number of parameters is tempting. However, when testing $H_0^*$ with any of the tests proposed in this paper, the hypothesis is rejected with a $p$-value of at most $10^{-10}$. The hypothesis of partial exchangeability with the above clusters thus seems to be too strong.

The left panel of Figure~\ref{fig:Taus} suggests why $H_0^*$ may not be adequate: the diagonal blocks corresponding to clusters $\mathcal{G}_2$ and $\mathcal{G}_5$ do not seem to be homogeneous. It is here that the more general hypothesis $H_0$ may be useful.  For example, we may wish to test that the entries in each of the diagonal blocks induced by the partition $\mathcal{G}$ are distinct, giving $10+1+28 =39$ possibly distinct values, while the entries in each of the off-diagonal blocks are the same, giving $15$ additional possibly distinct values. Under this assumption, the number of parameters in $\T$ is reduced from $153$ to $54$. This hypothesis can be formulated as $H_0$ in \eqref{eq:H0} with $\B \in \{0,1\}^{p \times L}$ with $L=54$ and $p=153$, which has exactly one entry equal to $1$ in each row. The exact form of $\B$ is reported in the Supplementary Material; for example, its first column has entries $B_{r1} = \mathbbm{1}\{(i_r,j_r) \in \mathcal{G}_1 \times \mathcal{G}_2 \}$ for $r \in \{1,\ldots, p\}$. Note that under $H_0$, there are only $12$ blocks with more than one entry, and the tests focus on these blocks only; the remaining $42$ blocks have one entry each and their contribution to any of the test statistics considered here is $0$.

To test $H_0$ with the above $\B$, we discarded the option $\SA=\Sh_{np}$, because the resulting tests may not hold their level, as discussed in Section~\ref{subsec:sim-study-level}. In order to assess the reliability of the tests using $\SA=(1/n)\I_p$, we generated $2500$ i.i.d. samples of size $n=65$ from the $18$-dimensional Normal distribution whose matrix of Kendall's taus is the matrix version $\hat{\bs{\Theta}}_{np}$ of  $\tt_{np} = \B\B^+ \th_{np}$. The matrix $\hat{\bs{\Theta}}_{np}$ is displayed in the right panel of Figure~\ref{fig:Taus}.  We then tested the hypothesis $H_0$ with this same $\B$, at significance level $\alpha=0.05$. For $E_{np}$ and $M_{np}$, we obtained sizes of $4.3\%$ and $4.5\%$, respectively, when using $\Sh_{np} = \Sh_{np}^{\rm P}$, and $3\%$ and $3.2\%$, respectively, when using $\Sh_{np} = \Sh_{np}^{\rm J}$. As these observed sizes are acceptable, we proceeded with applying the tests based on $E_{np}$ and $M_{np}$, respectively, with $\SA=(1/n)\I_p$ and $\Sh_{np} = \Sh_{np}^{\rm P}$. The $p$-values were $10.7$\% and $32.9$\% when $E_{np}$ and $M_{np}$ were used, respectively. In either case $H_0$ is not rejected at the $5$\% level and hence the data provide no evidence against this particular block structure. Exploiting the latter structure leads to the smoother estimator $\hat{\bs{\Theta}}_{np}$ of $\T$ shown in the right panel of Figure~\ref{fig:Taus}, which is likely more efficient than $\Th$ as suggested by the simulations in \cite{Perreault/Duchesne/Neslehova:2019}.

Finally, the formulation of the hypothesis in Eq.~\eqref{eq:H0} is sufficiently broad so that it can be used to test hypotheses about any submatrix of $\T$ individually. For example, we may wish to test whether the entries in block \#3 in the right panel of Figure~\ref{fig:Taus} are the same. The matrix $\B$ that is suited for this purpose is again reported in the Supplementary Material. Interestingly, the $p$-values of the tests based on $E_{np}$ and $M_{np}$, respectively, with $\SA=(1/n)\I_p$ and $\Sh_{np} = \Sh_{np}^{\rm P}$, are $0.3$\% and $0.4$\%, respectively. This reveals some differences in dependence between stations \#7 and \#8 located in Alaska, and stations \#2--\#6 located further south along the Pacific coast. This may be because cluster $\mathcal{G}_2$ is too heterogenous and could be divided, or because the clusters $\mathcal{G}_2$ and $\mathcal{G}_3$ should be merged into one group. Even more interestingly, perhaps, testing individual blocks one at the time may lead to more powerful procedures. We conjecture this because the tests proposed here lose their power with increasing $d$, as revealed by the simulation study in Section~\ref{subsec:sim-study-power}. However, proceeding this way would require adjustments for multiple testing and a careful balance between $d$ and the number of individual tests; this is left for future work.

\section{Discussion} \label{sec:conclusion}

We have provided a suite of new procedures  to test a wide class of hypotheses about the structure of Kendall correlation matrices. We have also worked out expressions that are easier to compute for a specific subset of these hypotheses. While rigorous theoretical investigations have shown that several combinations of test statistics, variance estimators and p-value approximations lead to asymptotically valid inferences, even in high dimension, simulations and empirical implementation have suggested the following practical guidelines.

First, even though their asymptotic distribution under the null is more complex, tests based on the statistics $E_{np}$ and $M_{np}$ computed with $\SA=(1/n)\I_p$ with jackknife p-value approximation tend to better hold their size while exhibiting good power. They are also easier to implement computationally for large $d$. Hence, we would recommend this combination as the most adequate procedure for testing $H_0$ among all methods that we have investigated. Second, if the more restricted null hypothesis $H_0^*$ is of interest, then in this case the use of the structured $\SA=\Sh_{np}$ also has good size and power properties, and the choice of the test statistic to use should therefore be driven by the departure from the null that one is trying to detect, with a preference for $M_{np}$ for applications in higher dimensions.

We have illustrated that the proposed tests can be used to induce some parsimony in a correlation matrix estimator or to rule out some potential parametric modeling avenues for the dependence structure. In future work, we intend to use these tests as the core ingredients of complex correlation structure learning algorithms.

It may be of interest to extend the methodology proposed here to other pair-wise rank-based measures of association, e.g., Spearman's rho. 
Apart from describing and estimating $\bs{\Sigma}_{np}$, this would however require new high-dimensional asymptotic results for $U$-statistics of higher order, or for more general rank statistics; of interest in this regard may be the theory developed, e.g., in \cite{Leung/Drton:2018} and \cite{Drton/Han/Shi:2020}.


%

\section*{Supplementary Material}

Appendices~\ref{app:proofs-prelim}--\ref{app:proofs-null} contain the proofs of the results in Sections~\ref{sec:preliminary-notions}--\ref{sec:null}.
Appendix~\ref{app:Sigma} discusses the estimators $\Sh_{np}^{\rm P}$ and $\Sh_{np}^{\rm J}$ of $\S_{np}$, as well as their structured equivalent $\Sb_{np}^{\rm P}$ and $\Sb_{np}^{\rm J}$.
Appendix~\ref{app:p-values} contains details about the p-value approximations used in the simulation study of Section~\ref{sec:sim-study}, while Appendix~\ref{app:sim} provides the empirical results of this latter study.
Appendices~\ref{app:local-alternatives} and \ref{app:application} provide additional details for Sections~\ref{sec:local-alternatives} and \ref{sec:application}, respectively.

\newpage

\spacingset{1.25} 

\begin{center}
\LARGE \bf Hypothesis tests for structured rank\\
correlation matrices\\
(Supplementary Material)
\end{center}

\subsection*{Description}\label{app:S0}

The following material supplements the content of the main text. 
Appendices~\ref{app:proofs-prelim}, \ref{app:exchangeability} and \ref{app:proofs-null} contain the proofs of the results in Sections~\ref{sec:preliminary-notions}, \ref{sec:exchangeability}, and \ref{sec:null}, respectively.
Appendix~\ref{app:Sigma} includes details about the estimators $\Sh_{np}^{\rm P}$ and $\Sh_{np}^{\rm J}$ of $\S_{np}$, as well as their structured equivalent $\Sb_{np}^{\rm P}$ and $\Sb_{np}^{\rm J}$; in particular, it contains a new result that eases the estimation of $\Sb_{np}^{\rm J}$ (Proposition~\ref{prop:sigma-exch}).
Appendix~\ref{app:sim-study} provides details about the numerical approximations of $p$-values
for the tests considered in the main paper and gives results of all simulation studies that were carried out. Appendix~\ref{app:local-alternatives} contains the proof of Theorem~\ref{thm:asymptotic-n-loc} and examples of asymptotic power curve calculations for the Normal and $t_4$ copula models. 
Appendix~\ref{app:application} provides additional details about the data application in Section~\ref{sec:application}.
Relevant code pertaining to the simulation study and the data application can be found on GitHub at \url{https://github.com/samperochkin/testing-tau}.

\clearpage

\tableofcontents

\clearpage
\appendix
\counterwithin{figure}{section}
\counterwithin{table}{section}
\counterwithin{equation}{section}
\counterwithin{theorem}{section}
\counterwithin{lemma}{section}
\counterwithin{proposition}{section}
\counterwithin{corollary}{section}
\counterwithin{remark}{section}

\section{Proofs from Section \ref{sec:preliminary-notions}} \label{app:proofs-prelim}

\subsection{Proof of Theorem~\ref{thm:Chen}}
First observe that, for any $r \in \{1,\dots,p\}$ and $\bs{x}, \bs{y} \in \mathbb{R}^d$,
\begin{align*}
\{\bs{P}_p \th_{np}\}_r = \frac{2}{n(n-1)} \sum_{1 \le \nu < \eta \le n} h_{r}^\prime(\bs{X}_\nu,\bs{X}_\eta),\qquad h_{r}^\prime(\bs{x},\bs{y}) = \sum_{s=1}^p P_{r s}\  h_s(\bs{x},\bs{y})
\end{align*}
where $h(\bs{x}, \bs{y})$ is as in Eq.~\eqref{eq:tau-hat}.
This representation makes it clear that $\bs{P}_p(\th_{np} - \t_p)$ is a centered $U$-statistic with H\'ajek projection $H_n'$ given, for any $\bs{x} \in \mathbb{R}^p$, by
$$
H_n^\prime = \frac{2}{n} \sum_{r=1}^n g^\prime(\bs{x}_r), \qquad g^\prime(\bs{x}) = \E\{h^\prime(\bs{x}, \bs{X})\} -\bs{P}_p \t_p = \bs{P}_p g(\bs{x}),
$$
where $g(\bs{x})$ is as in Eq.~\eqref{eq:Hajek}.
Let $\X^\prime$ be an independent copy of $\X$. For a real-valued random variable $Y$ and the function $\psi_1(z) = \exp(z) - 1$, denote by $\| Y \|_{\psi_1}$ the Orlicz norm of $Y$, viz. $\| Y \|_{\psi_1} = \inf \left\{ c  >0 : \E\{ \exp(|Y|/c)\} \leqslant 2 \right\}$. For all $n$, set $B_n^\prime = (2B_n)^2$.
The result follows from Theorem~2.1 of \cite{Chen:2018}, provided that (M.2) in Theorem~\ref{thm:Chen} in the main text implies that
\begin{align}
	&\max_{1 \leqslant k \leqslant p} \E \{ |h_k'(\X,\X') - (\bs{P}_{p} \t_p)_k|^{2+\ell} \} \leqslant (B_n^\prime)^{\ell}, \quad \ell \in \{1,2\}, \label{eq:proof-thm-Chen-1}\\
	&\max_{1 \leqslant k \leqslant p} \| h_k'(\X,\X') -(\bs{P}_{p} \t_p)_k\|_{\psi_1}  \leqslant B_n^\prime\label{eq:proof-thm-Chen-2}
\end{align}
both hold. These inequalities correspond to conditions (M.2) and (E.1) of \cite{Chen:2018}, respectively, and will be shown in turn. First note that for all $r \in \{1,\ldots, p\}$ and $\bs{x},\bs{y} \in \mathbb{R}^p$,
$|h_{r}(\bs{x},\bs{y}) - \tau_r | \leqslant 2$. This implies that, for all $r \in \{1,\dots,p\}$ and $\ell \in \{1,2\}$,
\begin{equation*} 
	\E\{| h_{r}'(\X,\X')- (\bs{P}_{p} \t_p)_r|^{2+\ell}\} \leqslant \E\Big\{\Big(\sum_{s=1}^p |P_{rs}| \ |h_s(\X,\X') - \tau_s|\Big)^{2+\ell}\Big\} \leqslant \Big(2\sum_{s=1}^p |P_{rs}|\Big)^{2+\ell}.
\end{equation*}
The last expression is at most $(B_n^\prime) ^\ell$, establishing the inequality \eqref{eq:proof-thm-Chen-1}.
To show that inequality \eqref{eq:proof-thm-Chen-2} holds, fix an arbitrary  $r \in \{1,\dots,p\}$ and set $c =  4\sum_{s=1}^p |P_{rs}|$. Then
\begin{equation*} 
	\E\Big[\exp\Big\{c^{-1}| h_{r}'(\X,\X')-(\bs{P}_{p} \t_p)_r|\Big\}\Big] \leqslant e^{1/2} < 2,
\end{equation*}
so that $\| h_r'(\X,\X') -(\bs{P}_{p} \t_p)_r\|_{\psi_1} \leqslant 4\sum_{s=1}^p |P_{rs}| \leqslant 4 B_n \leqslant B_n^\prime$, since $B_n \geqslant 1$ by assumption.
\qed


\clearpage
\section{Proofs from Section~\ref{sec:exchangeability}} \label{app:exchangeability}

\subsection{Proof of Proposition~\ref{prop:eigen-Sigma}}
For the proof of Proposition~\ref{prop:eigen-Sigma}, introduce, for $k \in \{1,\ldots, d\}$ and $a_2 \neq b_2 \neq 0$ the vector $\bs{v}^k$ of length $p$ with components given, for all $r \in \{1,\ldots, p\}$, by
\begin{equation} \label{eq:eigenvector2}
	v_r^{k} = a_2\mathbbm{1}\{k \in (i_r,j_r)\} + b_2\mathbbm{1}\{k \not\in (i_r,j_r)\},
\end{equation}
where $(i_r,j_r) = \iota(r)$ is as defined in \eqref{eq:bijection}.
Furthermore, for $1 \leqslant i < j \leqslant d$ and $a_3 \neq b_3 \neq c_3 \neq 0$, let $\bs{v}^{ij}$ be the vector of length $p$ with components given, for all $r \in \{1,\ldots, p\}$, by
\begin{equation} \label{eq:eigenvector3}
	v_r^{ij} = a_3\mathbbm{1}\{|(i_r,j_r) \cap (i,j)| = 2\} + b_3\mathbbm{1}\{|(i_r,j_r) \cap (i,j)| = 1\} + c_3\mathbbm{1}\{|(i_r,j_r) \cap (i,j)| = 0\}.
\end{equation}
As a preliminary step, note the following auxiliary Lemma.
\begin{lemma}\label{lem:orthogonality}
For arbitrary $k \in \{1,\ldots, d\}$ and $1 \leqslant i < j \leqslant d$, let $\bs{v}^k$ be as in \eqref{eq:eigenvector2} with $b_2 \neq 0$ and $a_2=-\{(d-2)/2\}b_2$. Let also $\bs{v}^{ij}$ be as in \eqref{eq:eigenvector3} with $c_3 \neq 0$, $a_3=\{(d-2)(d-3)/2\}c_3$ and $b_3= -\{(d-3)/2\}c_3$. Then the vectors $\bs{1}_p$, $\bs{v}^k$ and $\bs{v}^{ij}$ are orthogonal.
\end{lemma}

\medskip
\noindent
\textsc{Proof of Lemma~\ref{lem:orthogonality}.}
Clearly, $\bs{v}^k$ is orthogonal to $\bs{1}_p$ since
$$
\sum_{r=1}^p v^k_r = a_2(d-1) + b_2\{p-(d-1)\} = \frac{d-1}{2} \{ 2a_2+b_2(d-2)\} =0.
$$
Similarly, $\bs{v}^{ij}$ is orthogonal to $\bs{1}_p$, because
$$
\sum_{r=1}^p v^{ij}_r = a_3 + 2(d-2)b_3+ \frac{c_3(d-2)(d-3)}{2}  = 0.
$$
It remains to show that $\bs{v}^{ij}$ is orthogonal to $\bs{v}^k$. Indeed, if $k \not \in \{i,j\}$,
\begin{align*}
\sum_{r=1}^p v^{ij}_r v^k_r & = a_3b_2 + 2(d-3)b_3b_2 + 2b_3a_2 + (d-3)c_3a_2 + \frac{(d-3)(d-4)}{2} c_3b_2\\
&= c_3b_2\Biggl\{ \frac{(d-2)(d-3)}{2} - (d-3)^2 + \frac{(d-3)(d-4)}{2}\Biggr\} = 0,
\end{align*}
while if $k \in \{i,j\}$,
\begin{align*}
\sum_{r=1}^p v^{ij}_r v^k_r & = a_3a_2 + (d-2)b_3a_2 + (d-2)b_3b_2 +  \frac{(d-2)(d-3)}{2} c_3b_2\\
&= c_3b_2\Biggl\{ -\frac{(d-2)^2(d-3)}{4} +\frac{(d-2)^2(d-3)}{4}- \frac{(d-2)(d-3)}{2} + \frac{(d-2)(d-3)}{2}\Biggr\} = 0.
\end{align*}
This concludes the proof of Lemma~\ref{lem:orthogonality}. \qed

\medskip
\noindent
\textsc{Proof of Proposition~\ref{prop:eigen-Sigma}.}
For $r,s \in \{1,\ldots, p\}$, let $c(r,s) = |\{i_r,j_r\}\cap\{i_s,j_s\}|$. With this notation, the eigenequations take the form
\begin{equation}\label{eq:eigen-equation1}
	0 = [\left( \SS_p - \delta \I_p \right) \bs{v}]_r = (s_2 - \delta) v_r + s_1 \Biggl( \sum_{\substack{s=1 \\ c(r,s) = 1}}^p v_s \Biggr) + s_0 \Biggl( \sum_{\substack{s=1 \\ c(r,s) = 0}}^p v_s \Biggr), \qquad r=1,\dots,p.
\end{equation}
{\it First eigenvalue.} From \eqref{eq:eigen-equation1}, one can easily derive that $\bs{v} = \bs{1}_p$ is an eigenvector of $\SS_p$ with the corresponding eigenvalue $\delta_{1,d}(s_0,s_1,s_2) = s_2 + 2(d-2)s_1 + (p-2d+3)s_0$. Consequently, the geometric multiplicity of $\delta_{1,p}$ is at least $1$.

\medskip
\noindent
{\it Second eigenvalue.} Next, let, for $k \in \{1,\dots,d\}$ and $a_2 \neq b_2 \neq 0$ the vector $\bs{v}^k$ be as in \eqref{eq:eigenvector2}.  Evaluating \eqref{eq:eigen-equation1} for $\bs{v} = \bs{v}^k$ reduces the set of equations to the following two, depending on whether $k \in \{i_r, j_r\}$ or not:
\begin{align*}
	0 &= ( s_2-\delta)a_2 + s_1 (d-2)(a_2+b_2) + s_0 \{p - 2(d-2) - 1\} b_2\\
	0 &=(s_2-\delta)b_2 + s_1 \{2a_2 + 2(d-3)b_2\} + s_0 \{(d-3)a_2 + (p - 3d + 6) b_2\}.
\end{align*}
These equations may be rewritten as
\begin{align*}
	0 &= \{s_2 + (d-4) s_1-(d-3)s_0-\delta\}a_2 + (1/2)\{2a_2+(d-2)b_2\}\{2s_1+(d-3)s_0\}\\
	0 &= \{s_2 + (d-4) s_1-(d-3)s_0-\delta\}b_2 + (1/2)\{2a_2+(d-2)b_2\}\{2s_1+(d-3)s_0\}.
\end{align*}
Consequently, $\delta=\delta_{2,d}(s_0,s_1,s_2) = s_2 + (d-4)s_1 - (d-3)s_0$,  $a_2 = -\{(d-2)/2\}b_2$, and an arbitrary $b_2 \neq 0$ are possible solutions of the above equations. This implies that $\delta_{2,d}$ is an eigenvalue of $\SS_p$ and that for each $k \in \{1,\ldots, d\}$, $\bs{v}^k$ defined in \eqref{eq:eigenvector2} with $a_2 = -\{(d-2)/2\}b_2$ is an eigenvector corresponding to $\delta_{2,d}$.

To determine the geometric multiplicity of $\delta_{2,d}$, it suffices to pick an arbitrary $b_2 \neq 0$, set $a_2 = -\{(d-2)/2\}b_2$ and consider $\bs{v}^k$, $k \in \{1,\ldots, d\}$ as in \eqref{eq:eigenvector2}, all with the same coefficients $a_2, b_2$. Because $\sum_{k=1}^d \bs{v}^k = \bs{0}_p$, these vectors are linearly dependent. However, it turns out that $\bs{v}^1,\ldots, \bs{v}^{d-1}$ are linearly independent, i.e., that for any $w_1,\ldots, w_{d-1} \in \mathbb{R}$,  $\sum_{k=1}^{d-1} w_k\bs{v}^k = \bs{0}_p$ implies that $w_1 = \ldots = w_{d-1} =0$. To show this, let $\bar w = \sum_{k=1}^{d-1} w_k$ and note that for each $r \in \{1,\ldots, p\}$,
$$
\sum_{k=1}^{d-1} w_k v^k_r = a_2(w_{i_r} + w_{j_r}) + b_2(\bar w - w_{i_r} - w_{j_r}).
$$
Hence for any $i_1 \neq i_2 \neq i_3 \in \{1,\ldots, d-1\}$, we have that $a_2(w_{i_1} + w_{i_2}) + b_2(\bar w - w_{i_1} -w_{i_2}) = 0$ and $a_2(w_{i_1} + w_{i_3}) + b_2(\bar w - w_{i_1} -w_{i_3}) = 0$. Subtracting these two equations from one another gives that $(b_2-a_2)(w_{i_3} - w_{i_2})=0$. Because $a_2 \neq b_2$, this implies that $w_{i_2} = w_{i_3}$. Because $i_1, i_2, i_3$ were arbitrary, $w_1 = \ldots = w_{d-1} =w$ for some $w \in \mathbb{R}$. However, the components of the vector $\sum_{k=1}^{d-1} w\bs{v}^k$ are all equal to $w \{ 2a_2 + b_2(d-3)\}$. Since $a_2 = -\{(d-2)/2\}b_2$, this term is zero if and only if $w=0$. The geometric multiplicity of $\delta_{2,d}$ is thus at least $d-1$.

\medskip
\noindent
{\it Third eigenvalue.} For $1\le i < j \le d$, and $a_3 \neq b_3 \neq c_3 \neq 0$, let $\bs{v}^{ij}$ be as in \eqref{eq:eigenvector3}.

Setting $\bs{v} = \bs{v}^{ij}$ in \eqref{eq:eigen-equation1} leads to the equations
\begin{align*}
	0 &= (s_2-\delta) a_3 + 2 s_1 (d-2) b_3 + s_0 (p-2d+3) c_3\\
	0 &= (s_2-\delta) b_3 + s_1 \{a_3 + (d-2)b_3 + (d-3)c_3\} + s_0 \{ (d-3)b_3 + (p-3d+6) c_3\}\\
	0 &= (s_2-\delta)c_3 + s_1\{ 4b_3 + 2(d-4)c_3\} + s_0\{a_3 + 2(d-4)b_3 + (p - 4d + 10) c_3\},
\end{align*}
which may be rewritten as
\begin{align*} 
	0 &= (s_2-2s_1+s_0-\delta) a_3 + \{a_3+b_3(d-2)\}(2s_1-s_0) + (1/2)\{2b_3 + c_3(d-3)\} (d-2) s_0 \\
	0 &= (s_2-2s_1+s_0-\delta) b_3 + \{a_3+b_3(d-2)\}s_1 + (1/2)\{2b_3 + c_3(d-3)\}\{2s_1+(d-4)s_0\}\\
	0 &= (s_2-2s_1+s_0-\delta)c_3 + \{a_3+b_3(d-2)\}s_0+  (1/2)\{2b_3 + c_3(d-3)\}\{4s_1+(d-6)s_0\}.
\end{align*}
Obviously, setting $\delta=\delta_{3} = s_2-2s_1+s_0$, $a_3=\{(d-2)(d-3)/2\}c_3$, $b_3= -\{(d-3)/2\}c_3$, and $c_3 \neq 0$ arbitrary solves these equations. Consequently, $\delta_{3}$ is an eigenvalue of $\SS_p$ and for any $1 \le i < j \le d$, $\bs{v}^{ij}$ in \eqref{eq:eigenvector3} with $a_3=\{(d-2)(d-3)/2\}c_3$ and $b_3= -\{(d-3)/2\}c_3$ and an arbitrary $c_3 \neq 0$ is its eigenvector.

To determine the geometric multiplicity of $\delta_{3}$, let $b_2 = -(d-1)/(d-2)$, $c_3 = d/(d-2)$, and set $a_2 = -\{(d-2)/2\}b_2$, $a_3=\{(d-2)(d-3)/2\}c_3$ and $b_3= -\{(d-3)/2\}c_3$. For $k \in \{1,\ldots, d\}$, let $\bs{v}^k$ be as in \eqref{eq:eigenvector2}, all with the same above coefficients $a_2, b_2$. Also, for $1 \le  i < j \le d$, let $\bs{v}^{ij}$ be as in \eqref{eq:eigenvector3}, again all with the same above coefficients $a_3, b_3, c_3$. Set $\bs{V}_1 = \bs{J}_p$, and let $\bs{V}_2$ and $\bs{V}_3$ be $p\times p$ matrices with $r$-th column equal to $\bs{v}^{i_r} + \bs{v}^{j_r}$ and $\bs{v}^{i_rj_r}$, respectively. It is easy to see that $\bs{V}_1 = \SS_p(1,1,1)$, as well as
\begin{align}
\bs{V}_2 & = \SS_p(2b_2, b_2 + a_2, 2a_2) = \SS_p(-2(d-1)/(d-2), (d-4)(d-1)/(d-2), (d-1)) \label{eq:V2}\\
\bs{V}_3 &= \SS_p(c_3, b_3, a_3) = \SS_p(d/(d-2), -d(d-3)/\{2(d-2)\}, d(d-3)/ 2).\label{eq:V3}
\end{align}
Hence, $\bs{V}_1 + \bs{V}_2 + \bs{V}_3 = \SS_p (0,0,p) = p \bs{I}_p$. From Lemma \ref{lem:orthogonality}, the intersection of the column spaces of $\bs{V}_{\ell}$ and $\bs{V}_{q}$ is $\bs{0}_p$ for any $1 \le \ell < q \le 3$. Because the matrices are symmetric, the same holds for the row spaces. Furthermore, the ranks of $\bs{V}_1$ and $\bs{V}_2$ equal $1$ and $d-1$, respectively; the latter follows from the above discussion of the geometric multiplicity of $\delta_{2,d}$. Because the rank of $\bs{V}_1 + \bs{V}_2 + \bs{V}_3 $ is obviously $p$, Theorem~2 in \cite{Marsaglia:1964} implies that the rank of $\bs{V}_3$ is $p-d$. Because the column vectors of $\bs{V}_3$ are eigenvectors corresponding to $\delta_{3}$, this means that the geometric multiplicity of $\delta_{3}$ is at least $p-d$.

\medskip
Because the geometric multiplicies of $\delta_{1,d}$, $\delta_{2,d}$, and $\delta_{3}$ are at least $1$, $d-1$ and $p-d$, respectively, they must in fact be equal to $1$, $d-1$ and $p-d$, respectively, and equal to the respective algebraic multiplicities.  Finally, note that it can happen that some (or all) of the eigenvalues $\delta_{1,d}$, $\delta_{2,d}$ or $\delta_{3}$ coincide. If this is the case, the geometric multiplicities of the eigenvalues that are equal simply add up because the eigenspaces of $\delta_{1,d}$, $\delta_{2,d}$ and $\delta_{3}$ are always orthogonal by Lemma \ref{lem:orthogonality}. \qed

\begin{remark} \label{rem:eigen-Sigma-2}
Although the proof of Proposition~\ref{prop:eigen-Sigma} assumes $d \geqslant 4$, the case $d=3$ can be worked out separately and turns out to be analogous, while the case $d=2$ is degenerate because $p=1$ and $\SS_1$ is a scalar. When $d=3$, $p=3$ and there is no $r,s \in\{1,\ldots, p\}$ such that $|\{ i_r, j_r\} \cap \{i_s, j_s\}| =0$, so that $s_0$ does not appear in $\SS_3$. In fact, $\SS_3 = s_1 \bs{J}_3 + (s_2 - s_1) \bs{I}_3$ and it easily follows that the eigenvalues are $\delta_{1,3} = s_2 + 2s_1$ and $\delta_{2,3} = s_2 -s_1$. The respective eigenspaces are spanned by the vectors $\bs{1}_3$, and $(1,-1/2,-1/2), (-1/2,1,-1/2), (-1/2,-1/2,1)$. Hence, the geometric as well as algebraic multiplicities of $\delta_{1,3}$ and $\delta_{2,3}$ are $1$ and $2$, respectively.
\end{remark}

\subsection{Proof of Proposition~\ref{prop:inverse}}
Use again the fact that the set $\Sset$ is the same as $\mathcal{S}_\mathcal{G}$ in Appendix~A.2 of \cite{Perreault/Duchesne/Neslehova:2019} when $\mathcal{G} = \{\{1,\ldots, d\}\}$. The claim that $\SS_p^{-1} \in \Sset$ is thus an immediate consequence of Proposition~A.2 therein. For the well-known relationship between the eigenvalues of $\SS_p$ and $\SS_p^{-1}$ see, e.g., Lemma~21.1.3 in \cite{Harville:1997}.  \qed

\subsection{Proof of Proposition~\ref{prop:pd}}
In view of Proposition~\ref{prop:eigen-Sigma} it suffices to show that $\delta_{\ell,d}(s_0,s_1,s_2) > 0$ for all $\ell \in \{1,2,3\}$ and $d \geqslant 2$ if and only if $s_1 \geqslant s_0 \geqslant 0$ and $s_2 -s_1 > s_1 -s_0$.

First suppose that $\delta_{\ell,d}(s_0,s_1,s_2) > 0$ for all $\ell \in \{1,2,3\}$ and $d \geqslant 2$. Then $\delta_{3}(s_0,s_1,s_2) = s_2 -2s_1 +s_0 > 0$ implies that $s_2 -s_1 > s_1 -s_0$. Furthermore, note that as $d \to \infty$, $\delta_{1,d}(s_0,s_1,s_2)/p \to s_0$ so that $s_0 \geqslant 0$ as otherwise $\delta_{1,d}(s_0,s_1,s_2) < 0$ for all  $d \geqslant d_1$ for some sufficiently large $d_1$. Similarly, observe that as $d \to \infty$, $\delta_{2,d}(s_0,s_1,s_2)/d \to s_1 -s_0$ which implies that $s_1 \geqslant s_0$ as otherwise $\delta_{2,d}(s_0,s_1,s_2) < 0$ for all $d \geqslant d_2$ for some sufficiently large $d_2$.

Conversely, assume that $s_1 \geqslant s_0 \geqslant 0$ and $s_2 -s_1 > s_1 -s_0$. Clearly, this implies that $s_2 > s_1 \geqslant s_0 \geqslant 0$ and also that $ \delta_{3}(s_0,s_1,s_2) = s_2 -2s_1 + s_0 > 0$ for any $d \geqslant 2$. For $d=2$, one thus has that $\delta_{1,2}(s_0,s_1,s_2) = s_2 > 0$ while $\delta_{2,2}(s_0,s_1,s_2) = \delta_{3}(s_0,s_1,s_2) = s_2 -2s_1 + s_0 > 0$. It remains to show that for an arbitrary fixed $d \geqslant 3$, $\delta_{1,d}(s_0,s_1,s_2) > 0$ and $\delta_{2,d}(s_0,s_1,s_2) > 0$. To this end, observe first that $\delta_{1,d}(s_0,s_1,s_2)-\delta_{2,d}(s_0,s_1,s_2) = d s_1 + \{d(d-3)/2\} s_0 \geqslant 0$. The claim thus follows from the fact that $\delta_{2,d}(s_0,s_1,s_2) = (s_2 -s_1) + (d-3)(s_1 -s_0)  \geqslant s_2 -s_1 > 0$.
\qed


\subsection{Auxiliary lemmas}

\begin{lemma}\label{lem:V2V3}
Suppose that $d \geqslant 4$. Then $\G^* - \G =(1/p) \bs{V}_2$   and $\bs{I}_p - \bs{\Gamma}^* = (1/p) \bs{V}_3$, where $\bs{V}_2$ and $\bs{V}_3$ are as in Eqs~\eqref{eq:V2} and \eqref{eq:V3}, respectively.
\end{lemma}

\medskip
\noindent
\textsc{Proof of Lemma~\ref{lem:V2V3}}.
Note that $\G = \SS_p(1,1,1)/p$ and that $\G^*$, $\bs{V}_2$ and $\bs{V}_3$ are in $\Sset$ as well, with entries given in Eqns \eqref{eq:Gamma-star}, \eqref{eq:V2} and \eqref{eq:V3} respectively.
The result follows from direct calculations, exploiting the identities $\SS_p(\bs{a})+\SS_p(\bs{b})=\SS_p(\bs{a}+\bs{b})$ and $\SS_p(\bs{a})/c=\SS_p(\bs{a}/c)$, $\bs{a},\bs{b} \in \mathbb{R}^3$, $c \in \mathbb{R}$.
\qed

\begin{lemma}\label{lem:Gamma-star}
Suppose that $d \geqslant 4$ and let $\bs{B}^*$ be a $p\times d$ matrix with entries $\B_{rk}^* = \mathbbm{1}\{k \in \{i_r,j_r\}, (i_r,j_r) = \iota(r)\}$ for $r \in \{1,\ldots, p\}$ and $k \in \{1,\ldots, d\}$. Then the following hold.
\begin{itemize}
\item[(a)] $\G^* = \bs{B}^* (\bs{B}^*)^{+}$, where $(\bs{B}^*)^{+}$ is the Moore-Penrose generalized inverse of $\bs{B}^*$ given by
\begin{equation} \label{eq:B-star-plus}
(\bs{B}^*)^{+} = \frac{1}{d-2} (\bs{B}^*)^\top  - \frac{1}{(d-1)(d-2)} \bs{J}_{d\times p},
\end{equation}
where $\bs{J}_{d\times p}$ denotes the $d \times p$ matrix of ones.
\item[(b)] $(\bs{I}_p -\G^*)(\G^* - \G) = \bs{0}_{p \times p}$, where  $\bs{0}_{p \times p}$ is the $p\times p$ matrix of zeros.
\end{itemize}
\end{lemma}

\medskip
\noindent
\textsc{Proof of Lemma~\ref{lem:Gamma-star}}. To show, in part (a), that $\G^* = \B^* (\B^*)^+$ with $(\B^*)^+$ as in Eq. \eqref{eq:B-star-plus}, first write
\begin{align*}
	\G^* = \B^* (\B^*)^+ = \frac{1}{d-2} \B^*(\bs{B}^*)^\top  - \frac{1}{(d-1)(d-2)} \B^* \bs{J}_{d\times p}.
\end{align*}
Because $\bs{J}_{d \times p}$ is a matrix of ones and $\B^*$ has only two non-zero entries in any given row, the latter term simplifies to $\{(d-1)(d-2)\}^{-1} \B^* \bs{J}_{d\times p} = 2 \{(d-1)(d-2)\}^{-1} \bs{J}_{p}$, and is thus in $\Sset$ since $\bs{J}_{p} = \SS_p(1,1,1)$.
Also, note that for all $r,s \in \{1,\dots,p\}$ the $(r,s)$-th entry of $\B^*(\bs{B}^*)^\top$ is given by $\{ \B^* (\B^*)^\top\}_{rs} = |\{i_r,j_r\} \cap \{i_s,j_s\}|$,
and thus that $\{1/(d-2)\} \B^*(\bs{B}^*)^\top = \{1/(d-2)\} \SS_p(0,1,2)$.
Using the identities stated in the proof of Lemma~\ref{lem:V2V3}, we then get that
\begin{equation} \label{eq:Gamma-star-proof}
	\G^* = \frac{1}{d-2}\SS_p(0,1,2) - \frac{2}{(d-1)(d-2)}\SS_p(1,1,1) = \SS_p(c,b,a),
\end{equation}
where $a=2/(d-1)$, $b = (d-3)/\{(d-1)(d-2)\}$ and $c = -2/\{(d-1)(d-2)\}$, as required.

To show that the matrix $(\B^*)^+$ is indeed the Moore-Penrose pseudoinverse of $\B^*$, we need to show that $\B^*(\B^*)^+\B^* = \B^*$, $(\B^*)^+\B^*(\B^*)^+=(\B^*)^+$ and that both $\B^*(\B^*)^+$ and $(\B^*)^+\B^*$ are symmetric.

To verify that $\B^*(\B^*)^+\B^* = \B^*$, multiply Eq. \eqref{eq:B-star-plus} on each side by $\B^*$ to get
\begin{align*} 
	\B^*(\B^*)^+\B^* = \frac{1}{d-2} \B^* \left\{ \bs{R}_1 - \frac{1}{d-1} \bs{R}_2\right\}, \qquad  \bs{R}_1 = (\B^*)^{\top} \B^*,\ \bs{R}_2 = \bs{J}_{d \times p} \B^*.
\end{align*}
Direct calculations show that the $d \times d$ matrix $\bs{R}_1$ has entries given by, for all $k,\ell \in \{1,\dots,d\}$,
\begin{equation*}
	\{\bs{R}_1\}_{k\ell} = \sum_{r = 1}^p \mathbbm{1}(k \in \{i_r,j_r\}) \times \mathbbm{1}(\ell \in \{i_r,j_r\}) = (d-1)\mathbbm{1}(k=\ell) + \mathbbm{1}(k \neq \ell),
\end{equation*}
so that $\bs{R}_1 = \bs{J}_d + (d-2)\I_d$.
Also note that $\bs{R}_2 = (d-1)\bs{J}_d$, since $\B^*$ has exactly $d-1$ non-zero entries on any given column.
Consequently, the sum $\bs{R}_1 - (d-1)^{-1}\bs{R}_2$ simplifies to
\begin{equation*}
	\bs{R}_1 - (d-1)^{-1}\bs{R}_2 = \bs{J}_d + (d-2)\I_d - \bs{J}_d = (d-2) \I_d.
\end{equation*}

To verify that $(\B^*)^+ \B^* (\B^*)^+ = (\B^*)^+$, first note that
\begin{equation*} 
	(\B^*)^+ \B^* (\B^*)^+ = \frac{1}{d-2} \left\{ \bs{R}_3 - \frac{1}{d-1} \bs{R}_4 \right\}, \qquad  \bs{R}_3 = (\B^*)^\top \G^*,\ \bs{R}_4 = \bs{J}_{d\times p} \G^*.
\end{equation*}
Because the entries in each column of $\G^*$ sum up to $1$, $\bs{R}_4 = \bs{J}_{d\times p}$.
A direct calculation gives that for all  $k \in \{1,\dots,d\}$, $r \in \{1,\dots,p\}$, and $a,b,c$ as in Eq. \eqref{eq:Gamma-star-proof}, $\{\bs{R}_3\}_{kr}=a + (d-2)b = 1$ if $k \in \{i_{r},j_{r}\}$ and $\{\bs{R}_3\}_{kr}=2b + (d-3)c = 0 $ otherwise. Hence $\bs{R}_3 = (\B^*)^\top$.

The symmetry of $\B^*(\B^*)^+$ and $(\B^*)^+\B^*$ is immediate from the fact that $\B^*(\B^*)^+ = \G^*$ and $(\B^*)^+\B^*= (d-2)^{-1}\{(\B^*)^\top\B^* - \bs{J}_d\}$.

To show part (b), we have from Lemma~\ref{lem:V2V3} that $(\bs{I}_p -\G^*)(\G^* - \G) = (1/p^2) \bs{V}_3\bs{V}_2$. However, $\bs{V}_2$ and $\bs{V}_3$ are symmetric matrices whose columns are orthogonal by Lemma~\ref{lem:orthogonality}.
This completes the proof.
\qed

\subsection{Proof of Proposition \ref{prop:decomposition}}

First note that when $d \geqslant 4$, then $\tt_{np}^*$ with entries given in Eq. \eqref{eq:tau-star} satisfies 
\begin{equation}\label{eq:prop:orthogonality}
(\th_{np} - \tt_{np}^*)^\top(\tt_{np}^* - \tt_{np}) = 0.
\end{equation}
This is an immediate consequence of part (b) of Lemma~\ref{lem:Gamma-star}.

Next, recall that $\SS_p^{-1/2}$ is given by $\SS_p^{-1/2} = \bs{V} \bs{\Delta}^{-1/2} \bs{V}^\top$, where $\bs{\Delta}^{-1/2}$ is a diagonal matrix whose entries are the inverse square root of the eigenvalues of $\SS_p$, and $\bs{V}$ is the matrix of orthonormal eigenvectors of $\SS_p$, and hence $\bs{V} \bs{V}^\top = \I_p$. Hence $\bs{V}$ is also the matrix of orthonormal eigenvectors of $\SS_p^{-1/2}$ and the diagonal entries of $\bs{\Delta}^{-1/2}$  are the corresponding eigenvalues.

To prove (a), recall from Lemma \ref{lem:V2V3} that $\tt_{np}^* - \tt_{np} = (1/p) \bs{V}_2 \th_{np}$. Because each column of $\bs{V}_2$ is a sum of two eigenvectors of $\SS_p^{-1/2}$ associated to $\delta_{2,d}^{-1/2}$, we have that
\begin{align*}
\SS_p^{-1/2} (\tt_{np}^* - \tt_{np}) = (1/p) \SS_p^{-1/2}\bs{V}_2 \th_{np} = (1/p)\delta_{2,d}^{-1/2} \bs{V}_2  \th_{np} = \delta_{2,d}^{-1/2}(\tt_{np}^* - \tt_{np}).
\end{align*}
as was to be shown.

To show part (b), note first that from Lemma \ref{lem:V2V3}, $\th_{np} - \tt_{np}^* = (1/p)\bs{V}_3\th_{np}$, where the column vectors of $\bs{V}_3$ are eigenvectors of  $\SS_p^{-1/2}$ corresponding to $\delta_{3}^{-1/2}$. Hence $\SS_p^{-1/2} (\th_{np} - \tt_{np}^*) = \delta_{3}^{-1/2}(\th_{np} - \tt_{np}^*)$, which, combined with part (a) and Eq. \eqref{eq:prop:orthogonality}, proves the claim.
\qed



\clearpage
\section{Proofs from Section~\ref{sec:null}} \label{app:proofs-null}

The following auxiliary result is needed for the proofs of Theorem~\ref{thm:asymptotic-n} and Proposition~\ref{prop:asymptotic-supremum}.
\begin{lemma} \label{lem:SSS}
Consider a $p \times p$ positive definite matrix $\SA_p$ and a $p \times p$ matrix $\bs{P}_p$ of rank $\rho$.
Let $\SA_p^\dag = \SA_{p}^{-1/2} \bs{P}_p \S_{p} \bs{P}_p^\top \SA_{p}^{-1/2}$ and $\SA_p^\ddag = \S_{p}^{1/2} \bs{P}_p^\top \SA_{p}^{-1} \bs{P}_p \S_{p}^{1/2}$, where $\S_p$ is a $p \times p$ positive definite and symmetric matrix. Then the following statements hold.
\begin{itemize}
	\item[(a)] $\SA_p^\dag$ and $\SA_p^\ddag$ share the same eigenvalues and they are of rank $\rho$.
	\item[(b)] If $\SA_p = \S_p$ and $\bs{P}_p = \I_p - \Gamma(\S_p)$ with $\Gamma$ is as in Eq.~\eqref{eq:Gamma}, then $\SA_p^\dag = \SA_p^\ddag = \I_p - \S_{p}^{-1/2} \Gamma(\S_{p}) \S_{p}^{1/2}$. Furthermore, $\SA_p^\dag $ is idempotent and of rank $p-L$.
\end{itemize}
\end{lemma}

\medskip
\noindent
\textsc{Proof of Lemma~\ref{lem:SSS}.}
(a) First write $\SA_p^\dag = \bs{R}\bs{R}^\top$ and $\SA_p^\ddag = \bs{R}^\top \bs{R}$, where $\bs{R} = \SA_p^{-1/2} \bs{P}_p \S_p^{1/2}$ is a $p\times p$ matrix. From Theorem~21.10.1 in \citep{Harville:1997} it then follows that $\SA_p^\dag$ and $\SA_p^\ddag$ have the same eigenvalues and consequently the same rank.
To show that $\mathrm{rank}(\SA_p^\dag) = \rho$, first recall that $\mathrm{rank}(\boldsymbol{R}^\top\boldsymbol{R}) = \mathrm{rank}(\boldsymbol{R})$ for any matrix $\boldsymbol{R}$ \citep[Cor.~7.4.5]{Harville:1997}, so that $\mathrm{rank}(\SA_p^\dag) = \mathrm{rank}(\SA_p^{-1/2} \bs{P}_p \S_p^{1/2})$.
By Eq.~(5.2) in Section~17.5 and Theorem~17.5.1 in \citet{Harville:1997}, we have that
\begin{equation} \label{eq:proof-rank-1}
	\mathrm{rank}(\SA_p^{-1/2} \bs{P}_p \S_p^{1/2}) \leqslant \min\{\mathrm{rank}(\SA_p^{-1/2}),\mathrm{rank}(\bs{P}_p),\mathrm{rank}(\S_p^{1/2})\} = \rho
\end{equation}
as well as
\begin{equation} \label{eq:proof-rank-2}
	\mathrm{rank}(\SA_p^{-1/2} \bs{P}_p \S_p^{1/2}) \geqslant \mathrm{rank}(\SA_p^{-1/2} \bs{P}_p) + \mathrm{rank}(\bs{P}_p \S_p^{1/2}) - \mathrm{rank}(\bs{P}_p),
\end{equation}
where, by Corollary~17.5.2 of \cite{Harville:1997},
$$
\mathrm{rank}(\SA_p^{-1/2} \bs{P}_p) = \mathrm{rank}(\bs{P}_p\S_p^{1/2}) = \mathrm{rank}(\bs{P}_p) = \rho.
$$
Combining Eqs.~\eqref{eq:proof-rank-1} and \eqref{eq:proof-rank-2} then gives that $\mathrm{rank}(\SA_p^{-1/2} \bs{P}_p \S_p^{1/2}) = \rho$.

(b) By construction, $\Gamma(\S_p)$, which is a function of $\B$ of Eq.~\eqref{eq:H0}, is idempotent and of rank equal to $\mathrm{rank}(\B) = L$.
In particular, this means that $\mathrm{rank}(\bs{P}_p)=p-L$ \cite[Lemma~18.4.2]{Harville:1997} and thus, by part (a), that $\mathrm{rank}(\SA_p^\dag) = p-L$.
Direct calculations show that $\SA_p^\dag = \SA_p^\ddag= \I_p - \S_{p}^{-1/2} \Gamma(\S_p) \S_p^{1/2}$ and that $\S_{p}^{-1/2} \Gamma(\S_p) \S_p^{1/2}$ is symmetric and idempotent.
\qed

\subsection{Proof of Theorem \ref{thm:asymptotic-n}}

First note that $\sqrt{n} \bs{P}_{np} \th_{np} = \sqrt{n} \bs{P}_{np} (\th_{np} - \t_p)$ by assumption.
Also note that the mapping $\bs{A} \mapsto \bs{A}^{-1}$ is continuous for non-singular matrices \citep{Stewart:1969} and the mapping $\bs{A} \mapsto \bs{A}^{1/2}$ is continuous on the set of positive definite matrices \citep[Eq.~(7.2.13), exercise~18, p.~411]{Horn/Johnson:1985}.
Consequently, the mapping $\bs{A} \mapsto \bs{A}^{-1/2}$ is continuous on the set of positive definite matrices.
Hence, by the Continuous Mapping Theorem, $\{n\SA_{np}\}^{-1/2}$ converges in probability to $\SA_{p}^{-1/2}$.  Furthermore, Eq.~\eqref{eq:asymptotic-tau-n} combined with Slutsky's Lemma yields that
\begin{align*}
	\SA_{np}^{-1/2} \bs{P}_{np} \th_{np} = (n \SA_{np})^{-1/2} \bs{P}_{np} \sqrt{n}(\th_{np} - \t_p)\rightsquigarrow \SA_{p}^{-1/2} \bs{P}_p \Z^*_p,
\end{align*}
where $\Z^*_p \sim \mathcal{N}_p(\bs{0}_p,\S_p)$.
Let $\bs{Z}_p = \SA_{p}^{-1/2} \bs{P}_p \bs{Z}_p^*$. Then $\bs{Z}_p \sim \mathcal{N}_p(\bs{0}_p, \SA_p^\dagger)$. Another application of the Continuous Mapping Theorem gives that $E_{np} \rightsquigarrow \| \Z_p \|_2^2$ and $M_{np} \rightsquigarrow \| \bs{Z}_p \|_{\infty}$ as $n \to \infty$.
The representation of $\| \Z_p \|_2^2$ as a weighted sum of $\chi^2$ distributions is well-known. For example, in Representation 3.1a.1 of Chapter 3 of \cite{Mathai/Provost:1992}, the weights are given by the eigenvalues of $\SA_p^\ddag= \S_{p}^{1/2} \bs{P}_p^\top \SA_{p}^{-1} \bs{P}_p \S_{p}^{1/2}$. By part (a) of Lemma~\ref{lem:SSS}, $\SA_p^\ddag$ and $\SA_p^\dag$ have the same eigenvalues.
\qed

\subsection{Proof of Proposition \ref{prop:asymptotic-euclidean}}
To prove part (a), let $\bs{P}_{np} = \I_p - \Gamma(\Sh_{np})$ and note that the mapping $\Gamma$ involves only inversions, transpositions and matrix multiplications, and is thus continuous.
By the definition of $\Gamma(\Sh_{np})$, given in Eq. \eqref{eq:Gamma}, we have that $\Gamma(\Sh_{np}) = \Gamma(n\Sh_{np})$, and so the Continuous Mapping Theorem implies that $\Gamma(\Sh_{np})$ converges in probability to $\Gamma(\S_{p})$ as $n \to \infty$.
Consequently, $\bs{P}_{np}$ converges in probability to the matrix $\bs{P}_p = \I_p - \Gamma(\S_p)$ as $n \to \infty$.
Because $\Gamma(\Sh_{np}) \t_p \in \mathcal{T}_p$ by construction and given that  $\t_p \in \mathcal{T}_p$ under $H_0$, we have that $\bs{P}_{np} \t_p = \bs{0}_p$. The main claim thus follows from Theorem \ref{thm:asymptotic-n}.

From Lemma~\ref{lem:SSS} (b), $\SA_p^\dag$ is idempotent and of rank $p-L$. Consequently, all its eigenvalues are in $\{0,1\}$ with exactly $p-L$ of them equal to one, see, e.g., Theorem 21.8.2 in \cite{Harville:1997}.

To prove part (b), note that $\Gamma(\I_p) = \Gamma\{(1/n)\I_p\} = \B\B^+$, which is symmetric and idempotent.
A direct application of Theorem~\ref{thm:asymptotic-n} with $\bs{P}_{np} = \bs{P}_p = \I_p - \B \B^+$ and $n\SA_{np} = \SA_{p} = \I_p$ implies that $E_{np} \rightsquigarrow \sum_{k=1}^m \lambda_k \chi_{\upsilon_k}^2$, where $\lambda_k$ is the $k$th distinct eigenvalue of
	$\SA_p^\dag = (\I_p - \B\B^+)\S_p(\I_p - \B\B^+)$.
To see that $\SA_p^\dag$ and $\S_p(\I_p - \B\B^+)$ share the same eigenvalues, let $\bs{R}_1 = (\I_p - \B\B^+)$ and $\bs{R}_2 = \S_p(\I_p - \B\B^+)$ and recall from Theorem~21.10.1 in \cite{Harville:1997} that the products $\SA_p^\dag = \bs{R}_1\bs{R}_2$ and $\bs{R}_2\bs{R}_1 = \S_p(\I_p - \B\B^+)$ have the same eigenvalues.
\qed

\subsection{Proof of Corollary~\ref{cor:asymptotic-euc-exchangeable}}

\begin{lemma}\label{lem:JGN}
Assume that $H_0^*$ holds with $\mathcal{G} = \{\{1,\ldots, d\}\}$. Then $\S_p$ and $\S_{np}$ are elements of $\Sset$. Furthermore, $\S_p = \SS_p(\sigma_0, \sigma_1, \sigma_2)$ and $\S_{np} = \SS_p(\sigma_{0n}, \sigma_{1n},  \sigma_{2n})$ where for each $\ell \in \{0,1,2\}$,
\begin{align*}
\sigma_\ell & = 16(\vartheta_{1,\ell} + \vartheta_{2,\ell} + \vartheta_{3,\ell} + \vartheta_{4,\ell}) -4(\beta+1)^2 \\
\sigma_{\ell n} & = \frac{16}{n(n-1)} \left\{ (n-2)(\vartheta_{1,\ell} + \vartheta_{2,\ell} + \vartheta_{3,\ell} + \vartheta_{4,\ell}) + \vartheta_{5,\ell} + \vartheta_{6,\ell}\right\} - \frac{2(2n-3)}{n(n-1)} (\beta+1)^2.
\end{align*}
In the above, $\beta = \tau_1 = \ldots = \tau_p$ and the expressions for $\vartheta_{1,\ell},\ldots, \vartheta_{6,\ell}$ for $\ell \in \{1,2,3\}$ are given in Eqs. \eqref{eq:sigma0}--\eqref{eq:sigma2} below.
\end{lemma}

\medskip
\noindent
\textsc{Proof of Lemma \ref{lem:JGN}.} Observe that the set $\Sset$ is in fact the same as $\mathcal{S}_\mathcal{G}$ in Appendix~A.2 of \cite{Perreault/Duchesne/Neslehova:2019} when $\mathcal{G} = \{\{1,\ldots, d\}\}$. The claim that
$\S_p,\S_{np}\in\Sset$ thus follows at once from Proposition~A.1 in the latter paper. The values of $\sigma_\ell$ and $ \sigma_{\ell n}$ for $\ell \in \{0,1,2\}$ may be calculated from the formulas  in \cite{Genest/Neslehova/BenGhorbal:2011} (see also Eqns. (A.1)--(A.3) in \cite{Perreault/Duchesne/Neslehova:2019}). To this end, suppose that $\bs{U} \sim C$ and for any $A \subset \{1,\ldots, d\}$, let $C_{|A|}$ be the distribution function of $(U_i : i \in A)$. Note that because $C$ is exchangeable, $C_{|A|}$ indeed only depends on the cardinality of $A$. Now fix some arbitrary $r,s \in \{1,\ldots, p\}$ and let $\iota(r) =(i_1,j_1)$ and $\iota(s) =(i_2,j_2)$. Suppose also that $|\{i_1,j_1\} \cap \{i_2, j_2\}| = \ell$ for some $\ell \in \{0,1,2\}$. From Eqs. (A.1) and (A.3) in \cite{Perreault/Duchesne/Neslehova:2019}, the $(r,s)$-th entry of $\S_p$ and $\S_{np}$ is given by $\sigma_\ell$ and $ \sigma_{\ell n}$, respectively, where
\begin{align*}
\sigma_\ell & = 16(\vartheta_{1,\ell} + \vartheta_{2,\ell} + \vartheta_{3,\ell} + \vartheta_{4,\ell}) -4(\beta+1)^2 \\
 \sigma_{\ell n} & = \frac{16}{n(n-1)} \left\{ (n-2)(\vartheta_{1,\ell} + \vartheta_{2,\ell} + \vartheta_{3,\ell} + \vartheta_{4,\ell}) + \vartheta_{5,\ell} + \vartheta_{6,\ell}\right\} - \frac{2(2n-3)}{n(n-1)} (\beta+1)^2
\end{align*}
as claimed.  From \cite{Genest/Neslehova/BenGhorbal:2011}, the coefficients appearing in these expressions may be calculated as follows. When $\ell =0$, one necessarily has that $d \geqslant 4$, so that
\begin{gather}\label{eq:sigma0}
\vartheta_{1,0} = {\rm E}\{C_{2}(U_{1}, U_{2}) C_{2}(U_{3}, U_{4})\}, \quad \vartheta_{2,0} =  {\rm E}\{\bar{C}_{2}(U_{1}, U_{2}) C_{2}(U_{3}, U_{4})\}, \\\notag
\vartheta_{3,0} = {\rm E}\{C_{2}(U_{1}, U_{2})\bar{C}_{2}(U_{3}, U_{4})\}, \quad \vartheta_{4,0} = {\rm E}\{\bar{C}_{2}(U_{1}, U_{2})\bar{C}_{2}(U_{3}, U_{4})\},\\
\vartheta_{5,0} = {\rm E}\{C_{4}(U_{1},U_{2}, U_{3}, U_{4})\}, \quad \vartheta_{6,0} = {\rm E}\{\tilde{C}_{4}(U_{1},U_{2}, U_{3}, U_{4})\},\notag
\end{gather}
and $\bar{C}_2$ denotes the survival function corresponding to $C_2$, while $\tilde{C}_{4} = C_{2} - 2C_{3} + C_{4}$. When $\ell=1$, then it must hold that $d \geqslant 3$. In this case,
\begin{gather} \label{eq:sigma1}
\vartheta_{1,1} = {\rm E}\{C_{2}(U_{1}, U_{2}) C_{2}(U_{2}, U_{3})\}, \quad \vartheta_{2,1} =  {\rm E}\{\bar{C}_{2}(U_{1}, U_{2}) C_{2}(U_{2}, U_{3})\}, \\\notag
\vartheta_{3,1} = {\rm E}\{C_{2}(U_{1}, U_{2})\bar{C}_{2}(U_{2}, U_{3})\}, \quad \vartheta_{4,1} = {\rm E}\{\bar{C}_{2}(U_{1}, U_{2})\bar{C}_{2}(U_{2}, U_{3})\},\\
\vartheta_{5,1} = {\rm E}\{C_{3}(U_{1},U_{2}, U_{3})\}, \quad \vartheta_{6,1} = 0.\notag
\end{gather}
Finally, when $\ell=2$, $d \geqslant 2$ and
\begin{gather}\label{eq:sigma2}
\vartheta_{1,2} = {\rm E}\{C_{2}(U_{1}, U_{2}) C_{2}(U_{1}, U_{2})\}, \quad \vartheta_{2,2} =  {\rm E}\{\bar{C}_{2}(U_{1}, U_{2}) C_{2}(U_{1}, U_{2})\}, \\\notag
\vartheta_{3,2} = {\rm E}\{C_{2}(U_{1}, U_{2})\bar{C}_{2}(U_{1}, U_{2})\}, \quad \vartheta_{4,2} = {\rm E}\{\bar{C}_{2}(U_{1}, U_{2})\bar{C}_{2}(U_{1}, U_{2})\},\\
\vartheta_{5,2} = {\rm E}\{C_{2}(U_{1},U_{2})\}, \quad \vartheta_{6,2} = 0.\notag
\end{gather}
This concludes the proof. \qed

\medskip
\noindent
\textsc{Proof of Corollary~\ref{cor:asymptotic-euc-exchangeable}}.
By Lemma~\ref{lem:JGN}, $\S_p = \SA_p(\sigma_0,\sigma_1,\sigma_2)$. Furthermore, under full exchangeability, $\B\B^+=\bs{J}_p/p$. Hence $\S_p(\I_p-\B\B^+) =  \SA_p(t_0,t_1,t_3)$, where
for $k \in \{0,1,2\}$, $t_k = \sigma_k - (1/p)\{\sigma_2 + \sigma_1 2(d-2) + \sigma_0 ( p-2d+3)\}$. By Proposition~\ref{prop:eigen-Sigma}, the eigenvalues of $\S_p(\I_p-\B\B^+)$ are  $\delta_{1,d}(t_0,t_1,t_2) =0$, $\delta_{2,d}(t_0,t_1,t_2) = \delta_{2,d}$, and $\delta_{3}(t_0,t_1,t_2) = \delta_3$, where $\delta_{2,d}$, and  $\delta_{3}$ are eigenvalues of $\S_p$ with geometric multiplicities $d-1$ and $p-d$, respectively. The claim then follows at once from Proposition~\ref{prop:asymptotic-euclidean}.
\qed

\subsection{Proof of Proposition~\ref{prop:asymptotic-supremum}}
For (a), note that, as argued in the proof of
Proposition~\ref{prop:asymptotic-euclidean}, the conditions for applying Theorem~\ref{thm:asymptotic-n} are satisfied.
By Lemma~\ref{lem:SSS}~(b), $\SA_p^\dag$ simplifies to $\SA_p^\dag = \I_p - \S_p^{-1/2} \Gamma(\S_p) \S_p^{1/2}$. Part (b) follows directly from Theorem~\ref{thm:asymptotic-n} with $\bs{P}_{np} = \bs{P}_p = \I_p - \B \B^+$ and $n\SA_{np} = \SA_{p} = \I_p$.
\qed

\subsection{Proof of Proposition~\ref{prop:P-sum-less-1}}
Fix an arbitrary $r \in \{1,\dots,p\}$ and let $\ell \in \{1,\dots,L\}$ be such that $B_{r\ell} \neq 0$. Also let $\zeta_{\ell}$ be the number of non-zero entries on the $\ell$-th column of $\B$, viz. $\zeta_{\ell} = \sum_{s=1}^p \mathbbm{1}(B_{s\ell} \neq 0)$.
Recall that $\B \B^+ = \Gamma(\I_p) = \B (\B^\top \B)^{-1} \B^\top$ and note that since $\B$ possesses a single non-zero element per row, $(\B^\top\B)^{-1}$ is a $L \times L$ diagonal matrix with $\ell$-th diagonal element given by $\gamma_{\ell}^{-1}$, where
$$
	\gamma_{\ell} = (\B^\top\B)_{\ell\ell} = \sum_{s=1}^p B_{s\ell}^2 \geqslant \zeta_\ell a^2.
$$
Consequently, for all $s \in \{1,\ldots,p\}$, $|(\B \B^+)_{rs}| = |B_{r\ell}||B_{s\ell} | \gamma_{\ell}^{-1} \leqslant c^2/(a^2 \zeta_\ell)\ \mathbbm{1}(B_{s\ell} \neq 0)$. Hence,
$$
	 \sum_{s=1}^p |P_{rs}| \leqslant 1 + \sum_{s=1}^p |(\B \B^+)_{rs}| \leqslant 1 + (c/a)^2 \zeta_\ell^{-1}\sum_{s=1}^p \mathbbm{1}(B_{s\ell} \neq 0) = 1 + (c/a)^2,
$$
which proves the claim.
\qed

\subsection{Proof of Proposition~\ref{prop:M1-exchangeable}}
From the proof of Proposition~\ref{prop:eigen-Sigma}, $(1/p)\bs{1}_p$ is an eigenvector of $\S_p$ associated to $\delta_{1,d}(\bs{\sigma})$.
Thus
$
	\bs{P}_p \S_p \bs{P}_p = \S_p - (1/p)\bs{J}_p \S_p - (1/p)\S_p \bs{J}_p + (1/p)^2 \bs{J}_p \S_p \bs{J}_p = \S_p - (1/p) \delta_{1,d} (\bs{\sigma}) \bs{J}_p.
$
In particular, all diagonal entries of $\bs{P}_p \S_p \bs{P}_p$ are equal to $\sigma_2 - (1/p)\delta_{1,d}(\bs{\sigma})$.

When $d=3$, Remark~\ref{rem:eigen-Sigma} implies that $\delta_{1,3} = \sigma_2 + 2\sigma_1$ and hence $\sigma_2 - (1/3)\delta_{1,3}(\bs{\sigma}) = (2/3)(\sigma_2-\sigma_1)$.
When $d \geqslant 4$,
 Proposition~\ref{prop:eigen-Sigma} yields that
$	(1/p)\delta_{1,d} (\bs{\sigma}) = 
	 (1/p)\{\sigma_2 - \sigma_1  + (2d-3)(\sigma_1 - \sigma_0)\} +\sigma_0$.
Proposition~\ref{prop:pd} implies that $\sigma_2-\sigma_1 > \sigma_1-\sigma_0 \geqslant 0$ and $\sigma_0 \geqslant 0$.
Since $(2d-3)/p$ is monotone decreasing in $d$ whenever $d \geqslant 3$, we have that $\sigma_2 - (1/p)\delta_{1,d} \ge (2/3)(\sigma_2 - \sigma_1)$. The latter expression is strictly positive by Proposition~\ref{prop:pd}.
\qed

\clearpage
\section{Estimation of the covariance matrix of Kendall's tau} \label{app:Sigma}

In this section, we present various estimators of $\S_{np}$.
The first option is the plug-in estimator of \cite{BenGhorbal/Genest/Neslehova:2009}, denoted here by $\Sh_{np}^{\rm P}$, which is also described in the Appendix of \cite{Perreault/Duchesne/Neslehova:2019}. Therein, it is explained that as $n \to \infty$, $n\Sh_{np}^{\rm P} \to \S_p$ in probability. The second option is a modified version of the jackknife estimator in \eqref{eq:sigma-jackknife}, which is constructed so that $n^2(n-1)/(n-2)^2 \Sh_{np}^{\rm J}$ is the jackknife estimator of $\S_{np}$, see Eq. (18) in \cite{Chen:2018}. Note also that $n\Sh_{np}^{\rm J}$ is the estimator in Eq.~(19) in the latter paper, as well as the estimator in Eq.~(2.6) in \cite{Rublik:2016}. The fact that $n\Sh_{np}^{\rm J}$ converges in probability to $\S_p$ as $n \to \infty$ follows directly from Theorem~6 in \cite{Arvesen:1969} concerning the consistency of the jackknife variance estimator for $U$-statistics, as in the proof of Theorem~2.1 in \cite{Rublik:2016}.

When testing the more restricted hypothesis $H_0^*$ of partial exchangeability, we can use constrained estimators of $\S_{np}$ that are in $\mathcal{S}_{\mathcal{G}}$.
One such, referred here as $\Sb_{np}^{\rm P}$, is a variant of the plug-in estimator $\Sh_{np}^{\rm P}$ and denoted $\bs{\tilde{\Sigma}}$ in Appendix A.3 of \cite{Perreault/Duchesne/Neslehova:2019}. As argued therein, $n\Sb_{np}^{\rm P} \to \S_p$ as $n\to \infty$ in probability under the hypothesis of partial exchangeability.
Alternatively, we can also use a structured version $ \Sb_{np}^{\rm J} $ of $\Sh_{np}^{\rm J}$ obtained by averaging out its entries over the blocks inherent to $\mathcal{S}_{\mathcal{G}}$, described in Appendix A.2 of \cite{Perreault/Duchesne/Neslehova:2019}. Under $H_0^*$, we again have that $n\Sb_{np}^{\rm J} \to \S_p$ as $n \to \infty$ in probability.

In the special case of full exchangeability, the calculation of $ \Sb_{np}^{\rm J} $ simplifies, as we now explain. This is advantageous computationally, particularly when $d$ is large. Because $ \Sb_{np}^{\rm J} \in \Sset$ by construction, its calculation reduces to that of the vector $\bs{\hat{\sigma}}_n^{\rm J} = (\hat\sigma_{0n}^{\rm J},\hat\sigma_{1n}^{\rm J},\hat\sigma_{2n}^{\rm J})$ for which $\Sb_{np}^{\rm J}=\SS_p(\bs{\hat{\sigma}}_n^{\rm J})$.

To this end, introduce, for each $\nu \in \{1,\dots,n\}$,
\begin{align*}
	\th_{np}^{(\nu)}= \frac{1}{n-1} \sum_{ \substack{\eta=1 \\ \eta \neq \nu}}^n h(\bs{X}_\nu,\bs{X}_\eta),
\end{align*}
so that from Eq. \eqref{eq:sigma-jackknife}, $\Sh_{np}^{\rm J} = (4/n^2) \sum_{\nu=1}^n (\th_{np}^{(\nu)} - \th_{np})(\th_{np}^{(\nu)} - \th_{np})^\top$. From Eq. \eqref{eq:S-cal}, $\hat\sigma_{n2}^{\rm J}$ is obtained by averaging all diagonal entries of $\Sh_{np}^{\rm J}$, viz.\begin{equation} \label{eq:sigmaJS-2}
\hat\sigma_{n2}^{\rm J} =\frac{4}{pn^2} \sum_{\nu=1}^n (\th_{np}^{(\nu)} - \th_{np})^\top (\th_{np}^{(\nu)} - \th_{np}).
\end{equation}
In order to calculate $\hat\sigma_{n0}^{\rm J}$ and $\hat\sigma_{n1}^{\rm J}$, let us first define the following two intermediate quantities
\begin{equation}\label{eq:sigmaJS-01}
	\hat\zeta_{0} = \frac{4}{n^2}\sum_{\nu=1}^n (\bar\tau_{np}^{(\nu)} - \bar\tau_{np})^2, \qquad  \hat\zeta_{1} = \frac{4}{dn^2}\sum_{i=1}^d\sum_{\nu=1}^n (\bar T_{ni}^{(\nu)} - \bar T_{ni})^2,
\end{equation}
where for $\nu \in \{1,\dots,n\}$, $i \in \{1,\dots,d\}$ and $\mathcal{R}_i$ as in Eq.~\eqref{eq:tau-bar},
\begin{align*} 
\bar\tau_{np}^{(\nu)} = (1/p)\bs{1}_p^\top \th_{np}^{(\nu)}, \qquad \bar T_{ni}^{(\nu)} = \frac{1}{d-1}\sum_{r \in \mathcal{R}_i} \hat\tau_{np,r}^{(\nu)}, \qquad  \bar T_{ni} = \frac{1}{d-1}\sum_{r \in \mathcal{R}_i} \hat\tau_{np,r}.
\end{align*}
The following proposition provides formulas for $\hat\sigma_{n0}^{\rm J}$ and $\hat\sigma_{n1}^{\rm J}$ that depend on $\hat\sigma_{n2}$, $\hat\zeta_{0}$ and $\hat\zeta_{1}$ only.
\begin{proposition} \label{prop:sigma-exch}
Let $\hat \sigma_{n2}^{\rm J}$ be as in Eq. \eqref{eq:sigmaJS-2} and $\hat\sigma_{n0}^{\rm J}$, $\hat\sigma_{n1}^{\rm J}$ be such that
\begin{align*}
	\hat\sigma_{n0}^{\rm J} = \frac{p\hat\zeta_{0} - 2(d-1)\hat\zeta_{1} + {\hat\sigma}_{n2}^{\rm J}}{{p-2d+3}} \qquad \text{and} \qquad \hat\sigma_{n1}^{\rm J} = \frac{(d-1)\hat\zeta_{1} - \hat\sigma_{n2}^{\rm J}}{d-2}.
\end{align*}
Then for $\bs{\hat\sigma}_{np}^{\rm J} = (\hat\sigma_{n0}^{\rm J},\hat\sigma_{n1}^{\rm J},\hat\sigma_{n2}^{\rm J})$ it holds that $\Sb_{np}^{\rm J} = \SS_p(\bs{\hat\sigma}_{np}^{\rm J})$.
\end{proposition}
By Proposition \ref{prop:sigma-exch} and the consistency of $\Sh_{np}^{\rm J}$, it follows that $n\hat{\bs{\sigma}}_{np}^{\rm J}$ is a consistent estimator of $\bs{\sigma}$ as $n \to \infty$, where $\bs{\sigma}$ is such that $\S_{p} = \SS_p(\bs{\sigma})$.

\begin{remark} \label{rem:pd}
In finite-samples, and in particular when $n$ is small, it can happen that $\Sh_{np} \in \{\Sh_{np}^{\rm P},\Sh_{np}^{\rm J}\}$ fails to be positive (semi)definite. When  $\Sh_{np}$ is positive semidefinite, as is always the case with $\Sh_{np}^{\rm J}$ for example, we use the Moore-Penrose pseudoinverses $\Sh_{np}^+$ and $(\Sh_{np}^{1/2})^+$ instead of $\Sh_{np}^{-1}$ and $\Sh_{np}^{-1/2}$, respectively.
When the estimate of $\S_{p}$ fails to be positive semidefinite, as is sometimes the case with $\Sh_{np}^{\rm P}$, we apply the eigenvalue method discussed by \cite{Rousseeuw/Molenberghs:1993}, i.e., we replace the negative eigenvalues of $\Sh_{np}^{\rm P}$ by zero, so that the resulting matrix is positive semidefinite.
\end{remark}

\medskip
\noindent
\textsc{Proof of Proposition~\ref{prop:sigma-exch}}.
Note that for each $r,s \in \{1,\ldots, p\}$, $\hat\Sigma_{rs}^{\rm J} = (4/n^2)\sum_{\nu=1}^n (\hat\tau_{nr}^{(\nu)} - \hat\tau_{nr})(\hat\tau_{ns}^{(\nu)} - \hat\tau_{ns})$.
From Eq. \eqref{eq:sigmaJS-01},
\small
\begin{equation*}
\hat\zeta_{1}  = \frac{4}{dn^2}\sum_{i=1}^d\sum_{\nu=1}^n (\bar T_{ni}^{(\nu)} - \bar T_{ni})^2 = \frac{1}{d(d-1)^2} \sum_{i=1}^d  \sum_{r,s \in \mathcal{R}_i} \hat\Sigma_{rs}^{\rm J} = \frac{1}{d(d-1)^2} \sum_{i=1}^d  \Bigl( \sum_{r \in \mathcal{R}_i} \hat\Sigma_{rr}^{\rm J} + \sum_{r,s \in \mathcal{R}_i, r \neq s} \hat\Sigma_{rs}^{\rm J}\Bigr).
\end{equation*}
\normalsize
Now for any fixed $r \in \{1,\ldots, p\}$, $\iota(r) = (i_r, j_r)$ and hence $r \in \mathcal{R}_{i_r}$, $r \in \mathcal{R}_{j_r}$, while $r \neq \mathcal{R}_i$ for all $i \not \in \{i_r, j_r\}$. This implies that
$$
\frac{1}{d(d-1)^2} \sum_{i=1}^d   \sum_{r \in \mathcal{R}_i} \hat\Sigma_{rr}^{\rm J}  = \frac{2p}{d(d-1)^2} \hat \sigma_{n2}^{\rm J} = \frac{1}{d-1} \hat \sigma_{n2}^{\rm J} .
$$
Furthermore, for any fixed $r \neq s \in \{1,\ldots, p\}$, set $\iota(r)= (i_r, j_r)$ and $\iota(s) = (i_s, j_s)$. Then $|\{i_r,j_r\} \cup  \{i_s,j_s\}| = 1$ if and only if $r, s \in \mathcal{R}_i$ for $i \in \{i_r,j_r\} \cup  \{i_s,j_s\}$ and $r,s \not\in \mathcal{R}_i$ otherwise. Hence,
$$
\frac{1}{d(d-1)^2} \sum_{i=1}^d  \sum_{r,s \in \mathcal{R}_i, r \neq s} \hat\Sigma_{rs}^{\rm J} = \frac{d(d-1)(d-2)}{d(d-1)^2}  \hat \sigma_{n1}^{\rm J} = \frac{(d-2)}{(d-1)}  \hat \sigma_{n1}^{\rm J}.
$$
Put together, $\hat\zeta_{1} = \{\hat\sigma_{n2}^{\rm J} + (d-2)\hat\sigma_{n1}^{\rm J}\}/(d-1)$ so that
\begin{equation} \label{eq:sigma1-hat-from-prime}
	 \hat\sigma_{n1}^{\rm J} = \frac{(d-1)\hat\zeta_{1} - \hat\sigma_{n2}^{\rm J}}{d-2}.
\end{equation}	
From Eq. \eqref{eq:sigmaJS-01} one also has that
\begin{align*}
\hat\zeta_0 & = \frac{4}{n^2}\sum_{\nu=1}^n (\bar\tau_{np}^{(\nu)} - \bar\tau_{np})^2 = \frac{1}{p^2} \sum_{r,s=1}^p \hat\Sigma_{rs}^{\rm J} = \frac{1}{p^2} \{ p \hat\sigma_{n2}^{\rm J} + d(d-1)(d-2) \hat\sigma_{n1}^{\rm J} + p(p-2d+3) \hat\sigma_{n0}^{\rm J}\} \\
& = \frac{1}{p}  \{ \hat\sigma_{n2}^{\rm J} + 2(d-2) \hat\sigma_{n1}^{\rm J} + (p-2d+3) \hat\sigma_{n0}^{\rm J}\}.
\end{align*}
Substituting the value of $\hat\sigma_{n1}^{\rm J}$ given in \eqref{eq:sigma1-hat-from-prime} into the latter equation leads to
\begin{equation*}
	\hat\zeta_{0} = \frac{\hat{\sigma}_{n2}^{\rm J}}{p} + \frac{4(d-2)}{d(d-1)}\left(\frac{(d-1)\hat\zeta_{1} - \hat\sigma_{n2}^{\rm J}}{d-2} \right)  + \frac{p-2d+3}{p} \hat{\sigma}_{n0}^{\rm J}
	= \frac{-\hat{\sigma}_{n2}^{\rm J}}{p} +  \frac{4\hat\zeta_{1}}{d} + \frac{p-2d+3}{p} \hat{\sigma}_{n0}^{\rm J},
\end{equation*}
and solving for $\hat\sigma_{n0}^{\rm J}$ gives
\begin{equation*}
	\hat{\sigma}_{n0}^{\rm J} = \frac{p}{p-2d+3} \left( \hat\zeta_{0} - \frac{4\hat{\zeta}_1}{d}  + \frac{\hat{\sigma}_{n2}^{\rm J}}{p} \right) = \frac{p\hat\zeta_{0} - 2(d-1)\hat\zeta_{1} + \hat{\sigma}_{n2}^{\rm J}}{p-2d+3},
\end{equation*}
as claimed.
\qed

\clearpage


\clearpage
\section{Simulation study} \label{app:sim-study}

\subsection{Numerical approximations of p-values} \label{app:p-values}

In this section, we explain how the asymptotic results from Section~\ref{sec:null} are used to calculate approximate $p$-values for the tests proposed in Section~\ref{sec:tests}.
In what follows, $\Sh_{np}$ denotes a generic estimator of $\S_p$ which is assumed to be positive definite and such that $n \Sh_{np}$ converges elementwise in probability to $\S_p$ as $n \to \infty$; various estimators of $\S_p$ are described in Appendix~\ref{app:Sigma}.

For a given $\SA \in \{(1/n)\I_p, \Sh_{np}\}$, let $\G = \Gamma(\SA)$ be as in Eq.~\eqref{eq:Gamma} and $\Z_{p}^{(1)},\dots,\Z_{p}^{(N)}$ be i.i.d. with distribution $\mathcal{N}(\bs{0}_p,\SA_{np}^\dag)$, where $\SA_{np}^\dag = \SA^{-1/2}(\I_p - \G) \Sh_{np} (\I_p - \G)\SA^{-1/2}$.
In view of Propositions~\ref{prop:asymptotic-euclidean} and \ref{prop:asymptotic-supremum}, we compute approximate p-values for the tests based on $E_{np}$ in Eq.~\eqref{eq:euclidean} via
\begin{equation} \label{eq:pval-euc}
	\hat\alpha = \begin{cases}
		\Prob(\chi_{p-L}^2 > E_{np}) \quad & \text{if } \SA = \Sh_{np},\\
		\frac{1}{N} \sum_{\ell=1}^N \mathbbm{1}\left\{ \| \Z_{p}^{(\ell)} \|_2^2 >  E_{np} \right\} \quad & \text{if } \SA = (1/n)\I_p.
	\end{cases}
\end{equation}
Approximate $p$-values for tests based on $M_{np}$ in Eq.~\eqref{eq:supremum} are calculated by
\begin{equation} \label{eq:pval-sup}
	\hat\alpha = \frac{1}{N}\sum_{\ell=1}^N \mathbbm{1}\left\{\| \Z_p^{(\ell)}\|_{\infty} > M_{np} \right\}.
\end{equation}

To generate the i.i.d.\ replicates $\Z_p^{(1)},\dots,\Z_p^{(N)}$, we proceed as follows. Whenever the dimension $d < 50$, we draw these directly from the Normal distribution. This strategy becomes computationally infeasible for large values of $d$, because the dimension of $\Sh_{np}$ increases at a rate of $O(d^4)$.  When $d \geqslant 50$, $\S_p$ is estimated by the jackknife estimators  $\Sh_{np}^{\rm J}$ in Eq. \eqref{eq:sigma-jackknife} (under $H_0$) and its structured version $\Sb_{np}^{\rm J}$ (under $H_0^*$). We explain next how the explicit calculation of these jackknife estimators can be avoided altogether.

When tests of $H_0$ are used with the statistics based on $\SA = (1/n)\I_p$ and $\S_p$ is estimated by $\Sh_{np}^{\rm J}$ to obtain $\SA_{np}^\dag$, drawing i.i.d.\ observations from $\mathcal{N}(\bs{0}_p,\SA_{np}^\dag)$ is equivalent to employing the Gaussian multiplier bootstrap developed in Section~3.3 of \cite{Chen:2018}. This is because  $(\I_p - \B\B^+)\th_{np}$ is a $U$-statistic with kernel $(\I_p - \B\B^+)h$, as can be seen from the proof of Theorem~\ref{thm:Chen}. The replicates $\Z_p^{(1)},\dots,\Z_p^{(N)}$ are thus of the form
\begin{equation} \label{eq:bootstrap}
	\Z_p^{(\ell)} = \frac{2(\I_p - \B\B^+)}{\sqrt{n}(n-1)} \sum_{\nu=1}^n \left\{ \sum_{\eta \neq \nu} h(\X_\nu,\X_\eta) - \th_{np} \right\} w_\nu^{(\ell)},
\end{equation}
where $\bs{w}^{(\ell)} =  (w_1^{(\ell)},\dots,w_n^{(\ell)})$, $\ell \in \{1,\ldots, N\}$ are i.i.d.\  $\mathcal{N}_n(\bs{0}_n,\I_n)$ vectors.
From Theorem~3.6 of \cite{Chen:2018}, it follows that when $M_{np}$ in Eq. \eqref{eq:supremum} is used with $\SA = (1/n)\I_p$, then under $H_0$ and assuming that there exist constants $\underline{b},\bar{b} > 0$, $B_n = c \geqslant 1$ and $\gamma \in (0,1)$ such that (M.1), (M.2) and $c^2 \log^7(np) \leqslant n^{1-\gamma}$ hold, there exists a constant $\kappa(\underline{b}) > 0$ such that
\begin{equation*}
	\sup_{\alpha \in (0,1)} | \Prob\{M_{np} \leqslant q_{\alpha}^{\rm J}\} - \alpha| \leqslant \kappa(\underline{b})\times n^{-\gamma/6},
\end{equation*}
where $q_{\alpha}^{\rm J}$ is the conditional $\alpha$th quantile of $\| \Z_p\|_{\infty}$ given the data, where $\Z_p$ is as on the right-hand side of Eq. \eqref{eq:bootstrap} with $\bs{w}^{(\ell)} \equiv\bs{w} \sim \mathcal{N}_n(\bs{0}_n,\I_n)$.

When testing $H_0^*$ with $\mathcal{G} = \{\{1,\dots,d\}\}$ (full exchangeability) when $d \geqslant 50$, we make direct use of the estimator $\bs{\hat\sigma}_{np}^{\rm J}$ defined in Appendix~\ref{app:Sigma}, for which $\Sb_{np}^{\rm J}=\SS_p(\bs{\hat\sigma}_{np}^{\rm J})$ with $\SS_p(\cdot)$ as in Eq.~\eqref{eq:S-cal}.  Propositions~\ref{prop:eigen-Sigma} and \ref{prop:decomposition} allows us compute the statistics $E_{np}$ and $M_{np}$ with $\SA \in \{(1/n)\I_p, \Sh_{np}\}$ directly by using merely the eigenvalues of $\Sb_{np}^{\rm J}$.  To implement the $p$-value approximations in Eqs. \eqref{eq:pval-euc} and \eqref{eq:pval-sup} in a computationally feasible way, we proceed as follows. First,  \eqref{eq:pval-euc} can be used directly when $E_{np}$ is combined with $\SA = \Sh_{np} = \Sb_{np}^{\rm J}$. Secondly, for the test based on $E_{np}$ with $\SA = (1/n)\I_p$, we follow Corrolary~\ref{cor:asymptotic-euc-exchangeable} and replace $\| \Z_{p}^{(\ell)} \|_2^2$ in \eqref{eq:pval-euc} by $\delta_3(\bs{\hat\sigma}_{np}^{\rm J}) Y_1^{(\ell)} + \delta_{2,d}(\bs{\hat\sigma}_{np}^{\rm J}) Y_2^{(\ell)}$ for each $\ell \in \{1,\ldots, N\}$; here, 
$Y_1^{(\ell)}\sim \chi_{p-d}^2$ and $Y_2^{(\ell)} \sim \chi_{d-1}^2$,  $\ell \in \{1,\ldots, N\}$, are mutually independent and  $\delta_{2,d}$ and $\delta_{3}$ are 
as in Proposition~\ref{prop:eigen-Sigma}.
Third, for the tests based on $M_{np}$ with $\Sh_{np}$, we simply set $\Z_p^{(\ell)} = \bs{Y}_p^{(\ell)} - \mathbf{1}_p\bar{Y}^{(\ell)}$, where $\bar{Y}^{(\ell)} = (1/p) \sum_{r=1}^p Y_{p,r}^{(\ell)}$ and $\bs{Y}_p^{(\ell)} \sim \mathcal{N}_p(\bs{0}_p,\I_p)$.
Finally, for the tests based on $M_{np}$ with $\SA = \I_{p}$, observe that $\SA_{np}^\dag$ simplifies to $(\I_p - \G)\Sb_{np}^{\rm J}$ by Lemma~B.6 in \cite{Perreault/Duchesne/Neslehova:2019}; furthermore, $\G = \B\B^+= (1/p)\bs{J}_p$. We can thus 
 generate $\Z_p^{(\ell)}$, $\ell \in \{1,\ldots, N\}$, in Eq.~\eqref{eq:pval-sup} using the following proposition, whose proof is straightforward and hence omitted.

\begin{proposition} \label{prop:MC-H0-star}
Let $d \geqslant 4$, $\G = (1/p)\bs{J}_p$, $\G^*$ be as in Eq.~\eqref{eq:Gamma-star} and $\SS_p$ be a positive semidefinite matrix in $\Sset$ with at most three distinct eigenvalues $\delta_{1,d}$, $\delta_{2,d}$ and $\delta_{3}$, as defined in 
Proposition~\ref{prop:eigen-Sigma}. Also let $W_{r} \sim \mathcal{N}(0,1)$, $r = 1,\dots,p$, be independent random variables and define, for $i=1,\dots,d$,
$$
	\bar{W} = \frac{1}{p} \sum_{r=1}^p W_{r} \quad \text{and} \quad \bar{W}^{(i)} = \frac{1}{d-1} \sum_{r \in \mathcal{R}_{i}} W_{r}, \quad \text{where } \mathcal{R}_{i} = \{ s : i \in \{i_s,j_s\},\ (i_s,j_s) = \iota(s)\}.
$$
Then, the random vector $\bs{Y}_p^{(1)} = (Y_1^{(1)},\dots,Y_p^{(1)})$ whose $r$-th component, $r\in\{1,\dots,p\}$, is given by 
$$
	Y_r^{(1)} = \delta_{2,d}^{1/2}\left\{W_r - \frac{d-1}{d-2}(\bar{W}^{(i_r)} + \bar{W}^{(j_r)}) + \frac{d}{d-2} \bar{W} \right\}
$$
is such that $\bs{Y}_p^{(1)} \sim \mathcal{N}_p\{\bs{0}_p,(\I_p - \G^*)\SS_p\}$.
Similarly, the random vector $\bs{Y}_p^{(2)} = (Y_1^{(2)},\dots,Y_p^{(2)})$ whose $r$-th component, $r\in\{1,\dots,p\}$, is given by 
$$
Y_r^{(2)} = \delta_{3}^{1/2}\left\{\frac{d-1}{d-2}(\bar{W}^{(i_r)} + \bar{W}^{(j_r)}) - \frac{2(d-1)}{d-2} \bar{W} \right\}
$$
is such that $\bs{Y}_p^{(2)} \sim \mathcal{N}_p(\bs{0}_p,(\G^* - \G)\SS_p)$.
Furthermore, $\bs{Y}_p^{(1)} + \bs{Y}_p^{(2)} \sim \mathcal{N}_p\{\bs{0}_p,(\I_p - \G)\SS_p\}$.
\end{proposition}

\subsection{Simulation study results} \label{app:sim}

The tables in this section contain the simulation results.
The simulations involve samples from generated Normal, $t_4$ (Student $t$ with 4 degrees of freedom), Gumbel and Clayton copulas.
Unless otherwise stated, the entries of the tables for the Normal copula are based on $2500$ samples, while the entries of the tables for the three other copulas are based on $1000$ samples.
All tests were performed at the nominal 5\% level and Monte Carlo/bootstrap methods used 5000 replicates.

\subsubsection{Equicorrelation and exchangeability}

For this part of the simulation study, we consider the null hypotheses of equicorrelation ($H_0$ with $\B = \bs{1}_p$) and of full exchangeability ($H_0^*$ with $\mathcal{G} = \{\{ 1,\dots,d\}\}$).
We evaluate the performance of our tests on samples generated from the Normal, $t_4$, Gumbel, and Clayton copulas.
The null and alternative distributions are described in \eqref{eq:T-equi-null} and \eqref{eq:departure}, respectively; nested Archimedean copulas are used as alternatives in the case of the Gumbel and Clayton copulas.
We considered $n \in \{50,100,150,250\}$, $d\in \{5,15,25,50,100\}$; the copula parameter corresponds to $\tau \in \{0,0.3,0.6\}$ (excluding $\tau = 0$ for the Clayton  as  independence  is already included in the Normal model).

The following tables report the results of the simulation study for the tests of $H_0$ with $\B = \bs{1}_p$.
\begin{itemize}
\item Tables \ref{tab:sim-level-normal}--\ref{tab:sim-level-clayton}:
estimated sizes for the tests of $H_0$ using $\SA = (1/n)\I_p$ and $\SA =\Sh_{np}$.

\item Tables \ref{tab:sim-power-1-normal}--\ref{tab:sim-power-1-clayton}: estimated rejection rates of tests of $H_0$ using $\SA = (1/n)\I_{p}$;\\
single and column departures, $\Delta = 0.1$.

\item Tables \ref{tab:sim-power-2-normal}--\ref{tab:sim-power-2-clayton}: estimated rejection rates of tests of $H_0$ using $\SA = (1/n)\I_{p}$;\\
single and column departures, $\Delta = 0.2$.
\end{itemize}

The following tables report the results for the tests of $H_0^*$ with $\mathcal{G} = \{\{ 1,\dots,d\}\}$.
\begin{itemize}
\item Tables \ref{tab:sim-level-star-normal}--\ref{tab:sim-level-star-clayton}: estimated sizes for the tests of $H_0^*$ using $\SA = (1/n)\I_p,\Sh_{np}$.

\item Tables \ref{tab:sim-power-star-single-1-normal}--\ref{tab:sim-power-star-single-1-clayton}:
estimated rejection rates of tests of $H_0^*$ using $\SA = (1/n)\I_p$ and $\SA =\Sh_{np}$;\\
single departures, $\Delta = 0.1$.

\item Tables \ref{tab:sim-power-star-single-2-normal}--\ref{tab:sim-power-star-single-2-clayton}:
estimated rejection rates of tests of $H_0^*$ using $\SA = (1/n)\I_p$ and $\SA =\Sh_{np}$;\\
single departures, $\Delta = 0.2$.

\item Tables \ref{tab:sim-power-star-column-1-normal}--\ref{tab:sim-power-star-column-1-clayton}:
estimated rejection rates of tests of $H_0^*$ using $\SA = (1/n)\I_p$ and $\SA =\Sh_{np}$;\\
column departures, $\Delta = 0.1$.

\item Tables \ref{tab:sim-power-star-column-2-normal}--\ref{tab:sim-power-star-column-2-clayton}:
estimated rejection rates of tests of $H_0^*$ using $\SA = (1/n)\I_p$ and $\SA =\Sh_{np}$;\\
column departures, $\Delta = 0.2$.
\end{itemize}

\clearpage

\begin{table}[htbp]
 \captionsetup{width=1\linewidth,font=small,skip=0pt}
      \caption{Estimated sizes (in \%) for the tests of $H_0$ with $\B = \bs{1}_p$ performed at the nominal level 5\%. Each entry is based on $ 2500 $ samples of size $n$ in dimension $d$ drawn from a  Normal copula with Kendall's tau matrix $\bs{T}$ is as in Eq.~\eqref{eq:T-equi-null}.}
       \label{tab:sim-level-normal}
      \begin{center}
      \fontsize{8.75}{8.75}\selectfont
      \vskip-12pt
      \begin{tabular}{*{2}{l}*{12}{r}}
      \toprule
      \multicolumn{14}{c}{$E_{np}$ with $\SA = \Sh_{np}$}\\
      \midrule
      \multicolumn{2}{c}{} & & \multicolumn{3}{c}{$\tau = 0$} & & \multicolumn{3}{c}{$\tau = 0.3$} & & \multicolumn{3}{c}{$\tau = 0.6$}\\
      \cmidrule(lr){4-6}  \cmidrule(lr){8-10}  \cmidrule(lr){12-14}
      $\Sh_{np}$ & $d$\big|$n$ & & 50 & 150 & 250 & & 50 & 150 & 250 & & 50 & 150 & 250 \\
      \midrule
      \multirow{ 2}{*}{$\Sh_{np}^{\rm P}$} &  5 &&  39.5 & 13.4 & 10.1 && 34.6 & 12.8 & 9.1 && *34.6 & 10.3 & 7.2  \\
      & 15 &&   & *100 & 99.3 &&  &  & 98.8 &&  &  & 97.8  \\
      \cmidrule(lr){2-14}
      \multirow{ 2}{*}{$\Sh_{np}^{\rm J}$} &  5 &&  30.2 & 12 & 9.4 && 21.7 & 11 & 8.2 && 13.8 & 6.7 & 5.8 \\
      & 15 &&   & 100 & 99 &&  & 100 & 97 &&  & 100 & 87.6  \\
      \midrule
      \multicolumn{14}{c}{$M_{np}$ with $\SA = \Sh_{np}$}\\
      \midrule
      \multicolumn{2}{c}{} & & \multicolumn{3}{c}{$\tau = 0$} & & \multicolumn{3}{c}{$\tau = 0.3$} & & \multicolumn{3}{c}{$\tau = 0.6$}\\
      \cmidrule(lr){4-6}  \cmidrule(lr){8-10}  \cmidrule(lr){12-14}
      $\Sh_{np}$ & $d$\big|$n$ & & 50 & 150 & 250 & & 50 & 150 & 250 & & 50 & 150 & 250 \\
      \midrule
      \multirow{2}{*}{$\Sh_{np}^{\rm P}$}
      &  5 &&  31.1 & 10.4 & 9.1 && 29.6 & 10.1 & 8 && *28.9 & 9.4 & 6.8 \\
      & 15 &&   & *100 & 81.2 &&  &  & 68.6 &&  &  & 59.9 \\
      \cmidrule(lr){2-14}
      \multirow{2}{*}{$\Sh_{np}^{\rm J}$}
      &  5 &&  24 & 9.6 & 8.4 && 18.2 & 9.2 & 7.3 && 11 & 6.9 & 5.8 \\
      & 15 &&   & 99.9 & 77.5 &&  & 98.2 & 59.2 &&  & 92 & 39 \\
      \midrule
      \multicolumn{14}{c}{$E_{np}$ with $\SA = (1/n)\I_p$}\\
      \midrule
      \multicolumn{2}{c}{} & & \multicolumn{3}{c}{$\tau = 0$} & & \multicolumn{3}{c}{$\tau = 0.3$} & & \multicolumn{3}{c}{$\tau = 0.6$}\\
      \cmidrule(lr){4-6}  \cmidrule(lr){8-10}  \cmidrule(lr){12-14}
      $\Sh_{np}$ & $d$\big|$n$ & & 50 & 100 & 150 & & 50 & 100 & 150 & & 50 & 100 & 150 \\
      \midrule
      &  5 &&  5.4 & 5.3 & 6.1 && 5 & 4.1 & 5.4 && 4.2 & 4 & 4.6 \\
      $\Sh_{np}^{\rm P}$ & 15 &&  0.4 & 1.8 & 1.9 && 1.5 & 2.9 & 3.3 && 1.4 & 2.9 & 3.4 \\
      & 25 &&  0 & 0.2 & 0.4 && 0.6 & 1.1 & 2.5 && 0.3 & 0.7 & 1.7 \\
      \cmidrule(lr){2-14}
      &  5 &&  3.2 & 4 & 5.3 && 2.7 & 3.4 & 4.3 && 1.3 & 2.3 & 3.2 \\
      & 15 &&  0.1 & 0.7 & 1.2 && 0.6 & 1.4 & 2.4 && 0.2 & 1.4 & 1.7 \\
      $\Sh_{np}^{\rm J}$ & 25 &&  0 & 0 & 0.2 && 0.3 & 0.4 & 2 && 0 & 0.3 & 0.7 \\
      & 50 &&  0 & 0 & 0 && 0 & 0 & 0.4 && 0 & 0 & 0.1 \\
      & 100 &&  0 & 0 & 0 && 0 & 0 & 0 && 0 & 0 & 0 \\
      \midrule
      \multicolumn{14}{c}{$M_{np}$ with $\SA = (1/n)\I_p$}\\
      \midrule
      \multicolumn{2}{c}{} & & \multicolumn{3}{c}{$\tau = 0$} & & \multicolumn{3}{c}{$\tau = 0.3$} & & \multicolumn{3}{c}{$\tau = 0.6$}\\
      \cmidrule(lr){4-6}  \cmidrule(lr){8-10}  \cmidrule(lr){12-14}
      $\Sh_{np}$ & $d$\big|$n$ & & 50 & 100 & 150 & & 50 & 100 & 150 & & 50 & 100 & 150 \\
      \midrule
      &  5 &&  7.2 & 5.4 & 5.4 && 6.5 & 5.6 & 5.7 && 5.7 & 6.1 & 5 \\
      $\Sh_{np}^{\rm P}$ & 15 &&  3.7 & 5 & 5.2 && 3.6 & 5 & 4.4 && 2.9 & 3.5 & 5 \\
      & 25 &&  2.1 & 4.2 & 4.1 && 3.7 & 3.1 & 4.6 && 1.9 & 2.4 & 4 \\
      \cmidrule(lr){2-14}
      &  5 &&  5.2 & 4.4 & 5 && 3.7 & 4.4 & 5 && 2.5 & 4 & 3.8 \\
      & 15 &&  2.6 & 3.8 & 4.6 && 2.4 & 3.7 & 3.7 && 1.4 & 2 & 3.1 \\
      $\Sh_{np}^{\rm J}$ & 25 &&  1.5 & 3.6 & 3.6 && 2.4 & 2.4 & 3.8 && 1.1 & 1.7 & 3 \\
      & 50 &&  1.4 & 2.6 & 3.4 && 1.8 & 2.4 & 3.6 && 0.8 & 1 & 2 \\
      & 100 &&  1 & 1.8 & 2.3 && 1.6 & 2.6 & 3.3 && 0.2 & 1.4 & 1.5 \\
      \bottomrule
      \end{tabular}
      \end{center}
      \vskip-9pt
      \small
      Statistics: $E_{np}$ Euclidean norm-based statistic defined in Eq. \eqref{eq:euclidean}; $M_{np}$ supremum norm-based statistic defined in Eq. \eqref{eq:supremum}. Estimators: $\Sh_{np}^{\rm P}$ plug-in estimator; $\Sh_{np}^{\rm J}$ jackknife estimator. *The results marked by an asterisk were computed on at least 2000 simulations; the simulations for which $\SA$ was not positive definite were discarded. Blank entries correspond to cases where $\SA$ was positive definite less than 1\% of the times. In all other cases, $\SA$ was always positive definite. 
      \end{table}

            \begin{table}[htbp]
 \captionsetup{width=1\linewidth,font=small,skip=0pt}
      \caption{Estimated sizes (in \%) for the tests of $H_0$ with $\B = \bs{1}_p$ performed at the nominal level 5\%. Each entry is based on $ 1000 $ samples of size $n$ in dimension $d$ drawn from a  $t_4$ copula  with Kendall's tau matrix $\bs{T}$ is as in Eq.~\eqref{eq:T-equi-null}.}
       \label{tab:sim-level-t4}
      \begin{center}
      \fontsize{8.75}{8.75}\selectfont
      \vskip-12pt
      \begin{tabular}{*{2}{l}*{12}{r}}
      \toprule
      \multicolumn{14}{c}{$E_{np}$ with $\SA = \Sh_{np}$}\\
      \midrule
      \multicolumn{2}{c}{} & & \multicolumn{3}{c}{$\tau = 0$} & & \multicolumn{3}{c}{$\tau = 0.3$} & & \multicolumn{3}{c}{$\tau = 0.6$}\\
      \cmidrule(lr){4-6}  \cmidrule(lr){8-10}  \cmidrule(lr){12-14}
      $\Sh_{np}$ & $d$\big|$n$ & & 50 & 150 & 250 & & 50 & 150 & 250 & & 50 & 150 & 250 \\
      \midrule
      \multirow{ 2}{*}{$\Sh_{np}^{\rm P}$} &  5 &&  38.5 & 14.2 & 9.7 && 39.7 & 15.6 & 9.3 && *37.4 & 11.7 & 8.8  \\
      & 15 &&   &  & 100 &&  &  & 99.8 &&  &  & 99.2  \\
      \cmidrule(lr){2-14}
      \multirow{ 2}{*}{$\Sh_{np}^{\rm J}$} &  5 &&  30.9 & 12.8 & 9.5 && 29.2 & 14.2 & 8.6 && 15.8 & 9.1 & 8.1 \\
      & 15 &&   & 100 & 100 &&  & 100 & 99.4 &&  & 100 & 95.3  \\
      \midrule
      \multicolumn{14}{c}{$M_{np}$ with $\SA = \Sh_{np}$}\\
      \midrule
      \multicolumn{2}{c}{} & & \multicolumn{3}{c}{$\tau = 0$} & & \multicolumn{3}{c}{$\tau = 0.3$} & & \multicolumn{3}{c}{$\tau = 0.6$}\\
      \cmidrule(lr){4-6}  \cmidrule(lr){8-10}  \cmidrule(lr){12-14}
      $\Sh_{np}$ & $d$\big|$n$ & & 50 & 150 & 250 & & 50 & 150 & 250 & & 50 & 150 & 250 \\
      \midrule
      \multirow{2}{*}{$\Sh_{np}^{\rm P}$}
      &  5 &&  30.9 & 11.5 & 9.3 && 34.3 & 14.7 & 8.4 && *30.2 & 11.6 & 7.8 \\
      & 15 &&   &  & 91 &&  &  & 80.1 &&  &  & 65.6 \\
      \cmidrule(lr){2-14}
      \multirow{2}{*}{$\Sh_{np}^{\rm J}$}
      &  5 &&  24.4 & 10.4 & 8.9 && 24.6 & 12.7 & 7.9 && 13.5 & 8.6 & 6.5 \\
      & 15 &&   & 100 & 88.3 &&  & 99.8 & 71.4 &&  & 95.2 & 41.4 \\
      \midrule
      \multicolumn{14}{c}{$E_{np}$ with $\SA = (1/n)\I_p$}\\
      \midrule
      \multicolumn{2}{c}{} & & \multicolumn{3}{c}{$\tau = 0$} & & \multicolumn{3}{c}{$\tau = 0.3$} & & \multicolumn{3}{c}{$\tau = 0.6$}\\
      \cmidrule(lr){4-6}  \cmidrule(lr){8-10}  \cmidrule(lr){12-14}
      $\Sh_{np}$ & $d$\big|$n$ & & 50 & 100 & 150 & & 50 & 100 & 150 & & 50 & 100 & 150 \\
      \midrule
      &  5 &&  5.9 & 6 & 4.8 && 4.8 & 3.9 & 5.7 && 3.9 & 3.8 & 3.3 \\
      $\Sh_{np}^{\rm P}$ & 15 &&  0.1 & 0.2 & 1.1 && 0.4 & 1.2 & 1.9 && 0.6 & 1.2 & 1.7 \\
      & 25 &&  0 & 0 & 0 && 0.2 & 0.7 & 1.5 && 0.1 & 0.7 & 0.7 \\
      \cmidrule(lr){2-14}
      &  5 &&  3.4 & 4.9 & 4.2 && 3 & 3.3 & 4.6 && 1.6 & 2.5 & 2.8 \\
      & 15 &&  0 & 0.1 & 0.3 && 0.2 & 0.6 & 0.9 && 0.1 & 0.6 & 1.1 \\
      $\Sh_{np}^{\rm J}$ & 25 &&  0 & 0 & 0 && 0 & 0.4 & 0.9 && 0.1 & 0.2 & 0.2 \\
      & 50 &&  0 & 0 & 0 && 0 & 0 & 0.1 && 0 & 0.1 & 0 \\
      & 100 &&  0 & 0 & 0 && 0 & 0 & 0 && 0 & 0 & 0 \\
      \midrule
      \multicolumn{14}{c}{$M_{np}$ with $\SA = (1/n)\I_p$}\\
      \midrule
      \multicolumn{2}{c}{} & & \multicolumn{3}{c}{$\tau = 0$} & & \multicolumn{3}{c}{$\tau = 0.3$} & & \multicolumn{3}{c}{$\tau = 0.6$}\\
      \cmidrule(lr){4-6}  \cmidrule(lr){8-10}  \cmidrule(lr){12-14}
      $\Sh_{np}$ & $d$\big|$n$ & & 50 & 100 & 150 & & 50 & 100 & 150 & & 50 & 100 & 150 \\
      \midrule
      &  5 &&  6.2 & 7.1 & 5.5 && 7.1 & 5.2 & 6.6 && 5.7 & 3 & 5.3 \\
      $\Sh_{np}^{\rm P}$ & 15 &&  4.2 & 5.9 & 6 && 4.1 & 4.6 & 4.4 && 3.4 & 2.7 & 3.2 \\
      & 25 &&  2.8 & 4 & 4.3 && 2.3 & 4.3 & 4.1 && 2.5 & 2.9 & 3.6 \\
      \cmidrule(lr){2-14}
      &  5 &&  4.3 & 6.4 & 4.4 && 4.5 & 4.4 & 5.8 && 2.8 & 2.1 & 4.7 \\
      & 15 &&  3.2 & 5.2 & 5.2 && 3.1 & 2.9 & 4.1 && 2.1 & 1.6 & 2.4 \\
      $\Sh_{np}^{\rm J}$ & 25 &&  2.3 & 3.4 & 3.8 && 1.9 & 3.1 & 3.6 && 1 & 1.9 & 2.9 \\
      & 50 &&  1.5 & 3.5 & 4 && 2.3 & 3.1 & 4.2 && 1.1 & 2.3 & 3.6 \\
      & 100 &&  1 & 2 & 3.7 && 1.8 & 2.7 & 2.4 && 0.3 & 1.6 & 2.1 \\
      \bottomrule
      \end{tabular}
      \end{center}
      \vskip-9pt
      \small
      Statistics: $E_{np}$ Euclidean norm-based statistic defined in Eq. \eqref{eq:euclidean}; $M_{np}$ supremum norm-based statistic defined in Eq. \eqref{eq:supremum}. Estimators: $\Sh_{np}^{\rm P}$ plug-in estimator; $\Sh_{np}^{\rm J}$ jackknife estimator. *The results marked by an asterisk were computed on 994 simulations; the simulations for which $\SA$ was not positive definite were discarded. Blank entries correspond to cases where $\SA$ was positive definite less than 65\% of the times. In all other cases, $\SA$ was always positive definite.
      \end{table}

            \begin{table}[htbp]
 \captionsetup{width=1\linewidth,font=small,skip=0pt}
      \caption{Estimated sizes (in \%) for the tests of $H_0$ with $\B = \bs{1}_p$ performed at the nominal level 5\%. Each entry is based on $ 1000 $ samples of size $n$ in dimension $d$ drawn from a  Gumbel copula  with Kendall's tau matrix $\bs{T}$ is as in Eq.~\eqref{eq:T-equi-null}.}
       \label{tab:sim-level-gumbel}
      \begin{center}
      \fontsize{8.75}{8.75}\selectfont
      \vskip-12pt
      \begin{tabular}{*{2}{l}*{12}{r}}
      \toprule
      \multicolumn{14}{c}{$E_{np}$ with $\SA = \Sh_{np}$}\\
      \midrule
      \multicolumn{2}{c}{} & & \multicolumn{3}{c}{$\tau = 0$} & & \multicolumn{3}{c}{$\tau = 0.3$} & & \multicolumn{3}{c}{$\tau = 0.6$}\\
      \cmidrule(lr){4-6}  \cmidrule(lr){8-10}  \cmidrule(lr){12-14}
      $\Sh_{np}$ & $d$\big|$n$ & & 50 & 150 & 250 & & 50 & 150 & 250 & & 50 & 150 & 250 \\
      \midrule
      \multirow{ 2}{*}{$\Sh_{np}^{\rm P}$} &  5 &&  39.1 & 14.8 & 8.2 && 36.9 & 13.5 & 7.4 && *36.4 & 8 & 10  \\
      & 15 &&   & *100 & 99.2 &&  &  & 98.5 &&  &  & 98.1  \\
      \cmidrule(lr){2-14}
      \multirow{ 2}{*}{$\Sh_{np}^{\rm J}$} &  5 &&  30 & 13.4 & 7.4 && 22.9 & 11.7 & 6.6 && 15.4 & 5.3 & 8.1 \\
      & 15 &&   & 100 & 98.8 &&  & 100 & 97.2 &&  & 99.9 & 89.3  \\
      \midrule
      \multicolumn{14}{c}{$M_{np}$ with $\SA = \Sh_{np}$}\\
      \midrule
      \multicolumn{2}{c}{} & & \multicolumn{3}{c}{$\tau = 0$} & & \multicolumn{3}{c}{$\tau = 0.3$} & & \multicolumn{3}{c}{$\tau = 0.6$}\\
      \cmidrule(lr){4-6}  \cmidrule(lr){8-10}  \cmidrule(lr){12-14}
      $\Sh_{np}$ & $d$\big|$n$ & & 50 & 150 & 250 & & 50 & 150 & 250 & & 50 & 150 & 250 \\
      \midrule
      \multirow{2}{*}{$\Sh_{np}^{\rm P}$}
      &  5 &&  31 & 10.1 & 6.8 && 28.4 & 11.1 & 5.3 && *29.6 & 9.5 & 8 \\
      & 15 &&   & *99.9 & 79.1 &&  &  & 64.7 &&  &  & 54.4 \\
      \cmidrule(lr){2-14}
      \multirow{2}{*}{$\Sh_{np}^{\rm J}$}
      &  5 &&  22.5 & 9.2 & 6.2 && 19.1 & 9.9 & 4.8 && 13 & 7.2 & 6.4 \\
      & 15 &&   & 99.7 & 75 &&  & 96.4 & 53.6 &&  & 89.4 & 34.2 \\
      \midrule
      \multicolumn{14}{c}{$E_{np}$ with $\SA = (1/n)\I_p$}\\
      \midrule
      \multicolumn{2}{c}{} & & \multicolumn{3}{c}{$\tau = 0$} & & \multicolumn{3}{c}{$\tau = 0.3$} & & \multicolumn{3}{c}{$\tau = 0.6$}\\
      \cmidrule(lr){4-6}  \cmidrule(lr){8-10}  \cmidrule(lr){12-14}
      $\Sh_{np}$ & $d$\big|$n$ & & 50 & 100 & 150 & & 50 & 100 & 150 & & 50 & 100 & 150 \\
      \midrule
      &  5 &&  4.8 & 5.7 & 5.2 && 5.4 & 5.6 & 4.3 && 5.6 & 4.3 & 3.9 \\
      $\Sh_{np}^{\rm P}$ & 15 &&  0.4 & 1.7 & 2.4 && 1 & 3 & 3 && 1.2 & 1.7 & 4.1 \\
      & 25 &&  0 & 0 & 0.5 && 0.4 & 0.2 & 1 && 0 & 1.3 & 1.4 \\
      \cmidrule(lr){2-14}
      &  5 &&  3.1 & 4.1 & 4.5 && 3.7 & 4.4 & 3.7 && 3 & 3 & 3 \\
      & 15 &&  0.1 & 0.9 & 1.9 && 0.2 & 1.8 & 1.8 && 0 & 0.8 & 2.1 \\
      $\Sh_{np}^{\rm J}$ & 25 &&  0 & 0 & 0.4 && 0.2 & 0 & 0.8 && 0 & 0.5 & 0.6 \\
      & 50 &&  0 & 0 & 0 && 0 & 0.1 & 0 && 0 & 0 & 0.3 \\
      & 100 &&  0 & 0 & 0 && 0 & 0 & 0 && 0 & 0 & 0 \\
      \midrule
      \multicolumn{14}{c}{$M_{np}$ with $\SA = (1/n)\I_p$}\\
      \midrule
      \multicolumn{2}{c}{} & & \multicolumn{3}{c}{$\tau = 0$} & & \multicolumn{3}{c}{$\tau = 0.3$} & & \multicolumn{3}{c}{$\tau = 0.6$}\\
      \cmidrule(lr){4-6}  \cmidrule(lr){8-10}  \cmidrule(lr){12-14}
      $\Sh_{np}$ & $d$\big|$n$ & & 50 & 100 & 150 & & 50 & 100 & 150 & & 50 & 100 & 150 \\
      \midrule
      &  5 &&  5.4 & 6.1 & 5.1 && 6.9 & 7.5 & 4.9 && 6.9 & 5.6 & 4.4 \\
      $\Sh_{np}^{\rm P}$ & 15 &&  3.9 & 5.3 & 4.3 && 3.7 & 4.3 & 6.1 && 4.1 & 3.8 & 4.8 \\
      & 25 &&  2.3 & 3.9 & 4.4 && 3.3 & 5.3 & 3.8 && 2.4 & 2.5 & 3.2 \\
      \cmidrule(lr){2-14}
      &  5 &&  3.6 & 4.7 & 4.5 && 4.7 & 6.4 & 3.8 && 4.6 & 4.2 & 4 \\
      & 15 &&  2.3 & 4.2 & 3.9 && 2.5 & 3.4 & 4.6 && 1.8 & 2.5 & 3.4 \\
      $\Sh_{np}^{\rm J}$ & 25 &&  2 & 3 & 3.8 && 2.3 & 4.3 & 3 && 1.7 & 1.6 & 2.3 \\
      & 50 &&  2.1 & 2.6 & 3.2 && 2.3 & 3.3 & 3.2 && 0.6 & 2.3 & 1.9 \\
      & 100 &&  0.4 & 2 & 1.8 && 1.3 & 2 & 2.9 && 0.5 & 1.8 & 2.2 \\
      \bottomrule
      \end{tabular}
      \end{center}
      \vskip-9pt
      \small
      Statistics: $E_{np}$ Euclidean norm-based statistic defined in Eq. \eqref{eq:euclidean}; $M_{np}$ supremum norm-based statistic defined in Eq. \eqref{eq:supremum}. Estimators: $\Sh_{np}^{\rm P}$ plug-in estimator; $\Sh_{np}^{\rm J}$ jackknife estimator. *The results marked by an asterisk were computed on at least 802 simulations; the simulations for which $\SA$ was not positive definite were discarded. Blank entries correspond to cases where $\SA$ was positive definite less than 1\% of the times. In all other cases, $\SA$ was always positive definite.
      \end{table}

            \begin{table}[htbp]
 \captionsetup{width=1\linewidth,font=small,skip=0pt}
      \caption{Estimated sizes (in \%) for the tests of $H_0$ with $\B = \bs{1}_p$ performed at the nominal level 5\%. Each entry is based on $ 1000 $ samples of size $n$ in dimension $d$ drawn from a  Clayton copula  with Kendall's tau matrix $\bs{T}$ is as in Eq.~\eqref{eq:T-equi-null}.}
       \label{tab:sim-level-clayton}
      \begin{center}
      \fontsize{8.75}{8.75}\selectfont
      \vskip-12pt
      \begin{tabular}{*{2}{l}*{12}{r}}
      \toprule
      \multicolumn{14}{c}{$E_{np}$ with $\SA = \Sh_{np}$}\\
      \midrule
      \multicolumn{2}{c}{} & & \multicolumn{3}{c}{$\tau = 0$} & & \multicolumn{3}{c}{$\tau = 0.3$} & & \multicolumn{3}{c}{$\tau = 0.6$}\\
      \cmidrule(lr){4-6}  \cmidrule(lr){8-10}  \cmidrule(lr){12-14}
      $\Sh_{np}$ & $d$\big|$n$ & & 50 & 150 & 250 & & 50 & 150 & 250 & & 50 & 150 & 250 \\
      \midrule
      \multirow{ 2}{*}{$\Sh_{np}^{\rm P}$} &  5 &&   &  &  && 33.1 & 11 & 8 && *30.9 & 9.8 & 8  \\
      & 15 &&   &  &  &&  &  & 98.9 &&  &  & 95.1  \\
      \cmidrule(lr){2-14}
      \multirow{ 2}{*}{$\Sh_{np}^{\rm J}$} &  5 &&   &  &  && 20.9 & 8.6 & 6.6 && 12.4 & 6.9 & 6.8 \\
      & 15 &&   &  &  &&  & 100 & 97.3 &&  & 100 & 81.3  \\
      \midrule
      \multicolumn{14}{c}{$M_{np}$ with $\SA = \Sh_{np}$}\\
      \midrule
      \multicolumn{2}{c}{} & & \multicolumn{3}{c}{$\tau = 0$} & & \multicolumn{3}{c}{$\tau = 0.3$} & & \multicolumn{3}{c}{$\tau = 0.6$}\\
      \cmidrule(lr){4-6}  \cmidrule(lr){8-10}  \cmidrule(lr){12-14}
      $\Sh_{np}$ & $d$\big|$n$ & & 50 & 150 & 250 & & 50 & 150 & 250 & & 50 & 150 & 250 \\
      \midrule
      \multirow{2}{*}{$\Sh_{np}^{\rm P}$}
      &  5 &&   &  &  && 26.9 & 9.9 & 8 && *27.3 & 8.8 & 7.1 \\
      & 15 &&   &  &  &&  &  & 68.6 &&  &  & 45.5 \\
      \cmidrule(lr){2-14}
      \multirow{2}{*}{$\Sh_{np}^{\rm J}$}
      &  5 &&   &  &  && 18.3 & 8.6 & 7 && 12.2 & 6.8 & 5.9 \\
      & 15 &&   &  &  &&  & 98.2 & 59.6 &&  & 85.2 & 27.1 \\
      \midrule
      \multicolumn{14}{c}{$E_{np}$ with $\SA = (1/n)\I_p$}\\
      \midrule
      \multicolumn{2}{c}{} & & \multicolumn{3}{c}{$\tau = 0$} & & \multicolumn{3}{c}{$\tau = 0.3$} & & \multicolumn{3}{c}{$\tau = 0.6$}\\
      \cmidrule(lr){4-6}  \cmidrule(lr){8-10}  \cmidrule(lr){12-14}
      $\Sh_{np}$ & $d$\big|$n$ & & 50 & 100 & 150 & & 50 & 100 & 150 & & 50 & 100 & 150 \\
      \midrule
      &  5 &&   &  &  && 4.2 & 5.5 & 3.3 && 3.1 & 4.1 & 4.2 \\
      $\Sh_{np}^{\rm P}$ & 15 &&   &  &  && 1.5 & 2.8 & 3.4 && 1.2 & 2.9 & 3.3 \\
      & 25 &&   &  &  && 0.6 & 1.5 & 1.1 && 0.4 & 1.1 & 2.3 \\
      \cmidrule(lr){2-14}
      &  5 &&   &  &  && 2.8 & 4.2 & 2.9 && 1.4 & 2.7 & 3.5 \\
      & 15 &&   &  &  && 0.7 & 1.5 & 2.3 && 0.3 & 1.7 & 2 \\
      $\Sh_{np}^{\rm J}$ & 25 &&   &  &  && 0.1 & 0.9 & 0.8 && 0.1 & 0.2 & 1.6 \\
      & 50 &&   &  &  && 0 & 0.3 & 0.1 && 0 & 0.2 & 0.1 \\
      & 100 &&   &  &  && 0 & 0 & 0 && 0 & 0 & 0 \\
      \midrule
      \multicolumn{14}{c}{$M_{np}$ with $\SA = (1/n)\I_p$}\\
      \midrule
      \multicolumn{2}{c}{} & & \multicolumn{3}{c}{$\tau = 0$} & & \multicolumn{3}{c}{$\tau = 0.3$} & & \multicolumn{3}{c}{$\tau = 0.6$}\\
      \cmidrule(lr){4-6}  \cmidrule(lr){8-10}  \cmidrule(lr){12-14}
      $\Sh_{np}$ & $d$\big|$n$ & & 50 & 100 & 150 & & 50 & 100 & 150 & & 50 & 100 & 150 \\
      \midrule
      &  5 &&   &  &  && 6.1 & 5.8 & 4.1 && 5.5 & 5.4 & 5.2 \\
      $\Sh_{np}^{\rm P}$ & 15 &&   &  &  && 3.3 & 4.3 & 3.7 && 2.5 & 3.8 & 3.9 \\
      & 25 &&   &  &  && 3.6 & 4.1 & 3.6 && 1.7 & 3.3 & 4.1 \\
      \cmidrule(lr){2-14}
      &  5 &&   &  &  && 3.8 & 5.2 & 3.5 && 2.6 & 4.7 & 4 \\
      & 15 &&   &  &  && 1.7 & 3.6 & 3.3 && 1.5 & 2.6 & 3.1 \\
      $\Sh_{np}^{\rm J}$ & 25 &&   &  &  && 2.8 & 3.5 & 2.9 && 1.1 & 2.4 & 3.7 \\
      & 50 &&   &  &  && 2 & 2.4 & 3.4 && 0.9 & 2.5 & 2.3 \\
      & 100 &&   &  &  && 1.3 & 2.4 & 4.2 && 0.8 & 1.5 & 1.8 \\
      \bottomrule
      \end{tabular}
      \end{center}
      \vskip-9pt
      \small
      Statistics: $E_{np}$ Euclidean norm-based statistic defined in Eq. \eqref{eq:euclidean}; $M_{np}$ supremum norm-based statistic defined in Eq. \eqref{eq:supremum}. Estimators: $\Sh_{np}^{\rm P}$ plug-in estimator; $\Sh_{np}^{\rm J}$ jackknife estimator. *The results marked by an asterisk were computed on 999 simulations; the simulations for which $\SA$ was not positive definite were discarded. Blank entries (whenever $\tau \neq 0$) correspond to cases where $\SA$ was positive definite less than 1\% of the times. In all other cases, $\SA$ was always positive definite.
      \end{table}

            \begin{table}[htbp]
 \captionsetup{width=1\linewidth,font=small,skip=0pt}
    \caption{Estimated rejection rates (in \%) of tests of $H_0$ with $\B = \bs{1}_p$ and $\SA = (1/n)\I_{p}$, performed at nominal level $5$\%. Each entry is based on $ 2500 $ $n \times d$ datasets drawn from a  Normal copula  with Kendall's tau matrix $\bs{T}_\Delta$ in Eq.~\eqref{eq:departure} (a, single dep.) or (b, column dep.) with $\Delta= 0.1 $; $\bs{T}$ is as in Eq.~\eqref{eq:T-equi-null}.}
       \label{tab:sim-power-1-normal}
      \begin{center}
      \fontsize{8.75}{8.75}\selectfont
      \vskip-12pt
      \begin{tabular}{*{2}{l}*{12}{r}}
      \toprule
      \multicolumn{14}{c}{$E_{np}$ with $\SA = (1/n)\I_p$ for single departures ($\Delta =  0.1 $)}\\
      \midrule
      \multicolumn{2}{c}{} & & \multicolumn{3}{c}{$\tau = 0$} & & \multicolumn{3}{c}{$\tau = 0.3$} & & \multicolumn{3}{c}{$\tau = 0.6$}\\
      \cmidrule(lr){4-6}  \cmidrule(lr){8-10}  \cmidrule(lr){12-14}
      $\Sh_{np}$ & $d$\big|$n$ & & 50 & 100 & 150 & & 50 & 100 & 150 & & 50 & 100 & 150 \\
      \midrule
      &  5 &&  9.4 & 10.8 & 17.4 && 10.6 & 17.7 & 26.5 && 22.6 & 53.8 & 77.6 \\
      $\Sh_{np}^{\rm P}$ & 15 &&  1 & 2.2 & 3.9 && 1.8 & 4 & 4.9 && 1.7 & 5.4 & 10 \\
      & 25 &&  0 & 0.2 & 0.4 && 0.4 & 1.4 & 2.6 && 0.4 & 1.9 & 3.9 \\
      \cmidrule(lr){2-14}
      &  5 &&  6.8 & 9.1 & 15.6 && 6.6 & 14.5 & 23.8 && 11.1 & 43.6 & 72.8 \\
      & 15 &&  0.4 & 1.2 & 2.1 && 0.8 & 1.7 & 3.4 && 0.4 & 2.6 & 5.8 \\
      $\Sh_{np}^{\rm J}$ & 25 &&  0 & 0 & 0.2 && 0.1 & 0.6 & 2 && 0.2 & 0.7 & 2 \\
      & 50 &&  0 & 0 & 0 && 0 & 0.2 & 0.3 && 0 & 0 & 0.3 \\
      & 100 &&  0 & 0 & 0 && 0 & 0 & 0 && 0 & 0 & 0 \\
      \midrule
      \multicolumn{14}{c}{$M_{np}$ with $\SA = (1/n)\I_p$ for single departures ($\Delta =  0.1 $)}\\
      \midrule
      &  5 &&  10.8 & 13.2 & 19.8 && 13 & 23.4 & 34.2 && 32.8 & 72.2 & 90.3 \\
      $\Sh_{np}^{\rm P}$ & 15 &&  4.5 & 7 & 9.5 && 4.7 & 7.8 & 13.8 && 5.3 & 33.2 & 68.2 \\
      & 25 &&  2.7 & 4.4 & 6.7 && 3.2 & 5 & 7.4 && 2.4 & 21.2 & 52.7 \\
      \cmidrule(lr){2-14}
      &  5 &&  7.8 & 11.4 & 18.8 && 9.1 & 20.3 & 32 && 21.7 & 67.4 & 88.1 \\
      & 15 &&  3.4 & 5.6 & 7.9 && 3.4 & 5.8 & 11.4 && 2.6 & 26.9 & 63 \\
      $\Sh_{np}^{\rm J}$ & 25 &&  2.2 & 3.3 & 6.1 && 2.2 & 4.1 & 6.1 && 1.4 & 17 & 48 \\
      & 50 &&  1.1 & 2.4 & 4 && 1.6 & 2.3 & 4.4 && 0.8 & 6.2 & 27.8 \\
      & 100 &&  0.6 & 2.2 & 3.2 && 1.6 & 3 & 4 && 0.6 & 2.4 & 14.2 \\
      \midrule
      \multicolumn{14}{c}{$E_{np}$ with $\SA = (1/n)\I_p$ for column departures ($\Delta =  0.1 $)}\\
      \midrule
      \multicolumn{2}{c}{} & & \multicolumn{3}{c}{$\tau = 0$} & & \multicolumn{3}{c}{$\tau = 0.3$} & & \multicolumn{3}{c}{$\tau = 0.6$}\\
      \cmidrule(lr){4-6}  \cmidrule(lr){8-10}  \cmidrule(lr){12-14}
      $\Sh_{np}$ & $d$\big|$n$ & & 50 & 100 & 150 & & 50 & 100 & 150 & & 50 & 100 & 150 \\
      \midrule
      &  5 &&  17 & 30 & 44.4 && 25.3 & 50.5 & 68.4 && 62.3 & 93.5 & 99.4 \\
      $\Sh_{np}^{\rm P}$ & 15 &&  7.6 & 34.8 & 60 && 17 & 52.5 & 74 && 54.5 & 96.1 & 99.8 \\
      & 25 &&  2.2 & 18.6 & 45 && 9.2 & 38 & 64.4 && 37 & 92.2 & 99.6 \\
      \cmidrule(lr){2-14}
      &  5 &&  13.4 & 26.7 & 41.7 && 18.8 & 46.5 & 65.7 && 49.5 & 91 & 99.2 \\
      & 15 &&  4.5 & 27.8 & 54.3 && 11.1 & 44.9 & 70.3 && 41 & 93.4 & 99.7 \\
      $\Sh_{np}^{\rm J}$ & 25 &&  1.2 & 15.1 & 39.4 && 5.9 & 31.2 & 58.4 && 24 & 86.7 & 99.4 \\
      & 50 &&  0 & 1.6 & 10.8 && 0.4 & 10.8 & 35.2 && 3.4 & 59.9 & 95.1 \\
      & 100 &&  0 & 0 & 0.5 && 0 & 1.2 & 8.9 && 0 & 21.8 & 74 \\
      \midrule
      \multicolumn{14}{c}{$M_{np}$ with $\SA = (1/n)\I_p$ for column departures ($\Delta =  0.1 $)}\\
      \midrule
      \multicolumn{2}{c}{} & & \multicolumn{3}{c}{$\tau = 0$} & & \multicolumn{3}{c}{$\tau = 0.3$} & & \multicolumn{3}{c}{$\tau = 0.6$}\\
      \cmidrule(lr){4-6}  \cmidrule(lr){8-10}  \cmidrule(lr){12-14}
      $\Sh_{np}$ & $d$\big|$n$ & & 50 & 100 & 150 & & 50 & 100 & 150 & & 50 & 100 & 150 \\
      \midrule
      &  5 &&  14.3 & 20.2 & 30.3 && 20 & 35.4 & 50.2 && 47.5 & 82.4 & 95.4 \\
      $\Sh_{np}^{\rm P}$ & 15 &&  10.3 & 25.9 & 39.2 && 18.6 & 41.8 & 63.9 && 52.4 & 93.7 & 99.4 \\
      & 25 &&  8.9 & 22.8 & 38.5 && 16.7 & 40.8 & 63.4 && 50.6 & 93.5 & 99.7 \\
      \cmidrule(lr){2-14}
      &  5 &&  10.6 & 17.6 & 28.5 && 13.7 & 30.9 & 47.5 && 33.8 & 76 & 94 \\
      & 15 &&  8 & 21.9 & 35.8 && 14.2 & 37 & 60.2 && 42 & 90.7 & 99.2 \\
      $\Sh_{np}^{\rm J}$ & 25 &&  7 & 20.8 & 36 && 13.4 & 37.6 & 60.4 && 42.3 & 91.1 & 99.5 \\
      & 50 &&  4.3 & 16.4 & 36.4 && 10.8 & 37.1 & 61.8 && 32.5 & 88.3 & 99.5 \\
      & 100 &&  3.2 & 13.9 & 31.1 && 8.7 & 33.3 & 57.1 && 29.5 & 87.5 & 99.4 \\
      \bottomrule
      \end{tabular}
      \end{center}
      \vskip-9pt
      \small
      Statistics: $E_{np}$ Euclidean norm-based statistic defined in Eq.~\eqref{eq:euclidean}; $M_{np}$ supremum norm-based statistic defined in Eq.~\eqref{eq:supremum}. Estimators: $\Sh_{np}^{\rm P}$ plug-in estimator; $\Sh_{np}^{\rm J}$ jackknife estimator.
      \end{table}

            \begin{table}[htbp]
 \captionsetup{width=1\linewidth,font=small,skip=0pt}
    \caption{Estimated rejection rates (in \%) of tests of $H_0$ with $\B = \bs{1}_p$ and $\SA = (1/n)\I_{p}$, performed at nominal level $5$\%. Each entry is based on $ 1000 $ $n \times d$ datasets drawn from a  $t_4$ copula  with Kendall's tau matrix $\bs{T}_\Delta$ in Eq.~\eqref{eq:departure} (a, single dep.) or (b, column dep.) with $\Delta= 0.1 $; $\bs{T}$ is as in Eq.~\eqref{eq:T-equi-null}.}
       \label{tab:sim-power-1-t4}
      \begin{center}
      \fontsize{8.75}{8.75}\selectfont
      \vskip-12pt
      \begin{tabular}{*{2}{l}*{12}{r}}
      \toprule
      \multicolumn{14}{c}{$E_{np}$ with $\SA = (1/n)\I_p$ for single departures ($\Delta =  0.1 $)}\\
      \midrule
      \multicolumn{2}{c}{} & & \multicolumn{3}{c}{$\tau = 0$} & & \multicolumn{3}{c}{$\tau = 0.3$} & & \multicolumn{3}{c}{$\tau = 0.6$}\\
      \cmidrule(lr){4-6}  \cmidrule(lr){8-10}  \cmidrule(lr){12-14}
      $\Sh_{np}$ & $d$\big|$n$ & & 50 & 100 & 150 & & 50 & 100 & 150 & & 50 & 100 & 150 \\
      \midrule
      &  5 &&  8.4 & 9.4 & 13.5 && 8.7 & 14.1 & 18.9 && 13.4 & 34.2 & 61.8 \\
      $\Sh_{np}^{\rm P}$ & 15 &&  0.1 & 0.7 & 1.6 && 1.2 & 2.3 & 3.6 && 0.8 & 1.8 & 5.2 \\
      & 25 &&  0 & 0 & 0 && 0.8 & 1.1 & 1 && 0.2 & 0.9 & 1.1 \\
      \cmidrule(lr){2-14}
      &  5 &&  5 & 7.7 & 11.9 && 5.8 & 11.7 & 17 && 6.7 & 27.4 & 57 \\
      & 15 &&  0 & 0.5 & 1.1 && 0.5 & 1.2 & 3.1 && 0.1 & 1.5 & 3.8 \\
      $\Sh_{np}^{\rm J}$ & 25 &&  0 & 0 & 0 && 0.4 & 0.9 & 0.8 && 0 & 0.3 & 0.6 \\
      & 50 &&  0 & 0 & 0 && 0 & 0 & 0 && 0 & 0 & 0 \\
      & 100 &&  0 & 0 & 0 && 0 & 0 & 0 && 0 & 0 & 0 \\
      \midrule
      \multicolumn{14}{c}{$M_{np}$ with $\SA = (1/n)\I_p$ for single departures ($\Delta =  0.1 $)}\\
      \midrule
      &  5 &&  10.1 & 12.1 & 17.5 && 10.6 & 18.8 & 29.8 && 24.1 & 55.3 & 81.1 \\
      $\Sh_{np}^{\rm P}$ & 15 &&  4.8 & 7.6 & 8.2 && 3.8 & 6.2 & 10.8 && 4.3 & 17.8 & 43.9 \\
      & 25 &&  3.9 & 4.9 & 4.9 && 3.8 & 5.1 & 7.2 && 3.5 & 8.8 & 30 \\
      \cmidrule(lr){2-14}
      &  5 &&  7.8 & 10.9 & 15.6 && 7.9 & 17 & 28.1 && 17.5 & 49.4 & 78.4 \\
      & 15 &&  3.5 & 5.9 & 7.1 && 2.5 & 4.9 & 9.6 && 2.5 & 14 & 40.1 \\
      $\Sh_{np}^{\rm J}$ & 25 &&  2.8 & 4.3 & 4.4 && 2.7 & 4.4 & 6.2 && 2.5 & 6.8 & 27.1 \\
      & 50 &&  1.4 & 2.4 & 3.3 && 1.1 & 3.4 & 4.5 && 0.9 & 2.1 & 10.4 \\
      & 100 &&  0.7 & 1.6 & 3 && 1.7 & 3.4 & 3.7 && 0.7 & 1.9 & 5.1 \\
      \midrule
      \multicolumn{14}{c}{$E_{np}$ with $\SA = (1/n)\I_p$ for column departures ($\Delta =  0.1 $)}\\
      \midrule
      \multicolumn{2}{c}{} & & \multicolumn{3}{c}{$\tau = 0$} & & \multicolumn{3}{c}{$\tau = 0.3$} & & \multicolumn{3}{c}{$\tau = 0.6$}\\
      \cmidrule(lr){4-6}  \cmidrule(lr){8-10}  \cmidrule(lr){12-14}
      $\Sh_{np}$ & $d$\big|$n$ & & 50 & 100 & 150 & & 50 & 100 & 150 & & 50 & 100 & 150 \\
      \midrule
      &  5 &&  11.4 & 24.9 & 37.4 && 18.2 & 39.6 & 56.1 && 48.8 & 86 & 96.4 \\
      $\Sh_{np}^{\rm P}$ & 15 &&  4.3 & 20 & 42.1 && 10.7 & 33 & 57 && 38.5 & 84.5 & 98.4 \\
      & 25 &&  1.6 & 7.8 & 28.2 && 4.5 & 24.1 & 43.7 && 17.3 & 72.1 & 94.9 \\
      \cmidrule(lr){2-14}
      &  5 &&  8.5 & 22.4 & 35 && 13.5 & 36.9 & 53.7 && 39.1 & 83.8 & 95.7 \\
      & 15 &&  2.7 & 15.6 & 38.4 && 7 & 26.6 & 51.2 && 27.9 & 78.3 & 97.4 \\
      $\Sh_{np}^{\rm J}$ & 25 &&  0.9 & 6 & 24.7 && 3.4 & 20.3 & 40.3 && 10.4 & 66.2 & 93.4 \\
      & 50 &&  0 & 0.3 & 2.5 && 0.2 & 2.5 & 16.2 && 1.4 & 31.2 & 76.7 \\
      & 100 &&  0 & 0 & 0 && 0 & 0 & 1.5 && 0 & 5.2 & 31.3 \\
      \midrule
      \multicolumn{14}{c}{$M_{np}$ with $\SA = (1/n)\I_p$ for column departures ($\Delta =  0.1 $)}\\
      \midrule
      \multicolumn{2}{c}{} & & \multicolumn{3}{c}{$\tau = 0$} & & \multicolumn{3}{c}{$\tau = 0.3$} & & \multicolumn{3}{c}{$\tau = 0.6$}\\
      \cmidrule(lr){4-6}  \cmidrule(lr){8-10}  \cmidrule(lr){12-14}
      $\Sh_{np}$ & $d$\big|$n$ & & 50 & 100 & 150 & & 50 & 100 & 150 & & 50 & 100 & 150 \\
      \midrule
      &  5 &&  10.7 & 18.7 & 23.7 && 16.7 & 29.3 & 36.9 && 38.7 & 71.6 & 87.8 \\
      $\Sh_{np}^{\rm P}$ & 15 &&  8.5 & 19.3 & 31 && 16.6 & 31.4 & 51.3 && 43.1 & 83 & 97 \\
      & 25 &&  7.4 & 18.1 & 30.8 && 12.7 & 34.5 & 53 && 35.3 & 80.6 & 97.2 \\
      \cmidrule(lr){2-14}
      &  5 &&  8.7 & 16.6 & 22.7 && 11.7 & 26 & 34.7 && 28.3 & 65.3 & 85.2 \\
      & 15 &&  6.6 & 16.7 & 28.1 && 13.6 & 28.2 & 48.3 && 35.6 & 79 & 95.6 \\
      $\Sh_{np}^{\rm J}$ & 25 &&  6 & 16.3 & 28.2 && 10.2 & 31.3 & 51.1 && 29.2 & 77.4 & 96.9 \\
      & 50 &&  4.3 & 13.1 & 23.4 && 10 & 26.3 & 47.4 && 25.7 & 78.1 & 97.5 \\
      & 100 &&  3.1 & 9.6 & 21.9 && 7.3 & 23 & 45.3 && 24.2 & 74.2 & 95.1 \\
      \bottomrule
      \end{tabular}
      \end{center}
      \vskip-9pt
      \small
      Statistics: $E_{np}$ Euclidean norm-based statistic defined in Eq.~\eqref{eq:euclidean}; $M_{np}$ supremum norm-based statistic defined in Eq.~\eqref{eq:supremum}. Estimators: $\Sh_{np}^{\rm P}$ plug-in estimator; $\Sh_{np}^{\rm J}$ jackknife estimator.
      \end{table}

            \begin{table}[htbp]
 \captionsetup{width=1\linewidth,font=small,skip=0pt}
    \caption{Estimated rejection rates (in \%) of tests of $H_0$ with $\B = \bs{1}_p$ and $\SA = (1/n)\I_{p}$, performed at nominal level $5$\%. Each entry is based on $ 1000 $ $n \times d$ datasets drawn from a  Gumbel copula  with Kendall's tau matrix $\bs{T}_\Delta$ in Eq.~\eqref{eq:departure} (a, single dep.) or (b, column dep.) with $\Delta= 0.1 $; $\bs{T}$ is as in Eq.~\eqref{eq:T-equi-null}.}
       \label{tab:sim-power-1-gumbel}
      \begin{center}
      \fontsize{8.75}{8.75}\selectfont
      \vskip-12pt
      \begin{tabular}{*{2}{l}*{12}{r}}
      \toprule
      \multicolumn{14}{c}{$E_{np}$ with $\SA = (1/n)\I_p$ for single departures ($\Delta =  0.1 $)}\\
      \midrule
      \multicolumn{2}{c}{} & & \multicolumn{3}{c}{$\tau = 0$} & & \multicolumn{3}{c}{$\tau = 0.3$} & & \multicolumn{3}{c}{$\tau = 0.6$}\\
      \cmidrule(lr){4-6}  \cmidrule(lr){8-10}  \cmidrule(lr){12-14}
      $\Sh_{np}$ & $d$\big|$n$ & & 50 & 100 & 150 & & 50 & 100 & 150 & & 50 & 100 & 150 \\
      \midrule
      &  5 &&  7.4 & 13.1 & 16.6 && 10.9 & 16.5 & 24.4 && 16.8 & 41.2 & 67.7 \\
      $\Sh_{np}^{\rm P}$ & 15 &&  0.5 & 3 & 3.8 && 1 & 3.8 & 4.8 && 1 & 5.1 & 7.1 \\
      & 25 &&  0 & 0.2 & 0.8 && 0.1 & 1.2 & 1.7 && 0.6 & 0.8 & 2.9 \\
      \cmidrule(lr){2-14}
      &  5 &&  5.4 & 10.3 & 15.1 && 6.3 & 13.9 & 22.3 && 8.8 & 33.3 & 63.1 \\
      & 15 &&  0.2 & 1.5 & 2 && 0.3 & 2.2 & 3.1 && 0.2 & 3.2 & 4.3 \\
      $\Sh_{np}^{\rm J}$ & 25 &&  0 & 0.1 & 0.4 && 0 & 0.7 & 0.9 && 0.1 & 0.2 & 1.4 \\
      & 50 &&  0 & 0 & 0 && 0 & 0 & 0 && 0 & 0 & 0.3 \\
      & 100 &&  0 & 0 & 0 && 0 & 0 & 0 && 0 & 0 & 0 \\
      \midrule
      \multicolumn{14}{c}{$M_{np}$ with $\SA = (1/n)\I_p$ for single departures ($\Delta =  0.1 $)}\\
      \midrule
      &  5 &&  10 & 15.6 & 19.8 && 13.1 & 21.3 & 33.4 && 29.9 & 63.3 & 85 \\
      $\Sh_{np}^{\rm P}$ & 15 &&  4.1 & 8.1 & 11 && 3.3 & 8.5 & 13.3 && 4.8 & 22.7 & 55.9 \\
      & 25 &&  2.8 & 4.9 & 7 && 3.3 & 5.2 & 9.8 && 2 & 11.5 & 35.7 \\
      \cmidrule(lr){2-14}
      &  5 &&  6.8 & 13.3 & 18.4 && 9.6 & 18.2 & 31.1 && 19.6 & 57.1 & 82.3 \\
      & 15 &&  2.8 & 6.1 & 9.4 && 2 & 6.9 & 11.1 && 2.5 & 17.6 & 50.7 \\
      $\Sh_{np}^{\rm J}$ & 25 &&  2.3 & 4 & 6 && 2.2 & 4.4 & 8.8 && 1.2 & 9.1 & 32.4 \\
      & 50 &&  1.7 & 2.5 & 3.2 && 1.6 & 2 & 4.8 && 0.7 & 3.2 & 15 \\
      & 100 &&  0.6 & 1.7 & 3.3 && 1 & 2.1 & 3.4 && 0.5 & 1.6 & 7.8 \\
      \midrule
      \multicolumn{14}{c}{$E_{np}$ with $\SA = (1/n)\I_p$ for column departures ($\Delta =  0.1 $)}\\
      \midrule
      \multicolumn{2}{c}{} & & \multicolumn{3}{c}{$\tau = 0$} & & \multicolumn{3}{c}{$\tau = 0.3$} & & \multicolumn{3}{c}{$\tau = 0.6$}\\
      \cmidrule(lr){4-6}  \cmidrule(lr){8-10}  \cmidrule(lr){12-14}
      $\Sh_{np}$ & $d$\big|$n$ & & 50 & 100 & 150 & & 50 & 100 & 150 & & 50 & 100 & 150 \\
      \midrule
      &  5 &&  16.5 & 31.4 & 49.7 && 22.8 & 45.9 & 62.6 && 51.7 & 87.2 & 97.4 \\
      $\Sh_{np}^{\rm P}$ & 15 &&  9.8 & 33.1 & 59.6 && 14 & 46.1 & 69.1 && 44.9 & 85.7 & 98.1 \\
      & 25 &&  2.4 & 19.6 & 48.1 && 5.4 & 33.2 & 59.4 && 27.1 & 79.7 & 96.8 \\
      \cmidrule(lr){2-14}
      &  5 &&  12.7 & 28.4 & 47 && 17.7 & 42.6 & 60.6 && 40.8 & 83 & 96.9 \\
      & 15 &&  6.8 & 27.1 & 54.4 && 8.6 & 38.7 & 65.4 && 32.4 & 80.2 & 97.1 \\
      $\Sh_{np}^{\rm J}$ & 25 &&  1.5 & 16.2 & 42.6 && 3.4 & 27.3 & 53.9 && 17.6 & 73.8 & 95.7 \\
      & 50 &&  0.2 & 2.4 & 14.7 && 0.3 & 8.2 & 27.2 && 2.5 & 44.3 & 82.2 \\
      & 100 &&  0 & 0 & 0.7 && 0 & 0.8 & 5.6 && 0.1 & 11.9 & 49.7 \\
      \midrule
      \multicolumn{14}{c}{$M_{np}$ with $\SA = (1/n)\I_p$ for column departures ($\Delta =  0.1 $)}\\
      \midrule
      \multicolumn{2}{c}{} & & \multicolumn{3}{c}{$\tau = 0$} & & \multicolumn{3}{c}{$\tau = 0.3$} & & \multicolumn{3}{c}{$\tau = 0.6$}\\
      \cmidrule(lr){4-6}  \cmidrule(lr){8-10}  \cmidrule(lr){12-14}
      $\Sh_{np}$ & $d$\big|$n$ & & 50 & 100 & 150 & & 50 & 100 & 150 & & 50 & 100 & 150 \\
      \midrule
      &  5 &&  16 & 22.1 & 35.5 && 19.1 & 32.4 & 42.1 && 40.1 & 74.1 & 89.6 \\
      $\Sh_{np}^{\rm P}$ & 15 &&  12.1 & 25.8 & 40.4 && 14.7 & 41.2 & 60.3 && 45 & 82.2 & 96.3 \\
      & 25 &&  10.2 & 23.4 & 41.3 && 12 & 38.6 & 61.5 && 36.7 & 80.2 & 97.5 \\
      \cmidrule(lr){2-14}
      &  5 &&  11.8 & 19.3 & 32.7 && 14.3 & 28.5 & 39.2 && 26.8 & 67.1 & 87.3 \\
      & 15 &&  9.1 & 23.2 & 38.1 && 10.4 & 36.8 & 57.4 && 35.7 & 77.9 & 95.4 \\
      $\Sh_{np}^{\rm J}$ & 25 &&  8.2 & 21.5 & 39.7 && 9.7 & 35 & 59.4 && 31.2 & 76.2 & 96.8 \\
      & 50 &&  5.9 & 21.1 & 36.2 && 8.7 & 33.2 & 53.1 && 23.8 & 73.4 & 94.8 \\
      & 100 &&  4.3 & 18.5 & 36.7 && 7.6 & 29.4 & 54 && 19.3 & 68.6 & 95 \\
      \bottomrule
      \end{tabular}
      \end{center}
      \vskip-9pt
      \small
      Statistics: $E_{np}$ Euclidean norm-based statistic defined in Eq.~\eqref{eq:euclidean}; $M_{np}$ supremum norm-based statistic defined in Eq.~\eqref{eq:supremum}. Estimators: $\Sh_{np}^{\rm P}$ plug-in estimator; $\Sh_{np}^{\rm J}$ jackknife estimator.
      \end{table}

            \begin{table}[htbp]
 \captionsetup{width=1\linewidth,font=small,skip=0pt}
    \caption{Estimated rejection rates (in \%) of tests of $H_0$ with $\B = \bs{1}_p$ and $\SA = (1/n)\I_{p}$, performed at nominal level $5$\%. Each entry is based on $ 1000 $ $n \times d$ datasets drawn from a  Clayton copula  with Kendall's tau matrix $\bs{T}_\Delta$ in Eq.~\eqref{eq:departure} (a, single dep.) or (b, column dep.) with $\Delta= 0.1 $; $\bs{T}$ is as in Eq.~\eqref{eq:T-equi-null}.}
       \label{tab:sim-power-1-clayton}
      \begin{center}
      \fontsize{8.75}{8.75}\selectfont
      \vskip-12pt
      \begin{tabular}{*{2}{l}*{12}{r}}
      \toprule
      \multicolumn{14}{c}{$E_{np}$ with $\SA = (1/n)\I_p$ for single departures ($\Delta =  0.1 $)}\\
      \midrule
      \multicolumn{2}{c}{} & & \multicolumn{3}{c}{$\tau = 0$} & & \multicolumn{3}{c}{$\tau = 0.3$} & & \multicolumn{3}{c}{$\tau = 0.6$}\\
      \cmidrule(lr){4-6}  \cmidrule(lr){8-10}  \cmidrule(lr){12-14}
      $\Sh_{np}$ & $d$\big|$n$ & & 50 & 100 & 150 & & 50 & 100 & 150 & & 50 & 100 & 150 \\
      \midrule
      &  5 &&   &  &  && 10 & 17.1 & 22.4 && 18.7 & 42.3 & 64.3 \\
      $\Sh_{np}^{\rm P}$ & 15 &&   &  &  && 2.2 & 3.1 & 5.5 && 1.9 & 5.2 & 7.2 \\
      & 25 &&   &  &  && 0.2 & 1.2 & 1.5 && 0.7 & 1 & 2.9 \\
      \cmidrule(lr){2-14}
      &  5 &&   &  &  && 7.1 & 14.4 & 19.9 && 11.6 & 36.1 & 60.5 \\
      & 15 &&   &  &  && 1.1 & 2.2 & 3.9 && 0.5 & 3.1 & 4.1 \\
      $\Sh_{np}^{\rm J}$ & 25 &&   &  &  && 0 & 0.7 & 1 && 0.2 & 0.6 & 1.2 \\
      & 50 &&   &  &  && 0 & 0.1 & 0.2 && 0 & 0 & 1 \\
      & 100 &&   &  &  && 0 & 0 & 0 && 0 & 0 & 0 \\
      \midrule
      \multicolumn{14}{c}{$M_{np}$ with $\SA = (1/n)\I_p$ for single departures ($\Delta =  0.1 $)}\\
      \midrule
      &  5 &&   &  &  && 11.6 & 21.2 & 31.1 && 29.4 & 62.3 & 82.6 \\
      $\Sh_{np}^{\rm P}$ & 15 &&   &  &  && 3.5 & 6.4 & 13.8 && 4.4 & 22.5 & 51.4 \\
      & 25 &&   &  &  && 2.3 & 3.9 & 6 && 2.5 & 12.1 & 31.9 \\
      \cmidrule(lr){2-14}
      &  5 &&   &  &  && 8.6 & 18.8 & 28.9 && 20.4 & 56 & 80.5 \\
      & 15 &&   &  &  && 2.5 & 4.6 & 12.6 && 2.6 & 17.4 & 47.1 \\
      $\Sh_{np}^{\rm J}$ & 25 &&   &  &  && 2 & 3.4 & 5.5 && 1.4 & 10.2 & 28.4 \\
      & 50 &&   &  &  && 1.3 & 2 & 3.7 && 1.2 & 3.3 & 13.2 \\
      & 100 &&   &  &  && 1.1 & 2.4 & 2.4 && 0.7 & 2.5 & 7.7 \\
      \midrule
      \multicolumn{14}{c}{$E_{np}$ with $\SA = (1/n)\I_p$ for column departures ($\Delta =  0.1 $)}\\
      \midrule
      \multicolumn{2}{c}{} & & \multicolumn{3}{c}{$\tau = 0$} & & \multicolumn{3}{c}{$\tau = 0.3$} & & \multicolumn{3}{c}{$\tau = 0.6$}\\
      \cmidrule(lr){4-6}  \cmidrule(lr){8-10}  \cmidrule(lr){12-14}
      $\Sh_{np}$ & $d$\big|$n$ & & 50 & 100 & 150 & & 50 & 100 & 150 & & 50 & 100 & 150 \\
      \midrule
      &  5 &&   &  &  && 24.4 & 45.3 & 63.9 && 53.7 & 88.2 & 97.7 \\
      $\Sh_{np}^{\rm P}$ & 15 &&   &  &  && 14.4 & 43.5 & 64.8 && 43.2 & 86.9 & 98.2 \\
      & 25 &&   &  &  && 6.2 & 27.9 & 54.7 && 27.6 & 79.5 & 95.7 \\
      \cmidrule(lr){2-14}
      &  5 &&   &  &  && 19.6 & 41.4 & 62 && 43.9 & 84.7 & 96.8 \\
      & 15 &&   &  &  && 9.4 & 37 & 60.5 && 30.2 & 81.3 & 97.2 \\
      $\Sh_{np}^{\rm J}$ & 25 &&   &  &  && 3.9 & 22.9 & 50 && 17.6 & 71.2 & 94.8 \\
      & 50 &&   &  &  && 0.3 & 6.8 & 23.1 && 2.2 & 42.6 & 82.6 \\
      & 100 &&   &  &  && 0.1 & 0.8 & 5.5 && 0 & 12.5 & 51 \\
      \midrule
      \multicolumn{14}{c}{$M_{np}$ with $\SA = (1/n)\I_p$ for column departures ($\Delta =  0.1 $)}\\
      \midrule
      \multicolumn{2}{c}{} & & \multicolumn{3}{c}{$\tau = 0$} & & \multicolumn{3}{c}{$\tau = 0.3$} & & \multicolumn{3}{c}{$\tau = 0.6$}\\
      \cmidrule(lr){4-6}  \cmidrule(lr){8-10}  \cmidrule(lr){12-14}
      $\Sh_{np}$ & $d$\big|$n$ & & 50 & 100 & 150 & & 50 & 100 & 150 & & 50 & 100 & 150 \\
      \midrule
      &  5 &&   &  &  && 19.8 & 33.9 & 47.1 && 43.1 & 71.8 & 90.1 \\
      $\Sh_{np}^{\rm P}$ & 15 &&   &  &  && 15.4 & 36.3 & 55.6 && 41.5 & 82.6 & 95.6 \\
      & 25 &&   &  &  && 13.1 & 33.7 & 54.1 && 38.7 & 82.6 & 96.7 \\
      \cmidrule(lr){2-14}
      &  5 &&   &  &  && 14.7 & 28.7 & 44.8 && 31.3 & 65.7 & 88.4 \\
      & 15 &&   &  &  && 11.8 & 33 & 52.1 && 31.5 & 78.7 & 94.9 \\
      $\Sh_{np}^{\rm J}$ & 25 &&   &  &  && 10.6 & 30.6 & 50.9 && 31.3 & 79.6 & 95.9 \\
      & 50 &&   &  &  && 6.6 & 26.4 & 51.6 && 21.2 & 74.2 & 94.5 \\
      & 100 &&   &  &  && 6.1 & 25.2 & 44.5 && 17.9 & 70.8 & 95.5 \\
      \bottomrule
      \end{tabular}
      \end{center}
      \vskip-9pt
      \small
      Statistics: $E_{np}$ Euclidean norm-based statistic defined in Eq.~\eqref{eq:euclidean}; $M_{np}$ supremum norm-based statistic defined in Eq.~\eqref{eq:supremum}. Estimators: $\Sh_{np}^{\rm P}$ plug-in estimator; $\Sh_{np}^{\rm J}$ jackknife estimator.
      \end{table}

            \begin{table}[htbp]
 \captionsetup{width=1\linewidth,font=small,skip=0pt}
    \caption{Estimated rejection rates (in \%) of tests of $H_0$ with $\B = \bs{1}_p$ and $\SA = (1/n)\I_{p}$, performed at nominal level $5$\%. Each entry is based on $ 2500 $ $n \times d$ datasets drawn from a  Normal copula  with Kendall's tau matrix $\bs{T}_\Delta$ in Eq.~\eqref{eq:departure} (a, single dep.) or (b, column dep.) with $\Delta= 0.2 $; $\bs{T}$ is as in Eq.~\eqref{eq:T-equi-null}.}
       \label{tab:sim-power-2-normal}
      \begin{center}
      \fontsize{8.75}{8.75}\selectfont
      \vskip-12pt
      \begin{tabular}{*{2}{l}*{12}{r}}
      \toprule
      \multicolumn{14}{c}{$E_{np}$ with $\SA = (1/n)\I_p$ for single departures ($\Delta =  0.2 $)}\\
      \midrule
      \multicolumn{2}{c}{} & & \multicolumn{3}{c}{$\tau = 0$} & & \multicolumn{3}{c}{$\tau = 0.3$} & & \multicolumn{3}{c}{$\tau = 0.6$}\\
      \cmidrule(lr){4-6}  \cmidrule(lr){8-10}  \cmidrule(lr){12-14}
      $\Sh_{np}$ & $d$\big|$n$ & & 50 & 100 & 150 & & 50 & 100 & 150 & & 50 & 100 & 150 \\
      \midrule
      &  5 &&  21.2 & 44.7 & 66.1 && 34.8 & 72.5 & 92.8 && 91 & 100 & 100 \\
      $\Sh_{np}^{\rm P}$ & 15 &&  1.1 & 5.1 & 13.4 && 2.6 & 8.8 & 17 && 4.2 & 32.5 & 73.8 \\
      & 25 &&  0 & 0.4 & 1 && 0.4 & 2.5 & 5.8 && 0.6 & 6 & 15.1 \\
      \cmidrule(lr){2-14}
      &  5 &&  17.3 & 41 & 64.1 && 25.9 & 68 & 91.4 && 80.1 & 99.9 & 100 \\
      & 15 &&  0.6 & 2.6 & 9.5 && 1 & 5.2 & 11.8 && 0.8 & 17 & 58.8 \\
      $\Sh_{np}^{\rm J}$ & 25 &&  0 & 0.2 & 0.4 && 0.1 & 1.3 & 3.8 && 0 & 2.4 & 9.4 \\
      & 50 &&  0 & 0 & 0 && 0 & 0 & 0.4 && 0 & 0 & 0.5 \\
      & 100 &&  0 & 0 & 0 && 0 & 0 & 0 && 0 & 0 & 0 \\
      \midrule
      \multicolumn{14}{c}{$M_{np}$ with $\SA = (1/n)\I_p$ for single departures ($\Delta =  0.2 $)}\\
      \midrule
      &  5 &&  27.7 & 56.2 & 77.7 && 51.4 & 87.7 & 98.7 && 98.5 & 100 & 100 \\
      $\Sh_{np}^{\rm P}$ & 15 &&  9.3 & 33 & 59 && 15.6 & 60.4 & 89.6 && 84.4 & 100 & 100 \\
      & 25 &&  5.5 & 22.2 & 45.6 && 7.1 & 46.2 & 81.2 && 63.6 & 100 & 100 \\
      \cmidrule(lr){2-14}
      &  5 &&  23.1 & 53.3 & 76 && 42.8 & 85.9 & 98.3 && 97.1 & 100 & 100 \\
      & 15 &&  6.8 & 30.2 & 56.8 && 12 & 56.5 & 88 && 76.5 & 100 & 100 \\
      $\Sh_{np}^{\rm J}$ & 25 &&  4.5 & 20.3 & 43.6 && 5.6 & 43.1 & 79.6 && 53.6 & 100 & 100 \\
      & 50 &&  1.7 & 12.4 & 32.6 && 2.8 & 25.9 & 64.3 && 19.4 & 99.5 & 100 \\
      & 100 &&  0.8 & 5.8 & 19.3 && 1.6 & 13.2 & 45.5 && 4.6 & 97.6 & 100 \\
      \midrule
      \multicolumn{14}{c}{$E_{np}$ with $\SA = (1/n)\I_p$ for column departures ($\Delta =  0.2 $)}\\
      \midrule
      \multicolumn{2}{c}{} & & \multicolumn{3}{c}{$\tau = 0$} & & \multicolumn{3}{c}{$\tau = 0.3$} & & \multicolumn{3}{c}{$\tau = 0.6$}\\
      \cmidrule(lr){4-6}  \cmidrule(lr){8-10}  \cmidrule(lr){12-14}
      $\Sh_{np}$ & $d$\big|$n$ & & 50 & 100 & 150 & & 50 & 100 & 150 & & 50 & 100 & 150 \\
      \midrule
      &  5 &&  55.2 & 85.8 & 97 && 80.8 & 98.3 & 99.9 && 100 & 100 & 100 \\
      $\Sh_{np}^{\rm P}$ & 15 &&  55.8 & 95.1 & 99.7 && 80.2 & 99.6 & 100 && 99.9 & 100 & 100 \\
      & 25 &&  38.4 & 90.6 & 99.6 && 69.7 & 98.9 & 100 && 99.6 & 100 & 100 \\
      \cmidrule(lr){2-14}
      &  5 &&  49.4 & 83.9 & 96.6 && 75 & 98 & 99.9 && 99.8 & 100 & 100 \\
      & 15 &&  47.1 & 93.2 & 99.6 && 72.7 & 99.2 & 100 && 99.5 & 100 & 100 \\
      $\Sh_{np}^{\rm J}$ & 25 &&  31.4 & 87.9 & 99.4 && 59.8 & 98.4 & 100 && 99.1 & 100 & 100 \\
      & 50 &&  5.8 & 64.6 & 95.9 && 26.5 & 92.2 & 99.6 && 94 & 100 & 100 \\
      & 100 &&  0.2 & 19.8 & 72.4 && 4.5 & 67.2 & 97.4 && 66.6 & 100 & 100 \\
      \midrule
      \multicolumn{14}{c}{$M_{np}$ with $\SA = (1/n)\I_p$ for column departures ($\Delta =  0.2 $)}\\
      \midrule
      \multicolumn{2}{c}{} & & \multicolumn{3}{c}{$\tau = 0$} & & \multicolumn{3}{c}{$\tau = 0.3$} & & \multicolumn{3}{c}{$\tau = 0.6$}\\
      \cmidrule(lr){4-6}  \cmidrule(lr){8-10}  \cmidrule(lr){12-14}
      $\Sh_{np}$ & $d$\big|$n$ & & 50 & 100 & 150 & & 50 & 100 & 150 & & 50 & 100 & 150 \\
      \midrule
      &  5 &&  40.3 & 69.3 & 86.8 && 66 & 94.1 & 99.1 && 99.1 & 100 & 100 \\
      $\Sh_{np}^{\rm P}$ & 15 &&  50.5 & 87.2 & 97.7 && 76.6 & 99 & 100 && 99.8 & 100 & 100 \\
      & 25 &&  50.5 & 89 & 99 && 77 & 98.9 & 100 && 99.8 & 100 & 100 \\
      \cmidrule(lr){2-14}
      &  5 &&  34.3 & 65.8 & 85.7 && 57 & 92.3 & 98.7 && 97.6 & 100 & 100 \\
      & 15 &&  45.4 & 84.8 & 97.4 && 71.3 & 98.6 & 100 && 99.6 & 100 & 100 \\
      $\Sh_{np}^{\rm J}$ & 25 &&  45.9 & 87.5 & 99 && 72.9 & 98.8 & 100 && 99.6 & 100 & 100 \\
      & 50 &&  43.3 & 88.6 & 98.9 && 67.9 & 99 & 100 && 99.6 & 100 & 100 \\
      & 100 &&  35.6 & 87.1 & 98.2 && 66.3 & 98.2 & 100 && 99.5 & 100 & 100 \\
      \bottomrule
      \end{tabular}
      \end{center}
      \vskip-9pt
      \small
      Statistics: $E_{np}$ Euclidean norm-based statistic defined in Eq.~\eqref{eq:euclidean}; $M_{np}$ supremum norm-based statistic defined in Eq.~\eqref{eq:supremum}. Estimators: $\Sh_{np}^{\rm P}$ plug-in estimator; $\Sh_{np}^{\rm J}$ jackknife estimator.
      \end{table}

            \begin{table}[htbp]
 \captionsetup{width=1\linewidth,font=small,skip=0pt}
    \caption{Estimated rejection rates (in \%) of tests of $H_0$ with $\B = \bs{1}_p$ and $\SA = (1/n)\I_{p}$, performed at nominal level $5$\%. Each entry is based on $ 1000 $ $n \times d$ datasets drawn from a  $t_4$ copula  with Kendall's tau matrix $\bs{T}_\Delta$ in Eq.~\eqref{eq:departure} (a, single dep.) or (b, column dep.) with $\Delta= 0.2 $; $\bs{T}$ is as in Eq.~\eqref{eq:T-equi-null}.}
       \label{tab:sim-power-2-t4}
      \begin{center}
      \fontsize{8.75}{8.75}\selectfont
      \vskip-12pt
      \begin{tabular}{*{2}{l}*{12}{r}}
      \toprule
      \multicolumn{14}{c}{$E_{np}$ with $\SA = (1/n)\I_p$ for single departures ($\Delta =  0.2 $)}\\
      \midrule
      \multicolumn{2}{c}{} & & \multicolumn{3}{c}{$\tau = 0$} & & \multicolumn{3}{c}{$\tau = 0.3$} & & \multicolumn{3}{c}{$\tau = 0.6$}\\
      \cmidrule(lr){4-6}  \cmidrule(lr){8-10}  \cmidrule(lr){12-14}
      $\Sh_{np}$ & $d$\big|$n$ & & 50 & 100 & 150 & & 50 & 100 & 150 & & 50 & 100 & 150 \\
      \midrule
      &  5 &&  17 & 34.8 & 52.5 && 24.3 & 55.2 & 81.6 && 76.8 & 99.8 & 100 \\
      $\Sh_{np}^{\rm P}$ & 15 &&  0.8 & 2.2 & 6.1 && 1.2 & 4.7 & 10.9 && 1.9 & 12.9 & 35.3 \\
      & 25 &&  0 & 0 & 0.6 && 0.1 & 1.3 & 2.3 && 0.5 & 1 & 5.1 \\
      \cmidrule(lr){2-14}
      &  5 &&  12.5 & 32 & 49.9 && 18.8 & 52.1 & 79.5 && 63.2 & 99.7 & 100 \\
      & 15 &&  0.2 & 1.1 & 4.2 && 0.5 & 2.8 & 8.8 && 0.7 & 6.8 & 26.8 \\
      $\Sh_{np}^{\rm J}$ & 25 &&  0 & 0 & 0.3 && 0 & 0.7 & 2 && 0.1 & 0.4 & 3 \\
      & 50 &&  0 & 0 & 0 && 0 & 0 & 0.2 && 0 & 0 & 0.2 \\
      & 100 &&  0 & 0 & 0 && 0 & 0 & 0 && 0 & 0 & 0 \\
      \midrule
      \multicolumn{14}{c}{$M_{np}$ with $\SA = (1/n)\I_p$ for single departures ($\Delta =  0.2 $)}\\
      \midrule
      &  5 &&  23.4 & 44 & 65 && 39.4 & 75.2 & 93.8 && 94.7 & 100 & 100 \\
      $\Sh_{np}^{\rm P}$ & 15 &&  7.9 & 24.4 & 46 && 11.2 & 44.7 & 76.7 && 58 & 99.7 & 100 \\
      & 25 &&  5.2 & 14.6 & 34.7 && 5.7 & 27.6 & 65 && 31.6 & 98.3 & 100 \\
      \cmidrule(lr){2-14}
      &  5 &&  20.7 & 41.8 & 63.9 && 32.8 & 72.1 & 93.3 && 91.7 & 100 & 100 \\
      & 15 &&  6.5 & 22.5 & 43.6 && 7.7 & 40 & 74.7 && 48.8 & 99.6 & 100 \\
      $\Sh_{np}^{\rm J}$ & 25 &&  4.4 & 13.7 & 33.2 && 4.7 & 25.6 & 63.2 && 24 & 97.9 & 100 \\
      & 50 &&  2.1 & 8.2 & 21.9 && 2 & 13.5 & 41.9 && 6.3 & 90 & 99.8 \\
      & 100 &&  1.8 & 4.6 & 14.2 && 2.1 & 7.1 & 28.4 && 1.2 & 75.9 & 99.8 \\
      \midrule
      \multicolumn{14}{c}{$E_{np}$ with $\SA = (1/n)\I_p$ for column departures ($\Delta =  0.2 $)}\\
      \midrule
      \multicolumn{2}{c}{} & & \multicolumn{3}{c}{$\tau = 0$} & & \multicolumn{3}{c}{$\tau = 0.3$} & & \multicolumn{3}{c}{$\tau = 0.6$}\\
      \cmidrule(lr){4-6}  \cmidrule(lr){8-10}  \cmidrule(lr){12-14}
      $\Sh_{np}$ & $d$\big|$n$ & & 50 & 100 & 150 & & 50 & 100 & 150 & & 50 & 100 & 150 \\
      \midrule
      &  5 &&  46.8 & 80.5 & 92.6 && 67.8 & 94.8 & 99.5 && 99.5 & 100 & 100 \\
      $\Sh_{np}^{\rm P}$ & 15 &&  42.4 & 86.3 & 97 && 65.8 & 96.7 & 99.8 && 98.8 & 100 & 100 \\
      & 25 &&  24.1 & 79.8 & 97.5 && 48.5 & 93.3 & 99.7 && 96.8 & 100 & 100 \\
      \cmidrule(lr){2-14}
      &  5 &&  43.7 & 78.7 & 91.7 && 62.2 & 94.2 & 99.5 && 98.6 & 100 & 100 \\
      & 15 &&  35.9 & 83.3 & 96.8 && 58.8 & 95.7 & 99.8 && 98.3 & 100 & 100 \\
      $\Sh_{np}^{\rm J}$ & 25 &&  18.3 & 76.6 & 96.9 && 42.1 & 91.6 & 99.6 && 95.4 & 100 & 100 \\
      & 50 &&  1.6 & 38.1 & 85.5 && 12.5 & 75.2 & 96.1 && 78.2 & 100 & 100 \\
      & 100 &&  0 & 4.6 & 36.1 && 0.9 & 33.4 & 80.4 && 39.2 & 99.7 & 100 \\
      \midrule
      \multicolumn{14}{c}{$M_{np}$ with $\SA = (1/n)\I_p$ for column departures ($\Delta =  0.2 $)}\\
      \midrule
      \multicolumn{2}{c}{} & & \multicolumn{3}{c}{$\tau = 0$} & & \multicolumn{3}{c}{$\tau = 0.3$} & & \multicolumn{3}{c}{$\tau = 0.6$}\\
      \cmidrule(lr){4-6}  \cmidrule(lr){8-10}  \cmidrule(lr){12-14}
      $\Sh_{np}$ & $d$\big|$n$ & & 50 & 100 & 150 & & 50 & 100 & 150 & & 50 & 100 & 150 \\
      \midrule
      &  5 &&  36 & 61.3 & 77.5 && 54.9 & 84.6 & 96.3 && 97.3 & 100 & 100 \\
      $\Sh_{np}^{\rm P}$ & 15 &&  46 & 79.8 & 93.3 && 66 & 95.9 & 99.7 && 98.7 & 100 & 100 \\
      & 25 &&  45.2 & 82.1 & 96.9 && 69.3 & 94.7 & 99.6 && 98.7 & 100 & 100 \\
      \cmidrule(lr){2-14}
      &  5 &&  31 & 58.7 & 76.5 && 47.5 & 82.2 & 95.3 && 94.2 & 100 & 100 \\
      & 15 &&  39.7 & 77.2 & 93 && 59.9 & 95 & 99.6 && 98.2 & 100 & 100 \\
      $\Sh_{np}^{\rm J}$ & 25 &&  41.2 & 80.8 & 96.6 && 65.1 & 94.3 & 99.6 && 98 & 100 & 100 \\
      & 50 &&  37.7 & 79.5 & 96.6 && 58.9 & 96.3 & 99.6 && 98.3 & 100 & 100 \\
      & 100 &&  30.8 & 76.9 & 97 && 54.9 & 95.7 & 99.5 && 97.9 & 100 & 100 \\
      \bottomrule
      \end{tabular}
      \end{center}
      \vskip-9pt
      \small
      Statistics: $E_{np}$ Euclidean norm-based statistic defined in Eq.~\eqref{eq:euclidean}; $M_{np}$ supremum norm-based statistic defined in Eq.~\eqref{eq:supremum}. Estimators: $\Sh_{np}^{\rm P}$ plug-in estimator; $\Sh_{np}^{\rm J}$ jackknife estimator.
      \end{table}

            \begin{table}[htbp]
 \captionsetup{width=1\linewidth,font=small,skip=0pt}
    \caption{Estimated rejection rates (in \%) of tests of $H_0$ with $\B = \bs{1}_p$ and $\SA = (1/n)\I_{p}$, performed at nominal level $5$\%. Each entry is based on $ 1000 $ $n \times d$ datasets drawn from a  Gumbel copula  with Kendall's tau matrix $\bs{T}_\Delta$ in Eq.~\eqref{eq:departure} (a, single dep.) or (b, column dep.) with $\Delta= 0.2 $; $\bs{T}$ is as in Eq.~\eqref{eq:T-equi-null}.}
       \label{tab:sim-power-2-gumbel}
      \begin{center}
      \fontsize{8.75}{8.75}\selectfont
      \vskip-12pt
      \begin{tabular}{*{2}{l}*{12}{r}}
      \toprule
      \multicolumn{14}{c}{$E_{np}$ with $\SA = (1/n)\I_p$ for single departures ($\Delta =  0.2 $)}\\
      \midrule
      \multicolumn{2}{c}{} & & \multicolumn{3}{c}{$\tau = 0$} & & \multicolumn{3}{c}{$\tau = 0.3$} & & \multicolumn{3}{c}{$\tau = 0.6$}\\
      \cmidrule(lr){4-6}  \cmidrule(lr){8-10}  \cmidrule(lr){12-14}
      $\Sh_{np}$ & $d$\big|$n$ & & 50 & 100 & 150 & & 50 & 100 & 150 & & 50 & 100 & 150 \\
      \midrule
      &  5 &&  22.2 & 46.2 & 67.7 && 32.9 & 70.3 & 89.8 && 84 & 99.8 & 100 \\
      $\Sh_{np}^{\rm P}$ & 15 &&  1.3 & 7 & 14.9 && 3.1 & 6.5 & 16.4 && 3.6 & 20.6 & 48.3 \\
      & 25 &&  0 & 0.4 & 1.6 && 0.1 & 1.6 & 4.3 && 0.9 & 3.2 & 9.7 \\
      \cmidrule(lr){2-14}
      &  5 &&  17.1 & 41.9 & 64.2 && 24.3 & 64.5 & 88.5 && 70.8 & 99.7 & 100 \\
      & 15 &&  0.2 & 3.5 & 10.1 && 1.3 & 2.9 & 11.3 && 0.8 & 11.7 & 35.7 \\
      $\Sh_{np}^{\rm J}$ & 25 &&  0 & 0.1 & 0.9 && 0.1 & 0.6 & 2.7 && 0 & 1.7 & 5.9 \\
      & 50 &&  0 & 0 & 0 && 0 & 0 & 0.1 && 0 & 0.1 & 0.5 \\
      & 100 &&  0 & 0 & 0 && 0 & 0 & 0 && 0 & 0 & 0 \\
      \midrule
      \multicolumn{14}{c}{$M_{np}$ with $\SA = (1/n)\I_p$ for single departures ($\Delta =  0.2 $)}\\
      \midrule
      &  5 &&  29.1 & 55.7 & 78.4 && 49.1 & 85.8 & 96.8 && 96.7 & 100 & 100 \\
      $\Sh_{np}^{\rm P}$ & 15 &&  10.4 & 33.5 & 59.7 && 16.6 & 57.3 & 89.5 && 71.6 & 100 & 100 \\
      & 25 &&  7.3 & 25.3 & 44.8 && 6.9 & 44.7 & 78.4 && 38.5 & 99.7 & 100 \\
      \cmidrule(lr){2-14}
      &  5 &&  24.5 & 53.4 & 77.2 && 41.6 & 83.7 & 95.9 && 94.7 & 100 & 100 \\
      & 15 &&  7.2 & 31.1 & 58 && 12.2 & 53.6 & 88.3 && 61.3 & 99.9 & 100 \\
      $\Sh_{np}^{\rm J}$ & 25 &&  6.4 & 23.9 & 43.2 && 4.5 & 42.2 & 77 && 30.7 & 99.6 & 100 \\
      & 50 &&  1.7 & 11.2 & 29.8 && 2.9 & 25.1 & 64.7 && 7.4 & 95.6 & 99.8 \\
      & 100 &&  1.3 & 6.4 & 23 && 1.6 & 14.2 & 46 && 1.3 & 85.3 & 99.9 \\
      \midrule
      \multicolumn{14}{c}{$E_{np}$ with $\SA = (1/n)\I_p$ for column departures ($\Delta =  0.2 $)}\\
      \midrule
      \multicolumn{2}{c}{} & & \multicolumn{3}{c}{$\tau = 0$} & & \multicolumn{3}{c}{$\tau = 0.3$} & & \multicolumn{3}{c}{$\tau = 0.6$}\\
      \cmidrule(lr){4-6}  \cmidrule(lr){8-10}  \cmidrule(lr){12-14}
      $\Sh_{np}$ & $d$\big|$n$ & & 50 & 100 & 150 & & 50 & 100 & 150 & & 50 & 100 & 150 \\
      \midrule
      &  5 &&  49.6 & 84.9 & 94.8 && 73.5 & 97.2 & 100 && 99.6 & 100 & 100 \\
      $\Sh_{np}^{\rm P}$ & 15 &&  52.4 & 94.8 & 99.3 && 73.6 & 98.2 & 100 && 99.3 & 100 & 100 \\
      & 25 &&  36 & 89.5 & 99 && 60 & 97.2 & 99.8 && 99.4 & 100 & 100 \\
      \cmidrule(lr){2-14}
      &  5 &&  43.9 & 83.5 & 94.2 && 68.5 & 97 & 100 && 99.4 & 100 & 100 \\
      & 15 &&  44.7 & 92.6 & 99.3 && 65.5 & 97.8 & 100 && 98.5 & 100 & 100 \\
      $\Sh_{np}^{\rm J}$ & 25 &&  30.3 & 87.3 & 98.8 && 51.5 & 96.4 & 99.7 && 97.7 & 100 & 100 \\
      & 50 &&  7.7 & 66.2 & 94.9 && 22.2 & 83.5 & 99 && 85.5 & 100 & 100 \\
      & 100 &&  0.5 & 21.6 & 71.4 && 3.1 & 53.2 & 92.8 && 52.2 & 99.9 & 100 \\
      \midrule
      \multicolumn{14}{c}{$M_{np}$ with $\SA = (1/n)\I_p$ for column departures ($\Delta =  0.2 $)}\\
      \midrule
      \multicolumn{2}{c}{} & & \multicolumn{3}{c}{$\tau = 0$} & & \multicolumn{3}{c}{$\tau = 0.3$} & & \multicolumn{3}{c}{$\tau = 0.6$}\\
      \cmidrule(lr){4-6}  \cmidrule(lr){8-10}  \cmidrule(lr){12-14}
      $\Sh_{np}$ & $d$\big|$n$ & & 50 & 100 & 150 & & 50 & 100 & 150 & & 50 & 100 & 150 \\
      \midrule
      &  5 &&  38.4 & 69.2 & 83.6 && 59.1 & 91 & 98.8 && 97.2 & 100 & 100 \\
      $\Sh_{np}^{\rm P}$ & 15 &&  48.3 & 87.1 & 96.5 && 71.6 & 97.4 & 99.9 && 99.4 & 100 & 100 \\
      & 25 &&  48.1 & 87.7 & 97.9 && 72.1 & 98 & 99.9 && 99.6 & 100 & 100 \\
      \cmidrule(lr){2-14}
      &  5 &&  32.3 & 65.8 & 82.1 && 51 & 89.1 & 98.2 && 94.6 & 100 & 100 \\
      & 15 &&  43.1 & 84.5 & 95.8 && 66.9 & 96.7 & 99.9 && 98.8 & 100 & 100 \\
      $\Sh_{np}^{\rm J}$ & 25 &&  44.2 & 85.6 & 97.7 && 68.8 & 97.6 & 99.9 && 99.2 & 100 & 100 \\
      & 50 &&  40.1 & 87.6 & 98.9 && 60.3 & 96.5 & 99.7 && 98 & 100 & 100 \\
      & 100 &&  40.2 & 86.3 & 98.1 && 56.5 & 97.2 & 99.9 && 97.1 & 100 & 100 \\
      \bottomrule
      \end{tabular}
      \end{center}
      \vskip-9pt
      \small
      Statistics: $E_{np}$ Euclidean norm-based statistic defined in Eq.~\eqref{eq:euclidean}; $M_{np}$ supremum norm-based statistic defined in Eq.~\eqref{eq:supremum}. Estimators: $\Sh_{np}^{\rm P}$ plug-in estimator; $\Sh_{np}^{\rm J}$ jackknife estimator.
      \end{table}

            \begin{table}[htbp]
 \captionsetup{width=1\linewidth,font=small,skip=0pt}
    \caption{Estimated rejection rates (in \%) of tests of $H_0$ with $\B = \bs{1}_p$ and $\SA = (1/n)\I_{p}$, performed at nominal level $5$\%. Each entry is based on $ 1000 $ $n \times d$ datasets drawn from a  Clayton copula  with Kendall's tau matrix $\bs{T}_\Delta$ in Eq.~\eqref{eq:departure} (a, single dep.) or (b, column dep.) with $\Delta= 0.2 $; $\bs{T}$ is as in Eq.~\eqref{eq:T-equi-null}.}
       \label{tab:sim-power-2-clayton}
      \begin{center}
      \fontsize{8.75}{8.75}\selectfont
      \vskip-12pt
      \begin{tabular}{*{2}{l}*{12}{r}}
      \toprule
      \multicolumn{14}{c}{$E_{np}$ with $\SA = (1/n)\I_p$ for single departures ($\Delta =  0.2 $)}\\
      \midrule
      \multicolumn{2}{c}{} & & \multicolumn{3}{c}{$\tau = 0$} & & \multicolumn{3}{c}{$\tau = 0.3$} & & \multicolumn{3}{c}{$\tau = 0.6$}\\
      \cmidrule(lr){4-6}  \cmidrule(lr){8-10}  \cmidrule(lr){12-14}
      $\Sh_{np}$ & $d$\big|$n$ & & 50 & 100 & 150 & & 50 & 100 & 150 & & 50 & 100 & 150 \\
      \midrule
      &  5 &&   &  &  && 29.1 & 67.2 & 88.3 && 82.2 & 99.6 & 100 \\
      $\Sh_{np}^{\rm P}$ & 15 &&   &  &  && 3 & 7.8 & 12.5 && 4.3 & 18 & 38.1 \\
      & 25 &&   &  &  && 0.6 & 2.9 & 4.4 && 1 & 4.1 & 9.5 \\
      \cmidrule(lr){2-14}
      &  5 &&   &  &  && 20.7 & 62.2 & 86.2 && 70.2 & 99.6 & 100 \\
      & 15 &&   &  &  && 1.1 & 5.3 & 9.8 && 1.4 & 9.4 & 28.6 \\
      $\Sh_{np}^{\rm J}$ & 25 &&   &  &  && 0.4 & 1.7 & 2.9 && 0.2 & 2 & 5.4 \\
      & 50 &&   &  &  && 0 & 0 & 0.4 && 0 & 0 & 0.5 \\
      & 100 &&   &  &  && 0 & 0 & 0 && 0 & 0 & 0.1 \\
      \midrule
      \multicolumn{14}{c}{$M_{np}$ with $\SA = (1/n)\I_p$ for single departures ($\Delta =  0.2 $)}\\
      \midrule
      &  5 &&   &  &  && 42.6 & 82.8 & 95.9 && 96 & 99.9 & 100 \\
      $\Sh_{np}^{\rm P}$ & 15 &&   &  &  && 12.4 & 55.1 & 84.7 && 66.1 & 99.8 & 100 \\
      & 25 &&   &  &  && 6.8 & 42.1 & 74.3 && 35.5 & 98.9 & 99.9 \\
      \cmidrule(lr){2-14}
      &  5 &&   &  &  && 35.3 & 81.4 & 95.4 && 92.6 & 99.9 & 100 \\
      & 15 &&   &  &  && 9.7 & 51.2 & 83.1 && 57.9 & 99.7 & 100 \\
      $\Sh_{np}^{\rm J}$ & 25 &&   &  &  && 5.5 & 38.9 & 73.1 && 27.7 & 98.5 & 99.9 \\
      & 50 &&   &  &  && 2.4 & 18.3 & 51.7 && 6.7 & 93.4 & 100 \\
      & 100 &&   &  &  && 1.6 & 10.1 & 37.8 && 1.3 & 80.6 & 99.9 \\
      \midrule
      \multicolumn{14}{c}{$E_{np}$ with $\SA = (1/n)\I_p$ for column departures ($\Delta =  0.2 $)}\\
      \midrule
      \multicolumn{2}{c}{} & & \multicolumn{3}{c}{$\tau = 0$} & & \multicolumn{3}{c}{$\tau = 0.3$} & & \multicolumn{3}{c}{$\tau = 0.6$}\\
      \cmidrule(lr){4-6}  \cmidrule(lr){8-10}  \cmidrule(lr){12-14}
      $\Sh_{np}$ & $d$\big|$n$ & & 50 & 100 & 150 & & 50 & 100 & 150 & & 50 & 100 & 150 \\
      \midrule
      &  5 &&   &  &  && 75.2 & 97.4 & 99.8 && 99.5 & 100 & 100 \\
      $\Sh_{np}^{\rm P}$ & 15 &&   &  &  && 72.9 & 98.1 & 99.9 && 99.9 & 100 & 100 \\
      & 25 &&   &  &  && 58.5 & 97.9 & 100 && 98.5 & 100 & 100 \\
      \cmidrule(lr){2-14}
      &  5 &&   &  &  && 70.1 & 97.2 & 99.8 && 99.2 & 100 & 100 \\
      & 15 &&   &  &  && 65.4 & 97.3 & 99.9 && 99.4 & 100 & 100 \\
      $\Sh_{np}^{\rm J}$ & 25 &&   &  &  && 49.2 & 97 & 99.8 && 96.6 & 100 & 100 \\
      & 50 &&   &  &  && 19.5 & 84.6 & 99.1 && 85.1 & 100 & 100 \\
      & 100 &&   &  &  && 3.4 & 49.9 & 94.5 && 47.9 & 99.8 & 100 \\
      \midrule
      \multicolumn{14}{c}{$M_{np}$ with $\SA = (1/n)\I_p$ for column departures ($\Delta =  0.2 $)}\\
      \midrule
      \multicolumn{2}{c}{} & & \multicolumn{3}{c}{$\tau = 0$} & & \multicolumn{3}{c}{$\tau = 0.3$} & & \multicolumn{3}{c}{$\tau = 0.6$}\\
      \cmidrule(lr){4-6}  \cmidrule(lr){8-10}  \cmidrule(lr){12-14}
      $\Sh_{np}$ & $d$\big|$n$ & & 50 & 100 & 150 & & 50 & 100 & 150 & & 50 & 100 & 150 \\
      \midrule
      &  5 &&   &  &  && 60.7 & 91.2 & 98.2 && 97.6 & 100 & 100 \\
      $\Sh_{np}^{\rm P}$ & 15 &&   &  &  && 70.1 & 95.9 & 99.9 && 99.4 & 100 & 100 \\
      & 25 &&   &  &  && 66.8 & 97.4 & 99.7 && 99.4 & 100 & 100 \\
      \cmidrule(lr){2-14}
      &  5 &&   &  &  && 52.5 & 88.9 & 97.9 && 95.1 & 100 & 100 \\
      & 15 &&   &  &  && 63.4 & 94.8 & 99.9 && 98.7 & 100 & 100 \\
      $\Sh_{np}^{\rm J}$ & 25 &&   &  &  && 62.5 & 97.2 & 99.7 && 98.9 & 100 & 100 \\
      & 50 &&   &  &  && 56.3 & 95.9 & 100 && 97.9 & 100 & 100 \\
      & 100 &&   &  &  && 54.3 & 95.6 & 99.7 && 97.3 & 100 & 100 \\
      \bottomrule
      \end{tabular}
      \end{center}
      \vskip-9pt
      \small
      Statistics: $E_{np}$ Euclidean norm-based statistic defined in Eq.~\eqref{eq:euclidean}; $M_{np}$ supremum norm-based statistic defined in Eq.~\eqref{eq:supremum}. Estimators: $\Sh_{np}^{\rm P}$ plug-in estimator; $\Sh_{np}^{\rm J}$ jackknife estimator.
      \end{table}

            \begin{table}[htbp]
 \captionsetup{width=1\linewidth,font=small,skip=0pt}
      \caption{Estimated sizes (in \%) for the tests of $H^*_0$ with $\mathcal{G} = \{\{1,\dots,d\}\}$ performed at the nominal level 5\%. Each entry is based on $ 2500 $ samples of size $n$ in dimension $d$ drawn from a  Normal copula  with Kendall's tau matrix $\bs{T}$ is as in Eq.~\eqref{eq:T-equi-null}.}
       \label{tab:sim-level-star-normal}
      \begin{center}
      \fontsize{8.75}{8.75}\selectfont
      \vskip-12pt
      \begin{tabular}{*{2}{l}*{12}{r}}
      \toprule
      \multicolumn{14}{c}{$E_{np}$ with $\SA = \Sh_{np}$}\\
      \midrule
      \multicolumn{2}{c}{} & & \multicolumn{3}{c}{$\tau = 0$} & & \multicolumn{3}{c}{$\tau = 0.3$} & & \multicolumn{3}{c}{$\tau = 0.6$}\\
      \cmidrule(lr){4-6}  \cmidrule(lr){8-10}  \cmidrule(lr){12-14}
      $\Sh_{np}$ & $d$\big|$n$ & & 50 & 100 & 150 & & 50 & 100 & 150 & & 50 & 100 & 150 \\
      \midrule
      &  5 &&  3.3 & 4.2 & 5.2 && 3.2 & 4.1 & 4.9 && 2.6 & 3.6 & 4 \\
      $\Sb_{np}^{\rm P}$ & 15 &&  4.4 & 4.8 & 4.1 && 3.5 & 4.6 & 4.4 && 3.8 & 4.6 & 4 \\
      & 25 &&  4.2 & 4.4 & 4.8 && 3.8 & 4.5 & 5.2 && 4.7 & 4 & 4.4 \\
      \cmidrule(lr){2-14}
      &  5 &&  5.4 & 5.3 & 6 && 4.7 & 5.2 & 5.6 && 2.7 & 3.6 & 3.9 \\
      & 15 &&  5.3 & 5.3 & 4.4 && 2.8 & 4.2 & 4.1 && 1.2 & 1.8 & 2.3 \\
      $\Sb_{np}^{\rm J}$ & 25 &&  5 & 4.6 & 4.8 && 1.9 & 2.8 & 3.6 && 0.1 & 0.5 & 1.8 \\
      & 50 &&  4.3 & 4.4 & 3.9 && 0.7 & 1 & 2.6 && 0 & 0.1 & 0.3 \\
      & 100 &&  2.9 & 3.6 & 4.1 && 0 & 0.3 & 0.9 && 0 & 0 & 0 \\
      \midrule
      \multicolumn{14}{c}{$M_{np}$ with $\SA = \Sh_{np}$}\\
      \midrule
      &  5 &&  4.4 & 4 & 4.7 && 3.6 & 4.5 & 5 && 3.4 & 4.7 & 4.3 \\
      $\Sb_{np}^{\rm P}$ & 15 &&  3.7 & 4.6 & 4.9 && 4.5 & 5.1 & 4.8 && 5.5 & 4.7 & 5.2 \\
      & 25 &&  3.1 & 4.6 & 4.3 && 5 & 4.3 & 4.5 && 5.7 & 5.4 & 5.4 \\
      \cmidrule(lr){2-14}
      &  5 &&  5.7 & 4.8 & 5 && 4.5 & 4.9 & 5.4 && 3 & 4.3 & 4.1 \\
      & 15 &&  3.7 & 4.7 & 5 && 3.6 & 4.6 & 4.2 && 3 & 3.2 & 4.3 \\
      $\Sb_{np}^{\rm J}$ & 25 &&  2.9 & 4.4 & 4.3 && 3.6 & 3.9 & 3.8 && 2.4 & 3.4 & 3.8 \\
      & 50 &&  3 & 3.1 & 3.4 && 3.5 & 4 & 4.3 && 2.5 & 4 & 4 \\
      & 100 &&  2.4 & 3.5 & 4.4 && 4.8 & 4.3 & 5.4 && 4.2 & 4.9 & 4.7 \\
      \midrule
      \multicolumn{14}{c}{$E_{np}$ with $\SA = (1/n)\I_p$}\\
      \midrule
      \multicolumn{2}{c}{} & & \multicolumn{3}{c}{$\tau = 0$} & & \multicolumn{3}{c}{$\tau = 0.3$} & & \multicolumn{3}{c}{$\tau = 0.6$}\\
      \cmidrule(lr){4-6}  \cmidrule(lr){8-10}  \cmidrule(lr){12-14}
      $\Sh_{np}$ & $d$\big|$n$ & & 50 & 100 & 150 & & 50 & 100 & 150 & & 50 & 100 & 150 \\
      \midrule
      &  5 &&  3.5 & 4.4 & 5.6 && 3.2 & 3.6 & 4.7 && 2.8 & 3.6 & 4.2 \\
      $\Sb_{np}^{\rm P}$ & 15 &&  4.4 & 5 & 4.2 && 3.5 & 4.8 & 4.7 && 3.6 & 4 & 4.5 \\
      & 25 &&  4.4 & 4.4 & 4.8 && 3.2 & 4.5 & 5.1 && 3.8 & 3.2 & 4.5 \\
      \cmidrule(lr){2-14}
      &  5 &&  5.4 & 5.2 & 6.2 && 4.6 & 4.2 & 5.3 && 2.6 & 3.2 & 4 \\
      & 15 &&  5.4 & 5.3 & 4.3 && 4.1 & 4.9 & 4.7 && 1.9 & 3.2 & 3.6 \\
      $\Sb_{np}^{\rm J}$ & 25 &&  4.9 & 4.6 & 4.9 && 3.4 & 4.4 & 5 && 1.5 & 1.4 & 3 \\
      & 50 &&  4.4 & 4.3 & 4.1 && 3 & 2.9 & 3.8 && 0.6 & 2 & 2.2 \\
      & 100 &&  2.9 & 3.7 & 4.2 && 2.5 & 3.4 & 4.1 && 0.4 & 0.7 & 1.3 \\
      \midrule
      \multicolumn{14}{c}{$M_{np}$ with $\SA = (1/n)\I_p$}\\
      \midrule
      \multicolumn{2}{c}{} & & \multicolumn{3}{c}{$\tau = 0$} & & \multicolumn{3}{c}{$\tau = 0.3$} & & \multicolumn{3}{c}{$\tau = 0.6$}\\
      \cmidrule(lr){4-6}  \cmidrule(lr){8-10}  \cmidrule(lr){12-14}
      $\Sh_{np}$ & $d$\big|$n$ & & 50 & 100 & 150 & & 50 & 100 & 150 & & 50 & 100 & 150 \\
      \midrule
      &  5 &&  4.4 & 4.1 & 4.7 && 3.8 & 4.3 & 5.2 && 4 & 4.9 & 4.5 \\
      $\Sb_{np}^{\rm P}$ & 15 &&  3.6 & 4.7 & 5.1 && 4.4 & 5.4 & 4.4 && 5 & 4.6 & 5.6 \\
      & 25 &&  2.9 & 4.7 & 4.2 && 5.2 & 4 & 5.2 && 5.6 & 5.1 & 5.4 \\
      \cmidrule(lr){2-14}
      &  5 &&  5.4 & 4.6 & 5.2 && 4.3 & 4.8 & 5.1 && 3.3 & 4.4 & 4.2 \\
      & 15 &&  3.7 & 4.8 & 4.8 && 4 & 5 & 4.5 && 3 & 3.4 & 4.6 \\
      $\Sb_{np}^{\rm J}$ & 25 &&  2.9 & 4.6 & 4.4 && 4.6 & 3.6 & 5 && 3.2 & 3.4 & 4.5 \\
      & 50 &&  3 & 3.4 & 3.8 && 5.1 & 4.6 & 4.8 && 4.2 & 4 & 4.8 \\
      & 100 &&  2.4 & 3.5 & 4.6 && 6.4 & 6.4 & 6 && 5.6 & 5.2 & 4.8 \\
      \bottomrule
      \end{tabular}
      \end{center}
      \vskip-9pt
      \small
      Statistics: $E_{np}$ Euclidean norm-based statistic defined in Eq. \eqref{eq:euclidean}; $M_{np}$ supremum norm-based statistic defined in Eq. \eqref{eq:supremum}. Estimators: $\Sb_{np}^{\rm P}$ structured plug-in estimator; $\Sb_{np}^{\rm J}$ structured jackknife estimator.
      \end{table}

            \begin{table}[htbp]
 \captionsetup{width=1\linewidth,font=small,skip=0pt}
      \caption{Estimated sizes (in \%) for the tests of $H^*_0$ with $\mathcal{G} = \{\{1,\dots,d\}\}$ performed at the nominal level 5\%. Each entry is based on $ 1000 $ samples of size $n$ in dimension $d$ drawn from a  $t_4$ copula  with Kendall's tau matrix $\bs{T}$ is as in Eq.~\eqref{eq:T-equi-null}.}
       \label{tab:sim-level-star-t4}
      \begin{center}
      \fontsize{8.75}{8.75}\selectfont
      \vskip-12pt
      \begin{tabular}{*{2}{l}*{12}{r}}
      \toprule
      \multicolumn{14}{c}{$E_{np}$ with $\SA = \Sh_{np}$}\\
      \midrule
      \multicolumn{2}{c}{} & & \multicolumn{3}{c}{$\tau = 0$} & & \multicolumn{3}{c}{$\tau = 0.3$} & & \multicolumn{3}{c}{$\tau = 0.6$}\\
      \cmidrule(lr){4-6}  \cmidrule(lr){8-10}  \cmidrule(lr){12-14}
      $\Sh_{np}$ & $d$\big|$n$ & & 50 & 100 & 150 & & 50 & 100 & 150 & & 50 & 100 & 150 \\
      \midrule
      &  5 &&  3.4 & 5.5 & 4.2 && 4.1 & 3.6 & 4.4 && 3.7 & 3.5 & 4.7 \\
      $\Sb_{np}^{\rm P}$ & 15 &&  3.4 & 3.8 & 4.8 && 3.9 & 5.7 & 5.4 && 3.3 & 4.2 & 4.9 \\
      & 25 &&  3.3 & 4.6 & 4.5 && 4.2 & 5.2 & 4.4 && 4.4 & 3.4 & 3.3 \\
      \cmidrule(lr){2-14}
      &  5 &&  6.5 & 7.1 & 4.9 && 6 & 4.7 & 5.5 && 3.8 & 3.5 & 4.9 \\
      & 15 &&  5.2 & 4.6 & 5.8 && 4 & 5.9 & 5.5 && 1.7 & 2.6 & 3.4 \\
      $\Sb_{np}^{\rm J}$ & 25 &&  6.5 & 6.1 & 5.9 && 4.3 & 4.9 & 4.3 && 0.9 & 2.1 & 1.9 \\
      & 50 &&  7.6 & 4.4 & 6.5 && 2.7 & 4.5 & 4.5 && 0 & 0.5 & 1.3 \\
      & 100 &&  12.6 & 6 & 6.2 && 1.4 & 2.2 & 2.4 && 0.1 & 0 & 0.1 \\
      \midrule
      \multicolumn{14}{c}{$M_{np}$ with $\SA = \Sh_{np}$}\\
      \midrule
      &  5 &&  3.4 & 5.9 & 4.3 && 3.3 & 4 & 5.9 && 4.2 & 3.2 & 4 \\
      $\Sb_{np}^{\rm P}$ & 15 &&  3.6 & 5.4 & 5 && 4 & 3.5 & 3.7 && 4.4 & 3.8 & 4.8 \\
      & 25 &&  2.9 & 3.7 & 4.1 && 3.3 & 5 & 3.1 && 4.6 & 4.8 & 3.7 \\
      \cmidrule(lr){2-14}
      &  5 &&  5 & 6.4 & 4.9 && 4.4 & 4.7 & 5.9 && 4 & 2.8 & 4.2 \\
      & 15 &&  4.1 & 5.6 & 5.5 && 4.1 & 3.6 & 3.9 && 2.9 & 3.2 & 4.1 \\
      $\Sb_{np}^{\rm J}$ & 25 &&  3.1 & 3.9 & 4.2 && 3.2 & 4.9 & 3 && 2.9 & 3.8 & 3.2 \\
      & 50 &&  3.2 & 3.9 & 4.5 && 3.9 & 4.5 & 4.6 && 5.2 & 3.3 & 5 \\
      & 100 &&  2.4 & 2.9 & 3 && 7 & 6.4 & 5.7 && 8.3 & 6.1 & 5.1 \\
      \midrule
      \multicolumn{14}{c}{$E_{np}$ with $\SA = (1/n)\I_p$}\\
      \midrule
      \multicolumn{2}{c}{} & & \multicolumn{3}{c}{$\tau = 0$} & & \multicolumn{3}{c}{$\tau = 0.3$} & & \multicolumn{3}{c}{$\tau = 0.6$}\\
      \cmidrule(lr){4-6}  \cmidrule(lr){8-10}  \cmidrule(lr){12-14}
      $\Sh_{np}$ & $d$\big|$n$ & & 50 & 100 & 150 & & 50 & 100 & 150 & & 50 & 100 & 150 \\
      \midrule
      &  5 &&  3.9 & 5.4 & 4.2 && 3.6 & 4.1 & 5.6 && 3.5 & 3.7 & 3.3 \\
      $\Sb_{np}^{\rm P}$ & 15 &&  3.4 & 3.7 & 5 && 3.5 & 4 & 4.3 && 3.6 & 4.1 & 4.1 \\
      & 25 &&  3.4 & 4.9 & 4.9 && 3.3 & 4 & 4.2 && 3.7 & 4.3 & 4.5 \\
      \cmidrule(lr){2-14}
      &  5 &&  6.3 & 6.7 & 5 && 5.5 & 4.3 & 5.9 && 4.1 & 4.3 & 3.6 \\
      & 15 &&  5 & 4.9 & 5.7 && 5.2 & 4.5 & 5 && 2.9 & 3.8 & 3.6 \\
      $\Sb_{np}^{\rm J}$ & 25 &&  6.4 & 6 & 5.7 && 4.4 & 4.4 & 4.7 && 3 & 3.6 & 4 \\
      & 50 &&  7.7 & 4.6 & 6.6 && 4.9 & 5.8 & 4.8 && 1.9 & 3.8 & 3.5 \\
      & 100 &&  12.8 & 6 & 6.3 && 4.9 & 4.6 & 6.2 && 1.2 & 2.9 & 2.5 \\
      \midrule
      \multicolumn{14}{c}{$M_{np}$ with $\SA = (1/n)\I_p$}\\
      \midrule
      \multicolumn{2}{c}{} & & \multicolumn{3}{c}{$\tau = 0$} & & \multicolumn{3}{c}{$\tau = 0.3$} & & \multicolumn{3}{c}{$\tau = 0.6$}\\
      \cmidrule(lr){4-6}  \cmidrule(lr){8-10}  \cmidrule(lr){12-14}
      $\Sh_{np}$ & $d$\big|$n$ & & 50 & 100 & 150 & & 50 & 100 & 150 & & 50 & 100 & 150 \\
      \midrule
      &  5 &&  3.5 & 5.8 & 3.9 && 4.1 & 4.1 & 5.6 && 3.4 & 2.5 & 5.1 \\
      $\Sb_{np}^{\rm P}$ & 15 &&  3.5 & 5.4 & 5.2 && 5 & 4.4 & 4.4 && 5.6 & 3.6 & 4.2 \\
      & 25 &&  2.9 & 3.9 & 4.2 && 3.3 & 5.1 & 4.5 && 5.9 & 5.2 & 4.5 \\
      \cmidrule(lr){2-14}
      &  5 &&  4.6 & 6.6 & 4.4 && 5.1 & 4.4 & 6 && 3.7 & 2.4 & 4.8 \\
      & 15 &&  3.8 & 5.5 & 5.6 && 5.1 & 4.4 & 4.8 && 4.1 & 3 & 3.7 \\
      $\Sb_{np}^{\rm J}$ & 25 &&  3 & 3.9 & 4.6 && 3.4 & 5.2 & 4.5 && 4.3 & 4.4 & 4.2 \\
      & 50 &&  3.3 & 3.9 & 4.4 && 5.2 & 6.3 & 6 && 6.7 & 6.5 & 4.2 \\
      & 100 &&  2.2 & 2.9 & 2.8 && 8.7 & 6.1 & 6.3 && 8.8 & 7.9 & 7.5 \\
      \bottomrule
      \end{tabular}
      \end{center}
      \vskip-9pt
      \small
      Statistics: $E_{np}$ Euclidean norm-based statistic defined in Eq. \eqref{eq:euclidean}; $M_{np}$ supremum norm-based statistic defined in Eq. \eqref{eq:supremum}. Estimators: $\Sb_{np}^{\rm P}$ structured plug-in estimator; $\Sb_{np}^{\rm J}$ structured jackknife estimator.
      \end{table}

            \begin{table}[htbp]
 \captionsetup{width=1\linewidth,font=small,skip=0pt}
      \caption{Estimated sizes (in \%) for the tests of $H^*_0$ with $\mathcal{G} = \{\{1,\dots,d\}\}$ performed at the nominal level 5\%. Each entry is based on $ 1000 $ samples of size $n$ in dimension $d$ drawn from a  Gumbel copula  with Kendall's tau matrix $\bs{T}$ is as in Eq.~\eqref{eq:T-equi-null}.}
       \label{tab:sim-level-star-gumbel}
      \begin{center}
      \fontsize{8.75}{8.75}\selectfont
      \vskip-12pt
      \begin{tabular}{*{2}{l}*{12}{r}}
      \toprule
      \multicolumn{14}{c}{$E_{np}$ with $\SA = \Sh_{np}$}\\
      \midrule
      \multicolumn{2}{c}{} & & \multicolumn{3}{c}{$\tau = 0$} & & \multicolumn{3}{c}{$\tau = 0.3$} & & \multicolumn{3}{c}{$\tau = 0.6$}\\
      \cmidrule(lr){4-6}  \cmidrule(lr){8-10}  \cmidrule(lr){12-14}
      $\Sh_{np}$ & $d$\big|$n$ & & 50 & 100 & 150 & & 50 & 100 & 150 & & 50 & 100 & 150 \\
      \midrule
      &  5 &&  3.1 & 4.3 & 4.6 && 3.1 & 4.1 & 4.3 && 4.2 & 3.2 & 2.8 \\
      $\Sb_{np}^{\rm P}$ & 15 &&  3.8 & 5.1 & 4.4 && 3.5 & 3.6 & 4.7 && 5.1 & 4.2 & 5.1 \\
      & 25 &&  3.7 & 4.7 & 5.5 && 4.2 & 3.4 & 3.1 && 4.3 & 3.7 & 5 \\
      \cmidrule(lr){2-14}
      &  5 &&  4.8 & 5.8 & 5.3 && 5.1 & 5 & 5.1 && 4.4 & 3.2 & 3 \\
      & 15 &&  4.5 & 5.5 & 5 && 3.4 & 3 & 4.2 && 1.6 & 1.9 & 3.2 \\
      $\Sb_{np}^{\rm J}$ & 25 &&  4.2 & 4.9 & 5.6 && 2.6 & 2.7 & 2.8 && 0.4 & 0.7 & 1.9 \\
      & 50 &&  2.9 & 5.4 & 5.1 && 1.5 & 2.1 & 2.8 && 0 & 0 & 0.3 \\
      & 100 &&  2.5 & 3.9 & 2.6 && 0.1 & 0.6 & 1.3 && 0 & 0 & 0 \\
      \midrule
      \multicolumn{14}{c}{$M_{np}$ with $\SA = \Sh_{np}$}\\
      \midrule
      &  5 &&  3.4 & 4.6 & 4.3 && 3.7 & 7 & 4.1 && 5.3 & 4.1 & 4.6 \\
      $\Sb_{np}^{\rm P}$ & 15 &&  3.7 & 5 & 4.2 && 5.6 & 5.2 & 5.5 && 4.8 & 5.1 & 4.3 \\
      & 25 &&  3.1 & 4.2 & 4.7 && 5 & 5.6 & 3.9 && 5.7 & 5.3 & 5.2 \\
      \cmidrule(lr){2-14}
      &  5 &&  4.1 & 5.3 & 4.9 && 4.6 & 6.9 & 4.5 && 4.8 & 4 & 4.4 \\
      & 15 &&  3.9 & 5.2 & 4.1 && 5.1 & 4.9 & 5.4 && 2.5 & 4.2 & 3.4 \\
      $\Sb_{np}^{\rm J}$ & 25 &&  2.9 & 4.2 & 4.4 && 4.4 & 5.3 & 3.6 && 3.1 & 2.8 & 3.9 \\
      & 50 &&  2.8 & 4 & 4.9 && 3.4 & 4.1 & 4.4 && 5.4 & 4.2 & 4.5 \\
      & 100 &&  2.9 & 3.3 & 3.5 && 5.9 & 5.4 & 4.8 && 10.1 & 7.5 & 7.7 \\
      \midrule
      \multicolumn{14}{c}{$E_{np}$ with $\SA = (1/n)\I_p$}\\
      \midrule
      \multicolumn{2}{c}{} & & \multicolumn{3}{c}{$\tau = 0$} & & \multicolumn{3}{c}{$\tau = 0.3$} & & \multicolumn{3}{c}{$\tau = 0.6$}\\
      \cmidrule(lr){4-6}  \cmidrule(lr){8-10}  \cmidrule(lr){12-14}
      $\Sh_{np}$ & $d$\big|$n$ & & 50 & 100 & 150 & & 50 & 100 & 150 & & 50 & 100 & 150 \\
      \midrule
      &  5 &&  3.3 & 4.2 & 4.4 && 4.1 & 4.8 & 4.2 && 4.9 & 3.9 & 3.5 \\
      $\Sb_{np}^{\rm P}$ & 15 &&  3.8 & 5.2 & 4.8 && 2.9 & 4.7 & 5.2 && 4.2 & 3.6 & 5.5 \\
      & 25 &&  3.6 & 4.6 & 5.6 && 4.4 & 4.1 & 3.6 && 3.9 & 5.1 & 3.3 \\
      \cmidrule(lr){2-14}
      &  5 &&  4.8 & 5.7 & 5 && 5.2 & 5.6 & 4.4 && 4.9 & 4.1 & 3.8 \\
      & 15 &&  4.6 & 5.5 & 5.1 && 3.4 & 4.9 & 5.2 && 3.4 & 2.9 & 4.9 \\
      $\Sb_{np}^{\rm J}$ & 25 &&  3.9 & 4.9 & 5.6 && 4.4 & 3.7 & 3.4 && 1.9 & 3.8 & 2.7 \\
      & 50 &&  3.1 & 5.4 & 5 && 2.1 & 2.4 & 4.5 && 0.9 & 2 & 1.7 \\
      & 100 &&  2.9 & 3.9 & 2.8 && 1.9 & 3 & 4 && 0.8 & 1.8 & 2.4 \\
      \midrule
      \multicolumn{14}{c}{$M_{np}$ with $\SA = (1/n)\I_p$}\\
      \midrule
      \multicolumn{2}{c}{} & & \multicolumn{3}{c}{$\tau = 0$} & & \multicolumn{3}{c}{$\tau = 0.3$} & & \multicolumn{3}{c}{$\tau = 0.6$}\\
      \cmidrule(lr){4-6}  \cmidrule(lr){8-10}  \cmidrule(lr){12-14}
      $\Sh_{np}$ & $d$\big|$n$ & & 50 & 100 & 150 & & 50 & 100 & 150 & & 50 & 100 & 150 \\
      \midrule
      &  5 &&  3.5 & 4.7 & 4.2 && 4.4 & 6.5 & 4 && 5.2 & 4.8 & 4.3 \\
      $\Sb_{np}^{\rm P}$ & 15 &&  3.8 & 5.3 & 4.1 && 5 & 4.3 & 5.7 && 7.1 & 5.3 & 5.9 \\
      & 25 &&  3.2 & 4.2 & 4.6 && 6.2 & 6.4 & 4 && 6.4 & 5.7 & 5 \\
      \cmidrule(lr){2-14}
      &  5 &&  4 & 5.1 & 4.9 && 5.3 & 6.8 & 4 && 5.2 & 4.7 & 4.2 \\
      & 15 &&  3.7 & 5 & 4.1 && 4.3 & 4.1 & 5.9 && 5.3 & 4.6 & 5.4 \\
      $\Sb_{np}^{\rm J}$ & 25 &&  3.2 & 4.3 & 4.6 && 5.6 & 6.4 & 4.1 && 4.8 & 4.4 & 4.5 \\
      & 50 &&  2.9 & 4 & 4.8 && 4.9 & 4.5 & 4.5 && 5.4 & 6.1 & 4.2 \\
      & 100 &&  2.9 & 3.3 & 4.1 && 6.6 & 3.9 & 5.6 && 8.6 & 9.4 & 7.8 \\
      \bottomrule
      \end{tabular}
      \end{center}
      \vskip-9pt
      \small
      Statistics: $E_{np}$ Euclidean norm-based statistic defined in Eq. \eqref{eq:euclidean}; $M_{np}$ supremum norm-based statistic defined in Eq. \eqref{eq:supremum}. Estimators: $\Sb_{np}^{\rm P}$ structured plug-in estimator; $\Sb_{np}^{\rm J}$ structured jackknife estimator.
      \end{table}

            \begin{table}[htbp]
 \captionsetup{width=1\linewidth,font=small,skip=0pt}
      \caption{Estimated sizes (in \%) for the tests of $H^*_0$ with $\mathcal{G} = \{\{1,\dots,d\}\}$ performed at the nominal level 5\%. Each entry is based on $ 1000 $ samples of size $n$ in dimension $d$ drawn from a  Clayton copula  with Kendall's tau matrix $\bs{T}$ is as in Eq.~\eqref{eq:T-equi-null}.}
       \label{tab:sim-level-star-clayton}
      \begin{center}
      \fontsize{8.75}{8.75}\selectfont
      \vskip-12pt
      \begin{tabular}{*{2}{l}*{12}{r}}
      \toprule
      \multicolumn{14}{c}{$E_{np}$ with $\SA = \Sh_{np}$}\\
      \midrule
      \multicolumn{2}{c}{} & & \multicolumn{3}{c}{$\tau = 0$} & & \multicolumn{3}{c}{$\tau = 0.3$} & & \multicolumn{3}{c}{$\tau = 0.6$}\\
      \cmidrule(lr){4-6}  \cmidrule(lr){8-10}  \cmidrule(lr){12-14}
      $\Sh_{np}$ & $d$\big|$n$ & & 50 & 100 & 150 & & 50 & 100 & 150 & & 50 & 100 & 150 \\
      \midrule
      &  5 &&   &  &  && 2.6 & 4.4 & 3.8 && 3.2 & 3.7 & 3.8 \\
      $\Sb_{np}^{\rm P}$ & 15 &&   &  &  && 3.8 & 4.6 & 6 && 4.8 & 4.1 & 3.7 \\
      & 25 &&   &  &  && 5 & 4.1 & 4.4 && 4.3 & 5.8 & 4 \\
      \cmidrule(lr){2-14}
      &  5 &&   &  &  && 4.5 & 5.6 & 4.2 && 2.9 & 3.7 & 3.9 \\
      & 15 &&   &  &  && 3.5 & 4.4 & 5.5 && 1.9 & 1.5 & 2.1 \\
      $\Sb_{np}^{\rm J}$ & 25 &&   &  &  && 2.8 & 3.3 & 3.8 && 0.6 & 1.2 & 1.4 \\
      & 50 &&   &  &  && 1.1 & 1.7 & 2.3 && 0 & 0 & 0.4 \\
      & 100 &&   &  &  && 0.3 & 0.4 & 1.2 && 0 & 0 & 0 \\
      \midrule
      \multicolumn{14}{c}{$M_{np}$ with $\SA = \Sh_{np}$}\\
      \midrule
      &  5 &&   &  &  && 3.1 & 4.4 & 3.9 && 4.2 & 4.5 & 4 \\
      $\Sb_{np}^{\rm P}$ & 15 &&   &  &  && 3.3 & 4.9 & 4.1 && 4.1 & 4.4 & 3.8 \\
      & 25 &&   &  &  && 5.2 & 4.6 & 4.3 && 6.9 & 5.5 & 5.6 \\
      \cmidrule(lr){2-14}
      &  5 &&   &  &  && 4.3 & 5.2 & 4 && 3.6 & 4.3 & 3.9 \\
      & 15 &&   &  &  && 2.9 & 4.6 & 3.7 && 2.7 & 3.5 & 3.1 \\
      $\Sb_{np}^{\rm J}$ & 25 &&   &  &  && 4.2 & 4.1 & 4.1 && 3.1 & 3.8 & 4.7 \\
      & 50 &&   &  &  && 4.2 & 4.6 & 5.9 && 4.6 & 6.4 & 4.5 \\
      & 100 &&   &  &  && 7.5 & 4.2 & 6.9 && 10 & 8.6 & 6.9 \\
      \midrule
      \multicolumn{14}{c}{$E_{np}$ with $\SA = (1/n)\I_p$}\\
      \midrule
      \multicolumn{2}{c}{} & & \multicolumn{3}{c}{$\tau = 0$} & & \multicolumn{3}{c}{$\tau = 0.3$} & & \multicolumn{3}{c}{$\tau = 0.6$}\\
      \cmidrule(lr){4-6}  \cmidrule(lr){8-10}  \cmidrule(lr){12-14}
      $\Sh_{np}$ & $d$\big|$n$ & & 50 & 100 & 150 & & 50 & 100 & 150 & & 50 & 100 & 150 \\
      \midrule
      &  5 &&   &  &  && 3.2 & 4.7 & 3.1 && 2 & 3.7 & 4.3 \\
      $\Sb_{np}^{\rm P}$ & 15 &&   &  &  && 3.7 & 4.6 & 5.2 && 3.4 & 4 & 4.7 \\
      & 25 &&   &  &  && 5.4 & 4.4 & 3.7 && 4.2 & 4 & 3.8 \\
      \cmidrule(lr){2-14}
      &  5 &&   &  &  && 4.1 & 5.5 & 3.1 && 2.7 & 3.8 & 4 \\
      & 15 &&   &  &  && 4.9 & 4.7 & 5.2 && 2.7 & 3.5 & 4.6 \\
      $\Sb_{np}^{\rm J}$ & 25 &&   &  &  && 6.2 & 4.3 & 3.6 && 2.9 & 2.6 & 3.4 \\
      & 50 &&   &  &  && 3.6 & 4 & 4.7 && 1.6 & 3.1 & 4.6 \\
      & 100 &&   &  &  && 3.8 & 4.1 & 4.9 && 0.9 & 2.1 & 2.8 \\
      \midrule
      \multicolumn{14}{c}{$M_{np}$ with $\SA = (1/n)\I_p$}\\
      \midrule
      \multicolumn{2}{c}{} & & \multicolumn{3}{c}{$\tau = 0$} & & \multicolumn{3}{c}{$\tau = 0.3$} & & \multicolumn{3}{c}{$\tau = 0.6$}\\
      \cmidrule(lr){4-6}  \cmidrule(lr){8-10}  \cmidrule(lr){12-14}
      $\Sh_{np}$ & $d$\big|$n$ & & 50 & 100 & 150 & & 50 & 100 & 150 & & 50 & 100 & 150 \\
      \midrule
      &  5 &&   &  &  && 3.5 & 5.2 & 3.4 && 3.3 & 4.9 & 4.3 \\
      $\Sb_{np}^{\rm P}$ & 15 &&   &  &  && 3.7 & 4.5 & 3.5 && 4.7 & 5.5 & 5.2 \\
      & 25 &&   &  &  && 5.8 & 5.3 & 4.8 && 6.3 & 5.3 & 6 \\
      \cmidrule(lr){2-14}
      &  5 &&   &  &  && 4.3 & 5.5 & 3.8 && 3.1 & 4.9 & 4.1 \\
      & 15 &&   &  &  && 3.7 & 4.6 & 3.8 && 3.6 & 4.8 & 4.7 \\
      $\Sb_{np}^{\rm J}$ & 25 &&   &  &  && 5.7 & 5.2 & 4.6 && 4.1 & 4.7 & 5.5 \\
      & 50 &&   &  &  && 6.8 & 6.3 & 5.6 && 6.6 & 5.6 & 6.5 \\
      & 100 &&   &  &  && 9.7 & 5.5 & 6.2 && 9.2 & 9.2 & 7.2 \\
      \bottomrule
      \end{tabular}
      \end{center}
      \vskip-9pt
      \small
      Statistics: $E_{np}$ Euclidean norm-based statistic defined in Eq. \eqref{eq:euclidean}; $M_{np}$ supremum norm-based statistic defined in Eq. \eqref{eq:supremum}. Estimators: $\Sb_{np}^{\rm P}$ structured plug-in estimator; $\Sb_{np}^{\rm J}$ structured jackknife estimator.
      \end{table}

            \begin{table}[htbp]
 \captionsetup{width=1\linewidth,font=small,skip=0pt}
\caption{Estimated rejection rates (in \%) of tests of $H_0^*$ with $\mathcal{G}=\{\{1,\ldots,d\}\}$ performed at nominal level $5$\%. Each entry is based on $ 2500 $ $n \times d$ datasets drawn from a  Normal copula  with Kendall's tau matrix $\bs{T}_{\Delta}$ in Eq.~\eqref{eq:departure} (i) with $\Delta =  0.1 $; $\bs{T}$ is as in Eq.~\eqref{eq:T-equi-null}.}
       \label{tab:sim-power-star-single-1-normal}
      \begin{center}
      \fontsize{8.75}{8.75}\selectfont
      \vskip-12pt
      \begin{tabular}{*{2}{l}*{12}{r}}
      \toprule
      \multicolumn{14}{c}{$E_{np}$ with $\SA = \Sh_{np}$ for  single  departures ($\Delta =  0.1 $)}\\
      \midrule
      \multicolumn{2}{c}{} & & \multicolumn{3}{c}{$\tau = 0$} & & \multicolumn{3}{c}{$\tau = 0.3$} & & \multicolumn{3}{c}{$\tau = 0.6$}\\
      \cmidrule(lr){4-6}  \cmidrule(lr){8-10}  \cmidrule(lr){12-14}
      $\Sh_{np}$ & $d$\big|$n$ & & 50 & 100 & 150 & & 50 & 100 & 150 & & 50 & 100 & 150 \\
      \midrule
      &  5 &&  6.6 & 9.2 & 16.1 && 7.6 & 19.4 & 30.5 && 22.9 & 63.6 & 87.4 \\
      $\Sb_{np}^{\rm P}$ & 15 &&  5.2 & 6.2 & 7.4 && 6.2 & 7.9 & 12.2 && 10 & 24.7 & 43.8 \\
      & 25 &&  4.7 & 5.8 & 5.8 && 4.5 & 6.2 & 9.1 && 7.6 & 15.1 & 27.2 \\
      \cmidrule(lr){2-14}
      &  5 &&  9.9 & 11 & 17.8 && 11 & 22 & 32.5 && 22.9 & 62.6 & 87 \\
      & 15 &&  6.7 & 6.7 & 8 && 4.8 & 7 & 11.2 && 2.8 & 13.9 & 33.4 \\
      $\Sb_{np}^{\rm J}$ & 25 &&  5.4 & 6.1 & 5.8 && 2.6 & 4.3 & 7.2 && 0.4 & 4.4 & 13.6 \\
      & 50 &&  3.9 & 4.8 & 4.9 && 0.9 & 2.1 & 3.1 && 0 & 0.4 & 1.8 \\
      & 100 &&  3.4 & 3 & 4.3 && 0.1 & 0.4 & 1.1 && 0 & 0 & 0 \\
      \midrule
      \multicolumn{14}{c}{$M_{np}$ with $\SA = \Sh_{np}$ for  single departures ($\Delta =  0.1 $)}\\
      \midrule
      &  5 &&  6.8 & 10.6 & 18.2 && 9.4 & 21.9 & 36 && 30.7 & 76.8 & 93.6 \\
      $\Sb_{np}^{\rm P}$ & 15 &&  4.3 & 6.5 & 8.5 && 6 & 13.6 & 23.4 && 21.8 & 67.8 & 91.2 \\
      & 25 &&  3.7 & 4.9 & 6.8 && 5.3 & 11 & 17.8 && 19 & 61.7 & 89.3 \\
      \cmidrule(lr){2-14}
      &  5 &&  8.8 & 11.8 & 19.3 && 10.8 & 23.6 & 37.4 && 29 & 75.6 & 93.3 \\
      & 15 &&  4.4 & 6.6 & 8.6 && 5.1 & 12.6 & 22.6 && 15.8 & 62.2 & 89.2 \\
      $\Sb_{np}^{\rm J}$ & 25 &&  3.5 & 4.6 & 6.8 && 4 & 10 & 17 && 12.7 & 56.7 & 87.7 \\
      & 50 &&  2.6 & 4.7 & 5.5 && 4 & 7 & 13 && 10.8 & 43.8 & 82.6 \\
      & 100 &&  2.3 & 3.8 & 4.3 && 4.2 & 5.6 & 9.9 && 8.2 & 37.7 & 74 \\
      \midrule
      \multicolumn{14}{c}{$E_{np}$ with $\SA = (1/n)\I_p$ for  single  departures ($\Delta =  0.1 $)}\\
      \midrule
      \multicolumn{2}{c}{} & & \multicolumn{3}{c}{$\tau = 0$} & & \multicolumn{3}{c}{$\tau = 0.3$} & & \multicolumn{3}{c}{$\tau = 0.6$}\\
      \cmidrule(lr){4-6}  \cmidrule(lr){8-10}  \cmidrule(lr){12-14}
      $\Sh_{np}$ & $d$\big|$n$ & & 50 & 100 & 150 & & 50 & 100 & 150 & & 50 & 100 & 150 \\
      \midrule
      &  5 &&  7.1 & 9.4 & 16.6 && 7.4 & 16.1 & 25.3 && 18.5 & 51.4 & 76.8 \\
      $\Sb_{np}^{\rm P}$ & 15 &&  5.5 & 6.2 & 7.4 && 4.2 & 5.8 & 6.5 && 5.6 & 7.9 & 12.8 \\
      & 25 &&  4.8 & 6 & 5.8 && 4 & 4.6 & 5.2 && 4.3 & 6.3 & 7.5 \\
      \cmidrule(lr){2-14}
      &  5 &&  9.4 & 10.9 & 17.7 && 9.6 & 17.6 & 26.2 && 17.1 & 49.2 & 75.8 \\
      & 15 &&  6.7 & 6.9 & 7.8 && 4.7 & 6.2 & 6.7 && 3.1 & 5.6 & 10.2 \\
      $\Sb_{np}^{\rm J}$ & 25 &&  5.4 & 6.1 & 6 && 4.1 & 4.3 & 5 && 1.8 & 4 & 5 \\
      & 50 &&  4.1 & 4.6 & 4.7 && 3.6 & 4.5 & 3.8 && 0.6 & 1.7 & 2.4 \\
      & 100 &&  3.5 & 3.1 & 4.4 && 2.4 & 3.8 & 3.2 && 0.2 & 0.9 & 1.9 \\
      \midrule
      \multicolumn{14}{c}{$M_{np}$ with $\SA = (1/n)\I_p$ for  single  departures ($\Delta =  0.1 $)}\\
      \midrule
      \multicolumn{2}{c}{} & & \multicolumn{3}{c}{$\tau = 0$} & & \multicolumn{3}{c}{$\tau = 0.3$} & & \multicolumn{3}{c}{$\tau = 0.6$}\\
      \cmidrule(lr){4-6}  \cmidrule(lr){8-10}  \cmidrule(lr){12-14}
      $\Sh_{np}$ & $d$\big|$n$ & & 50 & 100 & 150 & & 50 & 100 & 150 & & 50 & 100 & 150 \\
      \midrule
      &  5 &&  7 & 10.8 & 18.3 && 9 & 20.2 & 31.9 && 26.7 & 69.8 & 89.4 \\
      $\Sb_{np}^{\rm P}$ & 15 &&  4.4 & 6.6 & 8.6 && 5.7 & 8 & 13.7 && 9.5 & 36.4 & 70 \\
      & 25 &&  3.5 & 4.7 & 6.8 && 5.4 & 6.4 & 8.5 && 8.1 & 28.5 & 58.8 \\
      \cmidrule(lr){2-14}
      &  5 &&  8.6 & 12 & 18.9 && 9.9 & 20.9 & 32.6 && 23.8 & 68.2 & 88.9 \\
      & 15 &&  4.6 & 6.6 & 8.7 && 5.3 & 7.8 & 13.2 && 5.9 & 32.1 & 67.3 \\
      $\Sb_{np}^{\rm J}$ & 25 &&  3.6 & 4.5 & 6.9 && 4.9 & 6 & 7.9 && 4.4 & 23.6 & 54.8 \\
      & 50 &&  2.8 & 4.8 & 5.6 && 5.5 & 5.8 & 7.2 && 5 & 11.8 & 36.1 \\
      & 100 &&  2.4 & 4 & 4.4 && 6.2 & 5.5 & 6.6 && 5 & 7.8 & 26 \\
      \bottomrule
      \end{tabular}
      \end{center}
      \vskip-9pt
      \small
      Statistics: $E_{np}$ Euclidean norm-based statistic defined in Eq.~\eqref{eq:euclidean}; $M_{np}$ supremum norm-based statistic defined in Eq.~\eqref{eq:supremum}. Estimators: $\Sb_{np}^{\rm P}$ structured plug-in estimator; $\Sb_{np}^{\rm J}$ structured jackknife estimator.
      \end{table}

            \begin{table}[htbp]
 \captionsetup{width=1\linewidth,font=small,skip=0pt}
\caption{Estimated rejection rates (in \%) of tests of $H_0^*$ with $\mathcal{G}=\{\{1,\ldots,d\}\}$ performed at nominal level $5$\%. Each entry is based on $ 1000 $ $n \times d$ datasets drawn from a  $t_4$ copula  with Kendall's tau matrix $\bs{T}_{\Delta}$ in Eq.~\eqref{eq:departure} (i) with $\Delta =  0.1 $; $\bs{T}$ is as in Eq.~\eqref{eq:T-equi-null}.}
       \label{tab:sim-power-star-single-1-t4}
      \begin{center}
      \fontsize{8.75}{8.75}\selectfont
      \vskip-12pt
      \begin{tabular}{*{2}{l}*{12}{r}}
      \toprule
      \multicolumn{14}{c}{$E_{np}$ with $\SA = \Sh_{np}$ for  single  departures ($\Delta =  0.1 $)}\\
      \midrule
      \multicolumn{2}{c}{} & & \multicolumn{3}{c}{$\tau = 0$} & & \multicolumn{3}{c}{$\tau = 0.3$} & & \multicolumn{3}{c}{$\tau = 0.6$}\\
      \cmidrule(lr){4-6}  \cmidrule(lr){8-10}  \cmidrule(lr){12-14}
      $\Sh_{np}$ & $d$\big|$n$ & & 50 & 100 & 150 & & 50 & 100 & 150 & & 50 & 100 & 150 \\
      \midrule
      &  5 &&  5.9 & 8.1 & 12.6 && 7.6 & 15.4 & 24.2 && 16.3 & 46.9 & 73.5 \\
      $\Sb_{np}^{\rm P}$ & 15 &&  5.2 & 4.4 & 6 && 4.1 & 6.5 & 9.6 && 9.1 & 17.1 & 30.8 \\
      & 25 &&  3.4 & 5 & 5.2 && 4.7 & 7.2 & 6.9 && 6.1 & 11.5 & 18.8 \\
      \cmidrule(lr){2-14}
      &  5 &&  9.4 & 10.5 & 15 && 10.3 & 17.3 & 26.4 && 17.6 & 48.6 & 73.5 \\
      & 15 &&  7.3 & 6.6 & 7.2 && 4.7 & 6.6 & 9.8 && 4.8 & 11.8 & 25.4 \\
      $\Sb_{np}^{\rm J}$ & 25 &&  6.6 & 5.8 & 6.5 && 5.1 & 7 & 6.5 && 1.6 & 5.2 & 11 \\
      & 50 &&  6.6 & 8.5 & 6.3 && 3.2 & 4.2 & 4.7 && 0.2 & 0.7 & 2.5 \\
      & 100 &&  14.1 & 8.4 & 8 && 1 & 1.8 & 3.6 && 0 & 0.1 & 0.3 \\
      \midrule
      \multicolumn{14}{c}{$M_{np}$ with $\SA = \Sh_{np}$ for  single departures ($\Delta =  0.1 $)}\\
      \midrule
      &  5 &&  5.8 & 10 & 15.1 && 6.4 & 18.1 & 30.7 && 20.9 & 58.6 & 83.7 \\
      $\Sb_{np}^{\rm P}$ & 15 &&  3.8 & 6.2 & 7.4 && 5.2 & 8.6 & 15 && 14.9 & 45.4 & 76.3 \\
      & 25 &&  3.7 & 4.6 & 4.7 && 5 & 8.8 & 12.9 && 12.7 & 39.7 & 68.1 \\
      \cmidrule(lr){2-14}
      &  5 &&  8.4 & 11.1 & 16.2 && 8.9 & 19.2 & 32 && 21 & 58.4 & 84.1 \\
      & 15 &&  4.2 & 6.5 & 7.5 && 5.1 & 8.7 & 14.7 && 12.6 & 43.1 & 74.6 \\
      $\Sb_{np}^{\rm J}$ & 25 &&  3.9 & 4.7 & 4.9 && 5.1 & 8.4 & 12.6 && 8.7 & 37 & 66.5 \\
      & 50 &&  2.6 & 5.4 & 4.8 && 4.2 & 7.7 & 10.1 && 8.4 & 30 & 61 \\
      & 100 &&  2.2 & 3.1 & 4 && 6.2 & 4.5 & 6.2 && 10.5 & 24.1 & 52.4 \\
      \midrule
      \multicolumn{14}{c}{$E_{np}$ with $\SA = (1/n)\I_p$ for  single  departures ($\Delta =  0.1 $)}\\
      \midrule
      \multicolumn{2}{c}{} & & \multicolumn{3}{c}{$\tau = 0$} & & \multicolumn{3}{c}{$\tau = 0.3$} & & \multicolumn{3}{c}{$\tau = 0.6$}\\
      \cmidrule(lr){4-6}  \cmidrule(lr){8-10}  \cmidrule(lr){12-14}
      $\Sh_{np}$ & $d$\big|$n$ & & 50 & 100 & 150 & & 50 & 100 & 150 & & 50 & 100 & 150 \\
      \midrule
      &  5 &&  6.7 & 8.5 & 12.8 && 6.7 & 13.5 & 18.3 && 11.8 & 34.8 & 62.5 \\
      $\Sb_{np}^{\rm P}$ & 15 &&  5.8 & 4.4 & 6.4 && 3.6 & 4.4 & 5.7 && 4.3 & 6.5 & 9.6 \\
      & 25 &&  3.5 & 5 & 5.3 && 5.2 & 6.2 & 6 && 3.6 & 5.2 & 6.9 \\
      \cmidrule(lr){2-14}
      &  5 &&  9.7 & 10.6 & 14.3 && 9.8 & 16.2 & 20.9 && 13.5 & 34.6 & 62.6 \\
      & 15 &&  7.7 & 6.2 & 6.9 && 4.8 & 5.3 & 6.2 && 3.7 & 5.7 & 9.1 \\
      $\Sb_{np}^{\rm J}$ & 25 &&  6.6 & 5.8 & 6.8 && 6.2 & 6.8 & 6.5 && 2.2 & 4.3 & 5.6 \\
      & 50 &&  6.6 & 8.4 & 6.1 && 4.7 & 3.9 & 6.1 && 1.8 & 2.1 & 3.8 \\
      & 100 &&  14.4 & 8.4 & 8 && 4.2 & 3.9 & 6 && 1.2 & 2.8 & 3.2 \\
      \midrule
      \multicolumn{14}{c}{$M_{np}$ with $\SA = (1/n)\I_p$ for  single  departures ($\Delta =  0.1 $)}\\
      \midrule
      \multicolumn{2}{c}{} & & \multicolumn{3}{c}{$\tau = 0$} & & \multicolumn{3}{c}{$\tau = 0.3$} & & \multicolumn{3}{c}{$\tau = 0.6$}\\
      \cmidrule(lr){4-6}  \cmidrule(lr){8-10}  \cmidrule(lr){12-14}
      $\Sh_{np}$ & $d$\big|$n$ & & 50 & 100 & 150 & & 50 & 100 & 150 & & 50 & 100 & 150 \\
      \midrule
      &  5 &&  6.2 & 10.7 & 15.2 && 6.4 & 16.2 & 28.1 && 18.2 & 51.1 & 78.7 \\
      $\Sb_{np}^{\rm P}$ & 15 &&  3.9 & 6.6 & 7.5 && 4.6 & 5.8 & 10.6 && 7.2 & 21.6 & 46 \\
      & 25 &&  3.5 & 4.5 & 4.8 && 5.2 & 5.6 & 8.2 && 7.9 & 14.3 & 36.1 \\
      \cmidrule(lr){2-14}
      &  5 &&  7.9 & 11.1 & 16.1 && 8.5 & 17.4 & 28.8 && 19.3 & 50.1 & 79.1 \\
      & 15 &&  4.3 & 6.9 & 7.6 && 4.9 & 6.2 & 10.8 && 5.7 & 20.3 & 45.3 \\
      $\Sb_{np}^{\rm J}$ & 25 &&  4 & 4.8 & 4.8 && 5.3 & 5.7 & 8 && 6.2 & 12.2 & 34.1 \\
      & 50 &&  2.6 & 5.4 & 5.1 && 5 & 6.4 & 6.2 && 6.1 & 8.2 & 20.6 \\
      & 100 &&  2.3 & 3.1 & 4.1 && 5.7 & 5.5 & 7 && 8.7 & 8.2 & 14.7 \\
      \bottomrule
      \end{tabular}
      \end{center}
      \vskip-9pt
      \small
      Statistics: $E_{np}$ Euclidean norm-based statistic defined in Eq.~\eqref{eq:euclidean}; $M_{np}$ supremum norm-based statistic defined in Eq.~\eqref{eq:supremum}. Estimators: $\Sb_{np}^{\rm P}$ structured plug-in estimator; $\Sb_{np}^{\rm J}$ structured jackknife estimator.
      \end{table}

            \begin{table}[htbp]
 \captionsetup{width=1\linewidth,font=small,skip=0pt}
\caption{Estimated rejection rates (in \%) of tests of $H_0^*$ with $\mathcal{G}=\{\{1,\ldots,d\}\}$ performed at nominal level $5$\%. Each entry is based on $ 1000 $ $n \times d$ datasets drawn from a  Gumbel copula  with Kendall's tau matrix $\bs{T}_{\Delta}$ in Eq.~\eqref{eq:departure} (i) with $\Delta =  0.1 $; $\bs{T}$ is as in Eq.~\eqref{eq:T-equi-null}.}
       \label{tab:sim-power-star-single-1-gumbel}
      \begin{center}
      \fontsize{8.75}{8.75}\selectfont
      \vskip-12pt
      \begin{tabular}{*{2}{l}*{12}{r}}
      \toprule
      \multicolumn{14}{c}{$E_{np}$ with $\SA = \Sh_{np}$ for  single  departures ($\Delta =  0.1 $)}\\
      \midrule
      \multicolumn{2}{c}{} & & \multicolumn{3}{c}{$\tau = 0$} & & \multicolumn{3}{c}{$\tau = 0.3$} & & \multicolumn{3}{c}{$\tau = 0.6$}\\
      \cmidrule(lr){4-6}  \cmidrule(lr){8-10}  \cmidrule(lr){12-14}
      $\Sh_{np}$ & $d$\big|$n$ & & 50 & 100 & 150 & & 50 & 100 & 150 & & 50 & 100 & 150 \\
      \midrule
      &  5 &&  4.6 & 10.6 & 15.3 && 8.1 & 17.9 & 28 && 18.3 & 53 & 80 \\
      $\Sb_{np}^{\rm P}$ & 15 &&  4.7 & 6.7 & 7.6 && 4.3 & 8.1 & 11.8 && 9.2 & 23 & 38.1 \\
      & 25 &&  4.5 & 5.4 & 6.7 && 5.8 & 5.4 & 9.1 && 6.7 & 14.1 & 25 \\
      \cmidrule(lr){2-14}
      &  5 &&  7.1 & 13.2 & 16.7 && 11.4 & 19.7 & 29.7 && 19.1 & 53.1 & 80.1 \\
      & 15 &&  6.5 & 7.5 & 7.9 && 3.7 & 7.7 & 11.3 && 3.8 & 15.1 & 32.8 \\
      $\Sb_{np}^{\rm J}$ & 25 &&  5.2 & 5.4 & 6.7 && 3.2 & 4 & 7.6 && 0.5 & 5.3 & 13.3 \\
      & 50 &&  4.3 & 5.6 & 4.6 && 0.5 & 3.1 & 4.2 && 0 & 0.7 & 2.2 \\
      & 100 &&  4.1 & 4.3 & 4.1 && 0.3 & 0.7 & 1.4 && 0 & 0 & 0 \\
      \midrule
      \multicolumn{14}{c}{$M_{np}$ with $\SA = \Sh_{np}$ for  single departures ($\Delta =  0.1 $)}\\
      \midrule
      &  5 &&  5.9 & 12.3 & 18 && 9.2 & 20.1 & 34.9 && 26.6 & 65.6 & 88.6 \\
      $\Sb_{np}^{\rm P}$ & 15 &&  4 & 7 & 10.7 && 5 & 12.2 & 20.7 && 20.5 & 59.3 & 85.2 \\
      & 25 &&  3.4 & 5.2 & 7.5 && 4.5 & 10.1 & 18.3 && 20.4 & 54.9 & 80.7 \\
      \cmidrule(lr){2-14}
      &  5 &&  7.5 & 13.8 & 19.4 && 11.1 & 21.1 & 35.7 && 26.2 & 65.1 & 88.6 \\
      & 15 &&  4.1 & 7 & 10.7 && 4.4 & 10.8 & 20.4 && 15.6 & 54.9 & 83.8 \\
      $\Sb_{np}^{\rm J}$ & 25 &&  3.4 & 5.1 & 7.2 && 3.6 & 9.2 & 17.8 && 14.9 & 49.8 & 78.7 \\
      & 50 &&  3 & 5.4 & 4.1 && 4.5 & 6.7 & 12.2 && 12 & 43 & 72.4 \\
      & 100 &&  2 & 3.3 & 4.8 && 5.4 & 5.4 & 9.6 && 16 & 38.2 & 68.2 \\
      \midrule
      \multicolumn{14}{c}{$E_{np}$ with $\SA = (1/n)\I_p$ for  single  departures ($\Delta =  0.1 $)}\\
      \midrule
      \multicolumn{2}{c}{} & & \multicolumn{3}{c}{$\tau = 0$} & & \multicolumn{3}{c}{$\tau = 0.3$} & & \multicolumn{3}{c}{$\tau = 0.6$}\\
      \cmidrule(lr){4-6}  \cmidrule(lr){8-10}  \cmidrule(lr){12-14}
      $\Sh_{np}$ & $d$\big|$n$ & & 50 & 100 & 150 & & 50 & 100 & 150 & & 50 & 100 & 150 \\
      \midrule
      &  5 &&  5.4 & 10.9 & 15.9 && 7.8 & 14.7 & 24.1 && 14.1 & 40.7 & 67.3 \\
      $\Sb_{np}^{\rm P}$ & 15 &&  4.9 & 6.8 & 7.7 && 3.3 & 6.3 & 6.9 && 5.1 & 7.7 & 10.2 \\
      & 25 &&  4.6 & 5.4 & 6.8 && 3.4 & 4.6 & 6.7 && 5.7 & 3.9 & 7.1 \\
      \cmidrule(lr){2-14}
      &  5 &&  6.9 & 12.9 & 16.8 && 10 & 16.6 & 24.6 && 14.4 & 40.1 & 66.9 \\
      & 15 &&  6.3 & 7.3 & 8.2 && 3.7 & 6.2 & 7 && 3.7 & 6.6 & 9.6 \\
      $\Sb_{np}^{\rm J}$ & 25 &&  5.2 & 5.7 & 6.8 && 3.3 & 4.7 & 6.5 && 3.8 & 2.9 & 5.4 \\
      & 50 &&  4.6 & 5.6 & 4.1 && 2.7 & 3.4 & 4.7 && 1.3 & 3.2 & 3.5 \\
      & 100 &&  4.2 & 4.3 & 4.1 && 2.5 & 2.9 & 3.1 && 0.4 & 2 & 2.5 \\
      \midrule
      \multicolumn{14}{c}{$M_{np}$ with $\SA = (1/n)\I_p$ for  single  departures ($\Delta =  0.1 $)}\\
      \midrule
      \multicolumn{2}{c}{} & & \multicolumn{3}{c}{$\tau = 0$} & & \multicolumn{3}{c}{$\tau = 0.3$} & & \multicolumn{3}{c}{$\tau = 0.6$}\\
      \cmidrule(lr){4-6}  \cmidrule(lr){8-10}  \cmidrule(lr){12-14}
      $\Sh_{np}$ & $d$\big|$n$ & & 50 & 100 & 150 & & 50 & 100 & 150 & & 50 & 100 & 150 \\
      \midrule
      &  5 &&  6.2 & 12.9 & 18.4 && 9.3 & 18.3 & 31.8 && 23.2 & 59.7 & 83.7 \\
      $\Sb_{np}^{\rm P}$ & 15 &&  3.9 & 7.2 & 10.4 && 3.8 & 8.9 & 13.8 && 8.3 & 27.6 & 59.6 \\
      & 25 &&  3.3 & 5.2 & 7.4 && 5.6 & 7 & 10.5 && 8.2 & 18.4 & 43.7 \\
      \cmidrule(lr){2-14}
      &  5 &&  7.7 & 14.1 & 18.9 && 10.6 & 19 & 32.1 && 21.6 & 58.7 & 82.8 \\
      & 15 &&  4.1 & 7 & 10.8 && 3.7 & 8.5 & 13.6 && 6.3 & 24.7 & 57.1 \\
      $\Sb_{np}^{\rm J}$ & 25 &&  3.3 & 5.2 & 7 && 5.2 & 6.8 & 10.3 && 5.2 & 15.9 & 40 \\
      & 50 &&  2.9 & 5.4 & 4.3 && 3.9 & 4.7 & 7.1 && 6.1 & 9.7 & 24 \\
      & 100 &&  2.1 & 3.3 & 5 && 4.2 & 5.2 & 7.5 && 9 & 9.9 & 17.6 \\
      \bottomrule
      \end{tabular}
      \end{center}
      \vskip-9pt
      \small
      Statistics: $E_{np}$ Euclidean norm-based statistic defined in Eq.~\eqref{eq:euclidean}; $M_{np}$ supremum norm-based statistic defined in Eq.~\eqref{eq:supremum}. Estimators: $\Sb_{np}^{\rm P}$ structured plug-in estimator; $\Sb_{np}^{\rm J}$ structured jackknife estimator.
      \end{table}

            \begin{table}[htbp]
 \captionsetup{width=1\linewidth,font=small,skip=0pt}
\caption{Estimated rejection rates (in \%) of tests of $H_0^*$ with $\mathcal{G}=\{\{1,\ldots,d\}\}$ performed at nominal level $5$\%. Each entry is based on $ 1000 $ $n \times d$ datasets drawn from a  Clayton copula  with Kendall's tau matrix $\bs{T}_{\Delta}$ in Eq.~\eqref{eq:departure} (i) with $\Delta =  0.1 $; $\bs{T}$ is as in Eq.~\eqref{eq:T-equi-null}.}
       \label{tab:sim-power-star-single-1-clayton}
      \begin{center}
      \fontsize{8.75}{8.75}\selectfont
      \vskip-12pt
      \begin{tabular}{*{2}{l}*{12}{r}}
      \toprule
      \multicolumn{14}{c}{$E_{np}$ with $\SA = \Sh_{np}$ for  single  departures ($\Delta =  0.1 $)}\\
      \midrule
      \multicolumn{2}{c}{} & & \multicolumn{3}{c}{$\tau = 0$} & & \multicolumn{3}{c}{$\tau = 0.3$} & & \multicolumn{3}{c}{$\tau = 0.6$}\\
      \cmidrule(lr){4-6}  \cmidrule(lr){8-10}  \cmidrule(lr){12-14}
      $\Sh_{np}$ & $d$\big|$n$ & & 50 & 100 & 150 & & 50 & 100 & 150 & & 50 & 100 & 150 \\
      \midrule
      &  5 &&   &  &  && 8.8 & 19.1 & 26 && 19.8 & 54.7 & 79.8 \\
      $\Sb_{np}^{\rm P}$ & 15 &&   &  &  && 6.4 & 9.5 & 12.5 && 10 & 24.3 & 41.1 \\
      & 25 &&   &  &  && 5.8 & 7.9 & 7.5 && 6.3 & 15.6 & 23.7 \\
      \cmidrule(lr){2-14}
      &  5 &&   &  &  && 11.3 & 21.8 & 27.8 && 21.4 & 54.6 & 79.7 \\
      & 15 &&   &  &  && 5.3 & 8.5 & 11.7 && 4.2 & 18 & 33.1 \\
      $\Sb_{np}^{\rm J}$ & 25 &&   &  &  && 4 & 6.2 & 6.7 && 0.7 & 5.9 & 13.6 \\
      & 50 &&   &  &  && 1.4 & 1.8 & 3.7 && 0 & 0.5 & 2.8 \\
      & 100 &&   &  &  && 0.3 & 0.9 & 1.8 && 0 & 0 & 0.2 \\
      \midrule
      \multicolumn{14}{c}{$M_{np}$ with $\SA = \Sh_{np}$ for  single departures ($\Delta =  0.1 $)}\\
      \midrule
      &  5 &&   &  &  && 8.7 & 20.3 & 32.1 && 26.2 & 67 & 87.1 \\
      $\Sb_{np}^{\rm P}$ & 15 &&   &  &  && 6.2 & 10.5 & 24 && 21.5 & 60.1 & 87.2 \\
      & 25 &&   &  &  && 4.8 & 11.1 & 18.5 && 20.3 & 56.1 & 81.2 \\
      \cmidrule(lr){2-14}
      &  5 &&   &  &  && 10.3 & 21.3 & 33.1 && 26.3 & 66.5 & 86.8 \\
      & 15 &&   &  &  && 5.6 & 9.8 & 23.5 && 15.5 & 57.2 & 85.8 \\
      $\Sb_{np}^{\rm J}$ & 25 &&   &  &  && 3.4 & 10.4 & 16.8 && 14.7 & 52.5 & 79.8 \\
      & 50 &&   &  &  && 4.8 & 7.7 & 11.4 && 12.2 & 42.3 & 72.5 \\
      & 100 &&   &  &  && 5.8 & 7.7 & 9.4 && 13.6 & 35.8 & 69.7 \\
      \midrule
      \multicolumn{14}{c}{$E_{np}$ with $\SA = (1/n)\I_p$ for  single  departures ($\Delta =  0.1 $)}\\
      \midrule
      \multicolumn{2}{c}{} & & \multicolumn{3}{c}{$\tau = 0$} & & \multicolumn{3}{c}{$\tau = 0.3$} & & \multicolumn{3}{c}{$\tau = 0.6$}\\
      \cmidrule(lr){4-6}  \cmidrule(lr){8-10}  \cmidrule(lr){12-14}
      $\Sh_{np}$ & $d$\big|$n$ & & 50 & 100 & 150 & & 50 & 100 & 150 & & 50 & 100 & 150 \\
      \midrule
      &  5 &&   &  &  && 8 & 16 & 20.6 && 16.5 & 41.4 & 64.4 \\
      $\Sb_{np}^{\rm P}$ & 15 &&   &  &  && 5 & 5.5 & 7.6 && 4.9 & 8.7 & 9.9 \\
      & 25 &&   &  &  && 4.7 & 4.1 & 4.7 && 4.4 & 5.2 & 6.9 \\
      \cmidrule(lr){2-14}
      &  5 &&   &  &  && 10.4 & 17.8 & 22.5 && 17 & 41.4 & 63.9 \\
      & 15 &&   &  &  && 5.6 & 6.2 & 7.5 && 4.2 & 8.3 & 9 \\
      $\Sb_{np}^{\rm J}$ & 25 &&   &  &  && 5.3 & 4.2 & 4.9 && 3.4 & 4.5 & 5.7 \\
      & 50 &&   &  &  && 4.5 & 4.8 & 4.1 && 1.8 & 2.8 & 5 \\
      & 100 &&   &  &  && 3.9 & 4 & 4.9 && 1.1 & 3.2 & 2.7 \\
      \midrule
      \multicolumn{14}{c}{$M_{np}$ with $\SA = (1/n)\I_p$ for  single  departures ($\Delta =  0.1 $)}\\
      \midrule
      \multicolumn{2}{c}{} & & \multicolumn{3}{c}{$\tau = 0$} & & \multicolumn{3}{c}{$\tau = 0.3$} & & \multicolumn{3}{c}{$\tau = 0.6$}\\
      \cmidrule(lr){4-6}  \cmidrule(lr){8-10}  \cmidrule(lr){12-14}
      $\Sh_{np}$ & $d$\big|$n$ & & 50 & 100 & 150 & & 50 & 100 & 150 & & 50 & 100 & 150 \\
      \midrule
      &  5 &&   &  &  && 7.7 & 18.8 & 29 && 22.5 & 58.1 & 81.6 \\
      $\Sb_{np}^{\rm P}$ & 15 &&   &  &  && 4.6 & 7.1 & 14.6 && 8 & 26.1 & 54 \\
      & 25 &&   &  &  && 5 & 5.2 & 7.1 && 7.2 & 19 & 37.5 \\
      \cmidrule(lr){2-14}
      &  5 &&   &  &  && 9.2 & 19.9 & 29.8 && 22.4 & 57.2 & 81.1 \\
      & 15 &&   &  &  && 4.2 & 6.9 & 14.8 && 5.8 & 24.3 & 51.9 \\
      $\Sb_{np}^{\rm J}$ & 25 &&   &  &  && 4.3 & 4.8 & 6.8 && 4.9 & 16.5 & 35.1 \\
      & 50 &&   &  &  && 7.3 & 7.1 & 5.8 && 7.1 & 8.9 & 24 \\
      & 100 &&   &  &  && 7.7 & 8.1 & 6.7 && 10.8 & 10.6 & 16.4 \\
      \bottomrule
      \end{tabular}
      \end{center}
      \vskip-9pt
      \small
      Statistics: $E_{np}$ Euclidean norm-based statistic defined in Eq.~\eqref{eq:euclidean}; $M_{np}$ supremum norm-based statistic defined in Eq.~\eqref{eq:supremum}. Estimators: $\Sb_{np}^{\rm P}$ structured plug-in estimator; $\Sb_{np}^{\rm J}$ structured jackknife estimator.
      \end{table}

            \begin{table}[htbp]
 \captionsetup{width=1\linewidth,font=small,skip=0pt}
\caption{Estimated rejection rates (in \%) of tests of $H_0^*$ with $\mathcal{G}=\{\{1,\ldots,d\}\}$ performed at nominal level $5$\%. Each entry is based on $ 2500 $ $n \times d$ datasets drawn from a  Normal copula  with Kendall's tau matrix $\bs{T}_{\Delta}$ in Eq.~\eqref{eq:departure} (i) with $\Delta =  0.2 $; $\bs{T}$ is as in Eq.~\eqref{eq:T-equi-null}.}
       \label{tab:sim-power-star-single-2-normal}
      \begin{center}
      \fontsize{8.75}{8.75}\selectfont
      \vskip-12pt
      \begin{tabular}{*{2}{l}*{12}{r}}
      \toprule
      \multicolumn{14}{c}{$E_{np}$ with $\SA = \Sh_{np}$ for  single  departures ($\Delta =  0.2 $)}\\
      \midrule
      \multicolumn{2}{c}{} & & \multicolumn{3}{c}{$\tau = 0$} & & \multicolumn{3}{c}{$\tau = 0.3$} & & \multicolumn{3}{c}{$\tau = 0.6$}\\
      \cmidrule(lr){4-6}  \cmidrule(lr){8-10}  \cmidrule(lr){12-14}
      $\Sh_{np}$ & $d$\big|$n$ & & 50 & 100 & 150 & & 50 & 100 & 150 & & 50 & 100 & 150 \\
      \midrule
      &  5 &&  16.8 & 41.5 & 64.8 && 34.4 & 78.2 & 95.6 && 94.8 & 100 & 100 \\
      $\Sb_{np}^{\rm P}$ & 15 &&  7.7 & 13.8 & 21.7 && 12.7 & 31.6 & 55.5 && 48 & 95.7 & 100 \\
      & 25 &&  5.8 & 9.5 & 12.5 && 8.5 & 19 & 34 && 28.3 & 76.6 & 96.2 \\
      \cmidrule(lr){2-14}
      &  5 &&  21.9 & 45.6 & 67 && 40.8 & 80.2 & 96 && 94.2 & 100 & 100 \\
      & 15 &&  10.2 & 14.7 & 22.7 && 10.4 & 28.9 & 53.2 && 25.1 & 91.2 & 99.9 \\
      $\Sb_{np}^{\rm J}$ & 25 &&  6.6 & 9.8 & 12.8 && 4.9 & 14 & 28.4 && 4.4 & 52 & 90.9 \\
      & 50 &&  5.4 & 6.6 & 7 && 1.2 & 3.8 & 7.6 && 0 & 3.6 & 25.2 \\
      & 100 &&  3.6 & 4.1 & 4.8 && 0.1 & 0.8 & 2.1 && 0 & 0 & 0.5 \\
      \midrule
      \multicolumn{14}{c}{$M_{np}$ with $\SA = \Sh_{np}$ for  single departures ($\Delta =  0.2 $)}\\
      \midrule
      &  5 &&  20.8 & 52.5 & 75.1 && 46.9 & 88.4 & 98.7 && 98.7 & 100 & 100 \\
      $\Sb_{np}^{\rm P}$ & 15 &&  9 & 32.3 & 58.3 && 31.8 & 80 & 97.3 && 97.7 & 100 & 100 \\
      & 25 &&  6.6 & 22.8 & 46.2 && 26.2 & 76.6 & 96 && 96.4 & 100 & 100 \\
      \cmidrule(lr){2-14}
      &  5 &&  24.6 & 54.4 & 76.2 && 50.6 & 89 & 98.7 && 98.7 & 100 & 100 \\
      & 15 &&  9.1 & 32.3 & 58.1 && 28.9 & 79.2 & 97 && 96.4 & 100 & 100 \\
      $\Sb_{np}^{\rm J}$ & 25 &&  6.3 & 23.2 & 46.1 && 23.3 & 74.8 & 95.6 && 93.8 & 100 & 100 \\
      & 50 &&  3.9 & 15 & 34.5 && 19.1 & 67.3 & 92.5 && 88.8 & 100 & 100 \\
      & 100 &&  2.8 & 9.2 & 22.8 && 13.1 & 57.7 & 88.7 && 81.5 & 100 & 100 \\
      \midrule
      \multicolumn{14}{c}{$E_{np}$ with $\SA = (1/n)\I_p$ for  single  departures ($\Delta =  0.2 $)}\\
      \midrule
      \multicolumn{2}{c}{} & & \multicolumn{3}{c}{$\tau = 0$} & & \multicolumn{3}{c}{$\tau = 0.3$} & & \multicolumn{3}{c}{$\tau = 0.6$}\\
      \cmidrule(lr){4-6}  \cmidrule(lr){8-10}  \cmidrule(lr){12-14}
      $\Sh_{np}$ & $d$\big|$n$ & & 50 & 100 & 150 & & 50 & 100 & 150 & & 50 & 100 & 150 \\
      \midrule
      &  5 &&  17.6 & 42.2 & 64.3 && 29.6 & 70.4 & 92.4 && 88.8 & 100 & 100 \\
      $\Sb_{np}^{\rm P}$ & 15 &&  8.1 & 13.7 & 21.7 && 5.9 & 13.2 & 20.9 && 12.7 & 43.3 & 80.8 \\
      & 25 &&  5.8 & 9.6 & 12.5 && 4.4 & 8.2 & 10.7 && 6.8 & 16.9 & 26.3 \\
      \cmidrule(lr){2-14}
      &  5 &&  21.4 & 45 & 66.2 && 33.3 & 71.6 & 92.6 && 87 & 99.9 & 100 \\
      & 15 &&  10 & 14.9 & 22.5 && 6.3 & 13.2 & 20.8 && 6.6 & 34.3 & 75 \\
      $\Sb_{np}^{\rm J}$ & 25 &&  6.8 & 9.9 & 12.6 && 4.4 & 8 & 10.4 && 2.7 & 10.2 & 20.2 \\
      & 50 &&  5.2 & 6.4 & 6.8 && 3.4 & 4.4 & 6.9 && 0.8 & 3 & 5.4 \\
      & 100 &&  3.4 & 4 & 4.7 && 2.2 & 3.2 & 4.4 && 0.2 & 0.9 & 2 \\
      \midrule
      \multicolumn{14}{c}{$M_{np}$ with $\SA = (1/n)\I_p$ for  single  departures ($\Delta =  0.2 $)}\\
      \midrule
      \multicolumn{2}{c}{} & & \multicolumn{3}{c}{$\tau = 0$} & & \multicolumn{3}{c}{$\tau = 0.3$} & & \multicolumn{3}{c}{$\tau = 0.6$}\\
      \cmidrule(lr){4-6}  \cmidrule(lr){8-10}  \cmidrule(lr){12-14}
      $\Sh_{np}$ & $d$\big|$n$ & & 50 & 100 & 150 & & 50 & 100 & 150 & & 50 & 100 & 150 \\
      \midrule
      &  5 &&  21.6 & 52.7 & 75.4 && 43.4 & 86.1 & 98.3 && 97.7 & 100 & 100 \\
      $\Sb_{np}^{\rm P}$ & 15 &&  8.8 & 32.4 & 58.2 && 17 & 60.8 & 89.9 && 90 & 100 & 100 \\
      & 25 &&  6.6 & 23.2 & 46.1 && 11.3 & 49.6 & 82.4 && 80 & 100 & 100 \\
      \cmidrule(lr){2-14}
      &  5 &&  24.6 & 54.3 & 76.5 && 45.3 & 86.4 & 98.4 && 97.6 & 100 & 100 \\
      & 15 &&  9.1 & 32.3 & 58.1 && 16.3 & 60.2 & 89.7 && 84.9 & 100 & 100 \\
      $\Sb_{np}^{\rm J}$ & 25 &&  6.4 & 23.3 & 45.9 && 10 & 48.3 & 81.9 && 71.8 & 100 & 100 \\
      & 50 &&  4 & 14.8 & 34.5 && 7.9 & 32.7 & 69.3 && 46.8 & 99.8 & 100 \\
      & 100 &&  2.9 & 9.3 & 23 && 7 & 21.3 & 54.4 && 28.2 & 99.4 & 100 \\
      \bottomrule
      \end{tabular}
      \end{center}
      \vskip-9pt
      \small
      Statistics: $E_{np}$ Euclidean norm-based statistic defined in Eq.~\eqref{eq:euclidean}; $M_{np}$ supremum norm-based statistic defined in Eq.~\eqref{eq:supremum}. Estimators: $\Sb_{np}^{\rm P}$ structured plug-in estimator; $\Sb_{np}^{\rm J}$ structured jackknife estimator.
      \end{table}

            \begin{table}[htbp]
 \captionsetup{width=1\linewidth,font=small,skip=0pt}
\caption{Estimated rejection rates (in \%) of tests of $H_0^*$ with $\mathcal{G}=\{\{1,\ldots,d\}\}$ performed at nominal level $5$\%. Each entry is based on $ 1000 $ $n \times d$ datasets drawn from a  $t_4$ copula  with Kendall's tau matrix $\bs{T}_{\Delta}$ in Eq.~\eqref{eq:departure} (i) with $\Delta =  0.2 $; $\bs{T}$ is as in Eq.~\eqref{eq:T-equi-null}.}
       \label{tab:sim-power-star-single-2-t4}
      \begin{center}
      \fontsize{8.75}{8.75}\selectfont
      \vskip-12pt
      \begin{tabular}{*{2}{l}*{12}{r}}
      \toprule
      \multicolumn{14}{c}{$E_{np}$ with $\SA = \Sh_{np}$ for  single  departures ($\Delta =  0.2 $)}\\
      \midrule
      \multicolumn{2}{c}{} & & \multicolumn{3}{c}{$\tau = 0$} & & \multicolumn{3}{c}{$\tau = 0.3$} & & \multicolumn{3}{c}{$\tau = 0.6$}\\
      \cmidrule(lr){4-6}  \cmidrule(lr){8-10}  \cmidrule(lr){12-14}
      $\Sh_{np}$ & $d$\big|$n$ & & 50 & 100 & 150 & & 50 & 100 & 150 & & 50 & 100 & 150 \\
      \midrule
      &  5 &&  12.7 & 32.8 & 50.5 && 25.3 & 65.2 & 88.1 && 84.7 & 100 & 100 \\
      $\Sb_{np}^{\rm P}$ & 15 &&  6.5 & 10.4 & 15.9 && 10.8 & 23.3 & 41.8 && 33 & 83.9 & 98.6 \\
      & 25 &&  5.1 & 7.8 & 10.2 && 5.8 & 13.3 & 24.1 && 18.9 & 55.6 & 84.5 \\
      \cmidrule(lr){2-14}
      &  5 &&  19.6 & 37.1 & 53.4 && 32.1 & 68.9 & 89.4 && 86.5 & 100 & 100 \\
      & 15 &&  8.9 & 12.1 & 17.7 && 11.9 & 24.6 & 42.1 && 22.1 & 78.2 & 98.4 \\
      $\Sb_{np}^{\rm J}$ & 25 &&  8.5 & 10 & 11.7 && 5.6 & 12.7 & 23.5 && 7.2 & 40.9 & 77.4 \\
      & 50 &&  8.4 & 9.5 & 9 && 4.5 & 6.4 & 9.2 && 0.5 & 6.2 & 20.4 \\
      & 100 &&  13.8 & 9.9 & 9.3 && 2 & 3.7 & 4.8 && 0.1 & 0 & 1.9 \\
      \midrule
      \multicolumn{14}{c}{$M_{np}$ with $\SA = \Sh_{np}$ for  single departures ($\Delta =  0.2 $)}\\
      \midrule
      &  5 &&  18.7 & 40.2 & 63 && 34.8 & 76 & 95.9 && 94.7 & 100 & 100 \\
      $\Sb_{np}^{\rm P}$ & 15 &&  6.9 & 23 & 44.2 && 20.3 & 65.5 & 91.1 && 88.4 & 100 & 100 \\
      & 25 &&  5 & 14.8 & 34.6 && 17.8 & 56.9 & 87.8 && 87.3 & 99.8 & 100 \\
      \cmidrule(lr){2-14}
      &  5 &&  21.5 & 42.3 & 64.7 && 39 & 77.9 & 96.3 && 94.9 & 100 & 100 \\
      & 15 &&  7.7 & 23.5 & 44.4 && 20.5 & 64.8 & 91.1 && 86.1 & 100 & 100 \\
      $\Sb_{np}^{\rm J}$ & 25 &&  5.5 & 15.2 & 34.6 && 16.8 & 56.8 & 87.7 && 84.2 & 99.8 & 100 \\
      & 50 &&  2.8 & 8.9 & 23.1 && 10.4 & 47.7 & 80.3 && 71.9 & 100 & 100 \\
      & 100 &&  1.6 & 7.4 & 14.4 && 10.1 & 38.5 & 71.3 && 65.9 & 99.3 & 100 \\
      \midrule
      \multicolumn{14}{c}{$E_{np}$ with $\SA = (1/n)\I_p$ for  single  departures ($\Delta =  0.2 $)}\\
      \midrule
      \multicolumn{2}{c}{} & & \multicolumn{3}{c}{$\tau = 0$} & & \multicolumn{3}{c}{$\tau = 0.3$} & & \multicolumn{3}{c}{$\tau = 0.6$}\\
      \cmidrule(lr){4-6}  \cmidrule(lr){8-10}  \cmidrule(lr){12-14}
      $\Sh_{np}$ & $d$\big|$n$ & & 50 & 100 & 150 & & 50 & 100 & 150 & & 50 & 100 & 150 \\
      \midrule
      &  5 &&  13 & 33.3 & 51.4 && 21.1 & 54.3 & 81.1 && 76 & 99.8 & 100 \\
      $\Sb_{np}^{\rm P}$ & 15 &&  6.2 & 10.4 & 15.2 && 5 & 10.6 & 17.4 && 10.5 & 26.2 & 55.4 \\
      & 25 &&  5.5 & 7.7 & 10.2 && 4.6 & 7.5 & 8.5 && 6 & 11.6 & 18.9 \\
      \cmidrule(lr){2-14}
      &  5 &&  19.3 & 36.7 & 53.3 && 26.9 & 57.6 & 81.9 && 77 & 99.8 & 100 \\
      & 15 &&  8.8 & 12 & 17.4 && 6.9 & 11.9 & 18.5 && 8.1 & 23.2 & 52.3 \\
      $\Sb_{np}^{\rm J}$ & 25 &&  8.3 & 10 & 11.6 && 6.9 & 8.6 & 9.3 && 4.4 & 9.6 & 15.3 \\
      & 50 &&  8.2 & 9.6 & 8.8 && 5.5 & 4.4 & 4.9 && 1.8 & 5.7 & 5.9 \\
      & 100 &&  13.1 & 9.9 & 9.1 && 5.2 & 4.7 & 4.1 && 0.6 & 1.7 & 3.8 \\
      \midrule
      \multicolumn{14}{c}{$M_{np}$ with $\SA = (1/n)\I_p$ for  single  departures ($\Delta =  0.2 $)}\\
      \midrule
      \multicolumn{2}{c}{} & & \multicolumn{3}{c}{$\tau = 0$} & & \multicolumn{3}{c}{$\tau = 0.3$} & & \multicolumn{3}{c}{$\tau = 0.6$}\\
      \cmidrule(lr){4-6}  \cmidrule(lr){8-10}  \cmidrule(lr){12-14}
      $\Sh_{np}$ & $d$\big|$n$ & & 50 & 100 & 150 & & 50 & 100 & 150 & & 50 & 100 & 150 \\
      \midrule
      &  5 &&  18.8 & 40 & 63.1 && 30.5 & 72 & 92.9 && 92.8 & 100 & 100 \\
      $\Sb_{np}^{\rm P}$ & 15 &&  6.9 & 22.8 & 44 && 11.6 & 45.5 & 76.3 && 68.6 & 99.9 & 100 \\
      & 25 &&  4.8 & 14.9 & 34.6 && 7.4 & 29.9 & 67.3 && 53.5 & 99.1 & 100 \\
      \cmidrule(lr){2-14}
      &  5 &&  21.5 & 42.1 & 63.9 && 33.7 & 73.3 & 93.3 && 92.6 & 100 & 100 \\
      & 15 &&  7.5 & 23.1 & 44.6 && 11.6 & 45.2 & 76.2 && 63.5 & 99.8 & 100 \\
      $\Sb_{np}^{\rm J}$ & 25 &&  5.3 & 15 & 34.5 && 7.2 & 29.3 & 66.8 && 45.1 & 99 & 100 \\
      & 50 &&  3 & 9.1 & 23 && 6.6 & 19.2 & 48.8 && 25.7 & 96.8 & 100 \\
      & 100 &&  1.6 & 7.1 & 14.3 && 7.4 & 13.8 & 34.3 && 15.8 & 90.8 & 100 \\
      \bottomrule
      \end{tabular}
      \end{center}
      \vskip-9pt
      \small
      Statistics: $E_{np}$ Euclidean norm-based statistic defined in Eq.~\eqref{eq:euclidean}; $M_{np}$ supremum norm-based statistic defined in Eq.~\eqref{eq:supremum}. Estimators: $\Sb_{np}^{\rm P}$ structured plug-in estimator; $\Sb_{np}^{\rm J}$ structured jackknife estimator.
      \end{table}

            \begin{table}[htbp]
 \captionsetup{width=1\linewidth,font=small,skip=0pt}
\caption{Estimated rejection rates (in \%) of tests of $H_0^*$ with $\mathcal{G}=\{\{1,\ldots,d\}\}$ performed at nominal level $5$\%. Each entry is based on $ 1000 $ $n \times d$ datasets drawn from a  Gumbel copula  with Kendall's tau matrix $\bs{T}_{\Delta}$ in Eq.~\eqref{eq:departure} (i) with $\Delta =  0.2 $; $\bs{T}$ is as in Eq.~\eqref{eq:T-equi-null}.}
       \label{tab:sim-power-star-single-2-gumbel}
      \begin{center}
      \fontsize{8.75}{8.75}\selectfont
      \vskip-12pt
      \begin{tabular}{*{2}{l}*{12}{r}}
      \toprule
      \multicolumn{14}{c}{$E_{np}$ with $\SA = \Sh_{np}$ for  single  departures ($\Delta =  0.2 $)}\\
      \midrule
      \multicolumn{2}{c}{} & & \multicolumn{3}{c}{$\tau = 0$} & & \multicolumn{3}{c}{$\tau = 0.3$} & & \multicolumn{3}{c}{$\tau = 0.6$}\\
      \cmidrule(lr){4-6}  \cmidrule(lr){8-10}  \cmidrule(lr){12-14}
      $\Sh_{np}$ & $d$\big|$n$ & & 50 & 100 & 150 & & 50 & 100 & 150 & & 50 & 100 & 150 \\
      \midrule
      &  5 &&  16 & 42.8 & 64.9 && 30.7 & 73.4 & 91.4 && 89.2 & 100 & 100 \\
      $\Sb_{np}^{\rm P}$ & 15 &&  7.1 & 15.5 & 22.5 && 12.7 & 27.7 & 50.8 && 46.9 & 93.3 & 99.6 \\
      & 25 &&  5.8 & 9.1 & 10.6 && 9 & 16.7 & 27.4 && 24.1 & 70.4 & 92.7 \\
      \cmidrule(lr){2-14}
      &  5 &&  22 & 47.5 & 67.4 && 36.5 & 75.3 & 92.3 && 89.7 & 100 & 100 \\
      & 15 &&  10.1 & 16.7 & 23.6 && 12.1 & 26.6 & 49.5 && 28.8 & 88.7 & 99.2 \\
      $\Sb_{np}^{\rm J}$ & 25 &&  6.7 & 9.5 & 10.7 && 6.2 & 13.4 & 24.7 && 5 & 49.2 & 85.9 \\
      & 50 &&  5.4 & 5.3 & 8.4 && 1.7 & 4.6 & 10.4 && 0.2 & 5.3 & 28.1 \\
      & 100 &&  4.2 & 3.7 & 4.2 && 0.2 & 0.9 & 2.5 && 0 & 0 & 0.6 \\
      \midrule
      \multicolumn{14}{c}{$M_{np}$ with $\SA = \Sh_{np}$ for  single departures ($\Delta =  0.2 $)}\\
      \midrule
      &  5 &&  22.3 & 52.6 & 76.4 && 43.8 & 84.3 & 96.8 && 97 & 100 & 100 \\
      $\Sb_{np}^{\rm P}$ & 15 &&  9.6 & 32.2 & 58.8 && 29.6 & 72.3 & 96 && 93.6 & 100 & 100 \\
      & 25 &&  8.4 & 26.1 & 45.3 && 22.6 & 69.3 & 92.3 && 90.9 & 100 & 100 \\
      \cmidrule(lr){2-14}
      &  5 &&  25.3 & 54.2 & 78.4 && 46.6 & 85.1 & 97.4 && 96.8 & 100 & 100 \\
      & 15 &&  10.2 & 32.9 & 58.9 && 27.9 & 71.4 & 95.9 && 92.1 & 100 & 100 \\
      $\Sb_{np}^{\rm J}$ & 25 &&  8.2 & 26.2 & 45.1 && 21.2 & 67.7 & 92 && 87.1 & 100 & 100 \\
      & 50 &&  4.3 & 17.1 & 34.9 && 16 & 58.9 & 87.1 && 82.7 & 99.9 & 100 \\
      & 100 &&  2.4 & 10.6 & 26.6 && 13.6 & 50.8 & 81.6 && 72.7 & 99.9 & 100 \\
      \midrule
      \multicolumn{14}{c}{$E_{np}$ with $\SA = (1/n)\I_p$ for  single  departures ($\Delta =  0.2 $)}\\
      \midrule
      \multicolumn{2}{c}{} & & \multicolumn{3}{c}{$\tau = 0$} & & \multicolumn{3}{c}{$\tau = 0.3$} & & \multicolumn{3}{c}{$\tau = 0.6$}\\
      \cmidrule(lr){4-6}  \cmidrule(lr){8-10}  \cmidrule(lr){12-14}
      $\Sh_{np}$ & $d$\big|$n$ & & 50 & 100 & 150 & & 50 & 100 & 150 & & 50 & 100 & 150 \\
      \midrule
      &  5 &&  17.3 & 43 & 66 && 28.4 & 68.3 & 89 && 81.5 & 99.8 & 100 \\
      $\Sb_{np}^{\rm P}$ & 15 &&  7.4 & 15.5 & 22.5 && 8 & 13.5 & 23.9 && 11.9 & 32.7 & 59.1 \\
      & 25 &&  5.7 & 8.9 & 10.6 && 4.3 & 7.7 & 11.3 && 4.6 & 11.5 & 20.1 \\
      \cmidrule(lr){2-14}
      &  5 &&  22.3 & 46.4 & 67.5 && 32.7 & 70.8 & 90 && 81 & 99.8 & 100 \\
      & 15 &&  9.5 & 16.4 & 23 && 9 & 13.4 & 23.7 && 9.1 & 28.1 & 55.5 \\
      $\Sb_{np}^{\rm J}$ & 25 &&  6.6 & 9.4 & 10.5 && 3.8 & 7.2 & 10.7 && 3.4 & 7.7 & 17.2 \\
      & 50 &&  5.3 & 5.3 & 8.4 && 4.6 & 4 & 4.8 && 1.9 & 3.2 & 6.1 \\
      & 100 &&  4.3 & 3.9 & 4.2 && 2.4 & 3 & 4.2 && 0.9 & 1.9 & 2.9 \\
      \midrule
      \multicolumn{14}{c}{$M_{np}$ with $\SA = (1/n)\I_p$ for  single  departures ($\Delta =  0.2 $)}\\
      \midrule
      \multicolumn{2}{c}{} & & \multicolumn{3}{c}{$\tau = 0$} & & \multicolumn{3}{c}{$\tau = 0.3$} & & \multicolumn{3}{c}{$\tau = 0.6$}\\
      \cmidrule(lr){4-6}  \cmidrule(lr){8-10}  \cmidrule(lr){12-14}
      $\Sh_{np}$ & $d$\big|$n$ & & 50 & 100 & 150 & & 50 & 100 & 150 & & 50 & 100 & 150 \\
      \midrule
      &  5 &&  23 & 53.2 & 76.5 && 42 & 83.8 & 96 && 95.6 & 100 & 100 \\
      $\Sb_{np}^{\rm P}$ & 15 &&  9.8 & 32.7 & 58.9 && 18.9 & 57.4 & 89.9 && 79.8 & 100 & 100 \\
      & 25 &&  8 & 25.9 & 45.6 && 11.6 & 47.4 & 80 && 60 & 99.8 & 100 \\
      \cmidrule(lr){2-14}
      &  5 &&  25.7 & 54.4 & 77.5 && 44.1 & 84.4 & 96.3 && 95.4 & 100 & 100 \\
      & 15 &&  10.2 & 32.8 & 58.9 && 17.6 & 56.7 & 89.7 && 75.5 & 100 & 100 \\
      $\Sb_{np}^{\rm J}$ & 25 &&  7.8 & 25.9 & 44.9 && 10.4 & 46.4 & 80.2 && 51.9 & 99.8 & 100 \\
      & 50 &&  4 & 16.6 & 34.8 && 7 & 30.7 & 67.3 && 32.4 & 99.2 & 100 \\
      & 100 &&  2.2 & 10.4 & 26.2 && 4.9 & 22.1 & 54.5 && 17 & 95.9 & 100 \\
      \bottomrule
      \end{tabular}
      \end{center}
      \vskip-9pt
      \small
      Statistics: $E_{np}$ Euclidean norm-based statistic defined in Eq.~\eqref{eq:euclidean}; $M_{np}$ supremum norm-based statistic defined in Eq.~\eqref{eq:supremum}. Estimators: $\Sb_{np}^{\rm P}$ structured plug-in estimator; $\Sb_{np}^{\rm J}$ structured jackknife estimator.
      \end{table}

            \begin{table}[htbp]
 \captionsetup{width=1\linewidth,font=small,skip=0pt}
\caption{Estimated rejection rates (in \%) of tests of $H_0^*$ with $\mathcal{G}=\{\{1,\ldots,d\}\}$ performed at nominal level $5$\%. Each entry is based on $ 1000 $ $n \times d$ datasets drawn from a  Clayton copula  with Kendall's tau matrix $\bs{T}_{\Delta}$ in Eq.~\eqref{eq:departure} (i) with $\Delta =  0.2 $; $\bs{T}$ is as in Eq.~\eqref{eq:T-equi-null}.}
       \label{tab:sim-power-star-single-2-clayton}
      \begin{center}
      \fontsize{8.75}{8.75}\selectfont
      \vskip-12pt
      \begin{tabular}{*{2}{l}*{12}{r}}
      \toprule
      \multicolumn{14}{c}{$E_{np}$ with $\SA = \Sh_{np}$ for  single  departures ($\Delta =  0.2 $)}\\
      \midrule
      \multicolumn{2}{c}{} & & \multicolumn{3}{c}{$\tau = 0$} & & \multicolumn{3}{c}{$\tau = 0.3$} & & \multicolumn{3}{c}{$\tau = 0.6$}\\
      \cmidrule(lr){4-6}  \cmidrule(lr){8-10}  \cmidrule(lr){12-14}
      $\Sh_{np}$ & $d$\big|$n$ & & 50 & 100 & 150 & & 50 & 100 & 150 & & 50 & 100 & 150 \\
      \midrule
      &  5 &&   &  &  && 28.2 & 73.8 & 92.1 && 88.8 & 99.8 & 100 \\
      $\Sb_{np}^{\rm P}$ & 15 &&   &  &  && 12.3 & 30.1 & 50.9 && 46.3 & 92.4 & 99.9 \\
      & 25 &&   &  &  && 7.3 & 17.6 & 27.7 && 26.8 & 67.8 & 93.5 \\
      \cmidrule(lr){2-14}
      &  5 &&   &  &  && 34.3 & 76.6 & 92.7 && 89.9 & 99.8 & 100 \\
      & 15 &&   &  &  && 11 & 27.9 & 49.7 && 26.4 & 87 & 99.8 \\
      $\Sb_{np}^{\rm J}$ & 25 &&   &  &  && 4.8 & 13.9 & 25.4 && 7.5 & 47.1 & 86 \\
      & 50 &&   &  &  && 1.7 & 5.2 & 9.4 && 0.2 & 5.7 & 25.1 \\
      & 100 &&   &  &  && 0.1 & 1.5 & 2.1 && 0 & 0 & 0.8 \\
      \midrule
      \multicolumn{14}{c}{$M_{np}$ with $\SA = \Sh_{np}$ for  single departures ($\Delta =  0.2 $)}\\
      \midrule
      &  5 &&   &  &  && 38.8 & 83.6 & 96.7 && 95.6 & 99.9 & 100 \\
      $\Sb_{np}^{\rm P}$ & 15 &&   &  &  && 28.7 & 78.1 & 95 && 94.4 & 100 & 100 \\
      & 25 &&   &  &  && 25.2 & 73.9 & 94.2 && 90.9 & 99.9 & 100 \\
      \cmidrule(lr){2-14}
      &  5 &&   &  &  && 42.2 & 84.3 & 96.9 && 95.4 & 99.9 & 100 \\
      & 15 &&   &  &  && 27.3 & 76.8 & 94.8 && 92.5 & 100 & 100 \\
      $\Sb_{np}^{\rm J}$ & 25 &&   &  &  && 22.9 & 71.3 & 94 && 88.2 & 99.9 & 100 \\
      & 50 &&   &  &  && 18.3 & 60.6 & 89.9 && 78.8 & 99.9 & 100 \\
      & 100 &&   &  &  && 15.9 & 52.4 & 83.6 && 73.7 & 100 & 100 \\
      \midrule
      \multicolumn{14}{c}{$E_{np}$ with $\SA = (1/n)\I_p$ for  single  departures ($\Delta =  0.2 $)}\\
      \midrule
      \multicolumn{2}{c}{} & & \multicolumn{3}{c}{$\tau = 0$} & & \multicolumn{3}{c}{$\tau = 0.3$} & & \multicolumn{3}{c}{$\tau = 0.6$}\\
      \cmidrule(lr){4-6}  \cmidrule(lr){8-10}  \cmidrule(lr){12-14}
      $\Sh_{np}$ & $d$\big|$n$ & & 50 & 100 & 150 & & 50 & 100 & 150 & & 50 & 100 & 150 \\
      \midrule
      &  5 &&   &  &  && 24 & 65.7 & 87.1 && 80.2 & 99.7 & 100 \\
      $\Sb_{np}^{\rm P}$ & 15 &&   &  &  && 6.1 & 11.6 & 16.4 && 10.2 & 28.1 & 46.6 \\
      & 25 &&   &  &  && 4.3 & 8 & 8.9 && 5.5 & 12.2 & 16.6 \\
      \cmidrule(lr){2-14}
      &  5 &&   &  &  && 28.8 & 67.6 & 88.2 && 80.7 & 99.7 & 100 \\
      & 15 &&   &  &  && 6.8 & 12.3 & 17.3 && 8.1 & 24.3 & 44.3 \\
      $\Sb_{np}^{\rm J}$ & 25 &&   &  &  && 4.8 & 8.5 & 9 && 4.2 & 9.3 & 13.6 \\
      & 50 &&   &  &  && 5 & 5 & 5.1 && 1.3 & 4.3 & 5.9 \\
      & 100 &&   &  &  && 2.9 & 5 & 4.7 && 0.6 & 3.2 & 3.8 \\
      \midrule
      \multicolumn{14}{c}{$M_{np}$ with $\SA = (1/n)\I_p$ for  single  departures ($\Delta =  0.2 $)}\\
      \midrule
      \multicolumn{2}{c}{} & & \multicolumn{3}{c}{$\tau = 0$} & & \multicolumn{3}{c}{$\tau = 0.3$} & & \multicolumn{3}{c}{$\tau = 0.6$}\\
      \cmidrule(lr){4-6}  \cmidrule(lr){8-10}  \cmidrule(lr){12-14}
      $\Sh_{np}$ & $d$\big|$n$ & & 50 & 100 & 150 & & 50 & 100 & 150 & & 50 & 100 & 150 \\
      \midrule
      &  5 &&   &  &  && 34 & 81.2 & 95.5 && 94 & 99.9 & 100 \\
      $\Sb_{np}^{\rm P}$ & 15 &&   &  &  && 13.3 & 55.7 & 85 && 74.6 & 99.8 & 100 \\
      & 25 &&   &  &  && 10.7 & 45.9 & 76.5 && 55.3 & 99.2 & 100 \\
      \cmidrule(lr){2-14}
      &  5 &&   &  &  && 37.4 & 81.9 & 95.6 && 93.4 & 99.9 & 100 \\
      & 15 &&   &  &  && 13.3 & 54.9 & 84.4 && 70.2 & 99.8 & 100 \\
      $\Sb_{np}^{\rm J}$ & 25 &&   &  &  && 9.8 & 45.4 & 76.1 && 49.6 & 99.1 & 100 \\
      & 50 &&   &  &  && 8.1 & 27.9 & 59.5 && 30.9 & 97.2 & 100 \\
      & 100 &&   &  &  && 8.4 & 18 & 44.5 && 18.7 & 93.5 & 99.9 \\
      \bottomrule
      \end{tabular}
      \end{center}
      \vskip-9pt
      \small
      Statistics: $E_{np}$ Euclidean norm-based statistic defined in Eq.~\eqref{eq:euclidean}; $M_{np}$ supremum norm-based statistic defined in Eq.~\eqref{eq:supremum}. Estimators: $\Sb_{np}^{\rm P}$ structured plug-in estimator; $\Sb_{np}^{\rm J}$ structured jackknife estimator.
      \end{table}

            \begin{table}[htbp]
 \captionsetup{width=1\linewidth,font=small,skip=0pt}
\caption{Estimated rejection rates (in \%) of tests of $H_0^*$ with $\mathcal{G}=\{\{1,\ldots,d\}\}$ performed at nominal level $5$\%. Each entry is based on $ 2500 $ $n \times d$ datasets drawn from a  Normal copula  with Kendall's tau matrix $\bs{T}_{\Delta}$ in Eq.~\eqref{eq:departure} (i) with $\Delta =  0.1 $; $\bs{T}$ is as in Eq.~\eqref{eq:T-equi-null}.}
       \label{tab:sim-power-star-column-1-normal}
      \begin{center}
      \fontsize{8.75}{8.75}\selectfont
      \vskip-12pt
      \begin{tabular}{*{2}{l}*{12}{r}}
      \toprule
      \multicolumn{14}{c}{$E_{np}$ with $\SA = \Sh_{np}$ for  column  departures ($\Delta =  0.1 $)}\\
      \midrule
      \multicolumn{2}{c}{} & & \multicolumn{3}{c}{$\tau = 0$} & & \multicolumn{3}{c}{$\tau = 0.3$} & & \multicolumn{3}{c}{$\tau = 0.6$}\\
      \cmidrule(lr){4-6}  \cmidrule(lr){8-10}  \cmidrule(lr){12-14}
      $\Sh_{np}$ & $d$\big|$n$ & & 50 & 100 & 150 & & 50 & 100 & 150 & & 50 & 100 & 150 \\
      \midrule
      &  5 &&  10.8 & 23.6 & 37.2 && 14.4 & 36.2 & 55.6 && 42.4 & 86 & 97.6 \\
      $\Sb_{np}^{\rm P}$ & 15 &&  9.5 & 21.6 & 35 && 9.8 & 22.9 & 38.6 && 29.1 & 73.9 & 93.2 \\
      & 25 &&  7.4 & 14.6 & 21.8 && 7.3 & 14.8 & 24 && 19.6 & 55.4 & 81.6 \\
      \cmidrule(lr){2-14}
      &  5 &&  15.6 & 27.5 & 40.5 && 18.9 & 40.1 & 58.4 && 44.1 & 86.3 & 97.7 \\
      & 15 &&  11.5 & 24.3 & 37.6 && 7.6 & 21 & 37.6 && 13.2 & 63.4 & 90.5 \\
      $\Sb_{np}^{\rm J}$ & 25 &&  7.7 & 15.1 & 22.7 && 3.1 & 10.6 & 19.5 && 1.5 & 28.1 & 68.5 \\
      & 50 &&  4.2 & 7.4 & 12 && 0.4 & 2.6 & 6.2 && 0 & 0.7 & 10.7 \\
      & 100 &&  2.2 & 3.1 & 5.7 && 0 & 0.1 & 0.8 && 0 & 0 & 0.1 \\
      \midrule
      \multicolumn{14}{c}{$M_{np}$ with $\SA = \Sh_{np}$ for  column departures ($\Delta =  0.1 $)}\\
      \midrule
      &  5 &&  8.1 & 14 & 23.5 && 9.4 & 20.4 & 34.6 && 22.9 & 55.6 & 80.8 \\
      $\Sb_{np}^{\rm P}$ & 15 &&  6 & 11.5 & 17.7 && 7.8 & 13.5 & 22.2 && 19.4 & 48 & 72.4 \\
      & 25 &&  5.7 & 8.3 & 11.4 && 7.1 & 11.1 & 15.2 && 14.9 & 35.1 & 57.8 \\
      \cmidrule(lr){2-14}
      &  5 &&  11 & 16.9 & 26.5 && 11.4 & 23.6 & 37.8 && 25.6 & 62.2 & 85.6 \\
      & 15 &&  6.2 & 12 & 18.3 && 6.8 & 13.2 & 22.2 && 13.7 & 46 & 73.4 \\
      $\Sb_{np}^{\rm J}$ & 25 &&  5.7 & 8.3 & 11.7 && 5.4 & 10.1 & 14.4 && 8.5 & 29 & 55 \\
      & 50 &&  4.3 & 5.4 & 7.9 && 4.2 & 5.5 & 7.6 && 3.2 & 14.3 & 27.3 \\
      & 100 &&  3.5 & 5.2 & 5.6 && 4.9 & 5.9 & 6.4 && 3.3 & 6.6 & 12.2 \\
      \midrule
      \multicolumn{14}{c}{$E_{np}$ with $\SA = (1/n)\I_p$ for  column  departures ($\Delta =  0.1 $)}\\
      \midrule
      \multicolumn{2}{c}{} & & \multicolumn{3}{c}{$\tau = 0$} & & \multicolumn{3}{c}{$\tau = 0.3$} & & \multicolumn{3}{c}{$\tau = 0.6$}\\
      \cmidrule(lr){4-6}  \cmidrule(lr){8-10}  \cmidrule(lr){12-14}
      $\Sh_{np}$ & $d$\big|$n$ & & 50 & 100 & 150 & & 50 & 100 & 150 & & 50 & 100 & 150 \\
      \midrule
      &  5 &&  13.7 & 27.4 & 42.6 && 20.2 & 48.1 & 66.8 && 57.4 & 92.8 & 99.4 \\
      $\Sb_{np}^{\rm P}$ & 15 &&  18.5 & 44.4 & 66.4 && 24.1 & 56.9 & 76.2 && 68.5 & 97.4 & 99.8 \\
      & 25 &&  19.8 & 44.5 & 65.8 && 23.2 & 51.8 & 72.6 && 66.6 & 96.4 & 99.9 \\
      \cmidrule(lr){2-14}
      &  5 &&  16.9 & 30.3 & 44.8 && 23.4 & 49.9 & 68.2 && 56.9 & 92.4 & 99.3 \\
      & 15 &&  21.8 & 46.5 & 67.2 && 24 & 56.5 & 76 && 58 & 96 & 99.8 \\
      $\Sb_{np}^{\rm J}$ & 25 &&  21.8 & 45.7 & 66.3 && 21.4 & 50.1 & 71.9 && 50.4 & 93.8 & 99.6 \\
      & 50 &&  18.4 & 41.1 & 62.5 && 13.9 & 40.1 & 62.6 && 32.2 & 88 & 99.1 \\
      & 100 &&  13.5 & 30.7 & 52.2 && 8 & 26.2 & 44.7 && 11.5 & 68.4 & 95.1 \\
      \midrule
      \multicolumn{14}{c}{$M_{np}$ with $\SA = (1/n)\I_p$ for  column  departures ($\Delta =  0.1 $)}\\
      \midrule
      \multicolumn{2}{c}{} & & \multicolumn{3}{c}{$\tau = 0$} & & \multicolumn{3}{c}{$\tau = 0.3$} & & \multicolumn{3}{c}{$\tau = 0.6$}\\
      \cmidrule(lr){4-6}  \cmidrule(lr){8-10}  \cmidrule(lr){12-14}
      $\Sh_{np}$ & $d$\big|$n$ & & 50 & 100 & 150 & & 50 & 100 & 150 & & 50 & 100 & 150 \\
      \midrule
      &  5 &&  9.5 & 15.9 & 26.8 && 12 & 29.3 & 45.7 && 34.2 & 76.6 & 94 \\
      $\Sb_{np}^{\rm P}$ & 15 &&  9.6 & 24.2 & 37.3 && 20.8 & 43 & 64.8 && 62.8 & 95.3 & 99.5 \\
      & 25 &&  10.1 & 23.9 & 38.8 && 24.1 & 47.1 & 67 && 68.5 & 96.4 & 99.8 \\
      \cmidrule(lr){2-14}
      &  5 &&  11.8 & 18.2 & 28.5 && 15.3 & 32 & 48.1 && 36.5 & 77.8 & 94.4 \\
      & 15 &&  10.6 & 25 & 38.6 && 20.4 & 42.7 & 64.7 && 54.6 & 93.6 & 99.4 \\
      $\Sb_{np}^{\rm J}$ & 25 &&  10.2 & 24.2 & 39.2 && 22.4 & 45.2 & 66.2 && 59.8 & 94.8 & 99.7 \\
      & 50 &&  9.4 & 22.4 & 39 && 20.7 & 46.7 & 68.9 && 61.4 & 95.8 & 99.8 \\
      & 100 &&  7.8 & 19.5 & 36.2 && 22.9 & 45.7 & 67.1 && 59.7 & 94.8 & 99.6 \\
      \bottomrule
      \end{tabular}
      \end{center}
      \vskip-9pt
      \small
      Statistics: $E_{np}$ Euclidean norm-based statistic defined in Eq.~\eqref{eq:euclidean}; $M_{np}$ supremum norm-based statistic defined in Eq.~\eqref{eq:supremum}. Estimators: $\Sb_{np}^{\rm P}$ structured plug-in estimator; $\Sb_{np}^{\rm J}$ structured jackknife estimator.
      \end{table}

            \begin{table}[htbp]
 \captionsetup{width=1\linewidth,font=small,skip=0pt}
\caption{Estimated rejection rates (in \%) of tests of $H_0^*$ with $\mathcal{G}=\{\{1,\ldots,d\}\}$ performed at nominal level $5$\%. Each entry is based on $ 1000 $ $n \times d$ datasets drawn from a  $t_4$ copula  with Kendall's tau matrix $\bs{T}_{\Delta}$ in Eq.~\eqref{eq:departure} (i) with $\Delta =  0.1 $; $\bs{T}$ is as in Eq.~\eqref{eq:T-equi-null}.}
       \label{tab:sim-power-star-column-1-t4}
      \begin{center}
      \fontsize{8.75}{8.75}\selectfont
      \vskip-12pt
      \begin{tabular}{*{2}{l}*{12}{r}}
      \toprule
      \multicolumn{14}{c}{$E_{np}$ with $\SA = \Sh_{np}$ for  column  departures ($\Delta =  0.1 $)}\\
      \midrule
      \multicolumn{2}{c}{} & & \multicolumn{3}{c}{$\tau = 0$} & & \multicolumn{3}{c}{$\tau = 0.3$} & & \multicolumn{3}{c}{$\tau = 0.6$}\\
      \cmidrule(lr){4-6}  \cmidrule(lr){8-10}  \cmidrule(lr){12-14}
      $\Sh_{np}$ & $d$\big|$n$ & & 50 & 100 & 150 & & 50 & 100 & 150 & & 50 & 100 & 150 \\
      \midrule
      &  5 &&  7.4 & 20.3 & 30.5 && 10.3 & 29.7 & 42.5 && 33.4 & 76.8 & 92.3 \\
      $\Sb_{np}^{\rm P}$ & 15 &&  7.9 & 16.7 & 29.2 && 9 & 17.7 & 29.2 && 23.6 & 58.5 & 82 \\
      & 25 &&  7.3 & 11.4 & 20.7 && 5.9 & 12.5 & 17.4 && 12.5 & 40.2 & 64.7 \\
      \cmidrule(lr){2-14}
      &  5 &&  12.5 & 23.6 & 34.8 && 15.6 & 34.2 & 45.8 && 38.7 & 78.6 & 92.9 \\
      & 15 &&  13.4 & 20.1 & 32.6 && 9.9 & 17.9 & 29.8 && 14.3 & 53.1 & 80.3 \\
      $\Sb_{np}^{\rm J}$ & 25 &&  10.9 & 14.9 & 22.5 && 5.2 & 11.6 & 16.6 && 2.1 & 24.6 & 54 \\
      & 50 &&  8 & 9.8 & 15.2 && 1.6 & 5 & 8.4 && 0.1 & 2.7 & 12.1 \\
      & 100 &&  10.1 & 7.8 & 8.3 && 0.4 & 1.4 & 2.6 && 0 & 0.1 & 0.3 \\
      \midrule
      \multicolumn{14}{c}{$M_{np}$ with $\SA = \Sh_{np}$ for  column departures ($\Delta =  0.1 $)}\\
      \midrule
      &  5 &&  5.4 & 13 & 16.6 && 7.9 & 18.2 & 24.5 && 18.8 & 46.8 & 66.6 \\
      $\Sb_{np}^{\rm P}$ & 15 &&  6.1 & 9.8 & 11.8 && 7.6 & 10.7 & 18 && 15.2 & 35 & 56.6 \\
      & 25 &&  4.2 & 7.6 & 9.5 && 5.1 & 10.4 & 11.2 && 11.2 & 27.1 & 42.2 \\
      \cmidrule(lr){2-14}
      &  5 &&  9 & 16.1 & 20.1 && 10.6 & 21.6 & 27.6 && 21.7 & 53.8 & 73.1 \\
      & 15 &&  7.4 & 10.8 & 13.6 && 7.3 & 10.9 & 18.6 && 12.3 & 34 & 57.2 \\
      $\Sb_{np}^{\rm J}$ & 25 &&  4.4 & 8.4 & 9.6 && 4.3 & 10.3 & 11 && 6.9 & 23.9 & 40.8 \\
      & 50 &&  4.2 & 5.1 & 7.8 && 5.5 & 6.6 & 5.5 && 5.4 & 12.8 & 19.9 \\
      & 100 &&  2.5 & 3.7 & 5.7 && 7.9 & 5.2 & 6 && 10.6 & 7.7 & 9.9 \\
      \midrule
      \multicolumn{14}{c}{$E_{np}$ with $\SA = (1/n)\I_p$ for  column  departures ($\Delta =  0.1 $)}\\
      \midrule
      \multicolumn{2}{c}{} & & \multicolumn{3}{c}{$\tau = 0$} & & \multicolumn{3}{c}{$\tau = 0.3$} & & \multicolumn{3}{c}{$\tau = 0.6$}\\
      \cmidrule(lr){4-6}  \cmidrule(lr){8-10}  \cmidrule(lr){12-14}
      $\Sh_{np}$ & $d$\big|$n$ & & 50 & 100 & 150 & & 50 & 100 & 150 & & 50 & 100 & 150 \\
      \midrule
      &  5 &&  8.9 & 23.8 & 36.1 && 15 & 38.7 & 55.3 && 47 & 85.9 & 96.4 \\
      $\Sb_{np}^{\rm P}$ & 15 &&  16.3 & 34.5 & 54.6 && 22.2 & 42.9 & 63.4 && 58.9 & 90.9 & 99 \\
      & 25 &&  17.8 & 36 & 58 && 17.5 & 42 & 59.3 && 49.8 & 88.8 & 98.2 \\
      \cmidrule(lr){2-14}
      &  5 &&  12.8 & 25.1 & 38.5 && 19.6 & 40.9 & 57 && 49 & 86.3 & 96.4 \\
      & 15 &&  19.8 & 36.6 & 57.3 && 25.1 & 44.3 & 64.5 && 54.3 & 89.9 & 98.9 \\
      $\Sb_{np}^{\rm J}$ & 25 &&  22.4 & 40.3 & 61.2 && 20 & 42.9 & 59.6 && 40.9 & 86.4 & 97.8 \\
      & 50 &&  18.6 & 37 & 55.3 && 15.5 & 33.9 & 53.3 && 27 & 77.8 & 95.1 \\
      & 100 &&  18.2 & 31.2 & 48.1 && 10.3 & 24.2 & 37 && 14.8 & 61.2 & 86.1 \\
      \midrule
      \multicolumn{14}{c}{$M_{np}$ with $\SA = (1/n)\I_p$ for  column  departures ($\Delta =  0.1 $)}\\
      \midrule
      \multicolumn{2}{c}{} & & \multicolumn{3}{c}{$\tau = 0$} & & \multicolumn{3}{c}{$\tau = 0.3$} & & \multicolumn{3}{c}{$\tau = 0.6$}\\
      \cmidrule(lr){4-6}  \cmidrule(lr){8-10}  \cmidrule(lr){12-14}
      $\Sh_{np}$ & $d$\big|$n$ & & 50 & 100 & 150 & & 50 & 100 & 150 & & 50 & 100 & 150 \\
      \midrule
      &  5 &&  6.3 & 14.7 & 20.7 && 10.7 & 23.2 & 33.3 && 27.7 & 64 & 85.3 \\
      $\Sb_{np}^{\rm P}$ & 15 &&  7.2 & 17.3 & 28.5 && 18.2 & 32.6 & 51.8 && 55.4 & 87 & 97.6 \\
      & 25 &&  7.1 & 17.5 & 29.9 && 18 & 38.7 & 56.3 && 54.4 & 87.6 & 98.4 \\
      \cmidrule(lr){2-14}
      &  5 &&  9.4 & 17.2 & 22.8 && 12.5 & 26.5 & 35.8 && 32.4 & 66.9 & 86.3 \\
      & 15 &&  8.6 & 18.4 & 30.4 && 18.9 & 33.5 & 52.3 && 50.9 & 85.8 & 97.4 \\
      $\Sb_{np}^{\rm J}$ & 25 &&  8.1 & 18.5 & 31.3 && 18.1 & 38.6 & 56.5 && 48.2 & 86.4 & 98.3 \\
      & 50 &&  6.6 & 19 & 29.1 && 18.8 & 39.4 & 54.2 && 52.6 & 90.2 & 97.4 \\
      & 100 &&  5.8 & 14.4 & 28.1 && 19.2 & 37.4 & 54.2 && 52.8 & 87.9 & 97.4 \\
      \bottomrule
      \end{tabular}
      \end{center}
      \vskip-9pt
      \small
      Statistics: $E_{np}$ Euclidean norm-based statistic defined in Eq.~\eqref{eq:euclidean}; $M_{np}$ supremum norm-based statistic defined in Eq.~\eqref{eq:supremum}. Estimators: $\Sb_{np}^{\rm P}$ structured plug-in estimator; $\Sb_{np}^{\rm J}$ structured jackknife estimator.
      \end{table}

           \begin{table}[htbp]
 \captionsetup{width=1\linewidth,font=small,skip=0pt}
\caption{Estimated rejection rates (in \%) of tests of $H_0^*$ with $\mathcal{G}=\{\{1,\ldots,d\}\}$ performed at nominal level $5$\%. Each entry is based on $ 1000 $ $n \times d$ datasets drawn from a  Gumbel copula  with Kendall's tau matrix $\bs{T}_{\Delta}$ in Eq.~\eqref{eq:departure} (i) with $\Delta =  0.1 $; $\bs{T}$ is as in Eq.~\eqref{eq:T-equi-null}.}
       \label{tab:sim-power-star-column-1-gumbel}
      \begin{center}
      \fontsize{8.75}{8.75}\selectfont
      \vskip-12pt
      \begin{tabular}{*{2}{l}*{12}{r}}
      \toprule
      \multicolumn{14}{c}{$E_{np}$ with $\SA = \Sh_{np}$ for  column  departures ($\Delta =  0.1 $)}\\
      \midrule
      \multicolumn{2}{c}{} & & \multicolumn{3}{c}{$\tau = 0$} & & \multicolumn{3}{c}{$\tau = 0.3$} & & \multicolumn{3}{c}{$\tau = 0.6$}\\
      \cmidrule(lr){4-6}  \cmidrule(lr){8-10}  \cmidrule(lr){12-14}
      $\Sh_{np}$ & $d$\big|$n$ & & 50 & 100 & 150 & & 50 & 100 & 150 & & 50 & 100 & 150 \\
      \midrule
      &  5 &&  11.1 & 23.5 & 41.7 && 12.8 & 35.4 & 49.2 && 32.9 & 74.2 & 93.8 \\
      $\Sb_{np}^{\rm P}$ & 15 &&  12.5 & 23.2 & 41.2 && 7.8 & 23.3 & 36.2 && 21.2 & 53.1 & 77.4 \\
      & 25 &&  9.5 & 17.9 & 33.6 && 6.3 & 14.5 & 24.3 && 13 & 38.3 & 59.4 \\
      \cmidrule(lr){2-14}
      &  5 &&  16.7 & 28 & 44.7 && 16.9 & 38 & 51.9 && 36.3 & 75.5 & 94.6 \\
      & 15 &&  14.6 & 26 & 43.9 && 7.2 & 22 & 35.8 && 9.5 & 44.3 & 73.2 \\
      $\Sb_{np}^{\rm J}$ & 25 &&  10.6 & 18.8 & 35 && 2.3 & 11.5 & 21.3 && 1.3 & 18 & 44.7 \\
      & 50 &&  4.2 & 12.9 & 20.9 && 0.6 & 2.4 & 6 && 0 & 0.4 & 4.8 \\
      & 100 &&  1.8 & 4.6 & 9.5 && 0 & 0.4 & 1.4 && 0 & 0 & 0 \\
      \midrule
      \multicolumn{14}{c}{$M_{np}$ with $\SA = \Sh_{np}$ for  column departures ($\Delta =  0.1 $)}\\
      \midrule
      &  5 &&  8.7 & 15.9 & 27.3 && 9.3 & 18.9 & 26 && 16 & 46.6 & 67.9 \\
      $\Sb_{np}^{\rm P}$ & 15 &&  6.4 & 13.9 & 21.9 && 6.5 & 13.4 & 21.7 && 14.5 & 32.5 & 51.5 \\
      & 25 &&  5.1 & 9.8 & 13.9 && 4.9 & 10.1 & 12.8 && 12.3 & 23.8 & 38.7 \\
      \cmidrule(lr){2-14}
      &  5 &&  12.2 & 18.3 & 30.8 && 11.5 & 22.5 & 29.8 && 18.1 & 50.8 & 74.4 \\
      & 15 &&  6.9 & 14.6 & 23.5 && 5.4 & 13.2 & 22.4 && 10.2 & 31.1 & 51.2 \\
      $\Sb_{np}^{\rm J}$ & 25 &&  5.4 & 10.5 & 14.2 && 3.6 & 9 & 12.2 && 7.1 & 19.4 & 36.3 \\
      & 50 &&  5.2 & 6.1 & 8.2 && 4.1 & 5.8 & 6.4 && 6.3 & 10.5 & 17.1 \\
      & 100 &&  3.3 & 4.9 & 4.5 && 6.3 & 5.8 & 6.1 && 7.7 & 9.9 & 10.5 \\
      \midrule
      \multicolumn{14}{c}{$E_{np}$ with $\SA = (1/n)\I_p$ for  column  departures ($\Delta =  0.1 $)}\\
      \midrule
      \multicolumn{2}{c}{} & & \multicolumn{3}{c}{$\tau = 0$} & & \multicolumn{3}{c}{$\tau = 0.3$} & & \multicolumn{3}{c}{$\tau = 0.6$}\\
      \cmidrule(lr){4-6}  \cmidrule(lr){8-10}  \cmidrule(lr){12-14}
      $\Sh_{np}$ & $d$\big|$n$ & & 50 & 100 & 150 & & 50 & 100 & 150 & & 50 & 100 & 150 \\
      \midrule
      &  5 &&  13.4 & 29.5 & 48.3 && 19.6 & 44.4 & 61.8 && 49 & 86.5 & 97.4 \\
      $\Sb_{np}^{\rm P}$ & 15 &&  22.6 & 43.1 & 66.8 && 22.6 & 52.9 & 73.4 && 55.9 & 88.8 & 98.7 \\
      & 25 &&  24.3 & 45.7 & 70.8 && 20.3 & 49.5 & 71.2 && 52.4 & 88 & 98.3 \\
      \cmidrule(lr){2-14}
      &  5 &&  17.4 & 32.1 & 50.1 && 23.1 & 45.9 & 62.7 && 49.5 & 86.4 & 97.3 \\
      & 15 &&  24.6 & 45.2 & 67.9 && 23.3 & 53.2 & 73.2 && 51.8 & 87.3 & 98.5 \\
      $\Sb_{np}^{\rm J}$ & 25 &&  25.5 & 46.2 & 71.1 && 19 & 49.4 & 70.8 && 43 & 85.3 & 97.7 \\
      & 50 &&  18 & 49 & 67.4 && 14.1 & 38.2 & 59.1 && 27.5 & 73.5 & 94.5 \\
      & 100 &&  16.1 & 39.6 & 65.1 && 8.5 & 27.4 & 44.7 && 12.1 & 56.7 & 84.4 \\
      \midrule
      \multicolumn{14}{c}{$M_{np}$ with $\SA = (1/n)\I_p$ for  column  departures ($\Delta =  0.1 $)}\\
      \midrule
      \multicolumn{2}{c}{} & & \multicolumn{3}{c}{$\tau = 0$} & & \multicolumn{3}{c}{$\tau = 0.3$} & & \multicolumn{3}{c}{$\tau = 0.6$}\\
      \cmidrule(lr){4-6}  \cmidrule(lr){8-10}  \cmidrule(lr){12-14}
      $\Sh_{np}$ & $d$\big|$n$ & & 50 & 100 & 150 & & 50 & 100 & 150 & & 50 & 100 & 150 \\
      \midrule
      &  5 &&  9.7 & 18.5 & 30.6 && 12.2 & 26.9 & 37.6 && 26 & 66.7 & 86.6 \\
      $\Sb_{np}^{\rm P}$ & 15 &&  12.6 & 25.4 & 39.9 && 17 & 42.6 & 60.9 && 54 & 85.3 & 97.2 \\
      & 25 &&  12.9 & 24.9 & 42.7 && 20.7 & 43.4 & 65.3 && 57.8 & 88.1 & 98.7 \\
      \cmidrule(lr){2-14}
      &  5 &&  12.7 & 20.1 & 33 && 16.1 & 30 & 40.8 && 29.4 & 68.8 & 87.8 \\
      & 15 &&  13.4 & 26 & 40.7 && 17.5 & 42.8 & 61.6 && 51.2 & 83.9 & 96.7 \\
      $\Sb_{np}^{\rm J}$ & 25 &&  13.2 & 25.2 & 42.6 && 19.1 & 43.6 & 64.7 && 51 & 85.4 & 98.1 \\
      & 50 &&  12.3 & 27.2 & 44.5 && 21.4 & 42.6 & 64.3 && 51.6 & 87.3 & 97.5 \\
      & 100 &&  11.9 & 25.5 & 42 && 21.1 & 43.2 & 63.7 && 51.6 & 87.8 & 98.3 \\
      \bottomrule
      \end{tabular}
      \end{center}
      \vskip-9pt
      \small
      Statistics: $E_{np}$ Euclidean norm-based statistic defined in Eq.~\eqref{eq:euclidean}; $M_{np}$ supremum norm-based statistic defined in Eq.~\eqref{eq:supremum}. Estimators: $\Sb_{np}^{\rm P}$ structured plug-in estimator; $\Sb_{np}^{\rm J}$ structured jackknife estimator.
      \end{table}

            \begin{table}[htbp]
 \captionsetup{width=1\linewidth,font=small,skip=0pt}
\caption{Estimated rejection rates (in \%) of tests of $H_0^*$ with $\mathcal{G}=\{\{1,\ldots,d\}\}$ performed at nominal level $5$\%. Each entry is based on $ 1000 $ $n \times d$ datasets drawn from a  Clayton copula  with Kendall's tau matrix $\bs{T}_{\Delta}$ in Eq.~\eqref{eq:departure} (i) with $\Delta =  0.1 $; $\bs{T}$ is as in Eq.~\eqref{eq:T-equi-null}.}
       \label{tab:sim-power-star-column-1-clayton}
      \begin{center}
      \fontsize{8.75}{8.75}\selectfont
      \vskip-12pt
      \begin{tabular}{*{2}{l}*{12}{r}}
      \toprule
      \multicolumn{14}{c}{$E_{np}$ with $\SA = \Sh_{np}$ for  column  departures ($\Delta =  0.1 $)}\\
      \midrule
      \multicolumn{2}{c}{} & & \multicolumn{3}{c}{$\tau = 0$} & & \multicolumn{3}{c}{$\tau = 0.3$} & & \multicolumn{3}{c}{$\tau = 0.6$}\\
      \cmidrule(lr){4-6}  \cmidrule(lr){8-10}  \cmidrule(lr){12-14}
      $\Sh_{np}$ & $d$\big|$n$ & & 50 & 100 & 150 & & 50 & 100 & 150 & & 50 & 100 & 150 \\
      \midrule
      &  5 &&   &  &  && 15.4 & 31.7 & 50 && 33.6 & 73.2 & 93.2 \\
      $\Sb_{np}^{\rm P}$ & 15 &&   &  &  && 9.2 & 19.8 & 30.8 && 19.3 & 52.4 & 78.9 \\
      & 25 &&   &  &  && 7.5 & 12.5 & 17.6 && 13.8 & 37 & 58 \\
      \cmidrule(lr){2-14}
      &  5 &&   &  &  && 19.3 & 35 & 53.4 && 37.8 & 75.6 & 94 \\
      & 15 &&   &  &  && 7.5 & 18.9 & 30.5 && 8.9 & 44.4 & 75.7 \\
      $\Sb_{np}^{\rm J}$ & 25 &&   &  &  && 4 & 8.5 & 14.3 && 1.6 & 17.5 & 44.2 \\
      & 50 &&   &  &  && 0.7 & 2.8 & 5.9 && 0 & 0.5 & 6.5 \\
      & 100 &&   &  &  && 0 & 0.5 & 0.9 && 0 & 0 & 0 \\
      \midrule
      \multicolumn{14}{c}{$M_{np}$ with $\SA = \Sh_{np}$ for  column departures ($\Delta =  0.1 $)}\\
      \midrule
      &  5 &&   &  &  && 9.7 & 18.9 & 30.5 && 17.2 & 43 & 68.5 \\
      $\Sb_{np}^{\rm P}$ & 15 &&   &  &  && 7.6 & 12.8 & 16.2 && 13 & 31.4 & 53.4 \\
      & 25 &&   &  &  && 6.4 & 10.5 & 11.1 && 12.2 & 23 & 37 \\
      \cmidrule(lr){2-14}
      &  5 &&   &  &  && 13.4 & 21.9 & 33.6 && 21.6 & 49 & 74.8 \\
      & 15 &&   &  &  && 6.7 & 12.7 & 16.5 && 9.6 & 30.8 & 53.7 \\
      $\Sb_{np}^{\rm J}$ & 25 &&   &  &  && 4.5 & 9.6 & 10.4 && 6.9 & 20.1 & 34.8 \\
      & 50 &&   &  &  && 5.1 & 5.2 & 7.9 && 5.2 & 9.8 & 16 \\
      & 100 &&   &  &  && 9.1 & 7.5 & 6.6 && 8.8 & 9.8 & 10.2 \\
      \midrule
      \multicolumn{14}{c}{$E_{np}$ with $\SA = (1/n)\I_p$ for  column  departures ($\Delta =  0.1 $)}\\
      \midrule
      \multicolumn{2}{c}{} & & \multicolumn{3}{c}{$\tau = 0$} & & \multicolumn{3}{c}{$\tau = 0.3$} & & \multicolumn{3}{c}{$\tau = 0.6$}\\
      \cmidrule(lr){4-6}  \cmidrule(lr){8-10}  \cmidrule(lr){12-14}
      $\Sh_{np}$ & $d$\big|$n$ & & 50 & 100 & 150 & & 50 & 100 & 150 & & 50 & 100 & 150 \\
      \midrule
      &  5 &&   &  &  && 21.2 & 42.8 & 63.4 && 49.6 & 86.5 & 97.5 \\
      $\Sb_{np}^{\rm P}$ & 15 &&   &  &  && 20.8 & 48.2 & 68.9 && 54.9 & 89.5 & 98.5 \\
      & 25 &&   &  &  && 19.5 & 42.7 & 62.6 && 51.7 & 88.1 & 97.9 \\
      \cmidrule(lr){2-14}
      &  5 &&   &  &  && 24.5 & 45.3 & 64.2 && 51.1 & 86.9 & 97.3 \\
      & 15 &&   &  &  && 22.2 & 50.2 & 68.7 && 49.9 & 88.2 & 98.3 \\
      $\Sb_{np}^{\rm J}$ & 25 &&   &  &  && 20.4 & 43.4 & 62.5 && 42.2 & 84.7 & 97.4 \\
      & 50 &&   &  &  && 13.6 & 33 & 55 && 28.1 & 75.5 & 95.5 \\
      & 100 &&   &  &  && 8.7 & 21.6 & 36.2 && 14.9 & 54.7 & 84.7 \\
      \midrule
      \multicolumn{14}{c}{$M_{np}$ with $\SA = (1/n)\I_p$ for  column  departures ($\Delta =  0.1 $)}\\
      \midrule
      \multicolumn{2}{c}{} & & \multicolumn{3}{c}{$\tau = 0$} & & \multicolumn{3}{c}{$\tau = 0.3$} & & \multicolumn{3}{c}{$\tau = 0.6$}\\
      \cmidrule(lr){4-6}  \cmidrule(lr){8-10}  \cmidrule(lr){12-14}
      $\Sh_{np}$ & $d$\big|$n$ & & 50 & 100 & 150 & & 50 & 100 & 150 & & 50 & 100 & 150 \\
      \midrule
      &  5 &&   &  &  && 12.5 & 26.6 & 41.9 && 29.3 & 63.4 & 87.2 \\
      $\Sb_{np}^{\rm P}$ & 15 &&   &  &  && 17.5 & 37.1 & 56 && 49.6 & 85.6 & 96 \\
      & 25 &&   &  &  && 19.2 & 39.7 & 59.2 && 56.9 & 87.5 & 97.2 \\
      \cmidrule(lr){2-14}
      &  5 &&   &  &  && 16.4 & 29.9 & 45.3 && 33.8 & 66.8 & 88.8 \\
      & 15 &&   &  &  && 17.6 & 37.6 & 56.6 && 46.1 & 84.1 & 95.8 \\
      $\Sb_{np}^{\rm J}$ & 25 &&   &  &  && 18.8 & 39.3 & 58.7 && 52.1 & 85.7 & 97.2 \\
      & 50 &&   &  &  && 19.5 & 36.8 & 59 && 52.6 & 87.8 & 98.8 \\
      & 100 &&   &  &  && 23.1 & 41.2 & 57.4 && 54.2 & 87.8 & 98.2 \\
      \bottomrule
      \end{tabular}
      \end{center}
      \vskip-9pt
      \small
      Statistics: $E_{np}$ Euclidean norm-based statistic defined in Eq.~\eqref{eq:euclidean}; $M_{np}$ supremum norm-based statistic defined in Eq.~\eqref{eq:supremum}. Estimators: $\Sb_{np}^{\rm P}$ structured plug-in estimator; $\Sb_{np}^{\rm J}$ structured jackknife estimator.
      \end{table}

            \begin{table}[htbp]
 \captionsetup{width=1\linewidth,font=small,skip=0pt}
\caption{Estimated rejection rates (in \%) of tests of $H_0^*$ with $\mathcal{G}=\{\{1,\ldots,d\}\}$ performed at nominal level $5$\%. Each entry is based on $ 2500 $ $n \times d$ datasets drawn from a  Normal copula  with Kendall's tau matrix $\bs{T}_{\Delta}$ in Eq.~\eqref{eq:departure} (i) with $\Delta =  0.2 $; $\bs{T}$ is as in Eq.~\eqref{eq:T-equi-null}.}
       \label{tab:sim-power-star-column-2-normal}
      \begin{center}
      \fontsize{8.75}{8.75}\selectfont
      \vskip-12pt
      \begin{tabular}{*{2}{l}*{12}{r}}
      \toprule
      \multicolumn{14}{c}{$E_{np}$ with $\SA = \Sh_{np}$ for  column  departures ($\Delta =  0.2 $)}\\
      \midrule
      \multicolumn{2}{c}{} & & \multicolumn{3}{c}{$\tau = 0$} & & \multicolumn{3}{c}{$\tau = 0.3$} & & \multicolumn{3}{c}{$\tau = 0.6$}\\
      \cmidrule(lr){4-6}  \cmidrule(lr){8-10}  \cmidrule(lr){12-14}
      $\Sh_{np}$ & $d$\big|$n$ & & 50 & 100 & 150 & & 50 & 100 & 150 & & 50 & 100 & 150 \\
      \midrule
      &  5 &&  38.3 & 77.3 & 94.3 && 62.6 & 95.8 & 99.6 && 99 & 100 & 100 \\
      $\Sb_{np}^{\rm P}$ & 15 &&  30.2 & 72.3 & 92.2 && 46.9 & 90.5 & 99.2 && 97.8 & 100 & 100 \\
      & 25 &&  21.4 & 52.1 & 81.8 && 32 & 76.6 & 95.8 && 93.2 & 100 & 100 \\
      \cmidrule(lr){2-14}
      &  5 &&  49.3 & 81 & 95 && 69.8 & 97.1 & 99.6 && 99.4 & 100 & 100 \\
      & 15 &&  37.1 & 76.8 & 93.7 && 44.8 & 91 & 99.2 && 95.2 & 100 & 100 \\
      $\Sb_{np}^{\rm J}$ & 25 &&  21.6 & 54.4 & 83.6 && 18.9 & 71.4 & 95.2 && 70.9 & 100 & 100 \\
      & 50 &&  6.1 & 22.7 & 44.3 && 1.2 & 22.3 & 59.3 && 3.2 & 91.1 & 100 \\
      & 100 &&  0.8 & 4.7 & 10.8 && 0 & 0.4 & 6.1 && 0 & 1.2 & 60.9 \\
      \midrule
      \multicolumn{14}{c}{$M_{np}$ with $\SA = \Sh_{np}$ for  column departures ($\Delta =  0.2 $)}\\
      \midrule
      &  5 &&  21 & 48.1 & 72.3 && 30.4 & 71.6 & 92.6 && 70.3 & 99.8 & 100 \\
      $\Sb_{np}^{\rm P}$ & 15 &&  18.1 & 44.1 & 66.5 && 32.2 & 69.5 & 89.5 && 83.3 & 100 & 100 \\
      & 25 &&  14 & 33.2 & 52.6 && 24.4 & 55.6 & 80.2 && 80 & 99.8 & 100 \\
      \cmidrule(lr){2-14}
      &  5 &&  31 & 58.5 & 80.4 && 45.6 & 82.6 & 95.9 && 91.2 & 100 & 100 \\
      & 15 &&  22.2 & 50.3 & 72.2 && 34.2 & 74.1 & 92 && 89.1 & 100 & 100 \\
      $\Sb_{np}^{\rm J}$ & 25 &&  14.9 & 36 & 57.6 && 23.2 & 57.6 & 83.5 && 78.7 & 100 & 100 \\
      & 50 &&  8.4 & 19.8 & 32.6 && 12.3 & 33.7 & 58.4 && 52.5 & 97.8 & 100 \\
      & 100 &&  7.3 & 9.5 & 16.4 && 7.2 & 16.7 & 31.5 && 20.2 & 84.4 & 99.2 \\
      \midrule
      \multicolumn{14}{c}{$E_{np}$ with $\SA = (1/n)\I_p$ for  column  departures ($\Delta =  0.2 $)}\\
      \midrule
      \multicolumn{2}{c}{} & & \multicolumn{3}{c}{$\tau = 0$} & & \multicolumn{3}{c}{$\tau = 0.3$} & & \multicolumn{3}{c}{$\tau = 0.6$}\\
      \cmidrule(lr){4-6}  \cmidrule(lr){8-10}  \cmidrule(lr){12-14}
      $\Sh_{np}$ & $d$\big|$n$ & & 50 & 100 & 150 & & 50 & 100 & 150 & & 50 & 100 & 150 \\
      \midrule
      &  5 &&  49.7 & 84.4 & 96.6 && 76.7 & 98.1 & 99.9 && 99.9 & 100 & 100 \\
      $\Sb_{np}^{\rm P}$ & 15 &&  66.9 & 96.4 & 99.8 && 85.4 & 99.7 & 100 && 100 & 100 & 100 \\
      & 25 &&  66.9 & 96.1 & 99.8 && 85 & 99.6 & 100 && 100 & 100 & 100 \\
      \cmidrule(lr){2-14}
      &  5 &&  55.1 & 85.8 & 97 && 79.5 & 98.3 & 99.9 && 99.9 & 100 & 100 \\
      & 15 &&  69.8 & 96.7 & 99.8 && 84.7 & 99.7 & 100 && 99.9 & 100 & 100 \\
      $\Sb_{np}^{\rm J}$ & 25 &&  69.2 & 96.4 & 99.8 && 82.9 & 99.6 & 100 && 99.9 & 100 & 100 \\
      & 50 &&  60.6 & 94.9 & 99.7 && 71.4 & 98.2 & 100 && 99.7 & 100 & 100 \\
      & 100 &&  47.6 & 87.4 & 98.6 && 53.1 & 95.3 & 99.7 && 97.3 & 100 & 100 \\
      \midrule
      \multicolumn{14}{c}{$M_{np}$ with $\SA = (1/n)\I_p$ for  column  departures ($\Delta =  0.2 $)}\\
      \midrule
      \multicolumn{2}{c}{} & & \multicolumn{3}{c}{$\tau = 0$} & & \multicolumn{3}{c}{$\tau = 0.3$} & & \multicolumn{3}{c}{$\tau = 0.6$}\\
      \cmidrule(lr){4-6}  \cmidrule(lr){8-10}  \cmidrule(lr){12-14}
      $\Sh_{np}$ & $d$\big|$n$ & & 50 & 100 & 150 & & 50 & 100 & 150 & & 50 & 100 & 150 \\
      \midrule
      &  5 &&  27.5 & 60.9 & 83 && 48.9 & 90.2 & 98.4 && 96.1 & 100 & 100 \\
      $\Sb_{np}^{\rm P}$ & 15 &&  50 & 86.5 & 97.6 && 80 & 99.2 & 100 && 99.9 & 100 & 100 \\
      & 25 &&  55 & 89.6 & 99.1 && 84.4 & 99.4 & 100 && 100 & 100 & 100 \\
      \cmidrule(lr){2-14}
      &  5 &&  35.6 & 66.8 & 85.9 && 59.2 & 93.1 & 98.9 && 98 & 100 & 100 \\
      & 15 &&  52.6 & 87.4 & 97.8 && 79.4 & 99.2 & 100 && 99.8 & 100 & 100 \\
      $\Sb_{np}^{\rm J}$ & 25 &&  55.7 & 90 & 99.1 && 83.2 & 99.4 & 100 && 99.8 & 100 & 100 \\
      & 50 &&  56.2 & 91.8 & 99.3 && 84.6 & 99.4 & 100 && 99.9 & 100 & 100 \\
      & 100 &&  55.6 & 92 & 99.2 && 85.2 & 99.4 & 100 && 99.9 & 100 & 100 \\
      \bottomrule
      \end{tabular}
      \end{center}
      \vskip-9pt
      \small
      Statistics: $E_{np}$ Euclidean norm-based statistic defined in Eq.~\eqref{eq:euclidean}; $M_{np}$ supremum norm-based statistic defined in Eq.~\eqref{eq:supremum}. Estimators: $\Sb_{np}^{\rm P}$ structured plug-in estimator; $\Sb_{np}^{\rm J}$ structured jackknife estimator.
      \end{table}

            \begin{table}[htbp]
 \captionsetup{width=1\linewidth,font=small,skip=0pt}
\caption{Estimated rejection rates (in \%) of tests of $H_0^*$ with $\mathcal{G}=\{\{1,\ldots,d\}\}$ performed at nominal level $5$\%. Each entry is based on $ 1000 $ $n \times d$ datasets drawn from a  $t_4$ copula  with Kendall's tau matrix $\bs{T}_{\Delta}$ in Eq.~\eqref{eq:departure} (i) with $\Delta =  0.2 $; $\bs{T}$ is as in Eq.~\eqref{eq:T-equi-null}.}
       \label{tab:sim-power-star-column-2-t4}
      \begin{center}
      \fontsize{8.75}{8.75}\selectfont
      \vskip-12pt
      \begin{tabular}{*{2}{l}*{12}{r}}
      \toprule
      \multicolumn{14}{c}{$E_{np}$ with $\SA = \Sh_{np}$ for  column  departures ($\Delta =  0.2 $)}\\
      \midrule
      \multicolumn{2}{c}{} & & \multicolumn{3}{c}{$\tau = 0$} & & \multicolumn{3}{c}{$\tau = 0.3$} & & \multicolumn{3}{c}{$\tau = 0.6$}\\
      \cmidrule(lr){4-6}  \cmidrule(lr){8-10}  \cmidrule(lr){12-14}
      $\Sh_{np}$ & $d$\big|$n$ & & 50 & 100 & 150 & & 50 & 100 & 150 & & 50 & 100 & 150 \\
      \midrule
      &  5 &&  33.3 & 71.6 & 87.7 && 49.2 & 89.4 & 98.8 && 97.1 & 100 & 100 \\
      $\Sb_{np}^{\rm P}$ & 15 &&  25.9 & 60.1 & 84.1 && 37.9 & 79.9 & 95.3 && 94.1 & 100 & 100 \\
      & 25 &&  16 & 42.9 & 68.2 && 23.5 & 62.8 & 87.7 && 84.8 & 99.9 & 100 \\
      \cmidrule(lr){2-14}
      &  5 &&  44.4 & 76.7 & 89.6 && 59.4 & 91.2 & 99 && 97.8 & 100 & 100 \\
      & 15 &&  36.6 & 67 & 87.5 && 40.9 & 83.1 & 95.9 && 90.4 & 100 & 100 \\
      $\Sb_{np}^{\rm J}$ & 25 &&  22.8 & 48.7 & 73.9 && 19.3 & 61.2 & 87.9 && 63.5 & 99.9 & 100 \\
      & 50 &&  10.9 & 24.8 & 40.3 && 5 & 25 & 52.6 && 4.8 & 85.9 & 99.6 \\
      & 100 &&  5.7 & 10.9 & 15.7 && 0.3 & 3.9 & 10.2 && 0 & 3.8 & 58.6 \\
      \midrule
      \multicolumn{14}{c}{$M_{np}$ with $\SA = \Sh_{np}$ for  column departures ($\Delta =  0.2 $)}\\
      \midrule
      &  5 &&  17.4 & 41.4 & 60.7 && 24.8 & 60.1 & 85.2 && 60.5 & 99 & 100 \\
      $\Sb_{np}^{\rm P}$ & 15 &&  16.7 & 34.5 & 53.4 && 26.8 & 56.2 & 77.1 && 73.6 & 99.3 & 100 \\
      & 25 &&  12.8 & 25.9 & 42.1 && 19 & 43.5 & 64.2 && 69.2 & 98.3 & 100 \\
      \cmidrule(lr){2-14}
      &  5 &&  27.6 & 52 & 69.6 && 38 & 70.6 & 90.2 && 83.9 & 99.9 & 100 \\
      & 15 &&  19.8 & 40.9 & 60.2 && 28.7 & 62.1 & 81.3 && 79.5 & 99.8 & 100 \\
      $\Sb_{np}^{\rm J}$ & 25 &&  14.3 & 29.3 & 46.5 && 18.9 & 45.7 & 66.4 && 68.5 & 99.1 & 100 \\
      & 50 &&  7.8 & 14.7 & 27.5 && 11.9 & 26.7 & 41.2 && 45.4 & 93.8 & 99.4 \\
      & 100 &&  6 & 10.8 & 11.7 && 9.9 & 14.1 & 20.4 && 20.5 & 70 & 93.7 \\
      \midrule
      \multicolumn{14}{c}{$E_{np}$ with $\SA = (1/n)\I_p$ for  column  departures ($\Delta =  0.2 $)}\\
      \midrule
      \multicolumn{2}{c}{} & & \multicolumn{3}{c}{$\tau = 0$} & & \multicolumn{3}{c}{$\tau = 0.3$} & & \multicolumn{3}{c}{$\tau = 0.6$}\\
      \cmidrule(lr){4-6}  \cmidrule(lr){8-10}  \cmidrule(lr){12-14}
      $\Sh_{np}$ & $d$\big|$n$ & & 50 & 100 & 150 & & 50 & 100 & 150 & & 50 & 100 & 150 \\
      \midrule
      &  5 &&  43.1 & 78.7 & 92 && 64.1 & 94.7 & 99.5 && 99.4 & 100 & 100 \\
      $\Sb_{np}^{\rm P}$ & 15 &&  62.3 & 90.5 & 98.6 && 77.3 & 97.8 & 99.9 && 99.8 & 100 & 100 \\
      & 25 &&  58.2 & 92.7 & 98.9 && 76.6 & 97 & 99.8 && 99.8 & 100 & 100 \\
      \cmidrule(lr){2-14}
      &  5 &&  48.6 & 81.5 & 93.1 && 69.8 & 95.1 & 99.5 && 99.4 & 100 & 100 \\
      & 15 &&  67.3 & 91.4 & 99 && 78.7 & 98 & 99.9 && 99.8 & 100 & 100 \\
      $\Sb_{np}^{\rm J}$ & 25 &&  64.3 & 94 & 99.3 && 76.5 & 96.9 & 99.8 && 99.4 & 100 & 100 \\
      & 50 &&  57.5 & 89.1 & 98.3 && 65.9 & 95.3 & 99.8 && 98.6 & 100 & 100 \\
      & 100 &&  47.6 & 83.8 & 96.6 && 52.1 & 90.8 & 99.2 && 94.7 & 100 & 100 \\
      \midrule
      \multicolumn{14}{c}{$M_{np}$ with $\SA = (1/n)\I_p$ for  column  departures ($\Delta =  0.2 $)}\\
      \midrule
      \multicolumn{2}{c}{} & & \multicolumn{3}{c}{$\tau = 0$} & & \multicolumn{3}{c}{$\tau = 0.3$} & & \multicolumn{3}{c}{$\tau = 0.6$}\\
      \cmidrule(lr){4-6}  \cmidrule(lr){8-10}  \cmidrule(lr){12-14}
      $\Sh_{np}$ & $d$\big|$n$ & & 50 & 100 & 150 & & 50 & 100 & 150 & & 50 & 100 & 150 \\
      \midrule
      &  5 &&  23.4 & 52.9 & 72.9 && 37.6 & 78.5 & 93.9 && 91.7 & 100 & 100 \\
      $\Sb_{np}^{\rm P}$ & 15 &&  42.6 & 78 & 93.2 && 68.9 & 96.1 & 99.7 && 99.6 & 100 & 100 \\
      & 25 &&  46.6 & 83 & 97 && 77.4 & 96.1 & 99.7 && 100 & 100 & 100 \\
      \cmidrule(lr){2-14}
      &  5 &&  32 & 59.4 & 76.2 && 49.4 & 83.3 & 95.5 && 94.8 & 100 & 100 \\
      & 15 &&  46.9 & 79.3 & 93.2 && 70.7 & 96.4 & 99.8 && 99.2 & 100 & 100 \\
      $\Sb_{np}^{\rm J}$ & 25 &&  49.7 & 83.8 & 97.3 && 77.7 & 96.3 & 99.7 && 99.8 & 100 & 100 \\
      & 50 &&  48.6 & 85 & 97.9 && 77.8 & 97.9 & 99.8 && 99.6 & 100 & 100 \\
      & 100 &&  45.8 & 85.8 & 97 && 78.6 & 97.9 & 99.9 && 100 & 100 & 100 \\
      \bottomrule
      \end{tabular}
      \end{center}
      \vskip-9pt
      \small
      Statistics: $E_{np}$ Euclidean norm-based statistic defined in Eq.~\eqref{eq:euclidean}; $M_{np}$ supremum norm-based statistic defined in Eq.~\eqref{eq:supremum}. Estimators: $\Sb_{np}^{\rm P}$ structured plug-in estimator; $\Sb_{np}^{\rm J}$ structured jackknife estimator.
      \end{table}

            \begin{table}[htbp]
 \captionsetup{width=1\linewidth,font=small,skip=0pt}
\caption{Estimated rejection rates (in \%) of tests of $H_0^*$ with $\mathcal{G}=\{\{1,\ldots,d\}\}$ performed at nominal level $5$\%. Each entry is based on $ 1000 $ $n \times d$ datasets drawn from a  Gumbel copula  with Kendall's tau matrix $\bs{T}_{\Delta}$ in Eq.~\eqref{eq:departure} (i) with $\Delta =  0.2 $; $\bs{T}$ is as in Eq.~\eqref{eq:T-equi-null}.}
       \label{tab:sim-power-star-column-2-gumbel}
      \begin{center}
      \fontsize{8.75}{8.75}\selectfont
      \vskip-12pt
      \begin{tabular}{*{2}{l}*{12}{r}}
      \toprule
      \multicolumn{14}{c}{$E_{np}$ with $\SA = \Sh_{np}$ for  column  departures ($\Delta =  0.2 $)}\\
      \midrule
      \multicolumn{2}{c}{} & & \multicolumn{3}{c}{$\tau = 0$} & & \multicolumn{3}{c}{$\tau = 0.3$} & & \multicolumn{3}{c}{$\tau = 0.6$}\\
      \cmidrule(lr){4-6}  \cmidrule(lr){8-10}  \cmidrule(lr){12-14}
      $\Sh_{np}$ & $d$\big|$n$ & & 50 & 100 & 150 & & 50 & 100 & 150 & & 50 & 100 & 150 \\
      \midrule
      &  5 &&  34.7 & 76.5 & 92 && 53.3 & 93.7 & 99.6 && 97.5 & 100 & 100 \\
      $\Sb_{np}^{\rm P}$ & 15 &&  33.2 & 76.5 & 94.2 && 39.4 & 85.3 & 97.3 && 91.7 & 100 & 100 \\
      & 25 &&  24.5 & 63 & 85.3 && 30.5 & 68.8 & 90.9 && 83.7 & 99.9 & 100 \\
      \cmidrule(lr){2-14}
      &  5 &&  44 & 81.2 & 93 && 63.2 & 94.9 & 99.7 && 98.7 & 100 & 100 \\
      & 15 &&  42.7 & 80.7 & 95 && 40.2 & 86.2 & 97.8 && 87.2 & 100 & 100 \\
      $\Sb_{np}^{\rm J}$ & 25 &&  26.8 & 66.8 & 87.1 && 20.6 & 65.2 & 90.4 && 49.6 & 99.7 & 100 \\
      & 50 &&  9.7 & 35.9 & 58.6 && 1.6 & 18 & 49.8 && 0.8 & 66.2 & 99.3 \\
      & 100 &&  1.8 & 9.6 & 22.3 && 0 & 0.5 & 5 && 0 & 0 & 17.4 \\
      \midrule
      \multicolumn{14}{c}{$M_{np}$ with $\SA = \Sh_{np}$ for  column departures ($\Delta =  0.2 $)}\\
      \midrule
      &  5 &&  18.6 & 47.1 & 68.7 && 27.7 & 68.7 & 88.6 && 61.2 & 98.9 & 100 \\
      $\Sb_{np}^{\rm P}$ & 15 &&  19.8 & 42.5 & 67.9 && 25.8 & 59.8 & 81.1 && 72.9 & 99.4 & 100 \\
      & 25 &&  14 & 36.6 & 57.3 && 21.6 & 45.4 & 72.6 && 65 & 98.5 & 100 \\
      \cmidrule(lr){2-14}
      &  5 &&  28.6 & 57.7 & 76.8 && 38.8 & 79 & 93.4 && 83.1 & 99.9 & 100 \\
      & 15 &&  24.2 & 49.6 & 74.8 && 29.4 & 64.8 & 86.1 && 78.4 & 99.7 & 100 \\
      $\Sb_{np}^{\rm J}$ & 25 &&  16.1 & 40.3 & 62.2 && 20.9 & 48.1 & 75.5 && 62.8 & 99 & 100 \\
      & 50 &&  8.8 & 21.5 & 34.5 && 11.7 & 24.9 & 44.4 && 35.7 & 89.2 & 99.2 \\
      & 100 &&  5 & 7.8 & 19.5 && 10.3 & 13.4 & 21.9 && 14.3 & 61.2 & 91.9 \\
      \midrule
      \multicolumn{14}{c}{$E_{np}$ with $\SA = (1/n)\I_p$ for  column  departures ($\Delta =  0.2 $)}\\
      \midrule
      \multicolumn{2}{c}{} & & \multicolumn{3}{c}{$\tau = 0$} & & \multicolumn{3}{c}{$\tau = 0.3$} & & \multicolumn{3}{c}{$\tau = 0.6$}\\
      \cmidrule(lr){4-6}  \cmidrule(lr){8-10}  \cmidrule(lr){12-14}
      $\Sh_{np}$ & $d$\big|$n$ & & 50 & 100 & 150 & & 50 & 100 & 150 & & 50 & 100 & 150 \\
      \midrule
      &  5 &&  44.6 & 83.9 & 94.5 && 70.7 & 97.1 & 100 && 99.6 & 100 & 100 \\
      $\Sb_{np}^{\rm P}$ & 15 &&  65.3 & 96.3 & 99.5 && 82.1 & 98.7 & 100 && 99.5 & 100 & 100 \\
      & 25 &&  67.8 & 95.6 & 99.4 && 78.7 & 99 & 99.9 && 100 & 100 & 100 \\
      \cmidrule(lr){2-14}
      &  5 &&  49.6 & 85 & 95 && 74.5 & 97.2 & 100 && 99.6 & 100 & 100 \\
      & 15 &&  67.2 & 96.4 & 99.5 && 82.1 & 98.7 & 100 && 99.5 & 100 & 100 \\
      $\Sb_{np}^{\rm J}$ & 25 &&  69.2 & 95.7 & 99.4 && 77.7 & 98.8 & 99.9 && 99.5 & 100 & 100 \\
      & 50 &&  66.7 & 95.8 & 99.8 && 67.3 & 97.1 & 99.7 && 98.9 & 100 & 100 \\
      & 100 &&  55.3 & 91.2 & 98.9 && 51.5 & 92.5 & 99.3 && 92.8 & 100 & 100 \\
      \midrule
      \multicolumn{14}{c}{$M_{np}$ with $\SA = (1/n)\I_p$ for  column  departures ($\Delta =  0.2 $)}\\
      \midrule
      \multicolumn{2}{c}{} & & \multicolumn{3}{c}{$\tau = 0$} & & \multicolumn{3}{c}{$\tau = 0.3$} & & \multicolumn{3}{c}{$\tau = 0.6$}\\
      \cmidrule(lr){4-6}  \cmidrule(lr){8-10}  \cmidrule(lr){12-14}
      $\Sh_{np}$ & $d$\big|$n$ & & 50 & 100 & 150 & & 50 & 100 & 150 & & 50 & 100 & 150 \\
      \midrule
      &  5 &&  25.7 & 60.7 & 80.1 && 41.6 & 86.4 & 97.3 && 91 & 100 & 100 \\
      $\Sb_{np}^{\rm P}$ & 15 &&  49.1 & 86.9 & 96.6 && 76.4 & 98 & 99.9 && 99.8 & 100 & 100 \\
      & 25 &&  53.9 & 88.9 & 98.1 && 80.1 & 98.5 & 99.9 && 99.9 & 100 & 100 \\
      \cmidrule(lr){2-14}
      &  5 &&  34 & 66.7 & 82.6 && 53.6 & 89.5 & 98.4 && 95 & 100 & 100 \\
      & 15 &&  51.3 & 87.7 & 96.8 && 76.6 & 98.1 & 99.9 && 99.5 & 100 & 100 \\
      $\Sb_{np}^{\rm J}$ & 25 &&  54.6 & 88.8 & 98.1 && 78.9 & 98.4 & 99.9 && 99.9 & 100 & 100 \\
      & 50 &&  59.7 & 91.5 & 99.1 && 78.2 & 99 & 99.9 && 99.3 & 100 & 100 \\
      & 100 &&  54.3 & 90.7 & 98.7 && 76.9 & 98.8 & 99.9 && 99.9 & 100 & 100 \\
      \bottomrule
      \end{tabular}
      \end{center}
      \vskip-9pt
      \small
      Statistics: $E_{np}$ Euclidean norm-based statistic defined in Eq.~\eqref{eq:euclidean}; $M_{np}$ supremum norm-based statistic defined in Eq.~\eqref{eq:supremum}. Estimators: $\Sb_{np}^{\rm P}$ structured plug-in estimator; $\Sb_{np}^{\rm J}$ structured jackknife estimator.
      \end{table}

            \begin{table}[htbp]
 \captionsetup{width=1\linewidth,font=small,skip=0pt}
\caption{Estimated rejection rates (in \%) of tests of $H_0^*$ with $\mathcal{G}=\{\{1,\ldots,d\}\}$ performed at nominal level $5$\%. Each entry is based on $ 1000 $ $n \times d$ datasets drawn from a  Clayton copula  with Kendall's tau matrix $\bs{T}_{\Delta}$ in Eq.~\eqref{eq:departure} (i) with $\Delta =  0.2 $; $\bs{T}$ is as in Eq.~\eqref{eq:T-equi-null}.}
       \label{tab:sim-power-star-column-2-clayton}
      \begin{center}
      \fontsize{8.75}{8.75}\selectfont
      \vskip-12pt
      \begin{tabular}{*{2}{l}*{12}{r}}
      \toprule
      \multicolumn{14}{c}{$E_{np}$ with $\SA = \Sh_{np}$ for  column  departures ($\Delta =  0.2 $)}\\
      \midrule
      \multicolumn{2}{c}{} & & \multicolumn{3}{c}{$\tau = 0$} & & \multicolumn{3}{c}{$\tau = 0.3$} & & \multicolumn{3}{c}{$\tau = 0.6$}\\
      \cmidrule(lr){4-6}  \cmidrule(lr){8-10}  \cmidrule(lr){12-14}
      $\Sh_{np}$ & $d$\big|$n$ & & 50 & 100 & 150 & & 50 & 100 & 150 & & 50 & 100 & 150 \\
      \midrule
      &  5 &&   &  &  && 55.6 & 94.3 & 99.4 && 97 & 100 & 100 \\
      $\Sb_{np}^{\rm P}$ & 15 &&   &  &  && 36.9 & 84.1 & 96.3 && 90.7 & 100 & 100 \\
      & 25 &&   &  &  && 25.4 & 62.7 & 88.7 && 81.2 & 99.7 & 100 \\
      \cmidrule(lr){2-14}
      &  5 &&   &  &  && 64.9 & 95.4 & 99.6 && 98 & 100 & 100 \\
      & 15 &&   &  &  && 36 & 85.3 & 96.8 && 85 & 100 & 100 \\
      $\Sb_{np}^{\rm J}$ & 25 &&   &  &  && 15.7 & 58.1 & 88.2 && 47.8 & 99.3 & 100 \\
      & 50 &&   &  &  && 1.3 & 17.9 & 41.3 && 0.5 & 66.6 & 98.7 \\
      & 100 &&   &  &  && 0 & 0.3 & 4.3 && 0 & 0.2 & 21.1 \\
      \midrule
      \multicolumn{14}{c}{$M_{np}$ with $\SA = \Sh_{np}$ for  column departures ($\Delta =  0.2 $)}\\
      \midrule
      &  5 &&   &  &  && 26.4 & 66.9 & 87.4 && 60.1 & 98.5 & 100 \\
      $\Sb_{np}^{\rm P}$ & 15 &&   &  &  && 23.8 & 54.9 & 79 && 69.8 & 98.9 & 99.9 \\
      & 25 &&   &  &  && 19.9 & 42.4 & 67.7 && 64.1 & 97.4 & 100 \\
      \cmidrule(lr){2-14}
      &  5 &&   &  &  && 40.9 & 77.6 & 93.2 && 83.2 & 99.8 & 100 \\
      & 15 &&   &  &  && 26.1 & 61.6 & 83.3 && 76.3 & 99.5 & 100 \\
      $\Sb_{np}^{\rm J}$ & 25 &&   &  &  && 18.6 & 44 & 71.2 && 61.4 & 98.2 & 100 \\
      & 50 &&   &  &  && 10.5 & 26 & 38.9 && 34.3 & 87.3 & 98.9 \\
      & 100 &&   &  &  && 11.5 & 14.1 & 22.2 && 16.9 & 62.1 & 91.2 \\
      \midrule
      \multicolumn{14}{c}{$E_{np}$ with $\SA = (1/n)\I_p$ for  column  departures ($\Delta =  0.2 $)}\\
      \midrule
      \multicolumn{2}{c}{} & & \multicolumn{3}{c}{$\tau = 0$} & & \multicolumn{3}{c}{$\tau = 0.3$} & & \multicolumn{3}{c}{$\tau = 0.6$}\\
      \cmidrule(lr){4-6}  \cmidrule(lr){8-10}  \cmidrule(lr){12-14}
      $\Sh_{np}$ & $d$\big|$n$ & & 50 & 100 & 150 & & 50 & 100 & 150 & & 50 & 100 & 150 \\
      \midrule
      &  5 &&   &  &  && 71.4 & 97.4 & 99.8 && 99.3 & 100 & 100 \\
      $\Sb_{np}^{\rm P}$ & 15 &&   &  &  && 80.2 & 98.6 & 99.9 && 100 & 100 & 100 \\
      & 25 &&   &  &  && 77.3 & 99.1 & 100 && 99.6 & 100 & 100 \\
      \cmidrule(lr){2-14}
      &  5 &&   &  &  && 74.8 & 97.4 & 99.8 && 99.4 & 100 & 100 \\
      & 15 &&   &  &  && 80.8 & 98.7 & 99.9 && 100 & 100 & 100 \\
      $\Sb_{np}^{\rm J}$ & 25 &&   &  &  && 77 & 99.2 & 100 && 99.4 & 100 & 100 \\
      & 50 &&   &  &  && 65.7 & 96.3 & 99.7 && 98.1 & 100 & 100 \\
      & 100 &&   &  &  && 49.3 & 91.2 & 98.8 && 94.1 & 100 & 100 \\
      \midrule
      \multicolumn{14}{c}{$M_{np}$ with $\SA = (1/n)\I_p$ for  column  departures ($\Delta =  0.2 $)}\\
      \midrule
      \multicolumn{2}{c}{} & & \multicolumn{3}{c}{$\tau = 0$} & & \multicolumn{3}{c}{$\tau = 0.3$} & & \multicolumn{3}{c}{$\tau = 0.6$}\\
      \cmidrule(lr){4-6}  \cmidrule(lr){8-10}  \cmidrule(lr){12-14}
      $\Sh_{np}$ & $d$\big|$n$ & & 50 & 100 & 150 & & 50 & 100 & 150 & & 50 & 100 & 150 \\
      \midrule
      &  5 &&   &  &  && 44.5 & 86.1 & 97.5 && 91.3 & 100 & 100 \\
      $\Sb_{np}^{\rm P}$ & 15 &&   &  &  && 73.8 & 96 & 99.9 && 100 & 100 & 100 \\
      & 25 &&   &  &  && 76 & 98.6 & 99.8 && 99.8 & 100 & 100 \\
      \cmidrule(lr){2-14}
      &  5 &&   &  &  && 54.3 & 89.3 & 97.9 && 95.3 & 100 & 100 \\
      & 15 &&   &  &  && 74.9 & 96.1 & 99.9 && 99.8 & 100 & 100 \\
      $\Sb_{np}^{\rm J}$ & 25 &&   &  &  && 75.6 & 98.3 & 99.8 && 99.7 & 100 & 100 \\
      & 50 &&   &  &  && 77.3 & 98.1 & 100 && 99.4 & 100 & 100 \\
      & 100 &&   &  &  && 77.6 & 98.9 & 99.8 && 99.6 & 100 & 100 \\
      \bottomrule
      \end{tabular}
      \end{center}
      \vskip-9pt
      \small
      Statistics: $E_{np}$ Euclidean norm-based statistic defined in Eq.~\eqref{eq:euclidean}; $M_{np}$ supremum norm-based statistic defined in Eq.~\eqref{eq:supremum}. Estimators: $\Sb_{np}^{\rm P}$ structured plug-in estimator; $\Sb_{np}^{\rm J}$ structured jackknife estimator.
      \end{table}

      \clearpage 
      
      \subsubsection{Block equicorrelation and block exchangeability}
The following tables report the results of the simulation study involving Normal and $t_4$ replicates with Kendall's tau matrix $\T$ as in Eq.~\eqref{eq:T-block} with $c_{k\ell} = 0.4 - (0.15)|k-\ell|$ for all $k,\ell \in \{1,2,3\}$, as well as the associated alternatives (see \eqref{eq:departure}, \textit{single departure}).  Such a matrix $\T$ satisfies $H_0^*$ with $\mathcal{G} = \{\{1,\ldots, d_1\}, \{d_1+1,\ldots, d_1+d_2\}, \{d_1+d_2+1,\ldots, d\}\}$ and hence also $H_0$ with $\B$ the block-membership matrix described in Section~\ref{sec:2.3}. Note that when $d=6$ and the blocks are balanced, $H_0$ and $H_0^*$ in fact still hold, so the rejection rates are expected to fluctuate around the nominal level $5$\%. We sampled $n \times d$ datasets for all combinations of $(n,d) \in \{50,150,250\} \times \{5,15,25,50,100\}$.

\begin{itemize}
\item Tables~\ref{tab:sim-level-blocks-normal}--\ref{tab:sim-level-blocks-t4}:
estimated sizes for the tests of $H_0$ using $\SA = (1/n)\I_{p}$.

\item Tables~\ref{tab:sim-power-blocks-normal}--\ref{tab:sim-power-blocks-t4}:
estimated rejection rates of tests of $H_0$ using $\SA = (1/n)\I_{p}$;\\
single departure with $\Delta = 0.1$.

\item Tables~\ref{tab:sim-level-star-blocks-normal}--\ref{tab:sim-level-star-blocks-t4}:
estimated sizes for the tests of $H_0^*$ using $\SA = (1/n)\I_{p}$ and $\SA = \Sh_{np}$.

\item Tables~\ref{tab:sim-power-star-blocks-normal}--\ref{tab:sim-power-star-blocks-t4}:
estimated rejection rates of tests of $H_0^*$ using $\SA = (1/n)\I_{p}$ and $\SA = \Sh_{np}$;\\
single departure with $\Delta = 0.1$.
\end{itemize}

\clearpage

\begin{table}
 \captionsetup{width=1\linewidth,font=small}
      \caption{
      Estimated sizes (in \%) for the tests of $H_0$ with $\SA = (1/n)\I_{p}$, performed at the nominal level 5\%. Each entry is based on $ 2500 $ $n\times d$ datasets from a  Normal copula  with Kendall's tau matrix $\T$ as in Eq.~\eqref{eq:T-block} with $c_{k\ell} = 0.4 - (0.15)|k-\ell|$ for all $k,\ell \in \{1,\dots,3\}$.}
       \label{tab:sim-level-blocks-normal}
      \begin{center}
      \scriptsize
      \vskip-9pt
      \begin{tabular}{*{2}{l}*{3}{r}@{\hspace{0.7cm}}*{3}{r}*{1}{c}*{3}{r}@{\hspace{0.7cm}}*{3}{r}}
      \toprule
      & &  \multicolumn{6}{c}{$E_{np}$} & {\hspace{0.2cm}} & \multicolumn{6}{c}{$M_{np}$} \\
      \cmidrule(lr){3-8}\cmidrule(lr){10-15}
      &&  \multicolumn{3}{c}{balanced} & \multicolumn{3}{c}{unbalanced} & &  \multicolumn{3}{c}{balanced} & \multicolumn{3}{c}{unbalanced}\\
      \cmidrule(lr){3-8}\cmidrule(lr){10-15}
      $\Sh_{np}$ & $d$\big|$n$ & 50 & 150 & 250 & 50 & 150 & 250 && 50 & 150 & 250 & 50 & 150 & 250 \\
      \midrule
      & 6 & 6.6 & 5.5 & 5.8 & 5.8 & 4.8 & 4.7 && 6.5 & 5.7 & 5.1 & 6.1 & 4.6 & 4.9 \\
      $\Sh_{np}^{\rm P}$
      & 12 & 2.8 & 4.2 & 4.1 & 3 & 3.9 & 5 && 4.6 & 4.9 & 4.3 & 4.8 & 5 & 4.8 \\
      &18 & 1.2 & 3.1 & 3.7 & 1.4 & 3.2 & 4.5 && 3.2 & 4.5 & 4.2 & 3.5 & 4.6 & 5.1 \\
      \cmidrule(lr){2-15}
      &6& 3.8 & 4.7 & 5.2 & 3.4 & 3.6 & 4.1 && 4.4 & 5 & 4.3 & 4.1 & 4.2 & 4.1 \\
      $\Sh_{np}^{\rm J}$
      &12 & 1.5 & 2.7 & 3.3 & 1 & 2.6 & 4.2 && 2.8 & 4.2 & 3.9 & 3.1 & 4.3 & 4.2 \\
      &18& 0.4 & 1.8 & 2.7 & 0.5 & 2 & 3.4 && 2.2 & 3.5 & 3.6 & 2.6 & 3.8 & 4.5 \\
      \bottomrule
      \end{tabular}
      \end{center}
      \vskip-9pt
      \small
      Statistics: $E_{np}$ Euclidean norm-based statistic defined in Eq. \eqref{eq:euclidean}; $M_{np}$ supremum norm-based statistic defined in Eq. \eqref{eq:supremum}. Estimators: $\Sh_{np}^{\rm P}$ plug-in estimator; $\Sh_{np}^{\rm J}$ jackknife estimator.
      \end{table}

      \begin{table}
 \captionsetup{width=1\linewidth,font=small}
      \caption{
      Estimated sizes (in \%) for the tests of $H_0$ with $\SA = (1/n)\I_{p}$, performed at the nominal level 5\%. Each entry is based on $ 1000 $ $n\times d$ datasets from a  $t_4$ copula  with Kendall's tau matrix $\T$ as in Eq.~\eqref{eq:T-block} with $c_{k\ell} = 0.4 - (0.15)|k-\ell|$ for all $k,\ell \in \{1,\dots,3\}$.}
       \label{tab:sim-level-blocks-t4}
      \begin{center}
      \scriptsize
      \vskip-9pt
      \begin{tabular}{*{2}{l}*{3}{r}@{\hspace{0.7cm}}*{3}{r}*{1}{c}*{3}{r}@{\hspace{0.7cm}}*{3}{r}}
      \toprule
      & &  \multicolumn{6}{c}{$E_{np}$} & {\hspace{0.2cm}} & \multicolumn{6}{c}{$M_{np}$} \\
      \cmidrule(lr){3-8}\cmidrule(lr){10-15}
      &&  \multicolumn{3}{c}{balanced} & \multicolumn{3}{c}{unbalanced} & &  \multicolumn{3}{c}{balanced} & \multicolumn{3}{c}{unbalanced}\\
      \cmidrule(lr){3-8}\cmidrule(lr){10-15}
      $\Sh_{np}$ & $d$\big|$n$ & 50 & 150 & 250 & 50 & 150 & 250 && 50 & 150 & 250 & 50 & 150 & 250 \\
      \midrule
      & 6 & 4.3 & 7.5 & 5.2 & 4.8 & 5.3 & 4.9 && 5.4 & 7.6 & 4.5 & 7.3 & 5.4 & 3.7 \\
      $\Sh_{np}^{\rm P}$
      & 12 & 2.1 & 4 & 4.1 & 2.1 & 2.4 & 3.3 && 5.9 & 5.1 & 5.4 & 4.3 & 4.4 & 5 \\
      &18 & 0.9 & 1.8 & 3 & 1.2 & 1.8 & 3.1 && 4.6 & 5 & 4.3 & 4.5 & 4.9 & 4.8 \\
      \cmidrule(lr){2-15}
      &6& 3 & 6.8 & 4.8 & 2.4 & 4.5 & 4.5 && 3.9 & 6.7 & 4.2 & 4.4 & 4.8 & 3.4 \\
      $\Sh_{np}^{\rm J}$
      &12 & 0.7 & 2.8 & 3.7 & 1.1 & 2 & 2.5 && 3.6 & 4.4 & 4.7 & 2.5 & 3.5 & 4.4 \\
      &18& 0.5 & 1.5 & 1.6 & 0.7 & 1.4 & 2.9 && 3.4 & 4.2 & 3.8 & 3.6 & 3.9 & 4 \\
      \bottomrule
      \end{tabular}
      \end{center}
      \vskip-9pt
      \small
      Statistics: $E_{np}$ Euclidean norm-based statistic defined in Eq. \eqref{eq:euclidean}; $M_{np}$ supremum norm-based statistic defined in Eq. \eqref{eq:supremum}. Estimators: $\Sh_{np}^{\rm P}$ plug-in estimator; $\Sh_{np}^{\rm J}$ jackknife estimator.
      \end{table}

      \begin{table}
      \captionsetup{width=1\linewidth,font=small}
      \caption{Estimated rejection rates (in \%) of tests of $H_0$ with $\SA = (1/n)\I_{p}$, performed at nominal level $5$\%.
      Each entry is based on $ 2500 $ $n\times d$ datasets from a  Normal copula  with Kendall's tau matrix $\bs{T}_\Delta$ in Eq.~\eqref{eq:departure} (i) with $\Delta =  0.1 $ and $\T$ as in Eq.~\eqref{eq:T-block} with $c_{k\ell} = 0.4 - (0.15)|k-\ell|$ for all $k,\ell \in \{1,2,3\}$.}
       \label{tab:sim-power-blocks-normal}
      \begin{center}
      \scriptsize
      \vskip-9pt
      \begin{tabular}{*{2}{l}*{3}{r}@{\hspace{0.7cm}}*{3}{r}*{1}{c}*{3}{r}@{\hspace{0.7cm}}*{3}{r}}
      \toprule
      & &  \multicolumn{6}{c}{$E_{np}$ with $\SA=(1/n)\I_p$ for single dep.} & {\hspace{0.2cm}} & \multicolumn{6}{c}{$M_{np}$ with $\SA=(1/n)\I_p$ for single dep.} \\
      \cmidrule(lr){3-8}\cmidrule(lr){10-15}
      &&  \multicolumn{3}{c}{balanced} & \multicolumn{3}{c}{unbalanced} & &  \multicolumn{3}{c}{balanced} & \multicolumn{3}{c}{unbalanced}\\
      \cmidrule(lr){3-8}\cmidrule(lr){10-15}
      $\Sh_{np}$ & $d$\big|$n$ & 50 & 150 & 250 & 50 & 150 & 250 && 50 & 150 & 250 & 50 & 150 & 250 \\
      \midrule
      & 6 & 6 & 5.3 & 5.1 & 9.2 & 22.9 & 40.1 && 6.6 & 5.1 & 5.1 & 8.5 & 23.7 & 46.8 \\
      $\Sh_{np}^{\rm P}$
      & 12 & 3.6 & 7.7 & 11.7 & 3.9 & 7.5 & 13.6 && 4.4 & 11.4 & 28.3 & 6.6 & 19.6 & 46.6 \\
      &18 & 1.3 & 5.6 & 6.5 & 1.4 & 4.6 & 7.4 && 4.3 & 10 & 27.4 & 3.4 & 12.8 & 35.4 \\
      \cmidrule(lr){2-15}
      &6& 3.7 & 4.8 & 4.6 & 5.8 & 20.2 & 38.2 && 4.1 & 4.3 & 4.7 & 5.8 & 21.4 & 45.2 \\
      $\Sh_{np}^{\rm J}$
      &12 & 1.8 & 5.7 & 10 & 1.8 & 5.5 & 11.5 && 2.9 & 9.8 & 26.2 & 3.8 & 16.9 & 44.2 \\
      &18& 0.6 & 3.6 & 5.1 & 0.3 & 3.1 & 5.9 && 2.7 & 8 & 25.6 & 2.2 & 10.9 & 33.1 \\
      \bottomrule
      \end{tabular}
      \end{center}
      \vskip-9pt
      \small
      Statistics: $E_{np}$ Euclidean norm-based statistic defined in Eq. \eqref{eq:euclidean}; $M_{np}$ supremum norm-based statistic defined in Eq. \eqref{eq:supremum}. Estimators: $\Sh_{np}^{\rm P}$ plug-in estimator; $\Sh_{np}^{\rm J}$ jackknife estimator.
      \end{table}

      \begin{table}
      \captionsetup{width=1\linewidth,font=small}
      \caption{Estimated rejection rates (in \%) of tests of $H_0$ with $\SA = (1/n)\I_{p}$, performed at nominal level $5$\%.
      Each entry is based on $ 1000 $ $n\times d$ datasets from a  $t_4$ copula  with Kendall's tau matrix $\bs{T}_\Delta$ in Eq.~\eqref{eq:departure} (i) with $\Delta =  0.1 $ and $\T$ as in Eq.~\eqref{eq:T-block} with $c_{k\ell} = 0.4 - (0.15)|k-\ell|$ for all $k,\ell \in \{1,2,3\}$.}
       \label{tab:sim-power-blocks-t4}
      \begin{center}
      \scriptsize
      \vskip-9pt
      \begin{tabular}{*{2}{l}*{3}{r}@{\hspace{0.7cm}}*{3}{r}*{1}{c}*{3}{r}@{\hspace{0.7cm}}*{3}{r}}
      \toprule
      & &  \multicolumn{6}{c}{$E_{np}$ with $\SA=(1/n)\I_p$ for single dep.} & {\hspace{0.2cm}} & \multicolumn{6}{c}{$M_{np}$ with $\SA=(1/n)\I_p$ for single dep.} \\
      \cmidrule(lr){3-8}\cmidrule(lr){10-15}
      &&  \multicolumn{3}{c}{balanced} & \multicolumn{3}{c}{unbalanced} & &  \multicolumn{3}{c}{balanced} & \multicolumn{3}{c}{unbalanced}\\
      \cmidrule(lr){3-8}\cmidrule(lr){10-15}
      $\Sh_{np}$ & $d$\big|$n$ & 50 & 150 & 250 & 50 & 150 & 250 && 50 & 150 & 250 & 50 & 150 & 250 \\
      \midrule
      & 6 & 6 & 5.6 & 5.4 & 8.1 & 16.2 & 26.4 && 6.9 & 6.2 & 5 & 8.2 & 18.3 & 31 \\
      $\Sh_{np}^{\rm P}$
      & 12 & 2.2 & 5.5 & 7.1 & 2.5 & 5.9 & 9.6 && 4.9 & 9.7 & 18.4 & 6.1 & 14.7 & 32.6 \\
      &18 & 0.9 & 2.8 & 5.6 & 1.1 & 2.3 & 4.9 && 4 & 6.9 & 18.6 & 4.4 & 9.9 & 21.4 \\
      \cmidrule(lr){2-15}
      &6& 3.6 & 5.1 & 4.6 & 5.6 & 14.4 & 25 && 4.8 & 5.5 & 4.5 & 6.4 & 16.3 & 29.8 \\
      $\Sh_{np}^{\rm J}$
      &12 & 1 & 3.8 & 6.2 & 1.4 & 4.3 & 8.1 && 3.1 & 8.4 & 17.2 & 3.3 & 12.7 & 30.7 \\
      &18& 0.6 & 2.1 & 4.6 & 0.5 & 1.6 & 4.2 && 2.8 & 5.6 & 17.1 & 3 & 8.2 & 19.9 \\
      \bottomrule
      \end{tabular}
      \end{center}
      \vskip-9pt
      \small
      Statistics: $E_{np}$ Euclidean norm-based statistic defined in Eq. \eqref{eq:euclidean}; $M_{np}$ supremum norm-based statistic defined in Eq. \eqref{eq:supremum}. Estimators: $\Sh_{np}^{\rm P}$ plug-in estimator; $\Sh_{np}^{\rm J}$ jackknife estimator.
      \end{table}

      \begin{table}
\caption{Estimated sizes (in \%) for the tests of $H_0^*$ performed at the nominal level 5\%.
      Each entry is based on $ 2500 $ $n\times d$ datasets from a Normal copula with Kendall's tau matrix $\T$ as in Eq.~\eqref{eq:T-block} with $c_{k\ell} = 0.4 - (0.15)|k-\ell|$ for all $k,\ell \in \{1,\dots,3\}$.}
       \label{tab:sim-level-star-blocks-normal}
      \begin{center}
      \scriptsize
      \vskip-9pt
      \begin{tabular}{*{2}{l}*{3}{r}@{\hspace{0.7cm}}*{3}{r}*{1}{c}*{3}{r}@{\hspace{0.7cm}}*{3}{r}}
      \toprule
      & &  \multicolumn{6}{c}{$E_{np}$ with $\SA = \Sh_{np}$} & {\hspace{0.2cm}} & \multicolumn{6}{c}{$M_{np}$ with $\SA = \Sh_{np}$} \\
      \cmidrule(lr){3-8}\cmidrule(lr){10-15}
      &&  \multicolumn{3}{c}{balanced} & \multicolumn{3}{c}{unbalanced} & &  \multicolumn{3}{c}{balanced} & \multicolumn{3}{c}{unbalanced}\\
      \cmidrule(lr){3-8}\cmidrule(lr){10-15}
      $\Sh_{np}$ & $d$\big|$n$ & 50 & 150 & 250 & 50 & 150 & 250 && 50 & 150 & 250 & 50 & 150 & 250 \\
      \midrule
      & 6 & 0.4 & 3 & 3.3 & 1.2 & 3 & 3.6 && 1.6 & 4.1 & 4.1 & 2 & 3.6 & 4.2 \\
      $\Sb_{np}^{\rm P}$
      & 12 & 1.9 & 3.4 & 3.6 & 2.3 & 3.1 & 4.2 && 3 & 4.3 & 4 & 3.4 & 4.6 & 4.3 \\
      &18 & 2 & 3.2 & 4.2 & 3.1 & 3.6 & 5.2 && 3.2 & 4.1 & 4.8 & 3.9 & 4 & 4.6 \\
      \cmidrule(lr){2-15}
      &6& 4.8 & 5.1 & 5.2 & 4.5 & 4.7 & 4.6 && 5.2 & 5.4 & 4.9 & 4.5 & 4.8 & 4.7 \\
      $\Sb_{np}^{\rm J}$
      &12 & 3.1 & 4.2 & 4.1 & 3.1 & 3.6 & 4.6 && 3.9 & 4.6 & 4.1 & 4 & 4.7 & 4.3 \\
      &18& 1.8 & 2.7 & 3.8 & 2.2 & 3.2 & 5 && 2.9 & 4 & 4.7 & 3.5 & 3.9 & 4.2 \\
      \midrule
      & &  \multicolumn{6}{c}{$E_{np}$ with $\SA = (1/n)\I_{p}$} & {\hspace{0.2cm}} & \multicolumn{6}{c}{$M_{np}$ with $\SA = (1/n)\I_{p}$} \\
      \cmidrule(lr){3-8}\cmidrule(lr){10-15}
      &&  \multicolumn{3}{c}{balanced} & \multicolumn{3}{c}{unbalanced} & &  \multicolumn{3}{c}{balanced} & \multicolumn{3}{c}{unbalanced}\\
      \cmidrule(lr){3-8}\cmidrule(lr){10-15}
      $\Sh_{np}$ & $d$\big|$n$ & 50 & 150 & 250 & 50 & 150 & 250 && 50 & 150 & 250 & 50 & 150 & 250 \\
      \midrule
      & 6 & 3 & 4.2 & 5.1 & 2.8 & 3.8 & 4.1 && 3.7 & 4.8 & 4.4 & 3.8 & 4 & 4.2 \\
      $\Sb_{np}^{\rm P}$
      & 12 & 2.9 & 4.3 & 4.2 & 3.1 & 3.8 & 5.2 && 3.8 & 4.6 & 4 & 4.2 & 4.8 & 4.5 \\
      &18 & 2.9 & 4.4 & 4.8 & 2.9 & 4.4 & 5.6 && 3.6 & 4.6 & 4.2 & 4.7 & 4.5 & 5 \\
      \cmidrule(lr){2-15}
      &6& 4.3 & 4.9 & 5.4 & 4.2 & 4.4 & 4.3 && 4.6 & 5.1 & 4.6 & 4.5 & 4.2 & 4.4 \\
      $\Sb_{np}^{\rm J}$
      &12 & 3.6 & 4.4 & 4.3 & 3.6 & 4 & 5 && 4 & 4.5 & 4 & 4.2 & 4.6 & 4.4 \\
      &18& 3 & 4.2 & 4.8 & 3.4 & 4.2 & 5.6 && 3.3 & 4.3 & 4 & 4.2 & 4.5 & 4.9 \\
      \bottomrule
      \end{tabular}
      \end{center}
      \vskip-9pt
      \small
      Statistics: $E_{np}$ Euclidean norm-based statistic defined in Eq. \eqref{eq:euclidean}; $M_{np}$ supremum norm-based statistic defined in Eq. \eqref{eq:supremum}. Estimators: $\Sb_{np}^{\rm P}$ structured plug-in estimator; $\Sb_{np}^{\rm J}$ structured jackknife estimator.
      \end{table}

      \begin{table}
\caption{Estimated sizes (in \%) for the tests of $H_0^*$  performed at the nominal level 5\%.
      Each entry is based on $ 1000 $ $n\times d$ datasets from a $t_4$ copula with Kendall's tau matrix $\T$ as in Eq.~\eqref{eq:T-block} with $c_{k\ell} = 0.4 - (0.15)|k-\ell|$ for all $k,\ell \in \{1,\dots,3\}$.}
       \label{tab:sim-level-star-blocks-t4}
      \begin{center}
      \scriptsize
      \vskip-9pt
      \begin{tabular}{*{2}{l}*{3}{r}@{\hspace{0.7cm}}*{3}{r}*{1}{c}*{3}{r}@{\hspace{0.7cm}}*{3}{r}}
      \toprule
      & &  \multicolumn{6}{c}{$E_{np}$ with $\SA = \Sh_{np}$} & {\hspace{0.2cm}} & \multicolumn{6}{c}{$M_{np}$ with $\SA = \Sh_{np}$} \\
      \cmidrule(lr){3-8}\cmidrule(lr){10-15}
      &&  \multicolumn{3}{c}{balanced} & \multicolumn{3}{c}{unbalanced} & &  \multicolumn{3}{c}{balanced} & \multicolumn{3}{c}{unbalanced}\\
      \cmidrule(lr){3-8}\cmidrule(lr){10-15}
      $\Sh_{np}$ & $d$\big|$n$ & 50 & 150 & 250 & 50 & 150 & 250 && 50 & 150 & 250 & 50 & 150 & 250 \\
      \midrule
      & 6 & 1 & 3.3 & 2.5 & 0.8 & 3.6 & 3.4 && 1.5 & 4.2 & 3.3 & 2.3 & 4 & 3.6 \\
      $\Sb_{np}^{\rm P}$
      & 12 & 1.5 & 3.1 & 2.6 & 2 & 3.3 & 4.1 && 2.2 & 4.2 & 4.4 & 2.4 & 4 & 4.1 \\
      &18 & 2.9 & 3.8 & 4.1 & 2.1 & 4.6 & 3.3 && 2 & 3.5 & 4.2 & 3.6 & 4.5 & 3.3 \\
      \cmidrule(lr){2-15}
      &6& 4.7 & 5.5 & 3.6 & 4.8 & 5.8 & 5 && 4.8 & 6 & 4.6 & 5 & 5.3 & 4.4 \\
      $\Sb_{np}^{\rm J}$
      &12 & 3.8 & 4.1 & 3.4 & 3.7 & 4.9 & 4.9 && 3.2 & 4.9 & 4.8 & 2.8 & 4.6 & 4.2 \\
      &18& 4 & 4.2 & 4.3 & 3.5 & 5.1 & 3.6 && 2.1 & 3.7 & 4.2 & 4.2 & 4.5 & 3.4 \\
      \midrule
      & &  \multicolumn{6}{c}{$E_{np}$ with $\SA = (1/n)\I_{p}$} & {\hspace{0.2cm}} & \multicolumn{6}{c}{$M_{np}$ with $\SA = (1/n)\I_{p}$} \\
      \cmidrule(lr){3-8}\cmidrule(lr){10-15}
      &&  \multicolumn{3}{c}{balanced} & \multicolumn{3}{c}{unbalanced} & &  \multicolumn{3}{c}{balanced} & \multicolumn{3}{c}{unbalanced}\\
      \cmidrule(lr){3-8}\cmidrule(lr){10-15}
      $\Sh_{np}$ & $d$\big|$n$ & 50 & 150 & 250 & 50 & 150 & 250 && 50 & 150 & 250 & 50 & 150 & 250 \\
      \midrule
      & 6 & 2.5 & 6.3 & 5 & 2.2 & 4.7 & 4.6 && 3.1 & 6 & 4.2 & 4 & 4.7 & 3.6 \\
      $\Sb_{np}^{\rm P}$
      & 12 & 2.8 & 4.8 & 5.4 & 2.9 & 3.4 & 3.8 && 4.4 & 4.8 & 5 & 3.7 & 4.3 & 4.7 \\
      &18 & 4 & 3 & 4.2 & 3.3 & 4.2 & 5 && 5 & 5 & 4.1 & 5 & 4.8 & 4.4 \\
      \cmidrule(lr){2-15}
      &6& 3.5 & 7.2 & 4.9 & 4.3 & 5.4 & 4.9 && 4.2 & 6.9 & 4.4 & 4.9 & 5.3 & 3.7 \\
      $\Sb_{np}^{\rm J}$
      &12 & 4.3 & 5.4 & 5.8 & 4.7 & 3.8 & 4.1 && 5 & 4.9 & 5.3 & 4.1 & 4.2 & 4.5 \\
      &18& 5 & 3.5 & 4.3 & 4.9 & 4.5 & 5.2 && 5 & 4.9 & 4.3 & 4.9 & 5.1 & 5 \\
      \bottomrule
      \end{tabular}
      \end{center}
      \vskip-9pt
      \small
      Statistics: $E_{np}$ Euclidean norm-based statistic defined in Eq. \eqref{eq:euclidean}; $M_{np}$ supremum norm-based statistic defined in Eq. \eqref{eq:supremum}. Estimators: $\Sb_{np}^{\rm P}$ structured plug-in estimator; $\Sb_{np}^{\rm J}$ structured jackknife estimator.
      \end{table}

      \begin{table}
\caption{Estimated rejection rates (in \%) of tests of $H_0^*$ performed at nominal level $5$\%.
Each entry is based on $ 2500 $ $n\times d$ datasets from a  Normal copula  with Kendall's tau matrix  $\T_\Delta$ as in Eq.~$(\ref{eq:departure})$ (i) with $\Delta =  0.1 $ and $\T$ as in Eq.~$(\ref{eq:T-block})$ with $c_{k\ell} = 0.4 - (0.15)|k-\ell|$ for all $k,\ell \in \{1,\dots,3\}$.}
       \label{tab:sim-power-star-blocks-normal}
      \begin{center}
      \scriptsize
      \vskip-9pt
      \begin{tabular}{*{2}{l}*{3}{r}@{\hspace{0.7cm}}*{3}{r}*{1}{c}*{3}{r}@{\hspace{0.7cm}}*{3}{r}}
      \toprule
      & &  \multicolumn{6}{c}{$E_{np}$ with $\SA = \Sh_{np}$ for single dep.} & {\hspace{0.2cm}} & \multicolumn{6}{c}{$M_{np}$ with $\SA = \Sh_{np}$ for single dep.} \\
      \cmidrule(lr){3-8}\cmidrule(lr){10-15}
      &&  \multicolumn{3}{c}{balanced} & \multicolumn{3}{c}{unbalanced} & &  \multicolumn{3}{c}{balanced} & \multicolumn{3}{c}{unbalanced}\\
      \cmidrule(lr){3-8}\cmidrule(lr){10-15}
      $\Sh_{np}$ & $d$\big|$n$ & 50 & 150 & 250 & 50 & 150 & 250 && 50 & 150 & 250 & 50 & 150 & 250 \\
      \midrule
      & 6 & 0.5 & 2.4 & 3.4 & 2.6 & 25.8 & 51.4 && 1.5 & 3.4 & 3.7 & 3 & 28.7 & 56.8 \\
      $\Sb_{np}^{\rm P}$
      & 12 & 2.2 & 11.6 & 25 & 3.7 & 15.7 & 32.9 && 3.8 & 21 & 50.8 & 6.1 & 33.8 & 69.5 \\
      &18 & 3.8 & 11 & 19.6 & 3.7 & 12 & 21.5 && 5.2 & 23.6 & 57 & 5.8 & 31.1 & 67.6 \\
      \cmidrule(lr){2-15}
      &6& 4.8 & 5 & 4.8 & 10.4 & 33.6 & 57.4 && 4.8 & 4.9 & 4.8 & 10.7 & 34.7 & 60.9 \\
      $\Sb_{np}^{\rm J}$
      &12 & 4.2 & 15.1 & 28.8 & 5.4 & 17.2 & 34.5 && 6 & 25.3 & 54.2 & 7.4 & 34.6 & 69.8 \\
      &18& 3.3 & 10.4 & 19.2 & 2.9 & 10.7 & 20.4 && 5.7 & 24 & 57.6 & 5.3 & 30.9 & 67.2 \\
      \midrule
      & &  \multicolumn{6}{c}{$E_{np}$ with $\SA = (1/n)\I_{p}$ for single dep.} & {\hspace{0.2cm}} & \multicolumn{6}{c}{$M_{np}$ with $\SA = (1/n)\I_{p}$ for single dep.} \\
      \cmidrule(lr){3-8}\cmidrule(lr){10-15}
      &&  \multicolumn{3}{c}{balanced} & \multicolumn{3}{c}{unbalanced} & &  \multicolumn{3}{c}{balanced} & \multicolumn{3}{c}{unbalanced}\\
      \cmidrule(lr){3-8}\cmidrule(lr){10-15}
      $\Sh_{np}$ & $d$\big|$n$ & 50 & 150 & 250 & 50 & 150 & 250 && 50 & 150 & 250 & 50 & 150 & 250 \\
      \midrule
      & 6 & 2.5 & 4.7 & 4.3 & 5.5 & 21.3 & 39.3 && 3.5 & 4.2 & 4.6 & 5.8 & 22.4 & 45.9 \\
      $\Sb_{np}^{\rm P}$
      & 12 & 4 & 7.9 & 12 & 4.6 & 7.9 & 13.9 && 3.9 & 11.2 & 28 & 5.6 & 19.2 & 46.3 \\
      &18 & 3.4 & 6.9 & 7.9 & 3.6 & 5.8 & 8.8 && 5 & 10.1 & 27.4 & 4.1 & 13.2 & 35.8 \\
      \cmidrule(lr){2-15}
      &6& 4 & 4.8 & 4.5 & 7.6 & 22.1 & 39.6 && 4.6 & 4.4 & 4.6 & 6.8 & 22.6 & 45.7 \\
      $\Sb_{np}^{\rm J}$
      &12 & 4.5 & 8 & 12.1 & 5.3 & 7.9 & 14 && 3.7 & 10.9 & 27.7 & 5.7 & 18.4 & 45.9 \\
      &18& 3.6 & 6.8 & 7.7 & 3.5 & 5.8 & 8.6 && 4.8 & 9.6 & 27.1 & 3.8 & 12.5 & 34.8 \\
      \bottomrule
      \end{tabular}
      \end{center}
      \vskip-9pt
      \small
      Statistics: $E_{np}$ Euclidean norm-based statistic defined in Eq. \eqref{eq:euclidean}; $M_{np}$ supremum norm-based statistic defined in Eq. \eqref{eq:supremum}. Estimators: $\Sb_{np}^{\rm P}$ structured plug-in estimator; $\Sb_{np}^{\rm J}$ structured jackknife estimator.
      \end{table}

      \begin{table}
\caption{Estimated rejection rates (in \%) of tests of $H_0^*$ performed at nominal level $5$\%.
Each entry is based on $ 1000 $ $n\times d$ datasets from a  $t_4$ copula  with Kendall's tau matrix  $\T_\Delta$ as in Eq.~$(\ref{eq:departure})$ (i) with $\Delta =  0.1 $ and $\T$ as in Eq.~$(\ref{eq:T-block})$ with $c_{k\ell} = 0.4 - (0.15)|k-\ell|$ for all $k,\ell \in \{1,\dots,3\}$.}
       \label{tab:sim-power-star-blocks-t4}
      \begin{center}
      \scriptsize
      \vskip-9pt
      \begin{tabular}{*{2}{l}*{3}{r}@{\hspace{0.7cm}}*{3}{r}*{1}{c}*{3}{r}@{\hspace{0.7cm}}*{3}{r}}
      \toprule
      & &  \multicolumn{6}{c}{$E_{np}$ with $\SA = \Sh_{np}$ for single dep.} & {\hspace{0.2cm}} & \multicolumn{6}{c}{$M_{np}$ with $\SA = \Sh_{np}$ for single dep.} \\
      \cmidrule(lr){3-8}\cmidrule(lr){10-15}
      &&  \multicolumn{3}{c}{balanced} & \multicolumn{3}{c}{unbalanced} & &  \multicolumn{3}{c}{balanced} & \multicolumn{3}{c}{unbalanced}\\
      \cmidrule(lr){3-8}\cmidrule(lr){10-15}
      $\Sh_{np}$ & $d$\big|$n$ & 50 & 150 & 250 & 50 & 150 & 250 && 50 & 150 & 250 & 50 & 150 & 250 \\
      \midrule
      & 6 & 0.3 & 2.7 & 3.7 & 2.5 & 18.6 & 37.6 && 1.8 & 4 & 3.4 & 2.5 & 21.1 & 41 \\
      $\Sb_{np}^{\rm P}$
      & 12 & 3.3 & 9.4 & 18.2 & 3.5 & 13.7 & 22.4 && 2.3 & 13.8 & 35.9 & 4.2 & 26.4 & 51.4 \\
      &18 & 3.4 & 8.2 & 13.8 & 3.6 & 9.5 & 15.2 && 3 & 15.8 & 38.4 & 3.6 & 22.4 & 43.5 \\
      \cmidrule(lr){2-15}
      &6& 5.6 & 5.4 & 5.1 & 10.2 & 27.1 & 43.9 && 5.8 & 5.3 & 4.5 & 9.2 & 27.7 & 45.9 \\
      $\Sb_{np}^{\rm J}$
      &12 & 6.9 & 13.2 & 22.3 & 6.6 & 15.6 & 24.1 && 5 & 17.1 & 39.2 & 5.8 & 27.8 & 52.1 \\
      &18& 4.8 & 9.4 & 14.5 & 5 & 10.7 & 15.7 && 3.8 & 16.7 & 39.4 & 3.8 & 22.8 & 43.5 \\
      \midrule
      & &  \multicolumn{6}{c}{$E_{np}$ with $\SA = (1/n)\I_{p}$ for single dep.} & {\hspace{0.2cm}} & \multicolumn{6}{c}{$M_{np}$ with $\SA = (1/n)\I_{p}$ for single dep.} \\
      \cmidrule(lr){3-8}\cmidrule(lr){10-15}
      &&  \multicolumn{3}{c}{balanced} & \multicolumn{3}{c}{unbalanced} & &  \multicolumn{3}{c}{balanced} & \multicolumn{3}{c}{unbalanced}\\
      \cmidrule(lr){3-8}\cmidrule(lr){10-15}
      $\Sh_{np}$ & $d$\big|$n$ & 50 & 150 & 250 & 50 & 150 & 250 && 50 & 150 & 250 & 50 & 150 & 250 \\
      \midrule
      & 6 & 2.7 & 4.2 & 4.3 & 5.1 & 15 & 25.5 && 4 & 5.1 & 4.6 & 6.5 & 17.1 & 30.4 \\
      $\Sb_{np}^{\rm P}$
      & 12 & 3 & 6.7 & 9 & 3.7 & 7.9 & 10.9 && 3.5 & 9.6 & 18.5 & 4.8 & 14.7 & 31.9 \\
      &18 & 4.1 & 6.6 & 8.2 & 4.1 & 4.5 & 7.8 && 4.3 & 7.1 & 18.8 & 4.7 & 10 & 21.6 \\
      \cmidrule(lr){2-15}
      &6& 4.7 & 5.2 & 5.1 & 7.7 & 16 & 26.6 && 5.2 & 5.6 & 4.6 & 7.1 & 17.5 & 30.7 \\
      $\Sb_{np}^{\rm J}$
      &12 & 5.5 & 7.1 & 9.4 & 5.5 & 8.2 & 11.2 && 4.2 & 9.8 & 18.1 & 5.1 & 14.4 & 31.9 \\
      &18& 5.6 & 7.2 & 8.4 & 5.6 & 5.1 & 7.9 && 4.6 & 7.2 & 18.7 & 4.7 & 9.7 & 21.6 \\
      \bottomrule
      \end{tabular}
      \end{center}
      \vskip-9pt
      \small
      Statistics: $E_{np}$ Euclidean norm-based statistic defined in Eq. \eqref{eq:euclidean}; $M_{np}$ supremum norm-based statistic defined in Eq. \eqref{eq:supremum}. Estimators: $\Sb_{np}^{\rm P}$ structured plug-in estimator; $\Sb_{np}^{\rm J}$ structured jackknife estimator.
      \end{table}

\clearpage
\section{Additional material for Section~\ref{sec:local-alternatives}} \label{app:local-alternatives}


\subsection{Proof of Theorem~\ref{thm:asymptotic-n-loc}}

Suppose, without loss of generality, that $\bcth_0 + \boldsymbol{h}_n/\sqrt{n} \in \Theta$; this is true for all sufficiently large $n$ since $\Theta$ is assumed to be an open set. Let $P_n$  and $Q_n$ denote the distribution of the random sample $(\bs{U}_1,\ldots, \bs{U}_n)$  when $\bs{U}_\nu \sim C_{\bcth_0}$ and when $\bs{U}_\nu \sim C_{\bcth_0+\boldsymbol{h}_n/\sqrt{n}}$, respectively, for all $\nu \in \{1,\dots,n\}$. From Theorem 7.2 and Example 6.5 in \cite{vanderVaart:1998}, $P_n$ and $Q_n$ are mutually contiguous. 

From \eqref{eq:tau-u} and Theorem~7.2 in  \cite{vanderVaart:1998} it follows that under $P_n$ (i.e., under $H_0$), 
$$
\Bigl(\sqrt{n}(\th_{np}-\t_p),  \log \prod_{\nu=1}^n \frac{c_{\bcth_0+\boldsymbol{h}_n/\sqrt{n}}}{c_{\bcth_0}}(\bs{U}_\nu) \Bigr) \rightsquigarrow \mathcal{N}_{p+1} \left( \begin{pmatrix} \bs{0}_p \\ - \frac{1}{2} \boldsymbol{h}^\top {\I}_{\bcth_0} \boldsymbol{h} \end{pmatrix} , \begin{pmatrix} \bs{\Sigma}_p &  \bs{a}  \\ \bs{a}^\top &  \boldsymbol{h}^\top {\I}_{\bcth_0} \boldsymbol{h}  \end{pmatrix}\right),
$$
where ${\I}_{\bcth_0} = \E \dot \ell_{\bcth_0} (\bs{U}) \ell_{\bcth_0} (\bs{U})^\top$ with $\bs{U} \sim C_{\bcth_0}$; note that the existence of ${\I}_{\bcth_0} $ is also guaranteed by the latter theorem. Le Cam's Third Lemma \citep[Example 6.7]{vanderVaart:1998} then implies that under the sequence $Q_n$ of local alternatives (i.e., under $H_{1n}$),
\begin{equation}\label{eq:tau-alt}
\sqrt{n}(\th_{np}-\t_p) \rightsquigarrow \mathcal{N}_p(\bs{a}, \bs{\Sigma}_p),
\end{equation}
where the elements of the drift $\bs{a}$ are as given in Theorem~\ref{thm:asymptotic-n-loc}.

Next, because $P_n$ and $Q_n$ are mutually contiguous, $\bs{P}_{np}$ converges in probability under $H_{1n}$ to $\bs{P}_p$. Contiguity and the same arguments as in the  proof of Theorem 
\ref{thm:asymptotic-n} further imply that $\{n\SA_{np}\}^{-1/2}$ converges in probability to $\SA_{p}^{-1/2}$ under $H_{1n}$. Eq.~\eqref{eq:tau-alt} combined with Slutsky's Lemma yields that
\begin{align*}
	\SA_{np}^{-1/2} \bs{P}_{np} \th_{np} = (n \SA_{np})^{-1/2} \bs{P}_{np} \sqrt{n}(\th_{np} - \t_p)\rightsquigarrow \SA_{p}^{-1/2} \bs{P}_p \Z^*_p,
\end{align*}
where $\Z^*_p \sim \mathcal{N}_p(\bs{a},\S_p)$. This completes the proof.

\subsection{Examples}\label{app:local-alternatives-examples}

From Lemma~7.6 in \cite{vanderVaart:1998}, Assumption~\ref{ass:qm}  holds in particular when the map $\bcth \mapsto \sqrt{c_\bcth(\bs{u})}$ is continuously differentiable in $\bcth$ for every $\bm{u}$ and the elements of the matrix $I_\bcth = \E\{(\dot c_\bcth (\bs{U}) /c_\bcth(\bs{U}))(\dot c_\bcth^\top(\bs{U})/c_\bcth(\bs{U}))\}$ where $\bs{U}$ is distributed as $C_\bcth$ are well defined and continuous in $\bcth$. Here, $\dot c_\bcth$ denotes the vector of the first-order derivatives of $c_\bcth$ with respect to $\cth_1,\ldots, \cth_k$.  In this case, we further have that
$$
\dot \ell_\bcth = \frac{\partial}{\partial \bcth} \log c_\bcth = \frac{\dot c_\bcth}{c_\bcth}.
$$
For elliptical copulas, the parameter vector $\bcth$ is the vector of the entries of the correlation matrix above the main diagonal. From Proposition~7.34 in \cite{McNeil/Frey/Embrechts:2015}, we have that for $r \in \{1,\ldots, p\}$, $\cth_r = \sin(\pi\tau_r /2)$. For single and column departures, we thus have local alternatives of the form $\bcth_0 + \bs{h}_n^*/\sqrt{n}$, where for $r \in \{1,\ldots, p\}$, 
$$
h_{nr}^* = h_r \times \frac{\pi}{2} \cos\left (\frac{\pi\tau_{rn} }{2}\right)
$$
for some $\tau_{rn} \in [\tau_r, \tau_r + h_r/\sqrt{n}]$. As $n\to \infty$, $\bs{h}_n^* \to \bs{h}^*$ with $h_r^* = (h_r  \pi/2) \cos(\pi\tau_{r}/2)$.

\begin{example}\label{ex:normal} 
Consider the $d$-dimensional Normal copula with  correlation matrix $\R$, which is symmetric and positive definite. Denote the above-diagonal entries of $\R$ by $\bcth$, that is, $R_{i_rj_r} = \cth_r$ with $(i_r,j_r) = \iota(r)$ as in Eq.~\eqref{eq:bijection}, and note that for $\bs{u} \in [0,1]^d$ the density $c_{\bcth}$ of a $d$-dimensional Normal copula is given by
$$
	c_{\bcth}(\bs{u}) = \frac{1}{|\R|^{1/2}} \exp\left\{ - \frac{1}{2} \bs{x}^\top (\R^{-1} - \I_d) \bs{x} \right\},
$$
where $\bs{x} = \{\Phi^{-1}(u_1),\ldots,\Phi^{-1}(u_d)\}$ and $\Phi$ is the distribution function of the standard Normal distribution; see, e.g., \cite{Song:2000}.
In particular, for any $r \in\{1,\ldots,p\}$ and $\bs{u} \in [0,1]^d$, we have that $(\partial/\partial \cth_{r}) \log c_{\bcth}(\bs{u})$ is equivalent to $(\partial/\partial \cth_{r}) (-1/2)\{ \log |\R| - \bs{x}^\top \R^{-1}\bs{x} \}$ and hence,
by Eqs.~(8.12) and (8.18) in \cite{Harville:1997}, that
\begin{equation*}
	\{\dot{\ell}_{\bcth}(\bs{u})\}_r = - (\R^{-1})_{ij} + \frac{1}{2}\bs{x}^\top \R^{-1} \bs{E}^{(i,j)} \R^{-1} \bs{x}, \qquad \bs{x} = \{\Phi^{-1}(u_1),\ldots,\Phi^{-1}(u_d)\},
\end{equation*}
where $(i,j) = \iota(r)$ and $\mathbf{E}^{(i,j)} = (\partial/\partial \cth_r) \R$ is a $d \times d$ matrix with entries at positions $(i,j)$ and $(j,i)$ equal to one and all other entries equal to zero \citep[Eq.~(5.7)]{Harville:1997}.
\end{example}

\begin{example}\label{ex:student}
Next, consider the Student $t_{4}$ copula with correlation matrix $\R$; let again $\bcth$ denote the above-diagonal entries of $\R$. 
Starting from the explicit expression for the density of $t_4$ copulas given, e.g., in \cite{Demarta/McNeil:2005}, straightforward calculations show that, for any $r \in\{1,\ldots, p\}$,
$$
	\{\dot{\ell}_{\bcth}(\bs{u})\}_r = \frac{\partial}{\partial \cth_{r}} \left\{ -\frac{1}{2} \log |\R| - \frac{d+4}{2} \log ( 4 + \bs{x}^\top \R^{-1}\bs{x} ) \right\},
$$
where $\bs{x} = \{t_4^{-1}(u_1),\ldots,t_4^{-1}(u_d) \}$ and  $t_4^{-1}$ is the quantile function of the univariate standard Student $t_4$ distribution.
Using Eqs.~(8.12), (8.18) and (5.7) in \cite{Harville:1997} and the formula for the derivative of logarithms, we then obtain
$$
	\{\dot{\ell}_{\bcth}(\bs{u})\}_r = - (\R^{-1})_{ij} + \left(\frac{d+4}{2}\right) \frac{\bs{x}^\top \R^{-1} \bs{E}^{(i,j)} \R^{-1} \bs{x}}{4 + \bs{x}^\top \R^{-1}\bs{x}},\qquad \bs{x} = \{t_4^{-1}(u_1),\ldots,t_4^{-1}(u_d) \},
$$
where $(i,j) = \iota(r)$ and $\mathbf{E}^{(i,j)}$ is as in Example~\ref{ex:normal}.

\end{example}


\clearpage
\section{Additional material for Section~\ref{sec:application}} \label{app:application}

\defcitealias{PSMSL:2020}{PSMSL, 2020}

We used the RLR Monthly dataset of the Permanent Service for Mean Sea Level \citepalias{PSMSL:2020}, which consists of monthly averages, and focused on the month of February.
Out of the many stations included in the PSMSL dataset, we restricted ourselves to those located in a subset of $17$ countries in North and Central Americas.
We then narrowed down our search to those with $65$ consecutive years of observations up until $2018$; considering $2019$ had the effect of discarding the station in Trois-Rivi\`{e}res, QC (Canada), indexed $18$ in Figure~\ref{fig:stations}, which we considered particularly interesting for the application.
Going back to $1954$ allowed us to include the $d=18$ stations analyzed in Section~\ref{sec:application}; this seemed like an interesting place to stop.
We ended up with stations in the United States, Canada and Panama only.
The station names, as given by the PSMSL, are listed below; they are ordered according to their unique id in Figure~\ref{fig:stations}.

\small
\begin{center}
\spacingset{.5}
\begin{minipage}{.33\textwidth}
    \begin{enumerate}
        \item Honolulu
        \item San Francisco
        \item Crescent City
        \item Astoria (Tongue Point)
        \item Seattle
        \item Vancouver
    \end{enumerate}
\end{minipage}
\begin{minipage}{.25\textwidth}
    \begin{enumerate}
        \item[7.] Sitka
        \item[8.] Juneau
        \item[9.] Balboa
        \item[10.] Key West
        \item[11.] St. Peterburg
        \item[12.] Pensacola
    \end{enumerate}
\end{minipage}
\begin{minipage}{.4\textwidth}
    \begin{enumerate}
        \item[13.] Charleston I
        \item[14.] Sewells Point, Hampton Roads
        \item[15.] Kiptopeke Beach
        \item[16.] Lewes (Breakwater Harbor)
        \item[17.] Portland (Maine)
        \item[18.] Trois-Rivi\`{e}res
    \end{enumerate}
\end{minipage}
\end{center}

\normalsize
\noindent Two more American stations could have been added by considering $n=63$ ($1956-2018$); an analysis analogous to that of Section~\ref{sec:application} lead to similar results.

Most of the $d=18$ raw time series suggest a rise in February's mean sea levels with time.
To resolve this issue, we fitted a simple linear regression model to the series at each station, with the year of measurement as the explanatory variable.
Proceeding this way eliminated the trend as well as all ties among the observations. 
The fit at each station was adequate, and the residuals did not exhibit any significant auto-correlation; this was assessed both visually and by Ljung--Box tests at various lags \citep{Ljung/Box:1978}.
Among all the fitted linear regressions, only five yielded a negative slope, out of which only those corresponding to stations located at Sitka (\#7 -- AK, USA), Juneau (\#8 -- AK, USA) and Trois-Rivi\`{e}res (\#18 -- QC, Canada) were significant according to the standard t-test at significance level $0.05$.
In all five cases, we kept both the intercept and the linear coefficient.
Some examples of raw time series are given in Figure~\ref{fig:sea-levels}, along with their corresponding linear regression.
We then applied the methodology developed in this paper to the regression residuals. Note that although the residuals are not i.i.d, the work of \cite{Cote/Genest/Omelka:2019} shows that the asymptotic results derived here in the fixed $d$ setting still apply to residuals from regression models with Normal errors.

\begin{figure}
	\centering
	\includegraphics[width=1\textwidth]{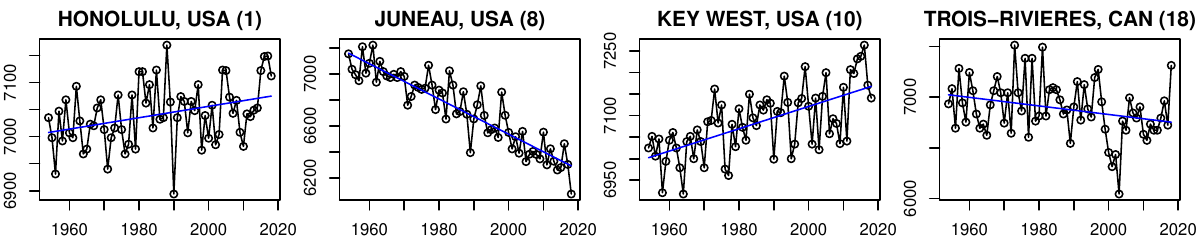}
\caption{Time series of February's mean sea level measured at four different locations from year $1954$ to $2018$, with corresponding linear regression, in blue, performed using time of measurement as explanatory variable.} \label{fig:sea-levels}
\end{figure}

Formally, the hypothesis considered in Section~\ref{sec:application} corresponds to $H_0$ in \eqref{eq:H0} with $p=153$, $L=54$ and a matrix $\B \in \{0,1\}^{p \times L}$ such that $\sum_{\ell=1}^L \B_{r\ell} = 1$ for all $r \in \{1,\dots,153\}$.
We used the first $15$ columns of $\B$ to record the entries of $\th_{np}$ that belong to each of the $15$ off-diagonal blocks shown in Figure~\ref{fig:Taus} (b), viz.
\begin{equation} \label{eq:application-B}
	B_{r\ell} =
	\begin{cases}
		\mathbbm{1}\{(i_r,j_r) \in \mathcal{G}_1 \times \mathcal{G}_2 \} & \text{if } \ell = 1\\
		\qquad \qquad \qquad \vdots\\
		\mathbbm{1}\{(i_r,j_r) \in \mathcal{G}_5 \times \mathcal{G}_6 \} & \text{if } \ell = 15
	\end{cases}
	\qquad r \in \{1,\dots,p\},\ \ell \in \{1,\dots,15\}.
\end{equation}
The rows corresponding to the $39$ entries of $\th_{np}$ that belong to a diagonal block are filled so that there is exactly one $1$ in each of the $L-15 = 39$ remaining columns of $\B$.
In other words, if you take $\B$ and remove its first $15$ columns and any row $r$ such that $(i_r,j_r) \in \mathcal{G}_{k} \times \mathcal{G}_{\ell}$ for some $1 \leqslant k < \ell \leqslant 6$, then, what is left is the identity matrix (or a permutation of it).

Finally, to complement the closing remarks of Section~\ref{subsec:application-results}, we report here the p-values obtained by individually testing whether the entries of a given block are all the same.
They can be found in Table~\ref{tab:application-results} along with the p-values associated to the global test.
The matrix $\B$ used for these tests is constructed in a similar fashion to that in \eqref{eq:application-B}.
More precisely, we first record the entries of $\th_{np}$ corresponding to the block of interest, say $\mathcal{G}_k \times \mathcal{G}_\ell$ ($1 \leqslant k < \ell \leqslant 6$), in the first column of $\B$, i.e. $B_{r1} = \mathbbm{1}\{(i_r,j_r) \in \mathcal{G}_k \times \mathcal{G}_\ell \}$ for all $r \in \{1,\ldots, p\}$.
Then, we fill the remaining $p - |\mathcal{G}_k \times \mathcal{G}_\ell|$ rows that corresponds to entries outside of $\mathcal{G}_k \times \mathcal{G}_\ell$ so that there is exactly one $1$ in each of $L - 1 = p - |\mathcal{G}_k \times \mathcal{G}_\ell|$ remaining columns. 

\vfill

\begin{table}[h]
\caption{P-values (\%) obtained from individually testing equi-correlation in each of the $12$ non-trivial blocks shown in Figure~\ref{fig:Taus} (b). The ID row provides the corresponding column of $\B$ or, alternatively, the corresponding block id as given in Figure~\ref{fig:Taus} (b). The last column reports the p-value obtained when testing $H_0$ with $\B$. Only the statistics $E_{np}$ and $M_{np}$ with $\SA = (1/n)\I_p$ were used.} \label{tab:application-results}
\centering
\small
\medskip
\begin{tabular}{rrrrrrrrrrrrrrr}
Block ID & 1 & 2 & 3 & 5 & 6 & 7 & 8 & 9 & 10 & 12 & 13 & 15 & * \\ 
\hline
$E_{np}$ & 10.2 & 60.3 & 0.4 & 46.3 & 21.8 & 3.5 & 20.0 & 32.1 & 36.6 & 60.1 & 59.1 & 88.1 & 10.7 \\ 
$M_{np}$ & 6.3 & 59.6 & 0.3 & 56.5 & 22.7 & 1.5 & 29.5 & 44.3 & 23.2 & 65.1 & 59.7 & 83.3 & 32.9 \\ 
\end{tabular}
\end{table}


\bibliographystyle{apalike}

\end{document}